\documentclass[a4paper,fleqn,usenatbib]{mnras}

\usepackage{newtxtext,newtxmath}
\usepackage{pdflscape}
\usepackage[T1]{fontenc}
\usepackage{ae,aecompl}
\usepackage{placeins}
\usepackage{enumitem}

\usepackage[dvipsnames]{xcolor}  

\usepackage{graphicx}	
\usepackage{amsmath}	
\usepackage{amssymb}	
\usepackage{soul}		
\setstcolor{red}
\setul{}{1mm}		        

\def\hi{\relax \ifmmode {\mbox H\,{\scshape i}}\else H\,{\scshape i}\fi}
\def\hii{\relax \ifmmode {\mbox H\,{\scshape ii}}\else H\,{\scshape ii}\fi}
\def\hei{\relax \ifmmode {\mbox He\,{\scshape i}}\else He\,{\scshape i}\fi}
\def\heii{\relax \ifmmode {\mbox He\,{\scshape ii}}\else He\,{\scshape ii}\fi}
\def\nii{\relax \ifmmode {\mbox N\,{\scshape ii}}\else N\,{\scshape ii}\fi}
\def\ni{\relax \ifmmode {\mbox N\,{\scshape i}}\else N\,{\scshape i}\fi}
\def\oi{\relax \ifmmode {\mbox O\,{\scshape i}}\else O\,{\scshape i}\fi}
\def\oii{\relax \ifmmode {\mbox O\,{\scshape ii}}\else O\,{\scshape ii}\fi}
\def\oiii{\relax \ifmmode {\mbox O\,{\scshape iii}}\else O\,{\scshape iii}\fi}
\def\sii{\relax \ifmmode {\mbox S\,{\scshape ii}}\else S\,{\scshape ii}\fi}
\def\siii{\relax \ifmmode {\mbox S\,{\scshape iii}}\else S\,{\scshape iii}\fi}
\def\ariii{\relax \ifmmode {\mbox Ar\,{\scshape iii}}\else Ar\,{\scshape iii}\fi}
\def\ariv{\relax \ifmmode {\mbox Ar\,{\scshape iv}}\else Ar\,{\scshape iv}\fi}
\def\neiii{\relax \ifmmode {\mbox Ne\,{\scshape iii}}\else Ne\,{\scshape iii}\fi}
\def\cliii{\relax \ifmmode {\mbox Cl\,{\scshape iii}}\else Cl\,{\scshape iii}\fi}
\def\feiii{\relax \ifmmode {\mbox Fe\,{\scshape iii}}\else Fe\,{\scshape iii}\fi}
\def\feii{\relax \ifmmode {\mbox Fe\,{\scshape ii}}\else Fe\,{\scshape ii}\fi}
\def\niqii{\relax \ifmmode {\mbox Ni\,{\scshape ii}}\else Ni\,{\scshape ii}\fi}
\def\cii{\relax \ifmmode {\mbox C\,{\scshape ii}}\else C\,{\scshape ii}\fi}
\def\mgi{\relax \ifmmode {\mbox Mg\,{\scshape i}}\else Mg\,{\scshape i}\fi}
\def\silii{\relax \ifmmode {\mbox Si\,{\scshape ii}}\else Si\,{\scshape ii}\fi}
\def\ha{\relax \ifmmode {\mbox H}\alpha\else H$\alpha$\fi}
\def\hb{\relax \ifmmode {\mbox H}\beta\else H$\beta$\fi}
\def\me{$^{-1}$}
\def\arcsec{\hbox{$^{\prime\prime}$}}

\def\deg{\hbox{$^{\circ}$}}
\defcitealias{paperII}{Paper~II}
 
\title[Bar effect on gas-phase abundance gradients. I. Chemical abundances]{Bar effect on gas-phase abundance gradients. I. Data sample and chemical abundances}

\author[A. Zurita et al.]{A. Zurita,$^{1,2}$\thanks{E-mail: azurita@ugr.es}
E. Florido,$^{1,2}$
F. Bresolin,$^{3}$
E. P\'erez-Montero$^{4}$
and I. P\'erez$^{1,2}$
\\
$^{1}$Dpto. de F\'\i sica y del Cosmos, Campus de Fuentenueva, Edificio Mecenas, Universidad de Granada, E18071--Granada, Spain\\
$^{2}$Instituto Carlos I de F\'\i sica Te\'orica y Computacional, Facultad de Ciencias, E18071--Granada, Spain\\
$^{3}$Institute for Astronomy, 2680 Woodlawn Drive, Honolulu, HI 96822, USA\\
$^{4}$Instituto de Astrof\' isica de Andaluc\' ia, Camino Bajo de Hu\'etor s/n, Aptdo. 3004, E18080-Granada, Spain\\
}

\date{Accepted XXX. Received YYY; in original form ZZZ}

\pubyear{2020}

\usepackage{etoolbox}
\makeatletter
\patchcmd\@combinedblfloats{\box\@outputbox}{\unvbox\@outputbox}{}{\errmessage{\noexpand patch failed}}
\makeatother

\begin{document}
\label{firstpage}
\pagerange{\pageref{firstpage}--\pageref{lastpage}}
\maketitle

\begin{abstract}
  Studies of gas-phase radial metallicity profiles in spirals published in the last decade have diminished the importance of galactic bars as agents that mix and flatten the  profiles, contradicting results obtained in the 1990s. We have collected a large sample of  2831 published \hii\ region emission-line fluxes in 51 nearby galaxies, including objects both with and without the presence of a bar, with the aim of revisiting the issue of whether bars affect  the radial metal distribution in spirals. In this first paper of a series of two, we present the galaxy and the \hii\ region samples. The methodology is homogeneous for the whole data sample and includes the derivation of \hii\ region chemical abundances,  structural parameters of  bars and discs, galactocentric distances, and radial abundance profiles. We have obtained O/H and N/O abundance ratios  from the $T_e$-based ({\em direct}) method for a sub-sample of 610 regions, and from a variety of strong-line methods for the whole \hii\ region sample. The strong-line methods have been evaluated in relation to the $T_e$-based one from both a comparison of the derived O/H and N/O abundances for individual \hii\ regions, and a comparison of the abundance gradients derived from both methodologies. The  median value and the standard deviation of the gradient distributions depend on the abundance method, and those based on the O3N2 indicator tend to flatten the steepest profiles, reducing the range of observed gradients.  A detailed analysis and discussion of the derived  O/H and N/O radial abundance gradients and y-intercepts for barred and unbarred galaxies is presented in the companion Paper II. The whole \hii\ region catalogue including emission-line fluxes, positions and derived abundances is made publicly available on the CDS VizieR facility, together with the  radial abundance gradients for all galaxies.
\end{abstract}

\begin{keywords}
ISM: abundances -- HII regions -- galaxies: spiral -- galaxies: ISM -- galaxies: abundances -- galaxies: structure
\end{keywords}


\section{Introduction}
The present-day distribution of metals in disc galaxies is a key diagnostic tool to understand their evolution. \hii\ regions are excellent tracers of this distribution in the gas-phase of spirals. Although the number of \hii\ regions does vary from galaxy to galaxy, typically hundreds of regions populate galaxy discs, having high luminosities concentrated in bright emission lines detectable up to large distances \citep[e.g.][]{n7479_hii,CedresCepa2002}, allowing to map the metal distribution across the face of galactic discs \citep[e.g.][]{bresolin2004,b12,pilyugin2014}. In the nearby universe, spectroscopic observations of \hii\ regions revealed, to first order, an exponential decrease in the relative abundance of oxygen to hydrogen  (normally used as a tracer for the metals) from the galaxy centre to the outer disc regions. The dependence of  12+$\log$(O/H) with galactocentric radius, normally termed radial {\em metallicity} profile, is then well parametrized by a straight line with a characteristic slope or {\em metallicity gradient.} Since their discovery \citep{Aller1942,Searle1971}, the present-day metallicity gradients have received much attention, as they are the result of the  galaxy evolution, where  a complex interplay between star-formation efficiency, infall of low metallicity gas, metal-enriched gas outflows,  interactions and mergers, stellar migration and gas flows within the disc, shapes the radial distribution of metals.

Radial gas flows can not by themselves produce the radial metallicity gradients observed in galaxies \citep{GoetzKoeppen1992}, but are known to be an efficient mechanism to change them \citep[e.g.][]{SchonrichBinney2009,SpitoniMatteucci2011,GrisoniSpitoniMatteucci2018}.
The gravitational potential of non-axisymmetric structures in disc galaxies, such as bars, is one of the most efficient mechanisms that can create large scale radial gas flows according to simulations \citep[e.g.][]{SW93,athanassoula03}. The induced gas-flows mix the gas and can flatten the metallicity gradient: assuming an initial negative metallicity profile, inwards gas flows dilute the higher metal content in the central regions, while outwards flows can enrich the more metal-poor outer disc areas.
 
In the 90s, \citet{vila-costas92}  were pioneering at investigating the properties of \hii\ regions in relation to the  properties of their host galaxies. In particular, they studied the  gas-phase radial metallicity gradients. In this, and in subsequent work \citep{Zaritsky94,Martin94,Dutil_Roy} a trend was reported for barred galaxies to show shallower metallicity gradients than unbarred galaxies, in agreement with the theoretical predictions. However, more recent work in which the galaxy samples have been considerably enlarged, the comparison of gas-phase metallicity profiles of barred and unbarred galaxies reveals no difference in the slope between the two types of spirals \citep{sanchez14,Sanchez-Menguiano2016,Zinchenko2019,Perez-Montero2016,Kaplan2016}.

Studies of the central gas-phase chemical abundances also yield contradictory results. For example, \citet{ellison} found larger central metallicity in barred galaxies with respect to unbarred ones.  \citet{considere} and \citet{dutil} reported a lower central metallicity in a sample of barred starbursts than in unbarred galaxies, while \citet{cacho} reported no difference in barred and unbarred galaxies. In a subsequent work, \citet{bulbos} found no difference in central metallicity but a enhanced N/O ratio in the centres of barred galaxies with respect to unbarred galaxies using Sloan Digital Sky Survey\footnote{\url{http://www.sdss.org}} (SDSS) spectra, as in \citet{ellison} and \citet{cacho}.

Apart from these contradicting results, there are important differences in methodology between the cited works, especially between the 90s works and  the most recent ones. On the one hand, in the 90s the chemical abundances were derived from integrated spectra or spectrophotometry of {\em individual} \hii\ regions, and authors used different calibrations of the R$_{23}$ parameter \citep{pagel79} or the [\oiii]/\hb\ and [\nii]/[\oiii] line ratios to derive their metallicities. On the other hand, more recent works have preferentially used empirical calibrations of the O3N2 and N2O2 parameters \citep[e.g.]{marino13}, the $R$ calibration \citep{PilyuginGrebel2016} or other model-based methods such as the HII-Chi-mistry code \citep{epm14}. Furthermore, the combination of spatial resolution of the IFU instruments used in these works, together with  the median distance of the sample galaxies, implies that these works are  based either on a spaxel-by-spaxel analysis or on integrated spectra of star-forming complexes, that normally include several {\em individual} \hii\ regions\footnote{The spatial element-resolution in CALIFA is 3\arcsec, that corresponds to $\sim1$~kpc \citep{sanchez14} at the average redshift of the survey.  The spatial resolution of VENGA is 5.6\arcsec, that corresponds to a median value of $\sim$387~pc at the median distance of their sample galaxies \citep[14.3 Mpc,][]{Kaplan2016}. Typical extragalactic \hii\ region diameters range $\sim90-800$~pc \citep[e.g.][]{n3359_hii, Oey2003}.}. This is also the case for the central abundances derived from SDSS spectra from  \citet{ellison}, \citet{cacho} and \citet{bulbos}, that include emission from the inner $\sim1-4$~kpc.

It is well known that different gas-phase metallicities are obtained depending on the strong-line method used \citep[e.g.][]{KewleyEllison2008,LopezSanchez-Esteban10}, but more importantly, this difference is not just a zero-point offset. Different methods also yield different radial metallicity gradients \citep{arellano-cordova,Bresolin09}. In the 90s, some of the largest data samples were based on  inhomogeneous compilations of slopes derived from previous authors with different calibrations \citep{Dutil_Roy}. In the more recent works, however, data analysis is homogeneous, but the improved instrumental sensitivity and  the lower spatial resolution imply  contamination from  the diffuse ionized gas  \citep[DIG, e.g.][]{zurita2000,oey2007}. The DIG has different physical conditions and ionizing spectra  compared to  \hii\ regions, and its inclusion on the extracted spectra can have a strong impact on the derived metallicities and, as a consequence, on the metallicity gradients \citep{zhang2017,sanders2017,vale-asari2019}.

Due to the most recent (mostly IFU-based) results, there might be  a new general and extended conviction among astronomers that the field is closed and bars have little impact on the gas-phase metallicity gradients. However, we believe there are enough reasons that justify a revision of the topic. Therefore, our aim is to  benefit from  the improved spectroscopic data sets available nowadays to make a compilation of emission-line measurements from  resolved \hii\ regions in nearby spirals. From these, we  perform a  homogeneous analysis of the data and revisit the topic of the influence of stellar bars on the gas-phase radial abundance gradient of O/H and N/O in spirals, aiming  to solve the controversy raised by previous works.

We divide our study of the bar effect on the gas-phase radial abundance profiles into two parts. The first one is included in this paper, and it is organized as follows. Sects.~\ref{sample} and \ref{hii_sample} present the galaxy and \hii\ region samples, respectively. The process to obtain physical properties and chemical abundances for the \hii\ regions, with the electron temperature-based  ({\em direct}) and strong-line methods is described in Sect.~\ref{abundances}, where we also perform a comparison between  the different methods employed. In Sect.~\ref{fits} we present the resulting radial abundance profiles. Finally, in Sect.~\ref{conclusiones} we present a summary and our conclusions.
The second part of the study is presented in the companion \citetalias{paperII}  \citep{paperII}, where we carefully analyse the derived radial metallicity gradients comparatively for (strongly and weakly) barred and unbarred galaxies. The results are compared with previous work, and discussed in the context of current knowledge on disc evolution and radial mixing induced by bars and spiral arms.

\section{Galaxy sample}
\label{sample}
The galaxy sample is shown in Table~\ref{tab:sample}. It comprises 51 nearby spiral galaxies (distances $<64$~Mpc). Our criterion for the sample selection was to include galaxies for which emission line ratios of \hii\ regions were available from previous publications, together with the corresponding information on their celestial coordinates (either absolute or relative to the galactic centre). We concentrated the search on spirals with inclination angles $i<70\deg$, with measurements for at least seven \hii\ regions covering a wide range in galactocentric distance so that more reliable radial profiles of the chemical abundances can be obtained.  As we want to minimize the amount of diffuse ionized gas contamination on the derived \hii\ region properties we gave priority to galaxies with resolved spectroscopic data. Therefore most of the spectroscopic data of our sample comes from long-slit  or fiber-fed spectroscopic campaigns of individual \hii\ regions. The median distance for the galaxy sample is $12\pm2$~Mpc, that yields an average spatial scale of 58~pc~arcsec\me\ \citep[c.f. $\sim350$~pc~arcsec\me\ for the CALIFA sample,][]{Walcher2014}. Typical extragalactic \hii\ region diameters range from $90-800$~pc \citep[e.g.][]{n3359_hii,Oey2003}. Our search is rather exhaustive, but it is probably not complete, and we can not exclude that we  have missed published data that meets the above-mentioned requirements. The distribution of sample galaxy properties is presented in Sect.~\ref{distrib_sample}.

There are other existing samples based on published \hii\ region data, as it is the case for the compilation presented in \citet{pilyugin2014}. Our sample is considerably smaller \citep[51 galaxies vs. 130 in][]{pilyugin2014} because of our more restrictive criteria for the compilation: (1) we only include spiral galaxies in the sample, while  \citet{pilyugin2014} also includes irregulars; (2) we prioritized the inclusion of data from resolved spectroscopy, but a considerable number of galaxies in the \citet{pilyugin2014} sample comes from integral field spectroscopy \citep[e.g. from][]{sanchez2012}; (3) we only compiled data for which both emission-line fluxes and \hii\ region coordinates were available, in order to recalculate both abundances and galactocentric distances with the same methodology, and (4) our minimum required number of regions per galaxy was set to seven, while according to figure~1 in \citet{pilyugin2014} this limit is smaller in their sample ($\sim$3 regions/galaxy). 
The latter requirement, together with the fact that our sample benefits from the publications on resolved \hii\ spectroscopy of the last few years, implies that we have a larger average number of \hii\ regions per galaxy, $\sim$56  against $\sim$29 in \citet{pilyugin2014}. Another advantage of our sample is the  availability of $T_e$-based abundance estimates, as we compiled auroral-line fluxes in addition to strong-line fluxes (see Sect. ~\ref{hii_sample}).

This work requires knowledge of several  galaxy structural parameters: (1) The disc inclination and position angle are necessary to compute \hii\ region deprojected galactocentric distances. The disc scale length (or equivalently the disc effective radius) is also needed to normalize distances and sizes. (2) Bar parameters (length, position angle and ellipticity) are also required. These  allow us to quantify bar strengths and to determine \hii\ region positions  in relation to the bar.  

In spite of the fact that the galaxies of the sample are nearby and well known targets, when structural parameters are available neither the methodology nor the  photometric bands employed by different authors are homogeneous. In order to ensure consistency in the data and in the corresponding analysis procedure, we  recalculated disc and bar structural parameters in a homogeneous way for the sample galaxies. The methodology and results are described in the following sections.
\begin{figure*}
\includegraphics[width=0.95\textwidth]{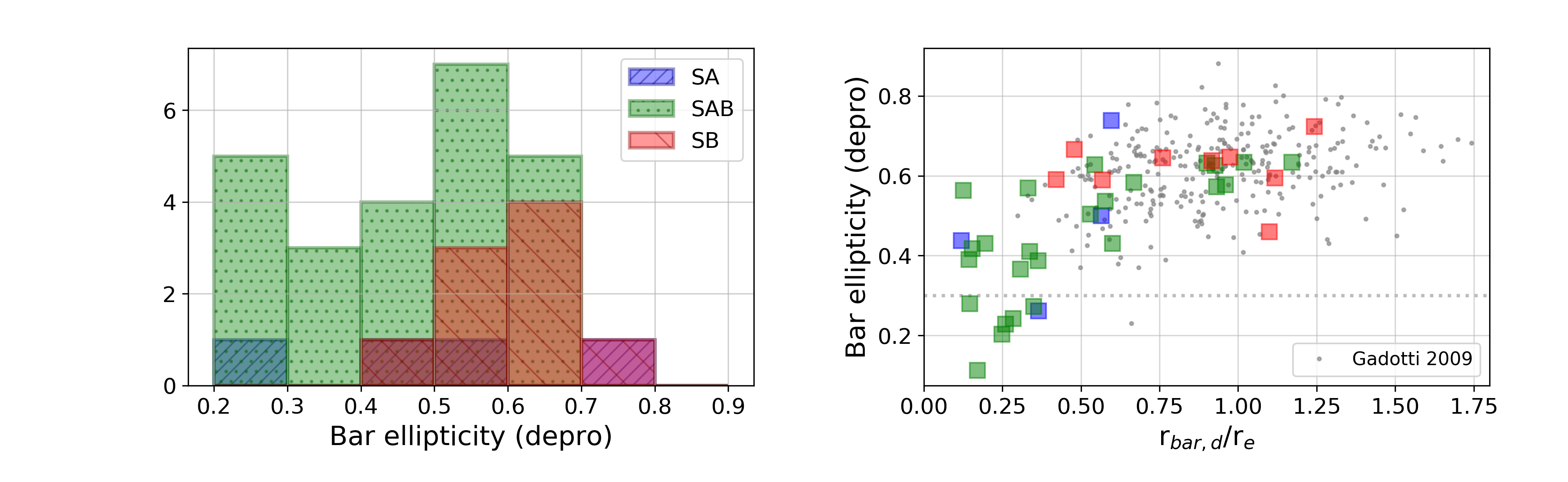}
\caption{{\em (Left)} Distribution of deprojected bar ellipticities ($e_{bar,d}$) for the three different RC3 visual bar classes: unbarred, SA galaxies (dashed blue), barred, SB (dashed orange) and intermediate or SAB (dotted green).  {\em (Right)}  Deprojected bar ellipticity as a function of the normalized (to the disc effective radius) bar radius. The straight dotted line marks the division between barred and unbarred galaxies. The  grey dots represent bar parameters for the face-on barred galaxies analysed by \citet{dimitrimorpho} for comparison.
\label{barsRC3}}
\end{figure*}

\subsection{Morphological analysis}
\label{morfo}
We compiled  broad-band images for all galaxies in the sample  (except for M31 and the Milky Way). All images are Sloan $r$-band or Johnson $R$-band images, except for NGC~1637 and NGC~1365, for which we used a $V$-band and $I$-band image, respectively. Most of the images come from the Sloan Digital Sky Survey  database, although a relevant number of them were obtained from the NASA/IPAC Extragalactic Database\footnote{\url{https://ned.ipac.caltech.edu/}}, and two galaxies (NGC~2336 and NGC~6384) were observed with CAFOS at the  Calar Alto Observatory 2.2m telescope in August 2018 through an $r$-band filter.
Table~\ref{referencias_imagenes} summarises the origin and photometric bands for these images.

The morphological characterization of the galaxies was performed through the widely used method of fitting ellipses to the image isophotes \citep[e.g.][]{wozniak95,aguirre00,marinova07} with the {\tt ellipse} task within IRAF\footnote{IRAF is distributed by the National Optical Astronomy Observatory, which is operated by the Association of Universities for Research in Astronomy (AURA) under a cooperative agreement with the National Science Foundation.}. For each galaxy the position of the nucleus was previously determined and fixed during the fit, whereas the position angle ($PA$) and the ellipticity ($e$) of the isophotal ellipses were set to be free parameters.

The disc scale length ($r_d$) was calculated from a fitting to the disc-dominated area of the surface brightness radial profile obtained with {\sc ellipse} to  a single exponential function. When a break in the surface brightness profile \citep[e.g.][]{MM13} was detected or we suspected could be present (e.g. for NGC~1058, NGC~3521, NGC~4258 and NGC~4303), we derived the disc scale length of the innermost part of the disc-dominated profile, excluding from the fit the outer disc. Therefore our derived $r_d$  values correspond to the inner disc scale length in these cases. The disc effective radius\footnote{The disc effective radius is the radius that contains half of the total disc integrated flux. For an exponential disc profile, $r_e=1.678 r_d$, with $r_d$ the disc scale length.}, $r_e$, was calculated from $r_d$. The disc $PA$ and ellipticity (and therefore the disc inclination $i$) were estimated from the average values in the disc-dominated region of the corresponding radial profiles. Table~\ref{tab:sample} shows the disc parameters obtained from our photometric analysis, except for the Milky Way\footnote{See brief explanation on the choice of parameters for the Milky Way in Appendix~\ref{notes_galaxies} (available online).} and M31 for which we used published values. For NGC~1365, NGC~1512 and NGC~2805 we have also partially used morphological parameters from the literature as compiled images were not suitable for the whole analysis. See Table~\ref{tab:sample} for the references.

\subsection{Bar parameters and classification}
\label{bar_class}
The bar parameters (length, position angle and ellipticity\footnote{The bar ellipticity, $e_{bar}$,  is related to the bar semi-axes by $e_{bar} = 1 - b/a$.}) were also determined from  the radial profiles obtained from the ellipse fits.  The bar radius was defined as the radius at which the ellipticity profile shows a local maximum value, $e_{bar}$, whereas the position angle remains approximately constant. 
This method has the advantage that it is simple to reproduce and to compare with published values in the literature, but see \citet{erwin} for a discussion on different methods to estimate the bar length and their systematic differences. In particular, the radius obtained from this procedure must be considered as a lower bound to the visual bar length, although the difference is not as large as initially thought \citep{SimonDiaz2016}. The bar position angle, $PA_{bar}$,  was  estimated from the average $PA$ within the bar radius.

Any oval feature with observed ellipticity above $\sim0.2$ was initially considered as a potential barred structure. The observed values for the  bar length and ellipticity were afterwards deprojected to their face-on values, using the analytical expressions in \citet{gadotti07} (in their appendix~A), considering the bar as a planar ellipse. Although this picture is a simplification of the real 3D shape of bars, the typical intrinsic flattening \citep[$\sim$0.34 on average --][]{jairo18} is similar to that of stellar discs. In addition, \citet{zou2014} used mock galaxies to quantify uncertainties in bar deprojected parameters and conclude that the 2D deprojection method used here is preferred over 1D methods, as it yields reasonably good uncertainties ($\sim10$\% for galaxy inclination angles $\lesssim60\deg$) in both bar length and ellipticity.

Disc galaxies of a given morphological type are generally classified into classes according to their bar visual prominence. Following the RC3 catalogue \citep{RC3} nomenclature, the standard classes are SB for strong bars, SA for unbarred galaxies and SAB for intermediate or weak bars. However, in this respect the RC3 bar type classification is known to be problematic and should be taken with caution \citep[see e.g.][]{marinova07,nair,erwin2018}. The classification has a certain degree of ambiguity. In fact, several galaxies classified as SA in RC3 (NGC~2541, NGC~1068, NGC~5194, NGC~4395)  contain a central oval distortion  \citep{chapelon99,scoville88,zaritsky93,buta_atlas}, 
while other galaxies classified as SAB (e.g. NGC~2903, NGC~925) have bars comparable in length and/or ellipticity to those in SB galaxies, or have no apparent bar, as is the case for NGC~2403. This can be better seen in Fig.~\ref{barsRC3} ({\em left}) \citep[also in Fig.~15 of][for a different sample]{marinova07}, where we plot for our sample galaxies the distribution of our estimated deprojected bar ellipticities ($e_{ bar,d}$) for the three different bar classes  separately, as recorded from the RC3 catalogue. It can be seen that SB galaxies (in orange) are all concentrated in the largest ellipticity range (as expected), with $e_{ bar,d}\geqslant0.4$, but SAB galaxies (in green) have significant overlap in ellipticity with SB galaxies and cover a wide range in  $e_{ bar,d}$ from 0.2 to around 0.7, and the four galaxies classified as SA (in blue), but having central oval distortions, have ellipticity values of up to $\sim0.7$.

The deprojected bar ellipticity is frequently used to quantify the bar strength \citep[e.g.][]{martin95,abraham00}, and correlates with the bar gravitational torque \citep[$Q_b$ --][]{SimonDiaz2016}. Therefore, in order to perform a more quantitative classification of bars, we have used the deprojected bar ellipticity to classify barred galaxies into: strongly barred (when $e_{bar,d} \geqslant 0.5$) and weakly barred (when $0.3 \leqslant  e_{bar,d} < 0.5$). Any galaxy with measured central oval distortion in the fitted isophotes, but for which the deprojected ellipticity is smaller than 0.3 was considered as unbarred. A similar criterion  was also adopted by several authors \citep[see e.g.][]{abraham99,marinova07,martin95}.

There is a relation between bar length and ellipticity, in the sense that very elliptical bars are also long. And for a given bar length, there is a minimum observed ellipticity. This can be seen in Fig.~\ref{barsRC3} ({\em right}). Our cutoff in ellipticity implies that we are considering as unbarred, galaxies with detected central oval structures of axial ratio $>$0.7 and with radius $\lesssim2.3$~kpc (or  $\lesssim0.4$~r$_e$).

In summary, after the bar classification described above, our sample comprises 22 strongly barred, 9 weakly barred and 20 unbarred galaxies, i.e. there is virtually equal representation of strongly barred and unbarred systems.

\subsection{Distribution of galaxy properties}
\label{distrib_sample}
Fig.~\ref{sample} shows the distribution of  absolute integrated B-band magnitude, morphological T-type, inclination angle and disc effective radius, separately for the unbarred, weakly barred and strongly barred galaxy  sub-samples, obtained as explained in Sect.~\ref{bar_class}. Median values are similar for all sub-samples within errors. We have performed a two-sample Anderson-Darling (hereinafter AD) test to the distributions of strongly barred and unbarred galaxies in all four  parameters. The AD P-values\footnote{The output of the AD test is a P-value or significance  level at which the null hypothesis can be rejected. A threshold level of $\sim5\%$ is usually adopted. Therefore, values below $5\%$ are statistically significant and indicate that the null hypothesis that both samples are drawn from the same parent distribution can be rejected.} are higher than 5\% in all cases, indicating that the distributions are not significantly different in strongly barred and unbarred galaxies.

Our galaxy sample is not intended to be complete in any parameter, but it is important to ensure that barred and unbarred galaxies cover a similar parameter space. The comparison of Fig.~\ref{sample} confirms that this is the case.

\FloatBarrier
\begin{figure}
\hspace{-0.5cm}
\includegraphics[width=0.55\textwidth]{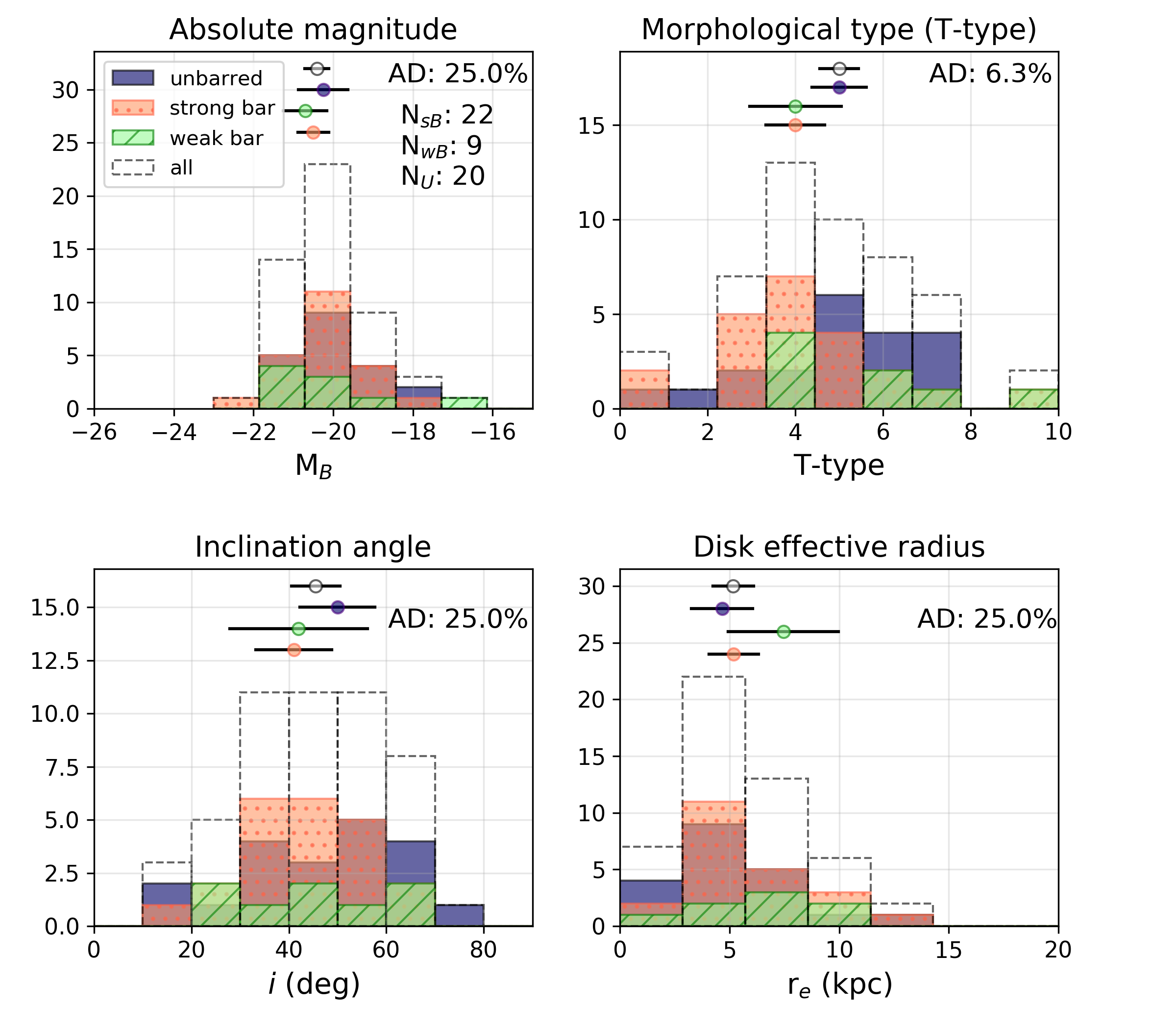}
\vspace{-0.3cm}
\caption{Histograms showing the distribution of absolute integrated B-band magnitude, morphological T-type, disc inclination angle and disc effective radius ($r_e$) for the  strongly barred (dotted orange), weakly barred (dashed green) and unbarred (blue) sub-samples of galaxies separately. The distribution for all galaxies is shown with a grey dashed-line. The circles in the upper side of the panels indicate the median value for each distribution and the horizontal error bar covers the 95\% confidence interval for the corresponding data value. The number of  galaxies in each sub-sample  and the P-value for the two-sample Anderson-Darling test (AD in \%) for the distribution of strongly barred and unbarred galaxies are also shown. The Anderson-Darling P-values are higher than 5\% in all cases, indicating that the distributions are not different in strongly barred and unbarred galaxies.\label{sample}}
\end{figure}
 
\begin{table*}
\begin{minipage}{1.05\textwidth}
\centering
\caption{Main properties of the galaxy sample: galaxy name (NGC number, except MW = Milky Way), morphological type, distance, absolute B-band magnitude, disc inclination ($i$), disc and bar position angles (PA, PA$_{bar}$), disc isophotal (r$_{25}$) and effective (r$_{e}$) radii,  observed bar radius or semi-major axis (r$_{bar}$), bar deprojected semi-major axis normalized to r$_{e}$, deprojected bar ellipticity ($e_{bar,d}$), nuclear activity, number of independent measurements of \hii\ regions in each galaxy (in parenthesis those that allow $T_e$-based abundance determinations).}
\label{tab:sample}
{\footnotesize
\begin{tabular}{lcccccccccccccc}
\hline
Galaxy & Type$^a$ & D   & M$_{B}$$^a$ & $i$ & PA  & PA$_{bar}$ & r$_{25}$$^a$ & r$_{e}$ & r$_{bar}$  &  r$_{bar,d}$/r$_{e}$ &  $e_{bar,d}$ & Nuc.      &  N \\
(NGC)       &       & (Mpc) & (mag)   & (deg)&(deg)& (deg)     & (kpc)         & (kpc)  &  (kpc)   &                     &          &   Act.$^b$   & \\
\hline
 MW        & .SBT4.. &   - & $-20.1$  &   -        &   -        &   -     & 26.8$^{c}$  & 4.5$\pm$0.4$^{d}$ & $5.0^{e}$ & 1.10 & 0.46$^{f}$ & - & 33 (32)\\
224$^{g}$  & .SAS3..  & 0.7 & $-21.0$  & 77         & 38         &   -       & 20.6       & 9.4$\pm$0.8   & 0.0 &   -   &   -   & Q     & 205 (22)\\
300       & .SAS7.. & 1.9 & $-17.9$  & 56$\pm$8   & 114$\pm$8  &   -       & 6.0        & 2.5$\pm$0.3   & 0.0 &   -   &   -   & -     & 42 (35)\\
598       & .SAS6.. & 0.8 & $-18.9$  & 53$\pm$8   & 22$\pm$15  &   -       & 8.6        & 3.7$\pm$0.3   & 0.0 &   -   &   -   & -     & 316 (120)\\
628       & .SAS5.. & 9.0 & $-20.0$  & 32$\pm$3   & 4$\pm$6    &   -       & 13.8       & 5.4$\pm$0.9   & 0.0 &   -   &   -   & -     & 210 (61)\\
925       & .SXS7.. & 9.3 & $-19.9$  & 54$\pm$6   & 120$\pm$6  & 110$\pm$3 & 14.2       & 8.3$\pm$0.8   & 4.8 & 0.60  & 0.43  & -     & 139 \\
1058      & .SAT5.. & 10.6 & $-18.6$ & 17$\pm$14  & 150$\pm$30 &   -       & 4.7        & 1.8$\pm$0.4   & 0.0 &   -   &   -   & S2    & 48 \\
1068      & RSAT3.. & 10.1 & $-20.6$ & 37$\pm$4   & 175$\pm$8  & 56$\pm$4  & 10.4       & 2.4$\pm$0.2   & 1.1 & 0.56  & 0.50  & S1h   & 22 \\
1097      & .SBS3.. & 16.0 & $-21.1$ & 45$\pm$5   & 117$\pm$10 & 148$\pm$5 & 21.7       & 8.8$\pm$0.2   & 8.7 & 1.12  & 0.59  & S3b   & 15 \\
1232      & .SXT5.. & 14.5 & $-20.4$ & 35$\pm$2   & 104$\pm$9  & 95$\pm$5  & 15.7       & 8.2$\pm$0.4   & 1.2 & 0.15  & 0.28  & -     & 34 (13)\\
1313      & .SBS7.. & 4.6  & $-19.0$ & 40$\pm$6   & 1$\pm$2    & 16$\pm$1  & 6.1        & 2.8$\pm$0.4   & 1.1 & 0.42  & 0.59  & -     & 45 (8)\\
1365$^{h}$& .SBS3.. & 19.6 & $-21.5$  & 41         & 220        & 85$\pm$1  & 32.0       & 11.4$\pm$1.9  & 9.5 & 0.97  & 0.65  & S1.8  & 87 (8)\\
1512$^{i}$& .SBR1.. & 12.0 & $-19.4$  & 49$\pm$5   & 80         & 45$\pm$2   & 15.6      & 5.5$\pm$0.8   & 4.1 & 0.92  & 0.64  & -     & 147 (8)\\
1637      & .SXT5.. & 12.0 & $-19.1$ & 39$\pm$1   & 36$\pm$2   & 76$\pm$5  & 7.0        & 3.0$\pm$0.1   & 1.4 & 0.53  & 0.50  & -     & 17 \\
1672      & .SBS3.. & 14.5 & $-20.6$ & 34$\pm$2   & 154$\pm$7  & 99$\pm$2  & 13.9       & 4.7$\pm$0.2   & 5.1 & 1.24  & 0.72  & S     & 17 \\
2336      & .SXR4.. & 33.0 & $-22.0$ & 56$\pm$1   & 179$\pm$5  & 122$\pm$2 & 34.0       & 11.5$\pm$1.3  & 4.6 & 0.67  & 0.58  & -     & 36 \\
2403      & .SXS6.. & 3.1  & $-19.0$ & 57$\pm$5   & 123$\pm$4  &   -       & 9.7        & 2.8$\pm$0.1   & 0.0 &   -   &   -   & -     & 77 (25)\\
2541      & .SAS6.. & 12.6 & $-18.9$ & 60$\pm$3   & 162$\pm$5  & 171$\pm$3 & 11.5       & 4.6$\pm$0.2   & 1.6 & 0.36  & 0.26  & -     & 19 \\
2805$^{j}$ & .SXT7.. & 28.0 & $-21.1$ & 36$\pm$2   & 123$\pm$3  & 114$\pm$20  & 25.7     & 12$\pm$2      & 2.1 & 0.17  & 0.11  & -     & 17 \\
2903      & .SXT4.. & 8.9  & $-20.6$ & 59$\pm$1   & 15$\pm$3   & 24$\pm$3  & 16.3       & 3.50$\pm$0.09 & 3.1 & 0.93  & 0.57  & -     & 53 (1)\\
2997      & .SXT5.. & 7.1  & $-19.9$ & 48$\pm$5   & 91$\pm$7   & 109$\pm$6 & 9.2        & 4.7$\pm$0.6   & 1.1 & 0.26  & 0.23  & -     & 22 (7)\\
3031      & .SAS2.. & 3.6  & $-20.4$ & 54$\pm$3   & 159$\pm$2  &   -       & 14.2       & 4.5$\pm$0.2   & 0.0 &   -   &   -   & Q     & 131 (29)\\
3184      & .SXT6.. & 14.4 & $-20.4$ & 16$\pm$3   & 148$\pm$10 & 82$\pm$4  & 15.5       & 7.7$\pm$1.1   & 1.9 & 0.25  & 0.20  & -     & 84 (25)\\
3227      & .SXS1P. & 20.6 & $-20.4$ & 66$\pm$1   & 152$\pm$3  & 160$\pm$3 & 16.1       & 6.7$\pm$0.7   & 1.8 & 0.35  & 0.27  & S1.5  & 9 \\
3310      & .SXR4P. & 18.7 & $-20.4$ & 32$\pm$5   & 172$\pm$13 & 31$\pm$20 & 8.4        & 4.9$\pm$1.2   & 0.6 & 0.13  & 0.58  & -     & 14 \\
3319      & .SBT6.. & 14.1 & $-19.6$ & 69$\pm$2   & 26$\pm$3   & 38$\pm$1  & 12.6       & 8.8$\pm$1.2   & 3.6 & 0.48  & 0.67  & -     & 13 \\
3344      & RSXR4.. & 9.8  & $-19.4$ & 27$\pm$5   & 150$\pm$30 & 179$\pm$1 & 10.1       & 3.50$\pm$0.09 & 1.1 & 0.34  & 0.41  & -     & 15 (1)\\
3351      & .SBR3.. & 10.5 & $-19.8$ & 43$\pm$2   & 12$\pm$2   & 113$\pm$2 & 11.3       & 4.7$\pm$0.4   & 3.3 & 0.96  & 0.58  & -     & 28 \\
3359      & .SBT5.. & 16.7 & $-20.4$ & 57$\pm$2   & 176$\pm$4  & 10$\pm$1  & 17.6       & 7.0$\pm$0.2   & 3.7 & 0.57  & 0.59  & -     & 35 (3)\\
3521      & .SXT4.. & 12.1 & $-21.1$ & 49$\pm$6   & 159$\pm$4  & 162$\pm$2 & 19.3       & 4.5$\pm$0.2   & 1.3 & 0.28  & 0.24  & -     & 13 \\
3621      & .SAS7.. & 7.2  & $-20.1$ & 64$\pm$4   & 163$\pm$2  &   -       & 12.9       & 4.2$\pm$0.4   & 0.0 &   -   &   -   & -     & 80 (12)\\
4254      & .SAS5.. & 14.4 & $-20.7$ & 39$\pm$4   & 41$\pm$10  &   -       & 11.3       & 4.8$\pm$0.2   & 0.0 &   -   &   -   & -     & 20 \\
4258      & .SXS4.. & 8.4  & $-21.1$ & 64$\pm$1   & 159$\pm$3  & 145$\pm$5 & 22.8       & 5.6$\pm$0.3   & 0.7 & 0.15  & 0.42  & S2    & 65 (6)\\
4303      & .SXT4.. & 14.5 & $-20.7$ & 30$\pm$5   & 58$\pm$20  & 2$\pm$1   & 13.6       & 3.0$\pm$0.6   & 2.5 & 0.93  & 0.63  & S2    & 22 \\
4321      & .SXS4.. & 17.2 & $-21.2$ & 27$\pm$3   & 31$\pm$10  & 99$\pm$4  & 18.6       & 5.9$\pm$0.2   & 5.4 & 1.02  & 0.63  & -     & 11 \\
4395      & .SAS9*. & 4.3  & $-17.6$ & 55$\pm$1   & 143$\pm$7  & 120$\pm$4 & 8.3        & 5.6$\pm$1.0   & 2.9 & 0.60  & 0.74  & S1.8  & 18 (5)\\
4625      & .SXT9P. & 9.5  & $-17.0$ & 21$\pm$6   & 160$\pm$10 & 21$\pm$6  & 3.0        & 0.9$\pm$0.1   & 0.3 & 0.36  & 0.39  & -     & 34 (1)\\
4651      & .SAT5.. & 29.1 & $-21.3$ & 49$\pm$2   & 79$\pm$5   &   -       & 16.8       & 7.6$\pm$0.4   & 0.0 &   -   &   -   & -     & 7 \\
4654      & .SXT6.. & 14.5 & $-20.1$ & 67$\pm$4   & 127$\pm$4  & 119$\pm$2 & 10.3       & 7.5$\pm$0.6   & 1.0 & 0.14  & 0.39  & -     & 7 \\
5194      & .SAS4P. & 8.0  & $-20.8$ & 42$\pm$8   & 38$\pm$10  & 138$\pm$3 & 13.0       & 7.7$\pm$0.2   & 0.7 & 0.12  & 0.44  & S2    & 104 (35)\\
5236      & .SXS5.. & 4.6  & $-20.3$ & 16$\pm$2   & 120$\pm40$ & 55$\pm$3  & 8.6        & 3.1$\pm$0.1   & 2.7 & 0.90  & 0.63  & -     & 94 (12)\\
5248      & .SXT4.. & 13.0 & $-19.9$ & 58$\pm$5   & 136$\pm$7  & 95$\pm$2  & 11.7       & 3.8$\pm$0.4   & 1.4 & 0.58  & 0.54  & -     & 11 \\
5457      & .SXT6.. & 6.7  & $-20.9$ & 36$\pm$8   & 57$\pm$30  & 79$\pm$3  & 28.1       & 8.6$\pm$0.2   & 1.6 & 0.19  & 0.43  & -     & 260 (137)\\
6384      & .SXR4.. & 29.7 & $-21.8$ & 46$\pm$9   & 29$\pm$4   & 36$\pm$1  & 26.6       & 11.4$\pm$1.9  & 3.5 & 0.31 & 0.37  & -     & 18 \\
6946      & .SXT6.. & 5.4  & $-20.9$ & 29$\pm$3   & 40$\pm$15  & 14$\pm$4  & 9.0        & 5.7$\pm$0.4   & 1.8 & 0.33  & 0.57  & -     & 10 (1)\\
7331      & .SAS3.. & 15.1 & $-21.5$ & 63$\pm$1   & 169$\pm$2  &   -       & 23.0       & 8.0$\pm$0.3   & 0.0 &   -   &   -   & -     & 16 \\
7518      & RSXR1.. & 34.8 & $-18.9$ & 32$\pm$6   & 41$\pm$7   & 124$\pm$2 & 7.2        & 3.20$\pm$0.08 & 3.2 & 1.17  & 0.63  & -     & 13 \\
7529$^{k}$ & RSA.4.. & 63.2 & $-19.6$ & 29$\pm$5   & 157$\pm$6  &   -       & 7.3        & 3.3$\pm$0.2   & 0.0 &   -   &   -   & -     & 30 \\
7591      & .SB.4.. & 53.2 & $-20.6$ & 61$\pm$2   & 142$\pm$6  & 18$\pm$3  & 15.0       & 6.2$\pm$0.5   & 2.5 & 0.76  & 0.64  & S     & 16 \\
7678      & .SXT5.. & 44.7 & $-21.1$ & 42$\pm$3   & 25$\pm$10  & 111$\pm$10& 16.5       & 6.8$\pm$0.6   & 2.8 & 0.54  & 0.63  & -     & 11 \\
7793      & .SAS7.. & 3.6  & $-18.4$ & 51$\pm$3   & 99$\pm$2   &   -       & 4.9        & 2.3$\pm$0.1   & 0.0 &   -   &   -   & -     & 41 (3)\\
\hline
\end{tabular}}
{\footnotesize
\begin{tabular}{ll}
$^{a}$ \citet[][RC3]{RC3}. M$_{B}$ obtained from the extinction         &   $^{g}$ Disc inclination and PA  from \citet{Corbelli2010}, r$_{e}$ from\\
corrected total blue magnitude and the distances in column 3, except  &   reported value for the disc scale  length in I-band by\\
  for the Milky Way \citep{Licquia2015}.                              &   \citet{Courteau2011}, adjusted to D = 744~kpc,  \\
$^{b}$ Nuclear activity according to \citet{veron-cetty06}.            &   r$_{d}$ = (5.6$\pm$0.5) kpc.\\
$^{c}$ From  \citet{Goodwin1998}, $r_{25}$=(26.8$\pm$1.1)~kpc.          &   $^{h}$ Disc inclination, PA and $r_d$ from \citet{zanmar-sanchez08}. \\
  $^{d}$ From the disc scale length reported by \citet{Licquia2016},   &   $^{i}$ Disc PA from \citet{koribalski09}.\\
 $r_d$=(2.71$\pm$0.22)~kpc.                                           &   $^{j}$ Disc inclination an PA from \citet{Erroz-Ferrer15}.\\
$^{e}$ From \citet{Wegg2015}, $r_{bar}$=(5.0$\pm$0.2)~kpc.              &   $^{k}$ Morphological type from the {\tt Hyperleda} database\\
$^{f}$ From \citet{Inma2011}.                                          &  \url{http://leda.univ-lyon1.fr/}.\\    
\end{tabular}}
\end{minipage}
\end{table*}

\section{H\,{\small II} region sample}
\label{hii_sample}

Our observational data come from a compilation of optical emission-line fluxes and positions of \hii\ regions from published papers. We then used a consistent method to obtain metallicities and deprojected galactocentric distances for all the regions.

Our criteria for the sample selection were described in Sect.~\ref{sample}. For these galaxies we collected \hii\ region emission-line fluxes and the corresponding information on their position within the host galaxies. It is a major compilation of published data from over 80 different papers. See Table~\ref{bibliografia_gal} for a list of references.


\subsection{H\,{\small II} region emission line fluxes}
The compilation includes fluxes, normalized to those of the H$\beta$ line, for the brightest emission lines ([\oii]$\,\lambda\lambda$3726,3729\footnote{The [\oii]$\,\lambda\lambda$3726,3729 is usually blended for the compiled observational data. Hereinafter it  will be referred to as [\oii]$\,\lambda$3727.},  [\oiii]$\,\lambda\lambda$4959,5007, [\nii]$\,\lambda\lambda$6548,6583, \ha, [\sii]$\,\lambda\lambda$6717,6731, [\siii]$\,\lambda\lambda$9069,9532). The auroral line fluxes ([\sii]$\,\lambda\lambda$4068,4076,  [\oiii]$\,\lambda$4363, [\nii]$\,\lambda$5755, [\siii]$\,\lambda$6312 and [\oii]$\,\lambda\lambda$7320,7330) were also compiled when available. These are necessary for a more accurate determination of chemical abundances via the $T_e-$based or {\em direct} method.

The original data is highly heterogeneous in a number of aspects, in particular in the data format for both positions and fluxes, in the number of emission lines provided by the different authors or in the flux error estimates. We adopted the following criteria:

\begin{enumerate}
\item When either only the brightest line of the [\oiii]$\,\lambda\lambda$4959,5007,  [\nii]$\,\lambda\lambda$6548,6583 or  [\siii]$\,\lambda\lambda$9069,9532 doublets is given, or the sum of the two line fluxes is given,  we calculated the flux of individual lines assuming the theoretical ratios: [\oiii]$\,\lambda$5007/[\oiii]$\,\lambda$4959 = 3,  [\nii]$\,\lambda$6583/[\nii]$\,\lambda$6548 = 3 and [\siii]$\,\lambda$9532/[\siii]$\,\lambda$9069 = 2.44  \citep[e.g.][]{Perez-Montero2017}.

\item The compilation contains observations of the same \hii\ regions by different authors. We retained all of them as independent observations. 

\item Emission line flux uncertainties were not always given by the authors. We used the measurements of different authors for the same \hii\ region to estimate errors in the compiled emission lines. Typical differences between  different authors are $\sim$15\% for  [\oiii]$\,\lambda\lambda$4959,5007 and [\nii]$\,\lambda\lambda$6548,6583,  $\sim$20\% for   [\sii]$\,\lambda\lambda$6717,6731, and around 30\% for [\oii]$\,\lambda$3727 and [\siii]$\,\lambda\lambda$9069,9532. For the auroral line fluxes, there was more uniformity among authors and we have compiled their reported uncertainties.
    


       
\item In most cases, the emission-line fluxes published by the different authors had already been corrected for internal extinction. In these cases, the provided corrected fluxes have been used. In a few cases such as the M31 data by \citet{sanders12}, and those for NGC~1672, NGC~5248 and NGC~1097 by \citet{storchi96},
only observed fluxes were given by the authors. In these particular cases we dereddened the lines using the \citet{ccm} extinction curve and either the $A_v$ values reported by the authors, or the observed \ha/\hb\ ratio by imposing a theoretical Balmer decrement \ha/\hb = 2.86 \citep{osterbrock}. 

\item We tried to be conservative and excluded from our compilation objects whose  data were identified as uncertain by the authors in their original papers, or that were likely to be uncertain (because no determination of the extinction coefficient was done and/or very few lines were detected). In addition, some \hii\ region data were discarded during the analysis process due to unphysical results that could be indicative of errors. In Appendix~\ref{notes_galaxies}, available online, we include specific notes for some of the galaxies.
\end{enumerate}

The final sample contains 2831 independent measurements of \hii\ regions belonging to 51 nearby spirals (including the Milky Way). For 709 of them, auroral lines fluxes for the determination of the $T_e$-based oxygen abundance are available.
\subsection{H\,{\small II} region location}
\label{gal_distances}
We have recalculated the deprojected radial positions for all the \hii\ regions from the celestial coordinates,  from offset positions from the galaxy centre, or from distances along a slit of a given centre and position angle, depending on the information given by the different authors in the original publications (see Table~\ref{bibliografia_gal}).
We have assumed thin discs and have employed the disc position angles and inclinations derived from our morphological analysis of the galaxies (or from the literature for a limited number of galaxies, see Table~\ref{tab:sample} and Sect.~\ref{morfo}). Our derived values for the disc effective radius (r$_{e}$) and the isophotal galaxy radius (r$_{25}$, from RC3), given in Table~\ref{tab:sample}, have been used to normalize the derived \hii\ region deprojected galactocentric distances.

\subsection{AGN contamination and circumnuclear regions}
\label{AGN}
Nuclear activity can potentially contaminate the spectra of the innermost \hii\ regions in AGNs, which  can therefore produce an ill determination of chemical abundances. In particular, enhanced [\nii]/\ha\ from AGNs could be misinterpreted as an enhanced nitrogen  abundance. Our galaxy sample contains 13 galaxies with nuclear activity (1 LINER, 10 Seyferts and 2 QSOs), according to \citet{veron-cetty06}. See column 13 in Table~\ref{tab:sample}.

\begin{figure*}
\includegraphics[width=1.0\textwidth]{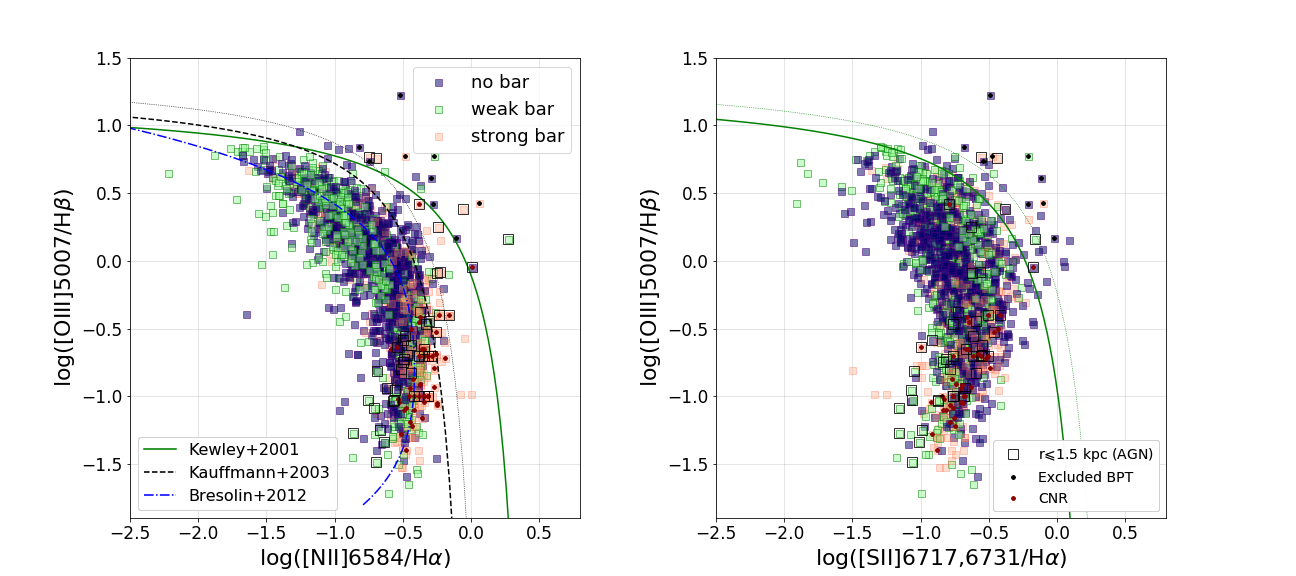}
\caption{[\oiii]$\,\lambda5007$/\hb\ vs. [\nii]$\,\lambda6583$/\hb\ (left) and  [\oiii]$\,\lambda5007$/\hb\ vs. [\sii]$\,\lambda6717,6731$/\hb\ (right) BPT  diagrams for the \hii\ regions compiled for this work.  The solid green and dashed black lines  separate the region for star-forming galaxies from the region for AGNs  according to \citet{kewley01} and \citet{kauffmann2003_AGN}, respectively. The dotted-dashed blue line shows the fit  obtained by \citet{b12} to the sequence outlined by a sample of Galactic and extragalactic \hii\ regions. The dotted-black and  dotted-green lines enclose the area in which regions located over the \citet{kauffmann2003_AGN} and \citet{kewley01} dividing lines, respectively, would still be compatible with star-forming, assuming an observational error of $\sim15\%$ for [\nii] and [\oiii], and 20\% for [\sii]. \hii\ regions located in weakly barred, strongly barred and unbarred galaxies are shown in green, orange, and blue, respectively. Overplotted black squares mark \hii\ regions located in galaxies containing an AGN at distances from the nucleus $\leqslant 1.5$~kpc. Overplotted red dots mark circumnuclear \hii\ regions. Black dots mark \hii\ regions that were excluded form the fittings for being simultaneously above the demarcation line by  \citet{kauffmann2003_AGN} in the [\oiii]$\,\lambda5007$/\hb\ vs. [\nii]$\,\lambda6583$/\hb\ diagram and the one by \citet{kewley01} in the [\oiii]$\,\lambda5007$/\hb\ vs. [\sii]$\,\lambda\lambda6717,6731$/\hb\ diagram.\label{BPT}}
\end{figure*}

Fig.~\ref{BPT} shows the BPT \citep{bpt} diagram for all the \hii\ regions of the sample. As we can see, the position of the \hii\ regions is  compatible with photoionization by massive stars in most cases. Only a small percentage of regions is above the demarcation line by \citet{kewley01}. Most of these regions are still compatible with photoionization if we take into account observational errors, and the majority belongs to normal (non-AGN) galaxies.

In any case,  to be on the safe side, we have been scrupulous with the data and have ignored in our radial abundance profile fits the innermost regions ($\lesssim$1.5~kpc) of active galaxies. This is the case for  NGC 1058, NGC~1365, NGC~1672, NGC~3227, NGC~4395, NGC~4395 and NGC~5194. \hii\ regions simultaneously  above the demarcation line by  \citet{kauffmann2003_AGN} in the [\oiii]$\,\lambda5007$/\hb\ vs. [\nii]$\,\lambda6583$/\hb\ diagram (left panel in  Fig.~\ref{BPT}) and the one by \citet{kewley01} in the [\oiii]$\,\lambda5007$/\hb\ vs. [\sii]$\,\lambda\lambda6717,6731$/\hb\ diagram (right), as well as circumnuclear \hii\ regions or {\em hotspots}  \citep[e.g.][]{KKB} in non-AGN galaxies (NGC~2903, NGC~2997, NGC~3351, NGC~4321, NGC~5236, NGC~5248, 52 regions in total) will not be considered for the determination of radial abundance gradients. We have opted however to calculate the corresponding strong-line abundances for these regions, in order to asses the behaviour of the different methods for these objects. 


We find a total of 89 regions that are either innermost \hii\ regions in AGNs, hotspots, or nebulae displaying signatures of shock excitation in the BPT diagrams. Five of them have auroral line measurements. We will further comment on this in Sect.~\ref{fits}.

\section{Physical properties and chemical abundances}
\label{abundances}
The determination of the electron temperature, $T_e$, is a necessary step to obtain accurate chemical abundances in \hii\ regions via the so-called $T_e$-based or $direct$ method. This is normally done by measuring a temperature sensitive emission line ratio (auroral-to-nebular), that requires the detection of the faint auroral lines, together with the corresponding nebular bright lines for the same ion (e.g [\oiii]$\,\lambda$4363/[\oiii]$\,\lambda\lambda$4959,5007 for the determination of $T_e$([\oiii])).

Our compilation includes auroral lines for a limited number of \hii\ regions (709 out of 2831 regions). Therefore, $T_e$ measurements and $T_e$-based abundances can  only be obtained for about 25\% of the regions (and reliably for just $\sim20$\%). For the rest of the \hii\ regions, only estimates based on strong emission lines can be obtained. For the regions with auroral line detection we have recalculated $T_e$-based abundances with an homogeneous methodology. This allows us to evaluate the different {\em strong-line abundances} in relation to the most accurate method for  \hii\ regions in nearby galaxies.
The methodology to obtain the electron temperature and density, and  $T_e$-based and {\em strong-line} chemical abundances is described in the following  subsections.
\subsection{Electron density and temperature}
\label{Te}
The physical conditions of the \hii\ regions have been derived from the corresponding collisionally  excited lines and the task {\tt temden} within the IRAF {\tt nebular} package \citep{nebular}, with updated values for the atomic parameters (collisional strengths and transition probabilities) as shown in Table~5 of \citet{Bresolin09}.

Our compilation of emission line fluxes includes the auroral lines [\sii]$\,\lambda\lambda$4068,4076,  [\oiii]$\,\lambda$4363, [\nii]$\,\lambda$5755, [\siii]$\,\lambda$6312 and [\oii]$\,\lambda\lambda$7320,7330 (referred to as [\oii]$\,\lambda\lambda$7325 in the following), when reported by the different authors. In total, the sample comprises 709 \hii\ regions with flux measurements for at least one of the auroral lines, of which we could derive the electron temperature (from at least one auroral line) for 639 \hii\ regions.

Density and temperature have been obtained simultaneously in an iterative process. The electron temperature ($T_e$) was initially set to 7500~K, and a first value for the electron density, $n_e$ was obtained from the [\sii]$\,\lambda$6717/[\sii]$\,\lambda$6731 line ratio. This initial value for $n_e$ is then used for a preliminary determination of the electron temperature and the ionic abundances (as described in the next section).  Using these initial estimates of the electron temperatures and ionic abundances, the electron density is recalculated, and  the reddening-corrected  fluxes of [\nii]$\,\lambda$5755 and [\oii]$\,\lambda$7325  are corrected for recombination contamination following \citet{liu2000}, where we have assumed N$^{2+}$/N$^+$  = O$^{2+}$/O$^+$. Regions with $T_e$ above the range of validity of \citet{liu2000} relations have not been corrected. Recombination corrections are small, and median values for the corrections are 3.0\% and 0.6\% for  [\oii]$\,\lambda$7325 and [\nii]$\,\lambda$5755, respectively. The ionic and total abundances were then re-calculated and the iterative process was then repeated until convergency was achieved.

The electron density has been determined for 1419 \hii\ regions. The vast majority (92\%) of them have electron densities below $\sim$200~cm$^{-3}$. For those regions with no reported [\sii]$\,\lambda\lambda$6717,6731 fluxes, or with  [\sii]$\,\lambda$6717/[\sii]$\,\lambda$6731 ratios above the low density limit we have adopted an electron density of 29.2~cm$^{-3}$ (the median value for the sample \hii\ regions). This particular choice of $n_e$ has virtually no effect on the subsequent temperature and abundance estimates, since we remain within the low-density regime.

All five values of the electron temperature $T_e$[\oiii], $T_e$[\oii], $T_e$[\nii], $T_e$[\sii] and $T_e$[\siii] have been measured in only 52 \hii\ regions, but in 377 \hii\ regions we derived at least two values of $T_e$, and at least one measurement of  $T_e$ was obtained for 639 \hii\ regions. In the calculation of the electron temperature we have discarded values with resulting uncertainties above 40\% and/or absolute errors larger than 4000~K. The absolute errors for  $T_e$ have been estimated from error propagation of the relevant line flux uncertainties. These are in the range $100-3800$~K for all determinations, with median values of $\sim$400-600~K for the five determinations of $T_e$.

Fig.~\ref{temp_rel} shows the relations obtained for electron temperatures for the different ionic species.  The derived values of $T_e$ support the relations predicted by photoionization models \citep{garnett92, stasinska82}. This is especially remarkable for the relation between $T_e$[\siii] and $T_e$[\oiii], and $T_e$[\siii]-$T_e$[\nii]. There is wide observational evidence supporting these theoretical relations \citep[e.g.][]{Perez-Montero2017,Croxall15, Bresolin09, Berg2020} but based on  smaller data samples.

\subsection{$T_e$-based ionic and total abundances}
\label{direct_abundances}
The N$^+$, O$^{+}$ and O$^{2+}$ ionic abundances have been calculated from the fluxes of collisionally  excited emission lines, assuming a two-zone scheme for the ionization structure of the \hii\ regions, with two zones of different $T_e$ where atomic species of similar ionization potential coexist. In view of the  relations for $T_e$ derived above for the different atomic species, we have adopted $T_e$([\nii]) as the temperature for the low-excitation region (N$^+$, O$^{+}$), and $T_e$([\oiii]) as the electron temperature for the highest ionization zone. The following criteria have been adopted when these temperatures were not available:
\begin{itemize}
\item When $T_e$[\nii] could not be  measured, the temperature for the low-excitation  zone has been obtained from the average value obtained from $T_e$[\oiii] and  $T_e$[\siii] and the inversion of the corresponding \citet{garnett92} relations, or from just one of them, depending on availability.
\item When $T_e$[\oiii] could not be measured, we have obtained the high-ionization zone temperature from either $T_e$[\siii] or $T_e$[\nii] and the inversion of the \citet{garnett92} relations, or from the average of the two resulting estimates of $T_e$[\oiii], when both $T_e$[\siii] and  $T_e$[\nii] were available.
\end{itemize}

Our measured values for $T_e$[\sii] and/or $T_e$[\oii] have only been employed when no other electron temperature was available. For these cases, either $T_e$[\oii] or the average of both $T_e$[\sii] and $T_e$[\oii] (when both were available) have been used as representative for the low-excitation zone, and the high-excitation zone temperature has been obtained from the inversion of Eq.~1 of \citet{garnett92}. When only $T_e$[\sii] was available, we have not reported ionic abundances, as this temperature yielded abnormal abundances in comparison with other \hii\ region abundances  in the same galaxy. This is the case for only 14 \hii\ regions: 3 in NGC~300, 2 from \citet{Pagel1979} and one region from \citet{Edmunds84}, and 11 in NGC~3184 from \citet{Berg2020}.

The final adopted values for the low and high ionization zones are available the corresponding electronic table available on CDS VizieR (via \url{https://vizier.u-strasbg.fr/viz-bin/VizieR)}. 

The task {\tt ionic} within the IRAF {\tt NEBULAR} package was employed for obtaining the ionic abundances of N$^+$, O$^{+}$ and O$^{2+}$. The task assumes a five-level atom approximation and yields the ionic abundance for the given zonal temperature, electron density and corresponding reddening-corrected emission line ratios ([\nii]$\,\lambda6584$/\hb, [\oii]$\,\lambda$3727/\hb\ and [\oiii]$\,\lambda5007$/\hb, for N$^+$, O$^{+}$ and O$^{2+}$ respectively). Absolute errors in the ionic abundances have been estimated from the propagation of the error in the adopted electron temperature. We have been very conservative with our calculations. In cases where the authors give auroral-line fluxes that are afterwards not used for temperature and/or abundance estimates, we have chosen not to use them, whether or not this is justified by the authors \citep[e.g.][]{Croxall15,croxall16,lin_m33}.

The total oxygen abundance is then obtained from the sum of O$^+$/H$^+$ and O$^{2+}$/H$^+$. We therefore assume that the amount of O$^{3+}$/H$^+$ is negligible, which might not be true for high excitation \hii\ regions. Photoionization  models predict a fraction of O$^{3+}$/O $\gtrsim1\%$ only in the highest excitation regions, those for which O$^{+}$/(O$^+$ + O$^{2+}$) $\lesssim 10\%$ \citep{Izotov06}. Only five \hii\ regions\footnote{M101\_N5451C, NGC598\_MA1, NGC598\_IC132, MW\_ NGC3603, MW\_Sh2-83} have low O$^{+}$/(O$^+$ + O$^{2+}$), with values in the range $7 - 9\%$. The high excitation line He~{\sc ii}~$\lambda4686$ is only detected in NGC598\_MA1 by \citet{Kehrig11}, and these  authors estimate O$^{3+}$ contributions that would increase $12 + \log$(O/H) by $\sim 0.06$~dex, that is smaller than our quoted uncertainty (0.19~dex, with an average of 0.12~dex for all regions). We can therefore safely assume O/H = O$^+$/H$^+$ +  O$^{2+}$/H$^+$ for all \hii\ regions.

For nitrogen we assume the usual relation N/O = N$^{+}$/O$^+$ \citep{peimbert69}, based on the similarity of the O and N ionization potential.
\begin{figure}
\hspace{-0.5cm}
\includegraphics[width=0.53\textwidth]{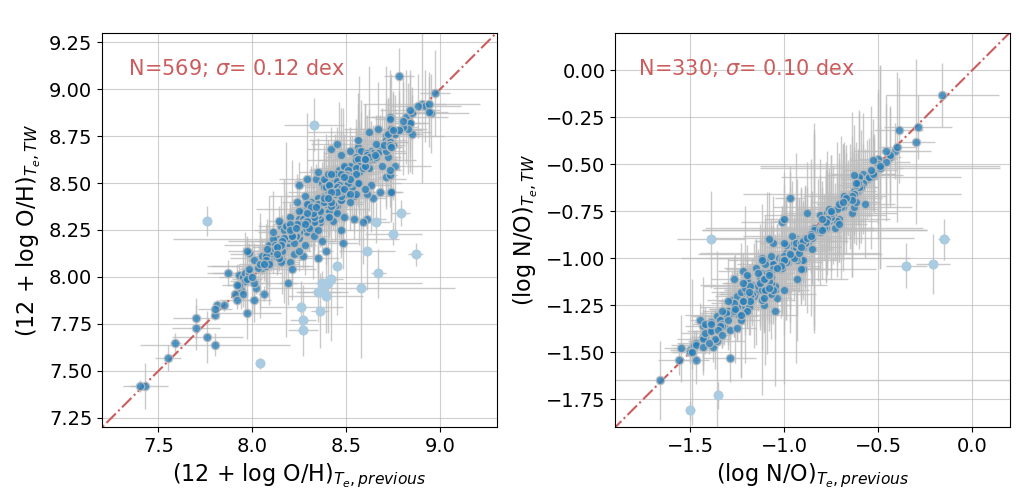}
\caption{Comparison between our derived  $T_e$-based O/H ({\em left}) and N/O ({\em right}) abundances, in the $y$-axis, with those obtained by other authors in previous studies ($x$-axis). The red dashed lines mark the 1:1 relation. Data points in light blue mark the regions with larger deviation from the 1:1 relation. \label{comparison_Te}}
\end{figure}

Fig.~\ref{comparison_Te} shows a comparison of our derived total oxygen ({\em left}) and nitrogen abundances ({\em right}) with those calculated by the original authors (Table~\ref{bibliografia_gal}). The agreement is good over the whole range of abundances, with a scatter of 0.12~dex and 0.09~dex rms for $12+\log$(O/H) and $\log$(N/O), respectively. Points deviating more than three times the scatter from the 1:1 relation have been marked in the plots with light blue symbols.  These constitute a small percentage of the total sample: 21 regions or 3.7\% of the total, and 8 regions or 4.7\% for O/H and N/O, respectively. These outliers  correspond to data collected from different authors, but $\sim60$\% comes from the work by \citet{Magrini10} and \citet{stanghellini10}, which suggests differences  in the methodology followed by us and by these authors  in the derivation of the total abundances.

\subsection{Strong-line abundance estimates}
\label{sl_abundances}
Direct or $T_e$-based abundances are only available for about 20\% of the \hii\ regions (610 out of 2831). We have then considered a number of different abundance determination methods based on strong-line flux ratios to estimate $12+\log$(O/H) and $\log$(N/O). As we are interested in the 
galactic radial profiles of the chemical abundances, a comparison is necessary, as there are well known offsets between the different methods, but more importantly, different methods can also yield different radial gradient slopes \citep[e.g.][]{Bresolin09,Moustakas_SINGS,arellano-cordova}. The methods used in this work have been selected among many others  because they are some of the most frequently used, and/or because they allow us to compare with previous studies:

\begin{itemize}
\item The HII-CHI-mistry method (hereinafter HCM, version 3.0) makes use of grids of photoionization models by \citet{epm14}  to calculate 12 + $\log$~(O/H), in addition to the N/O  abundance ratio. 

\item O3N2 = $\log$(([\oiii]$\,\lambda$5007/\hb)/([\nii]$\,\lambda6583$/\ha)) as calibrated by \citet{pp04}, and later by \citet{marino13}, hereinafter PP04 and M13, respectively. 

\item N2 = $\log$([\nii]$\,\lambda6583$/\ha) with the empirical calibration given by \citet{pp04}.

\item N2O2 = $\log$([\nii]$\,\lambda6583$/[\oii]$\,\lambda3727$) with the  empirical calibration of \citet{b07}, hereinafter B07.

\item R$_{23}$ = ([\oii]$\,\lambda3727$ + [\oiii]$\,\lambda\lambda$4959, 5007)/\hb, as calibrated from  theoretical model grids by \citet{m91}, hereinafter MG91, with the \citet{kkp99} parametrization (hereinafter KKP99).  The well-known disadvantage of this indicator is that it is double valued, i.e. the same value of  R$_{23}$ can yield two different values  for 12+$\log$(O/H), making it necessary to  use  secondary indicators to break the degeneracy. Initially we tried  the prescription of other authors to use the ratio [\nii]$\,\lambda6583$/[\oii]$\,\lambda3727$, by assigning the upper branch when this ratio is  larger than -1.2 \citep[e.g.][]{KewleyEllison2008}. However, this produces artificial discontinuities in some of the derived  radial abundance profiles (e.g. for NGC~2403, NGC~300, NGC~925, NGC~925). In addition, the [\nii]$\,\lambda6583$ emission-line flux was  not available for a considerable number of \hii\ regions. After a careful analysis of R$_{23}$ as a function of other oxygen abundance indicators,  and diagnostic emission-line ratios, we found it more reliable to use our derived HCM oxygen abundances for breaking the R$_{23}$ degeneracy: regions  with $12+\log$(O/H)$<7.9$ as estimated with HCM were assigned to the lower branch, and all the rest to the upper branch. For 2458 \hii\ regions with  R$_{23}$ measurements, 48 (2410) were assigned to the lower (upper) branch.

\item The $R$ calibration by \citet{PilyuginGrebel2016}, hereinafter PG16, that uses three bright emission-line ratios (R2=[\oii]$\lambda3727$/\hb, R3=[\oiii]$\,\lambda \lambda4959,5007$/\hb, and N2=[\nii]$\,\lambda\lambda6548,6583$/\hb) to determine $12+\log$(O/H) and $\log$(N/H).

\end{itemize}

For the N/O abundance ratio, in addition to HCM and the $R$ calibration, we have also used the empirical calibrations of N2O2 and N2S2 (=$\log$([\nii]$\,\lambda6583$/[\sii]$\,\lambda\lambda6717,6731$)) provided by \citet{pmc09}.

\subsection{Comparison between  T$_{\mbox{\lowercase{e}}}$- and strong-line-based  abundances}
\label{comparison}
We have employed the O/H and the N/O abundance ratios obtained from our direct estimate of $T_e$ (Sect.~\ref{direct_abundances}) to check the reliability of the strong-line methods used in this work (described in the previous section). There are several detailed evaluations of these methods in the literature,  that either compare $T_e$-based abundances with empirical calibrations \citep[e.g.][]{bresolin2004,Bresolin09,Perez-Montero2005,Yin2007,LopezSanchez-Esteban10,PilyuginGrebel2016} or different strong-line methods among themselves \citep[e.g.][]{Ho2015,RupkeKewleyChien2010,KewleyEllison2008}.

Our purpose here is to test the performance of the methods employed in this work in relation with the {\em direct} method for our specific sample of galactic and extragalactic \hii\ regions. The aim is to study possible systematic trends that could affect the radial abundance gradients, and that may be helpful at explaining our and previous (discrepant) results.
\begin{figure*}
\hspace{-0.6cm}\includegraphics[width=0.51\textwidth]{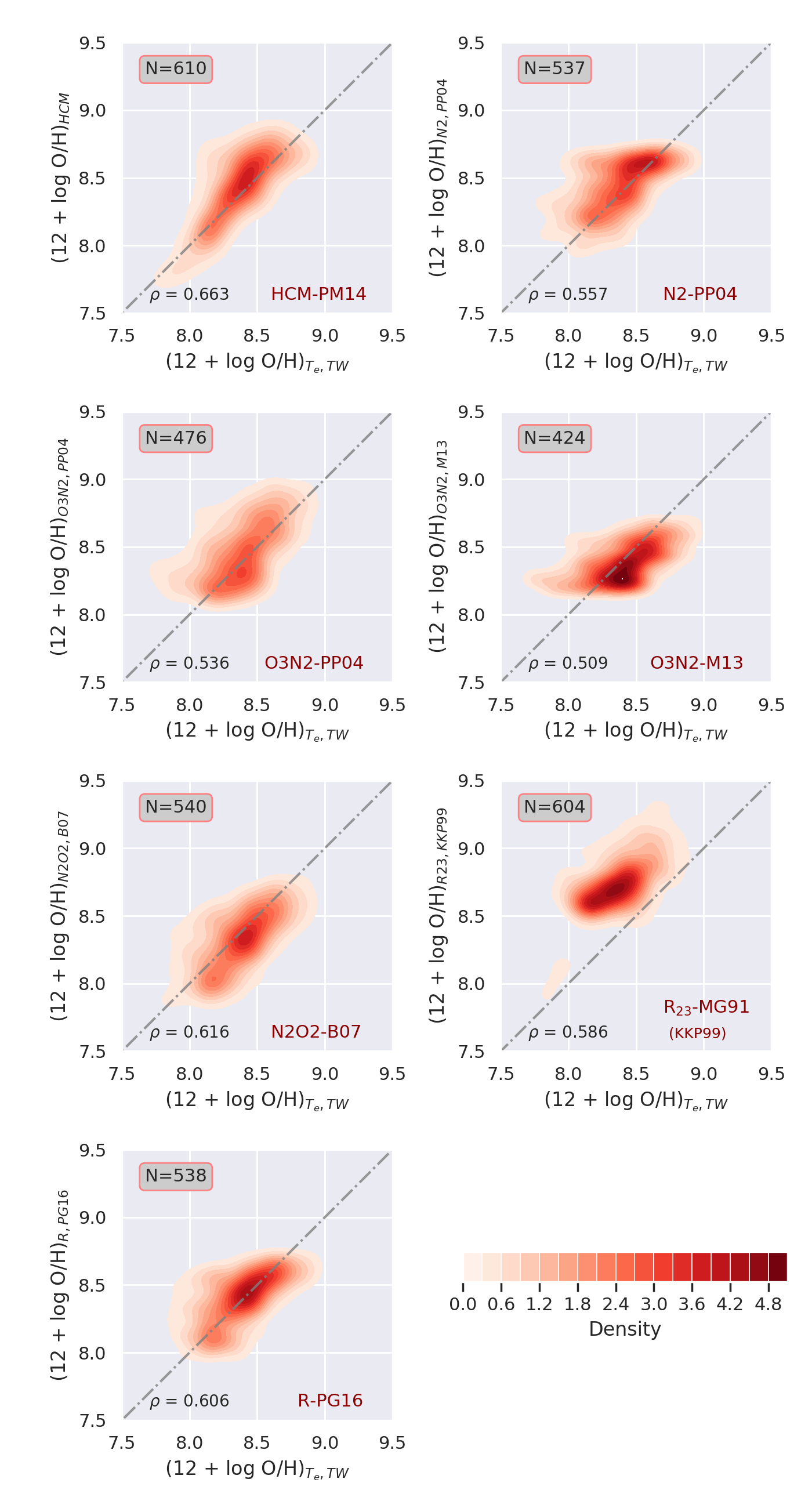}
\hspace{-0.1cm}\includegraphics[width=0.51\textwidth]{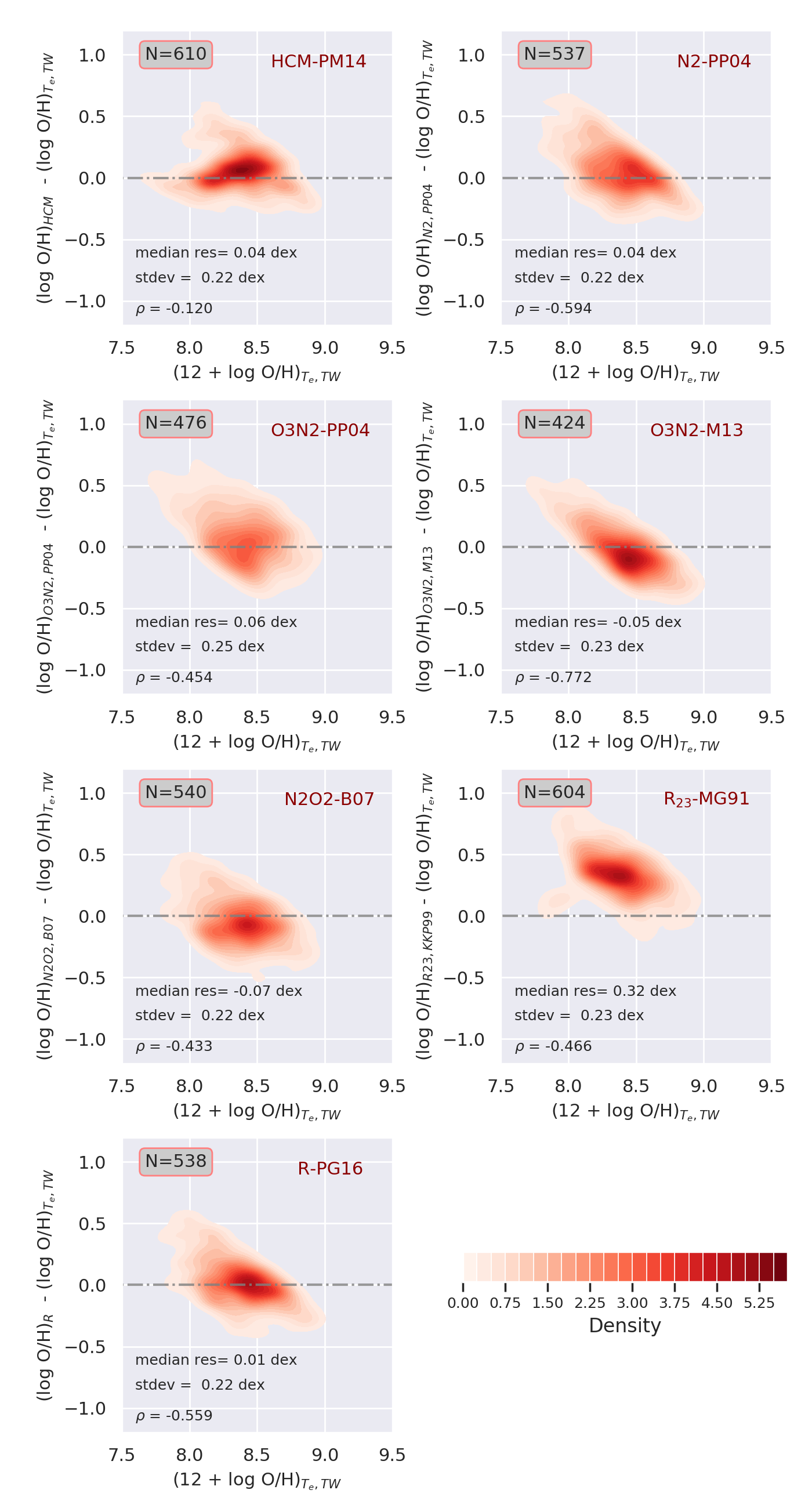}
\caption{{\em Left:} 2-D density plots showing a comparison of the total oxygen abundances obtained from the direct $T_e$-based method ($x$-axis) and those obtained from different strong-line methods ($y$-axis), as indicated in each panel (from left to right and from top to bottom: HCM, N2, O3N2-PP04, O3N2-M13, N2O2, R$_{23}$-MG91 and R-PG16). The dotted-dashed line marks the 1:1 relation. The Spearman's Rank correlation coefficient, $\rho$, is shown in all panels. The number of \hii\ region data used for each panel is indicated in the top-left corner. {\em Right:} Difference between the O/H abundances obtained from strong-lines and the corresponding $T_e$-based value ({\em residuals}) as a function of the $T_e$-based O/H abundances. The median value and the standard deviation (rms) for the residuals are shown in the lower-left corner of each panel, together with the  Spearman's Rank correlation coefficient.
\label{residuals_vs_O_abund}}
\end{figure*}

\begin{figure*}
\hspace{-0.8cm}\includegraphics[width=0.51\textwidth]{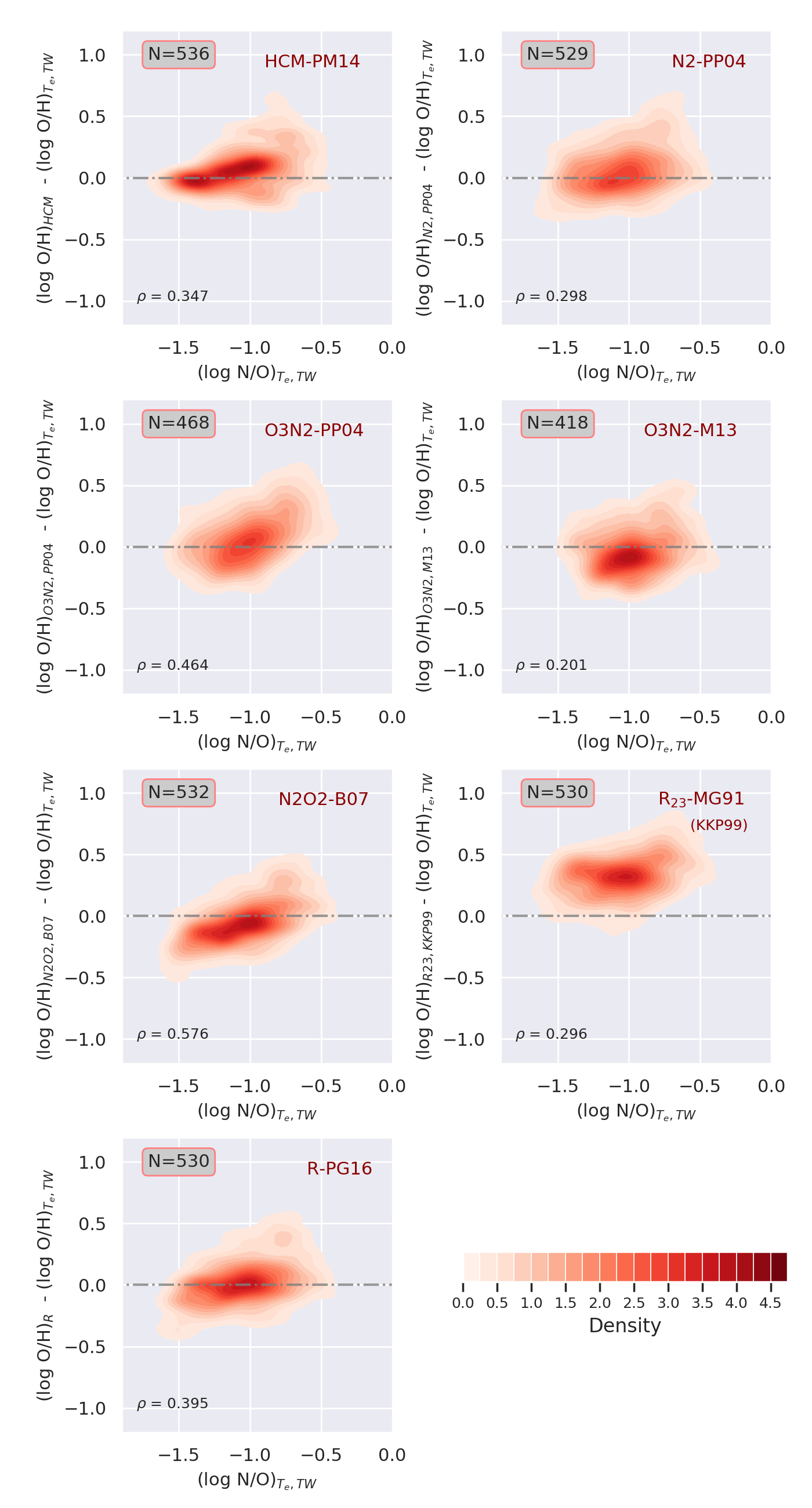}
\hspace{-0.1cm}\includegraphics[width=0.51\textwidth]{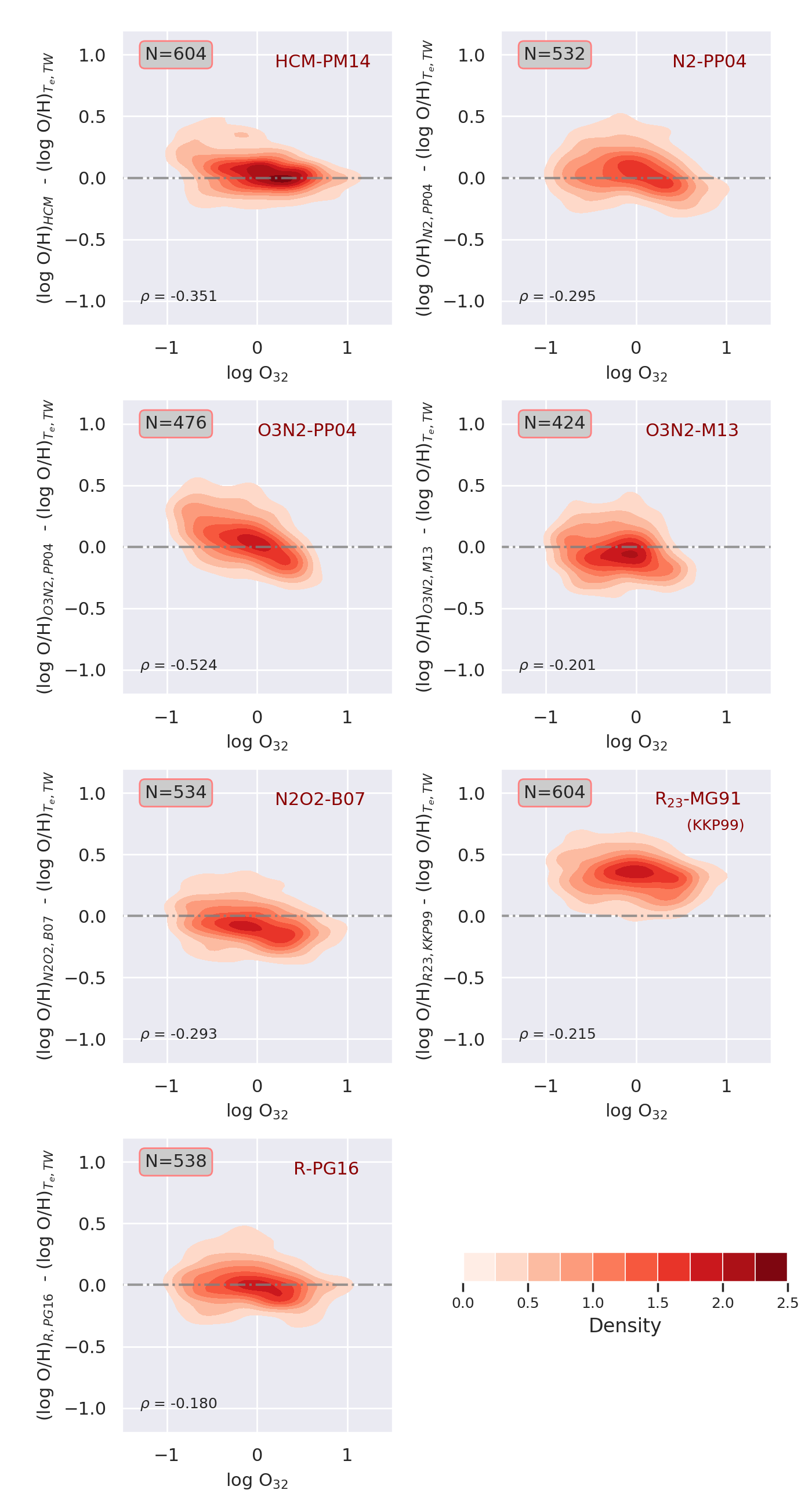}
\caption{Same as Fig.~\ref{residuals_vs_O_abund} but for the residuals against the $T_e$-based $\log$(N/O) abundances derived in this work ({\em left}) and
 $\log$O$_{32}$ ({\em right}) for the strong-line methods indicated in the top-right corner in each panel.\label{residuals_vs_NO_O32}}
\end{figure*}

The left-hand panel of Fig.~\ref{residuals_vs_O_abund} shows a comparison between the different strong-line metallicities and the  $T_e$-based ones. We have opted to show the comparison in the form of  2-D density plots as,  given the high overlapping in data points, scatter plots are more difficult to analyse. The same colour scale has been used for the six panels, where darker areas correspond to a higher concentration of data points. 
The number of regions used for each comparison is shown in the upper-left corner of each panel. This number is generally smaller than the total number of regions with $T_e$-based abundances (610) because for some of these regions either the relevant line ratios are not available or these are outside the validity range of the strong-line calibrations. The latter is the reason why the comparison with $T_e$-based  abundances for the two calibrations of the O3N2 parameter involves a different number of regions.
A first look at the plots shows the expected behaviour: a  positive correlation between strong-line abundances and $T_e$-based ones (Spearman's  coefficient, $\rho >0.5$ with all methods), and the well-known overestimation of the oxygen abundances obtained from R$_{23}$ (bottom-right panel) by $\sim$0.3~dex \citep[e.g.][]{Bresolin09,KewleyEllison2008}, as we have used a  calibration from photoionization models  \citep{m91}. The median value of the difference in $12+\log$(O/H) between the strong-line methods and the $T_e$-based abundances, that we will term {\em residuals}, are in all cases (except for $R_{23}$) smaller than the dispersion (rms deviation from equality), with the latter being pretty similar, $\sim$$0.22-0.25$~dex, in all cases.
However, important differences between  methods arise when we analyse the residuals as a function of 12+$\log$(O/H)$_{Te}$, $\log$(N/O)$_{Te}$, and the O32 (=[\oiii]$\,\lambda\lambda$5007,4959/[\oii]$\,\lambda$3727), the latter being  a proxy for the ionization parameter (right-hand panels of Figs.~\ref{residuals_vs_O_abund} and Fig.~\ref{residuals_vs_NO_O32}). In this respect,  the strong negative correlation ($\rho =-0.77$) of the residuals of the O3N2 method as calibrated by M13 with the $T_e$-based oxygen abundances is remarkable. This implies that oxygen abundances in the higher metallicity range are underestimated, with respect to $T_e$-based abundances by up to $\sim$$0.3-0.4$~dex, while the opposite occurs for the lower metallicity regions. That translates in a much lower range in  12+$\log$(O/H) for \hii\ region oxygen abundances obtained with this method, with respect to the range covered by their corresponding $T_e$-based abundances. This was already visible in the corresponding left-hand panel in Fig.~\ref{residuals_vs_O_abund}, where we can see that O3N2 abundances are all in the range 8.2-8.7~dex, while the $T_e$-based ones cover a wider range 7.7-8.9. We should recall  here that we have used this calibration only in the range of validity specified by the authors \citep{marino13}. For the N2 calibration by PP04, and the $R$ calibration by PG16 the residuals also show a moderate correlation ($\rho =-0.59$ and $-0.56$, respectively) with the $T_e$-based oxygen abundance. For N2O2, R$_{23}$ and O3N2 (as calibrated by PP04) only a weak dependence with the oxygen abundance is observed.

The left-hand panel of Fig.~\ref{residuals_vs_NO_O32} shows the same residuals in the oxygen abundances for all the strong-line methods but now as a function of the $T_e$-based N/O abundance. Here no method shows a significant dependence with N/O, except N2O2 for which we estimate a Spearman's coefficient of $\rho=0.58$. This behaviour is not unexpected and has been previously reported \citep{pmc09}, as N2O2 is also a good tracer of N/O. However, the observed dependence does not produce an overestimation of 12+$\log$(O/H) for regions of high N/O, but an  underestimation for regions with $\log$(N/O) below $\sim-0.9$, with respect to the ones obtained with the  $T_e$-based method. However, the slope in this plot is not very high and in
practice, this would imply  an underestimation of up to $\sim 0.2-0.3$ dex for the lowest N/O \hii\ regions ($\log$(N/O)$\lesssim-1.5$) that is close to the typical uncertainties in strong-line methods, with standard deviations of $\sim0.25$~dex. Surprisingly, no dependence is observed in the residuals of the N2 method with  $\log$(N/O). Finally, the residuals in $12+\log$(O/H) are plotted as a function of O32 in the right-hand panel of Fig.~\ref{residuals_vs_NO_O32}. Very weak dependences are observed for all methods except for O3N2 as calibrated by PP04  $\rho=-0.52$, a dependence already reported by \citet{Ho2015}.

In summary, overall, all the strong-line methods tested here globally reproduce $T_e$-based oxygen abundances in our sample of \hii\ regions (with the exception of R$_{23}$), but the residuals in  12+$\log$(O/H) correlate to some degree with $\log$(N/O) and/or  $\log$(O32). These correlations imply over- or under-estimations in  12+$\log$(O/H) depending on  the method. The correlation with O32 underlines the importance of the ionization parameter and its effects on strong line chemical abundances. The residuals are within typical uncertainties inherent to the methods, except for O3N2 as calibrated by M13, but can produce systematic effects in radial abundance gradients. As these methods are widely used, we will compute the radial metallicity  profiles for all the different strong-line methods mentioned in Sect.~\ref{sl_abundances}, in order to allow for comparisons with previous  results and to evaluate the influence of the methodology on  the O/H abundance we derive.

\begin{figure*}
  \includegraphics[width=\textwidth]{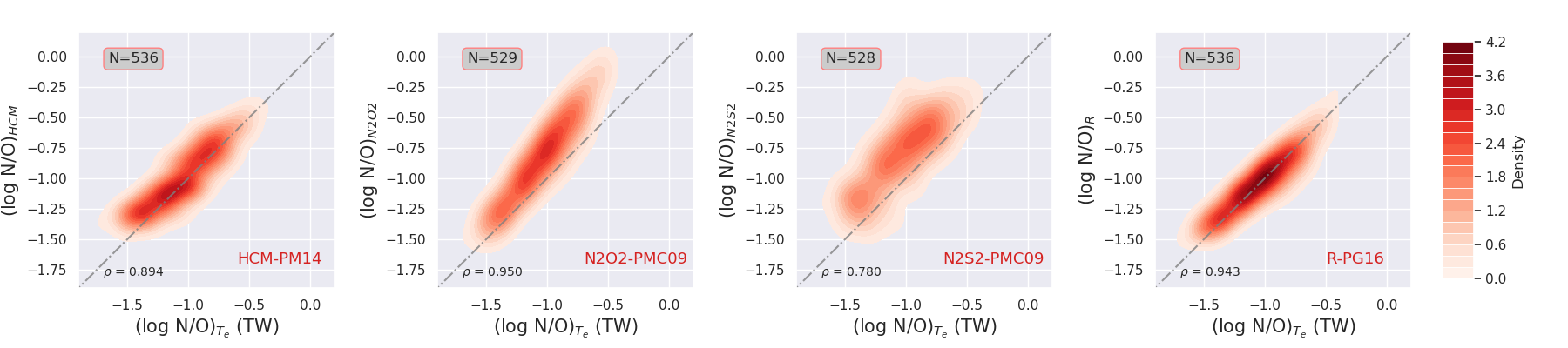}
  \includegraphics[width=\textwidth]{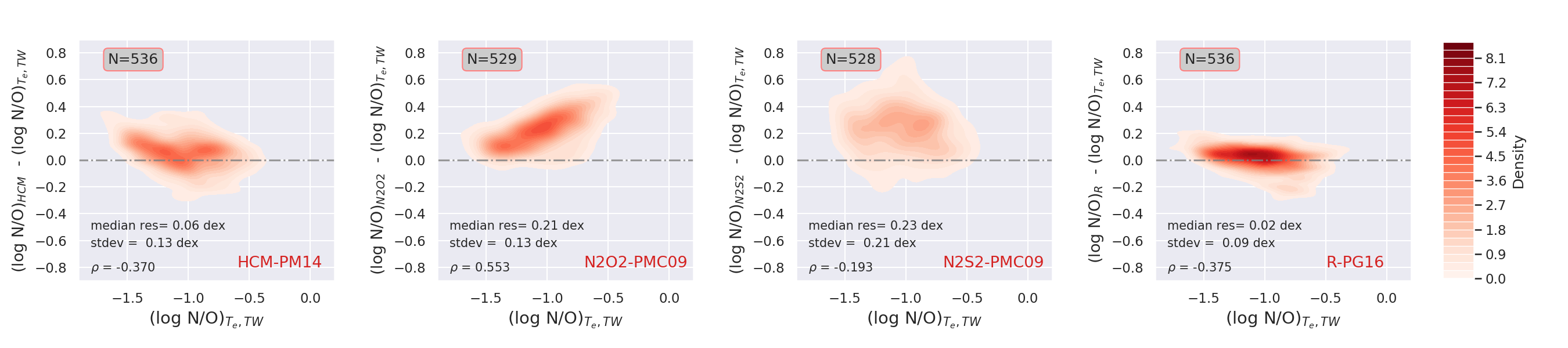}
  \includegraphics[width=\textwidth]{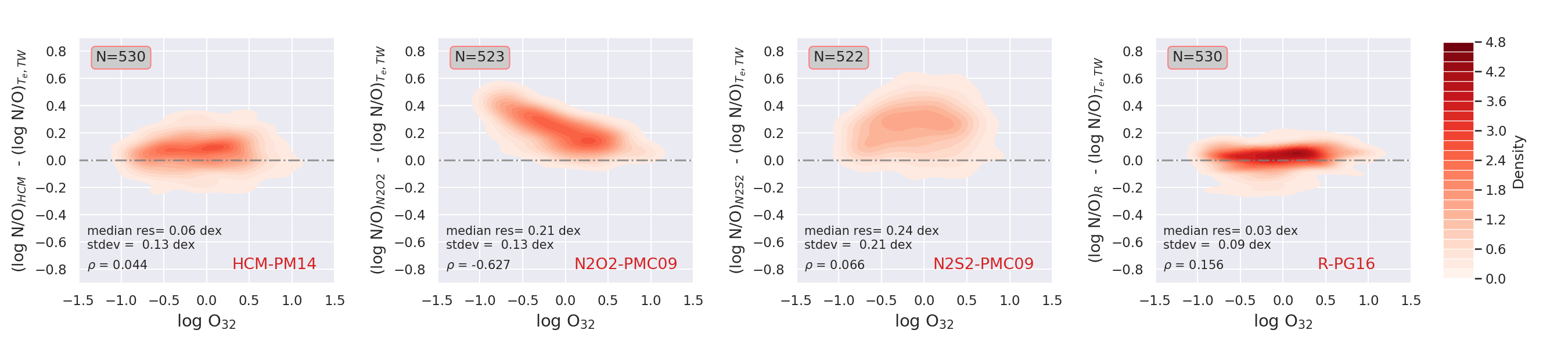}
  \caption{{\em Top:} Comparison of the $T_e$-based N/O abundances derived in this paper with those obtained from HCM, N2O2, N2S2 and the $R$ calibration. {\em Centre:} Differences between N/O strong-line and $T_e$-based N/O  abundances (from left to right: HCM, N2O2, N2S2, $R$) as a function of the $T_e$-based N/O abundance. {\em Bottom:}  Same residuals in  N/O abundance as a function of $\log$(O32) for the three methods. The median value and standard deviation (rms) for the residuals is shown in the lower-left corner of the central and bottom panels, together with the  Spearman's  Rank correlation coefficient.
    \label{comparison_NO_direct}}
\end{figure*}

The top panels of Fig.~\ref{comparison_NO_direct} show a comparison of the  $\log$(N/O) obtained with HCM, N2O2, N2S2 and the $R$ calibration with the values obtained from the $T_e$-based method for our sample of \hii\ regions. The corresponding residuals are plotted in the mid and lower panels as a function of $\log$(N/O) and $\log$(O32), respectively. The dispersion of the residuals is lower for the $R$ calibration, HCM and N2O2 ($\sim$0.09-0.13~dex, cf. 0.21~dex for N2S2), but the N/O abundance ratio obtained from N2O2 is systematically larger than the $T_e$-based N/O. In fact, the residuals of N2O2 show a positive correlation with $\log$(N/O)$_{T_e}$, with $\rho$ = 0.55, implying average overestimations in $\log$(N/O) as large as $\sim$0.5~dex for the \hii\ regions with the largest N/O abundance ratio ($\log$(N/O) $\sim -0.5$). The N2O2 method also yields residuals in $\log$(N/O)  that anti-correlate with $\log$(O32). The residuals obtained with HCM, N2S2 and the $R$ calibration do not show important correlations with $\log$(N/O) or $\log$(O32), with $\vert\rho\vert<0.4$, but the $R$ calibration and HCM are those that better match the $T_e$-based method, with low median residuals (<0.1~dex versus 0.23~dex for N2S2) as well as the low dispersion around the 1:1 relation already mentioned. Therefore, the N/O abundance ratios obtained with HCM  and with the $R$ calibration are those that better follow the $T_e$ scale for our \hii\ region sample.
In what follows we will then focus on the  $\log$(N/O) estimates from HCM and the $R$ calibration, although we will derive radial profile fits for all the methods. 
Previous works on radial abundance gradients in barred galaxies has been mostly concentrated on O/H, therefore there is not much work to compare with, with the exception of \citet{Perez-Montero2016}, that uses HCM.

\section{Radial abundance profiles}
\label{fits}
In order to derive the O/H and N/O radial abundance profiles for our sample of galaxies, we have used our recalculated deprojected galactocentric distances (Sect.~\ref{gal_distances}) and our derived oxygen and nitrogen abundances for all the \hii\ regions in the compilation, as described in Sects.~\ref{direct_abundances} and \ref{sl_abundances}.

The profiles are shown in Figs.~\ref{Afig1}-\ref{Afig54} in Appendix~\ref{perfiles} (available online) for all galaxies and for a few selected methods, where we also show an image for each galaxy and the location of the \hii\ regions. The figures contain different panels that show the $\log$~(N/O) radial profile (from HCM) and the metallicity  profiles obtained for five different strong-line methods: HCM, O3N2 (PP04), N2 (PP04), N2O2 (B07) and R$_{23}$ \citep[MG91, as parametrized by][]{kkp99}, as described in Sect.~\ref{abundances}.  Different colours and symbols have been employed for data from different authors as shown in the legend (upper-left corner of the galaxy image). \hii\  regions located within galactic bars have been marked with a dark grey edge. Black small open symbols represent $T_e$-based abundance estimates. Vertical lines in the radial profiles mark the location of $r_e$ (grey dashed line), r$_{25}$ (grey dotted line) and the  deprojected bar radius  (dark blue dashed-dotted line), for barred galaxies. Given the inhomogeneous nature of the data, we do not have the same emission lines for all regions, and therefore in some cases we could not obtain the profiles with all the methods due to the lack of the relevant line fluxes.
\begin{figure*}
\includegraphics[width=\textwidth, clip, trim = 0cm 1.8cm 0cm 0.0cm]{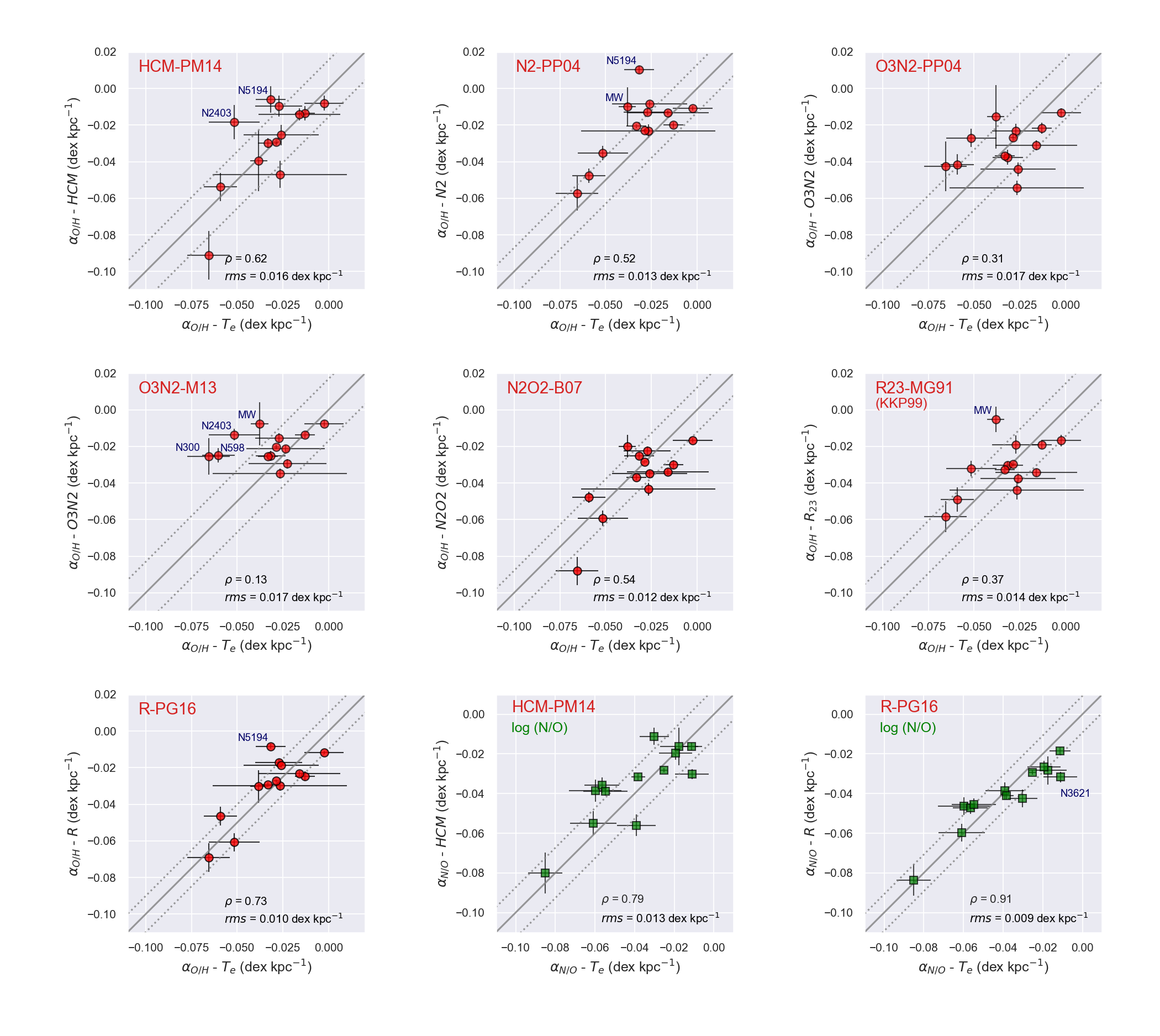}
\caption{Comparison of slopes obtained from different strong-line methods as a function of those derived from $T_e$-based abundances for the sub-sample of galaxies with reliable $T_e$-based abundance profiles. The first seven panels show the comparison for slopes of the oxygen abundance radial profile ($\alpha_{O/H}$, red circles). The slopes of the $\log$(N/O) radial profile ($\alpha_{N/O}$, green squares) are shown in the two bottom right panels. The solid grey line marks the 1:1 relation. The dotted lines show the rms dispersion about equality. The rms is also shown in the lower-right corner of each panel together with the Spearman's  correlation coefficient ($\rho$).  The galaxies with larger deviations from the 1:1 line are labelled.
\label{comparison_slopes_O_N}}
\end{figure*}

A first look at the profiles shows, as expected, a wide range of abundance profile slopes \citep[e.g.][]{Moustakas_SINGS,pilyugin2014} and radial breaks in the outer parts of some of them \citep[e.g.][]{b12,zahid,goddard11,croxall16}. In about 75\% of the galaxies our derived profiles cover the galactic disc beyond approximately r$_{25}$, while in 6 galaxies (12\% of the sample) the collected data cover only up to $\sim$0.6-0.8 times r$_{25}$. In order to characterize the abundance radial profiles we performed a least-squares linear fit to 12+$\log$(O/H) and $\log$(N/O) as a function of galactocentric radius ($R$), as parametrized by 12+$\log$(O/H)$ = b_{O/H} + \alpha_{O/H}~R$ and $\log$(N/O)$ = b_{N/O} + \alpha_{N/O}~R$, respectively. We performed an unweighted fit to the data, as flux error estimates are highly heterogeneous among the different authors and were not always provided. All available \hii\ regions in a galaxy were used for the linear fitting, except those described in Sect.~\ref{AGN} and regions with peculiar metallicities already identified in previous studies, as is the case of regions located in the {\em bridge} structure in NGC~1512 \citep{b12}. The regions excluded from the fit are marked in Figs.~\ref{Afig1} to \ref{Afig54} with a blue edge. In two  galaxies (NGC~2403 and NGC~2903) we excluded data from \citet{smith75} that systematically showed a larger deviation from the bulk of the data. A minimum of five data points per profile were required for the fitting \citep{Zaritsky94}. The slopes of the profiles normalized to $r_e$ and r$_{25}$ were also calculated.   

When the profiles showed evidence for a radial abundance gradient change or {\em break} after visual inspection, and/or the break has been reported by other authors \citep{bresolin09b,b12,zahid,goddard11,berg13,croxall16, Bresolin2019}, we have performed both a single and a double linear fit to the radial profile. The single linear fit to the data points is shown with a blue solid straight line in Figs.~\ref{Afig1}-\ref{Afig54}. The corresponding slope and $\chi_\nu^2$ values are shown, in the same colour, in the upper side for each panel. Double linear fits are shown with a dotted red line. The associated $\chi_\nu^2$ and the improvement with respect to the single linear fit (in parenthesis) are also shown in red. When this improvement is at least  10\%, the inner slope value is also given. This criterion in $\chi_\nu^2$ is somewhat arbitrary, but  it reflects numerically  the visual detection of breaks. In most cases the improvement in  $\chi_\nu^2$ is very significant ($>25\%$).

When  $T_e$-based abundances could be derived for a number of regions in a galaxy, we also derived the best linear fitting, and it is represented with a black dotted straight line in the profiles, with the slope and corresponding uncertainty also shown in black. Unfortunately, radial abundance profiles based on the $T_e$-based abundances could be reliably derived for only 13 galaxies. Although we can not base our study on these galaxies alone, this sub-sample offers the possibility of testing metallicity gradients derived from strong-line indicators (Sect.~\ref{comparison_slopes}). For four of the galaxies with $T_e$-based abundance profiles (NGC~628, NGC~3621, NGC~5154 and NGC~5457) we could attempt  a double linear fit, that results in a significant ($\chi_\nu^2$ improvement $>10\%$) break in the $\log$(N/O) radial profile. The fit is shown with dashed green lines and the corresponding inner slope is shown with the same colour in the corresponding plots.

Electronic tables with the radial abundance fit coefficients for all the methods, together with the corresponding correlation parameter,  $\chi_\nu^2$, and scatter in the fits are made publicly available on the CDS VizieR facility for all galaxies.

\subsection{Radial gas abundance  {\em breaks}}
\label{breaks}
The presence of a radial {\em break} or a change in the slope of the radial metallicity profile is a well known feature of galaxies having an extended disc of star formation \citep[see][for a review]{bresolin17}. For seven galaxies of our sample (NGC~1058, NGC~1512,  NGC~3359, NGC~3621, NGC~4625, NGC~5236, NGC~5457) radial breaks in the 12+$\log$(O/H) profile, at a galactocentric position close to the isophotal radius $r_{25}$, have been reported by previous work (see Table~\ref{bibliografia_gal} for the references). We confirm with our re-analysis of the compiled data the presence of such breaks in $12+\log$(O/H) via an improvement in $\chi_\nu^2$ in the double linear fit with respect to the single linear fit to the radial profile (see Section~\ref{fits} and Figs.~\ref{Afig1}-\ref{Afig54}). However, the break is not clearly detected with all the strong-line methods used in this work. In our sample, if a break is detected, it will probably be seen at least in the profiles derived from  N2O2 and R$_{23}$. NGC~3031 appears to show a break (in R$_{23}$ an possibly in the N2O2 profiles) at a radius of $\sim0.7\times$r$_{25}$, but to our knowledge, this has not been previously reported. In general, the break radius varies between 0.5 (for NGC~3359) and $1.4\times$r$_{25}$ (for NGC~5236) with an average of ($0.9\pm0.3$)$\times$r$_{25}$. The break radius changes among calibrations, but the maximum difference is $\sim25\%$ in a given galaxy.

The profile break was first detected in the $12+\log$(O/H) radial profiles of spirals, but a break in the $\log$(N/O) radial profile was also detected in  a few galaxies \citep[][]{bresolin09b,berg13,lopez-sanchez15,croxall16,Berg2020}.
In the case of  NGC~628 and NGC~5457 (M101), a break was detected in the radial $\log$(N/O) profile that was not seen in $12+\log$(O/H). All the galaxies in this sample presenting  radial breaks in the $12+\log$(O/H) profile, with one or several methods,  also show a break in the $\log$(N/O) profile with either HCM, or the $R$ calibration or with both of them, at approximately the same galactocentric radius. 
It is also interesting to note that for the four galaxies in which a radial break is detected in the $\log$(N/O) profile derived from the   T$_e$-based  abundances (NGC~628, NGC~3621, NGC~5194 and NGC~5457), such break is not detected in the T$_e$-based $12+\log$(O/H) profile. Given the lower dispersion in the  $\log$(N/O) profiles, the break seems to be more prominent and easily detected than in $12+\log$(O/H). However, a deeper analysis would be necessary.

\subsection{Comparison between $T_e$- and  strong-line-based slopes}
\label{comparison_slopes}
In Fig.~\ref{comparison_slopes_O_N} we compare $T_e$-based radial abundance gradient slopes with those  obtained from strong-line abundances for O/H (first seven panels, with red circles) and N/O (last two panels, with green squares). We have restricted the comparison to the better determined  $T_e$-based slopes, and therefore we consider only  galaxies with at least 10 \hii\ regions having  $T_e$-based abundances well distributed across their  discs, 13 galaxies in total\footnote{The T$_e$-based slopes for NGC~1512, NGC~2997 and NGC~4258, were not taken into account for this comparison because they were determined by a smaller number of data points and/or the data points are not well distributed across the disk and show a large dispersion.}. As the radial breaks are not always detected with all methods, we are comparing the slopes resulting from the single linear fit to all the radial abundance profiles.

Each panel in Fig.~\ref{comparison_slopes_O_N} shows the metallicity gradient as derived with a strong-line method, as a function of the slope derived with the  $T_e$-based method (in dex~kpc\me). The solid 1:1 line shows where the points would lie if the strong-line  and the $T_e$ methods  gave the same slope. The dotted lines mark the rms deviations from the 1:1 relation  indicated in the lower-right corner in each panel. The lowest rms values for the O/H slopes, $\alpha_{O/H}$, are found for the $R$ calibration, N2O2 and N2  (0.009-0.013 dex~kpc\me), while the highest rms is found for O3N2  (0.017~dex~kpc\me). We have labelled in each panel the galaxies with deviations from equality larger than the rms, considering error bars. NGC~5194 has a large deviation with HCM, N2 and R. In fact,  all galaxies exhibit a negative oxygen abundance gradient with all methods, with the exception of this galaxy, for which a positive gradient is obtained with  N2. This might be produced by the difficulty of strong-line methods in giving good estimates of metallicity for low-excitation \hii\ regions \citep{bresolin2004}. This is  evident in Fig.~\ref{n5194} where we can see that most of the strong-line methods yield low abundances for the innermost regions. 
The latter  are excluded from the fits (see Sect.~\ref{AGN}), and this does  not alter the metallicity gradient we derive except for the profiles obtained with N2 and HCM, for which  we found a systematic trend of decreasing oxygen abundances with decreasing distance to the galaxy centre from galactocentric distances as large as $\sim$4~kpc. The N2, HCM and R$_{23}$ methods are not able to estimate well the oxygen abundance of the \hii\ regions with low ionization degree in the Milky Way either \citep{Esteban18}, but in this particular case the slope remains negative, although the gradient is considerably shallower with N2 and R$_{23}$.

Figure~\ref{comparison_slopes_O_N} also shows a tendency for methods based on the O3N2 indicator to yield shallower oxygen abundance gradients as compared with the direct method. This is probably explained by the fact already discussed in Sect.~\ref{comparison} that oxygen abundances obtained from O3N2, notably with the M13 calibration, are restricted to a limited range in $12+\log$(O/H), smaller than the corresponding values obtained from the $T_e$-based method for the same regions. This effect, translated to the \hii\ region abundances in a given galaxy, may produce a smaller variation in O3N2 oxygen abundances across the galactic disc in the galaxy, that would imply a smaller derived radial abundance gradient. The importance of this method-induced flattening would of course depend on the average metallicity of the galaxy, but it would be expected to affect more those galaxies with larger gradients or with a larger range of {\em real} oxygen abundances across their discs. This is exactly what we see in the third and fourth panels in Fig.~\ref{comparison_slopes_O_N}, especially for the M13 calibration. Shallower slopes in radial abundance gradients derived using O3N2 were also found by \citet{BresolinKennicutt2015} in a sample of low surface brightness galaxies, and also by \citet{error-ferrer2019} for a sample of 38 spirals observed with MUSE.

Overall, the metallicity gradients derived with HCM,  N2, N2O2 and the R calibration show a moderate to strong correlation with the slopes derived with the $T_e$-based method. The Spearman's rank correlation coefficients for these methods are larger than $\sim$0.5, with the strongest correlation for R and HCM, with $\rho=0.73$ and $0.62$, respectively, for this sample of galaxies. The slopes derived with the O3N2 indicator as calibrated by M13, are the ones that more weakly correlate with the $T_e$-based ones ($\rho=0.13$). We are aware that the above comparison  must be interpreted with caution. First, it is based on a small data sample. Second, the number of data measurements that define the  $T_e$-based slope is typically lower (by even a factor of 10) than the number of measurements defining the strong-line method slope. This can give way to biases.  In spite of this, it is worth doing the comparison. The number of works in which strong-line methods are used to obtain radial abundance profiles in large samples of galaxies is rapidly increasing. 
Currently, this is probably one of the most complete data sets for comparing different metallicity estimates, and scrutinizing strong-line methods against the direct method  provides  a useful and  healthy check on their reliability until better data sets become  available.

The two bottom-right panels of Figure~\ref{comparison_slopes_O_N} show that the HCM and $R$ calibration slopes for the $\log$(N/O) radial profiles are in good agreement with the $T_e$-based gradients. The dispersion is similar to the one obtained from the $12+\log$(O/H), 0.013 and 0.009~dex~kpc\me, and the  Spearman coefficients, ($\rho=0.79$ and 0.91) indicate a strong correlation between  the two slopes for the two methods.


Therefore, based on both the comparison of the strong-line abundances with the $T_e$-based ones for individual \hii\ regions (Sect.~\ref{comparison}), and  the comparison of the radial abundance gradients, the $R$ calibration and the HCM methods are those that better match $T_e$-based $12+\log$(O/H) and $\log$(N/O) abundances in our sample. However, as we ultimately aim to understand the causes behind the contradicting results on the effect of bars on the metallicity gradients, and these have been derived from a variety of strong-line methods, we will continue using and analysing O/H and N/O results obtained from several methods. These are presented in the following sections.
\subsection{A characteristic O/H and N/O abundance gradient?}
Recent work on the gas-phase radial metallicity gradients in spirals has considerably enlarged the galaxy sample sizes, yielding important results, such  as the confirmation of an earlier proposal for the existence of a universal metallicity gradient in spirals when slopes are normalized to galaxy size \citep[e.g.][]{Zaritsky94,vila-costas92}. The result comes from the analysis of IFU spectra of different data samples: CALIFA in \citet{sanchez14,Sanchez-Menguiano2016} and MUSE-VLT in \citet[][]{Sanchez-Menguiano2018}, with the same metallicity indicator, O3N2, and  normalization of the gradients (disc effective radius). The latter work found $-0.10\pm0.03$~dex~r$_e^{-1}$. \citet{Ho2015} analysed  49 local field galaxies with the O3N2 and N2O2 calibrations, and also report the existence of a common slope  when metallicity gradients are normalized by the disc isophotal radius: $-0.39\pm0.18$~dex r$_{25}^{-1}$. Table~\ref{tab:universal_gradient} shows median or average slopes obtained by different authors for different galaxy samples and metallicity scales.

\begin{figure*}
\includegraphics[width=0.80\textwidth]{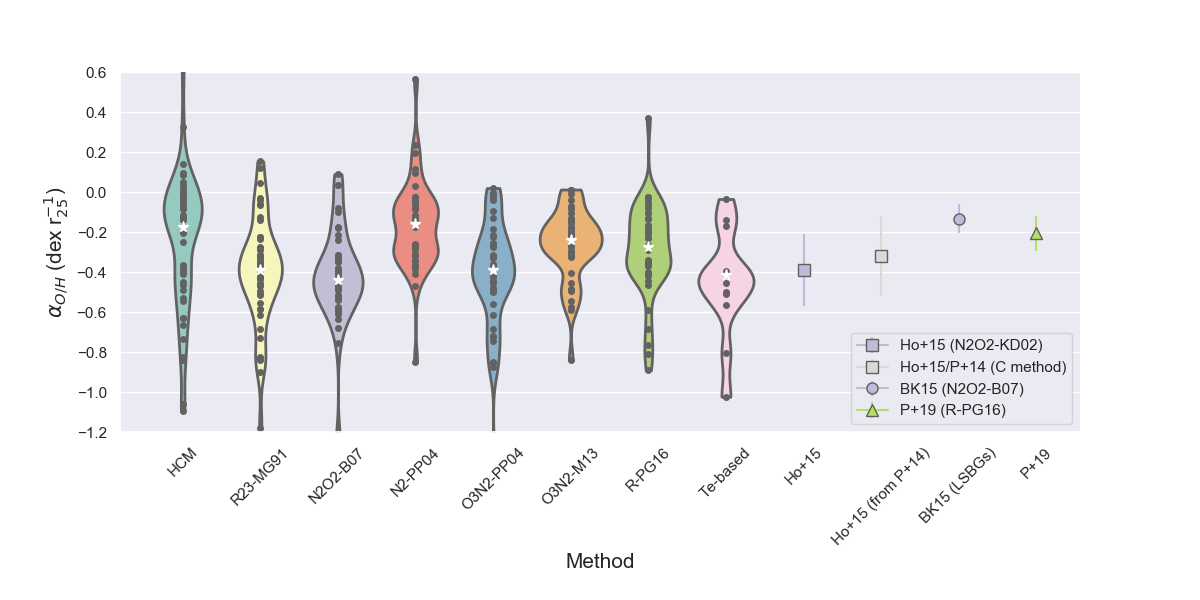}
\vspace{-0.3cm}
\includegraphics[width=0.80\textwidth]{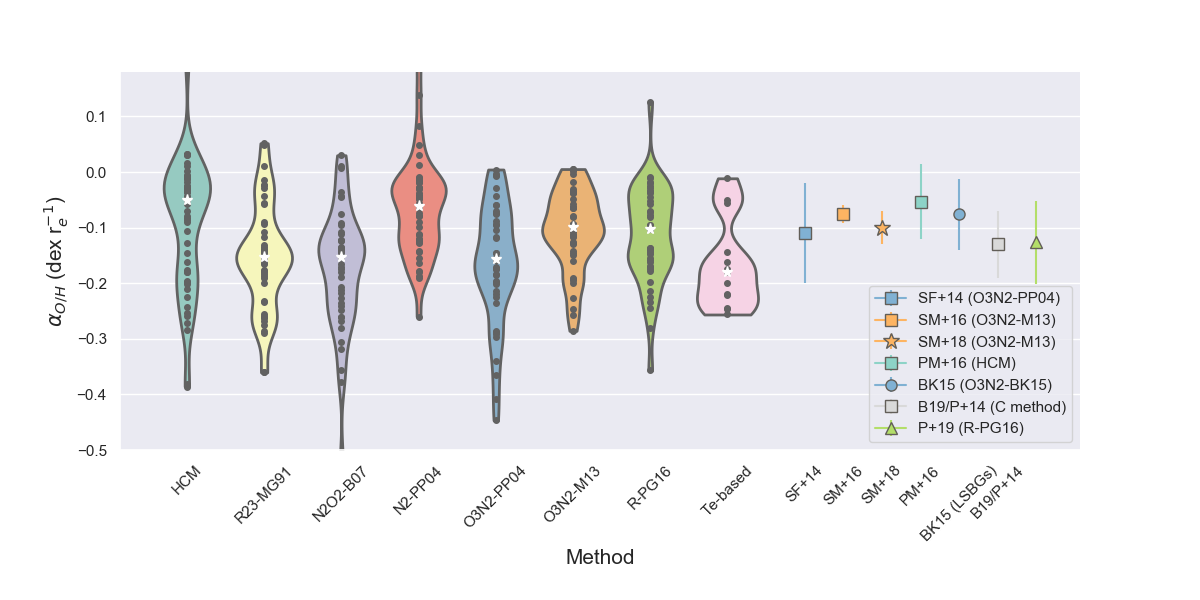}
\caption{Violin plots showing the distribution of the metallicity slopes derived with the different methods in units of dex~r$_{25}^{-1}$ ({\em top}) and
  dex~r$_{e}^{-1}$ ({\em bottom}). The grey points inside the violin plots represent the slope values for the individual galaxies of the sample, and the white star represents the median slope  for each method. For galaxies presenting metallicity breaks in their profiles we have considered the inner slope value.
Average or median values obtained by other authors are shown in the right hand side of each panel for comparison. These include \citet[Ho+15]{Ho2015}, \citet[P+14]{pilyugin2014}, \citet[BK15]{BresolinKennicutt2015}, \citet[SF+14]{sanchez14}, \citet[SM+16]{Sanchez-Menguiano2016}, \citet[SM+18]{Sanchez-Menguiano2018}, \citet[PM+16]{Perez-Montero2016},  \citet[B19]{Bresolin2019}, \citet[P+19]{P19}.\label{violin_plots}}
\end{figure*}

\begin{figure}
\includegraphics[width=0.4\textwidth]{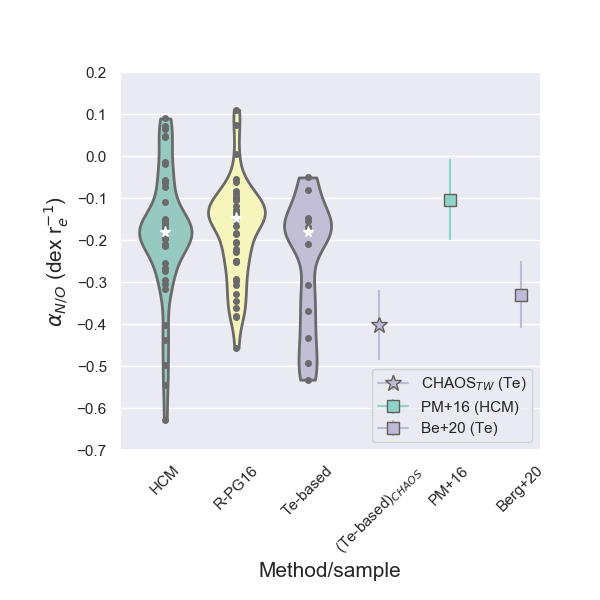}
\caption{Violin plot showing the distribution of the $\log$(N/O) abundance gradient derived with HCM, the $R$ calibration and the T$_e$-based method in units of dex~r$_{e}^{-1}$. For comparison we also show the median slope value derived from the T$_e$-based method in this work for the four CHAOS galaxies included in \citet{Berg2020}, violet star, and the median values derived by \citet{Perez-Montero2016} and \citet{Berg2020}.\label{violin_plots_NO}}
\end{figure}

The $\alpha_{O/H}$  and $\alpha_{N/O}$  median values  from our galaxy sample are shown in Table~\ref{tab:fits} for the different diagnostics employed in this paper, in units of dex~kpc$^{-1}$, dex~r$_{25}^{-1}$ and dex~r$_{e}^{-1}$. The number of galaxies used in the statistics varies between 42 and 51 depending on the strong-line method. We also show the median value derived for the 13 galaxies for which the T$_e$-based abundance profile is reliable (see Sect.~\ref{comparison_slopes}). The slope values are graphically shown in Figs.~\ref{violin_plots} and \ref{violin_plots_NO} for 12+$\log$(O/H) and $\log$(N/O), respectively. These contain violin plots with the distribution of slope values derived for the different methods (individual grey dots inside the violins), together with the probability density of the data at different slope values (the coloured violins themselves). The probability density is a kernel density estimate of the underlying distribution\footnote{We employed the  {\tt seaborn}  {\sc python} module.}.
Median values for the corresponding distributions are marked with a white star for each metallicity method. The two panels in Fig.~\ref{violin_plots} correspond to metallicity slopes normalized to r$_e$ and r$_{25}$ in the bottom and top panels, respectively. For comparison purposes, we also show the values reported by other authors from different samples and metallicity calibrations (Table~\ref{tab:universal_gradient}). Both in Table~\ref{tab:fits} and Figs.~\ref{violin_plots} and \ref{violin_plots_NO} we have considered the inner slope in the profiles with radial breaks, as explained in Sect.~\ref{fits}.

Fig.~\ref{violin_plots} nicely shows that the distribution of metallicity gradients is clearly dependent on the method in both the median value and the standard deviation. For all the strong-line methods we obtain a standard deviation in the slope distribution that range from $\sim$0.09 to $0.12$~dex~r$_{e}^{-1}$ or $\sim$0.22 to $0.33$~dex~r$_{25}^{-1}$, except for O3N2 as calibrated by  \citet{marino13}, for which it is systematically lower with the two normalizations (i.e. 0.07~dex~r$_{25}^{-1}$ and 0.17~dex~r$_{e}^{-1}$). This is not unexpected, given our analysis in Sects.~\ref{comparison} and \ref{comparison_slopes}, where our comparison of direct and strong-line methods showed a tendency for this calibration of the O3N2 index to flatten the steepest gradients, due to the small range in the O3N2-predicted metallicities in relation to the T$_e$-based ones (Fig.~\ref{residuals_vs_O_abund}). This implies that gradients derived with this method tend to cover a smaller range of values. Therefore, the O3N2 method, notably with the calibration performed by \citet{marino13}, favours the conclusion of the existence of a universal metallicity gradient, but the  dispersion in slope values  may  be biased by the method selection. 

 The HCM and N2 methods yield a shallower median slope than the rest of methods by an amount similar to the standard deviation of the distributions ($\sim$0.1~dex~r$_{e}^{-1}$ or $\sim$0.2~dex~r$_{25}^{-1}$). Regarding the median slope values, it is interesting to note that there is good agreement between the median slopes derived by us with a given method and the corresponding value obtained for a different sample by other authors with the same method.  The metallicity method employed seems to have a stronger effect on the differences in median values between works than differences in the galaxy sample and/or in the nature of the observational data. The largest difference is seen in the median slope in dex~r$_{25}^{-1}$ derived from the N2O2 scale between the sample of low surface brightness galaxies (LSBGs) in \citet{BresolinKennicutt2015}, violet circle in top panel of Fig.~\ref{violin_plots},  and our sample or the one by \citet{Ho2015}, both composed by high surface brightness galaxies. This difference between the slopes in low and high surface brightness galaxies, not clearly seen when slopes are normalized to r$_{e}^{-1}$ with the O3N2 scale, is deeply discussed in \citet{BresolinKennicutt2015}.

Recent work by \citet{Berg2020} questions the existence of a universal  gradient in $12+\log$(O/H). This conclusion comes from the diversity of gradients found with just four galaxies from the CHAOS project for which the slopes have been derived very reliably with the $T_e$-based method. Our  $T_e$-based slopes support their statement. However, \citet{Berg2020} find that the same four galaxies have a very similar value of $\alpha_{N/O}$ in terms of r$_e$, for gradients measured between 0.2 and 2 times r$_e$. We show in Fig.~\ref{violin_plots_NO} the distribution of $\alpha_{N/O}$  values for our sample galaxies  from the $T_e$-based  method and from both the HCM and the $R$ calibration.  The {\em rms} dispersion of the $\alpha_{N/O}$ value distributions are comparable to those for $\alpha_{O/H}$  ($\sigma=0.12$-$0.15$~dex~r$_{e}^{-1}$). Taking into account uncertainties, our median value for HCM also agrees with the one derived by \citet{Perez-Montero2016} with the same method but for a different and much larger sample of 201 galaxies from CALIFA, but \citet{Perez-Montero2016}  obtained a smaller dispersion of 0.096~dex~r$_{e}^{-1}$.
Our median value derived from the 13 galaxies with reliable T$_e$-based radial $\log$(N/O) abundance profiles is however considerably shallower than the characteristic value of $-0.33\pm0.08$~dex~r$_{e}^{-1}$ derived by \citet{Berg2020}. However, our median  $\alpha_{N/O}$ value for the four CHAOS galaxies\footnote{NGC~628, NGC~3184, NGC~5194 and NGC~5457.} included in \citet{Berg2020}, $-0.41\pm0.08$~dex~r$_{e}^{-1}$ (violet star in Fig.~\ref{violin_plots_NO}),  is closer to their reported characteristic value. Part of the discrepancy between our and their median value for these four galaxies might come from differences in the inner slope for NGC~5194, for which we derive a steeper slope of $-0.53\pm0.16$~dex~r$_{e}^{-1}$ than the cited authors do ($-0.17\pm0.11$~dex~r$_{e}^{-1}$), possibly due to different adopted values of r$_{e}$  for this galaxy.

Our galaxy sample was not designed to test the existence of a universal abundance gradient in spirals and might not be appropriate for this aim. However, the reasonable agreement of our data with values derived from previous studies, and the large measured dispersion of the slope distributions, might indicate that if a universal gradient exists, second order effects might be important in the modification of the characteristic gradient, and they might be relevant in producing the observed dispersion of slope values both for  $\alpha_{O/H}$ and  $\alpha_{N/O}$.

In \citetalias{paperII}  we will analyse the dependence of the radial abundance gradients of O/H and N/O with host galaxy properties, in particular with the presence of a galactic stellar bar.

\begin{table*}
\centering
\caption{Published values of the mean (or median) slopes of the $12 + \log$(O/H) and $\log$(N/O) radial abundance profiles from different
  data samples and methodologies as indicated.}
\label{tab:universal_gradient}
\begin{tabular}{ccl} 
\hline
 $\alpha_{O/H}$ &  Indicator and calibration$^{*}$ & Sample and reference \\
\hline
\hline
  $-0.11\pm0.09$~dex   r$_{e}^{-1}$   & O3N2 (PP04)    &  N=193 from the CALIFA sample, \citet{sanchez14}\\
  $-0.075\pm0.016$~dex r$_{e}^{-1}$   & O3N2 (M13)     &  N=122 from the CALIFA sample  \citet{Sanchez-Menguiano2016}\\
  $-0.10\pm0.03$~dex   r$_{e}^{-1}$   & O3N2 (M13)     &  N=102 from the AMUSING survey  \citet{Sanchez-Menguiano2018}\\
  $-0.076\pm0.064$~dex r$_{e}^{-1}$   & O3N2 (PP04)    &  N=10,  Low surface brightness galaxies  \citet{BresolinKennicutt2015}\\
  $-0.13\pm0.06$~dex r$_{e}^{-1}$     & C method (P14) &  N=46  from the \citet{pilyugin2014} sample, \citet{Bresolin2019}\\
  $-0.053\pm0.068$~dex r$_{e}^{-1}$   & HCM (PM14)     &  N=201 from the CALIFA sample,  \citet{Perez-Montero2016}\\
 $-0.13\pm0.07$~dex r$_{e}^{-1}$   & $R$ calibration (PG16) & N=147 from the MaNGA survey, calculated from data in \citet{P19}\\
\hline
  $-0.39\pm0.18$~dex r$_{25}^{-1}$    & N2O2 (KD02)    & N=49, Local field star-forming galaxies, \citet{Ho2015}\\
  $-0.133\pm0.072$~dex r$_{25}^{-1}$  & N2O2 (B07)     & N=10, Low surface brightness galaxies,  \citet{BresolinKennicutt2015} \\
  $-0.32\pm0.20$~dex r$_{25}^{-1}$    & C method (P14) & N=104 from the \citet{pilyugin2014} sample, \citet{Ho2015}\\
 $-0.21\pm0.09$~dex r$_{25}^{-1}$    & $R$ calibration (PG16) & N=147 from the MaNGA survey, calculated from data in \citet{P19}\\
\hline
 $\alpha_{N/O}$ &  Indicator and calibration$^{*}$ & Sample and reference \\
\hline
\hline
$-0.104\pm0.096$~dex   r$_{e}^{-1}$   & HCM (PM14)     &  N=201 from the CALIFA sample,  \citet{Perez-Montero2016}\\
$-0.33\pm0.08$~dex     r$_{e}^{-1}$   & T$_e$-based    &  N=4   (NGC~628, NGC~5194, NGC~5457, NGC~3184), \citet{Berg2020}\\
\hline
\end{tabular}
{\footnotesize
\begin{tabular}{l}

  $^{*}$ PP04: \citet{pp04}, M13:  \citet{marino13}, P14: \citet{pilyugin2014}, KD02: \citet{kewley02}, \\
  B07:  \citet{b07}, PM14: \citet{epm14}, PG16: \citet{PilyuginGrebel2016}\\ 
\end{tabular}}
\end{table*}

\begin{table*}
\centering
\caption{Median slope of the radial oxygen abundance gradient in different units for the galaxies analysed in this paper. The uncertainties show the standard deviation in the slope values.}
\label{tab:fits}
\begin{tabular}{lcccc} 
\hline
\multicolumn{5}{l}{$\log$(O/H) {\em vs.} R}\\  
\hline
\hline
Method$^{\dagger}$  &   $\alpha_{O/H}$   &  $\alpha_{O/H}$       &  $\alpha_{O/H}$     &    N  \\ 
                  &  (dex kpc$^{-1}$)  &  (dex r$_{25}^{-1}$)  & (dex r$_{e}^{-1}$)   &     \\ 
\hline
HCM  (PM14)      & $-0.01\pm0.03$   & $-0.17\pm0.33$  & $-0.05\pm0.12$    & 51\\
R23  (M91)         & $-0.03\pm0.03$   & $-0.39\pm0.26$  & $-0.15\pm0.10$    & 48\\
N2O2 (B07)        & $-0.03\pm0.03$   & $-0.44\pm0.26$  & $-0.15\pm0.11$    & 43\\  
N2   (PP04)         & $-0.01\pm0.02$   & $-0.16\pm0.22 $ & $-0.06\pm0.09$    & 44\\
O3N2 (PP04)      & $-0.03\pm0.03$   & $-0.39\pm0.27$  & $-0.16\pm0.11$    & 44\\ 
O3N2 (M13)       & $-0.02\pm0.02$   & $-0.24\pm0.18$  & $-0.10\pm0.07$    & 44\\ 
R (PG16)            & $-0.02\pm0.02$   & $-0.27\pm0.23$  & $-0.10\pm0.09$    & 42\\ 
T$_e$-based$^{*}$ & $-0.03\pm0.02$   & $-0.41\pm0.25$  & $-0.17\pm0.08$    & 13\\ 
\hline
\multicolumn{5}{l}{$\log$(N/O) {\em vs.} R}\\
\hline
\hline
Method              &  $\alpha_{N/O}$    &  $\alpha_{N/O}$     & $\alpha_{N/O}$         & N\\ 
                    &  (dex kpc$^{-1}$) &  (dex r$_{25}^{-1}$)  & (dex r$_{e}^{-1}$)     & \\
\hline
HCM  (PM14)          & $-0.04\pm0.04$   &  $-0.5\pm0.3 $     & $ -0.18\pm0.15$      & 45 \\
R (PG16)                 & $-0.04\pm0.02$   &  $-0.4\pm0.2$      & $-0.15\pm0.12$       & 42\\  
T$_e$-based$^{*}$ & $-0.05\pm0.02$   &  $-0.5\pm0.3 $     & $ -0.18\pm0.15$      & 13 \\
\hline
\end{tabular}
{\footnotesize
\begin{center}
  \begin{tabular}{l}    
  $^{\dagger}$ Acronyms as follow: PM14: \citet{epm14}, M91: \citet{m91},  B07: \citet{b07} \\
     PP04: \citet{pp04}, M13:  \citet{marino13}, PG16: \citet{PilyuginGrebel2016}\\
  $^{*}$ Median value for the galaxies with more reliable determination of the gradients (i.e. NGC~1512, NGC~2997 \\
   and NGC~4258 were excluded).\\
  \end{tabular}
\end{center}
}
\end{table*}

\section{Summary and conclusions}
\label{conclusiones} 
This paper is the first of a series of two, devoted to revisit the issue of the effect of stellar bars on the radial distribution of metals in the gas-phase of spiral galaxies, as an attempt to solve the existing discrepancies found in the literature. For that aim we have made a compilation of published emission-line fluxes and positions of over 2800 \hii\ regions, that  belong to a sample of 51 nearby (D < 64~Mpc) spiral galaxies. We have used an homogeneous methodology to derive: (a) the structural parameters for bars and discs from broad-band imaging, and (b) O/H  and N/O abundance ratios from the {\em direct} or $T_e$-based method for a sub-sample of regions (610), and from a variety of strong-line methods (based on the $R$ calibration, R$_{23}$, N2O2, O3N2, N2  and HCM for $12+\log$(O/H), and for $\log$(N/O) on N2O2, N2S2, the $R$ calibration and HCM). The abundances derived from the strong-line methods have been compared to the more reliable $T_e$-based ones for the subsample of regions for which {\em direct} abundances could be calculated. Radial $12+\log$(O/H) and $\log$(N/O) abundance profiles have been derived for all galaxies from re-calculated deprojected \hii\ region galactocentric distances for all the abundance methods and have been analysed. Our main results and conclusions are summarised below:
\begin{itemize}
\item The RC3 visual classification of bars should be taken with caution. In our sample, the 'AB' class contains galaxies with a wide range in bar ellipticity (a proxy for bar-strength). Using this parameter, our sample contains 22 strongly barred, 9 weakly barred and 20 unbarred galaxies.

\item Overall, all strong-line methods tested here reproduce the $T_e$-based abundances (except for the well-known offset for the R$_{23}$ method calibrated from photoionization
models from \citet{m91}). However, we find that the residuals correlate to some extent  with  $12+\log$(O/H)$_{T_e}$, $\log$(N/O)$_{T_e}$  and/or $\log$(O32) for some of the methods. Special caution must be taken with the O3N2 method, notably with the calibration performed by \citet{marino13}, that yields a smaller range in metallicities than the $T_e$-based one.

\item The above result translates in a tendency for this method to flatten the radial metallicity profiles of the galaxies with the steepest gradients in our sample. This is clearly seen in our comparison of slopes derived from strong-line methods with those obtained from the  T$_e$-based method for a subsample of 13 galaxies with reliable determination of the T$_e$-based metallicity gradient.

\item Based on both the comparison of $T_e$-based with strong-line metallicities for individual \hii\ regions, and the comparison of the  metallicity gradients derived from both methodologies,  the $R$ calibration, HCM and N2O2 yield a better agreement with the T$_e$-based scale for our \hii\ region sample.

\item For $\log$(N/O) the HCM method and the $R$ calibration are those that better match the $T_e$-based scale for individual \hii\ regions. These also reproduce well the $T_e$-based radial  $\log$(N/O) slopes.

\item We confirm the presence of breaks in the radial abundance profiles of the galaxies analysed, in agreement with previous work by different authors.  However, the breaks in the $12+\log$(O/H) are not clearly detected with all the strong-line methods. In our sample, if a break is detected, the N2O2 and R$_{23}$ methods are more prone to show it.

\item Most of the galaxies with a break in their metallicity profile (with one or several methods)  show also a break in their $\log$(N/O) radial profile at a similar radius. The break radius (in either  the $12+\log$(O/H) or the $\log$(N/O) profiles) is typically in the range between 0.5 to 1.4 times r$_{25}$  with an average of ($0.9\pm0.3$)$\times$r$_{25}$.

\item We have analysed the median slope and dispersion values derived for $12+\log$(O/H) and $\log$(N/O) for the different methods. Both the median and the dispersion depend on the selected method to derive the abundances, but are in reasonable agreement with results derived by other authors from different data samples. The O3N2 method as calibrated by \citet{marino13} yields a smaller dispersion than the rest of methods with the two normalizations of the profiles (with r$_e$ and r$_{25}$), that supports the proposal of a universal metallicity gradient. However,  the low measured dispersion in the slope distribution might be biased by the method selection and the associated problems with O3N2 metallicity scale mentioned above (in comparison with the $T_e$-based abundances).
\end{itemize}

In \citetalias{paperII} we analyse the derived radial abundance gradients comparatively for (strongly and weakly) barred and unbarred galaxies. The results are compared with previous work, and discussed in the context of current knowledge on disc evolution and radial mixing induced by bars and spiral arms.


\section*{Acknowledgements}
We kindly thank all the authors who have previously published the \hii\ data that form the basis for this work, and Simon Verley for his invaluable help with {\sc python}. 
We also acknowledge the anonymous referee for his/her suggestions that improved the clarity of the manuscript.
We thank Calar Alto Observatory for allocation of director's discretionary time to this programme. AZ, EF and IP acknowledge support from the Spanish {\em Ministerio de Economia y Competitividad} and {\em FEDER programme} via grant AYA2017-84897-P and from the Junta de Andalucia local government through the FQM-108 project. 
EPM acknowledges  funding from the Spanish MINECO project  Estallidos 6 AYA2016-79724-C4 and  the Spanish Science Ministry "Centro de Excelencia Severo Ochoa Program under grant SEV-2017-0709''.
This research made use of Astropy,\footnote{http://www.astropy.org} a community-developed core {\sc python} package for Astronomy \citep{astropy:2013, astropy:2018}. This research has made use of the VizieR catalogue access tool, CDS, Strasbourg, France (DOI : 10.26093/cds/vizier). The original description  of the VizieR service was published in 2000, A\&AS 143, 23. This research has made use of the NASA/IPAC Extragalactic Database (NED) which is operated by the Jet Propulsion Laboratory, California Institute of Technology, under contract with the National Aeronautics and Space Administration. 

\section*{Data Availability}
The data underlying this article are available in the article, in its online supplementary material and on the CDS VizieR facility (via https://vizier.u-strasbg.fr/viz-bin/VizieR).



\bibliographystyle{mnras}
\bibliography{referencias} 


\newpage
\appendix

\noindent{\Large\bf Supporting information}

\setcounter{page}{0}
\pagenumbering{arabic}
\setcounter{page}{1}
\section{References for HII region data}
Table~\ref{bibliografia_gal}  contains the references for the published work from which we have compiled the  \hii\ region emission-line fluxes and positions for each galaxy. We have highlighted in boldface the papers that report radial abundance profile {\em breaks} in the corresponding galaxies (see Sect.~\ref{breaks} for further details).
\begin{table*}
\centering
\small
\caption{List of papers from which the \hii\ region emission-line fluxes and positions have been compiled. Boldface references highlight
papers reporting radial abundance breaks in the corresponding galaxies.}
\label{bibliografia_gal}
\begin{tabular}{ll}
\hline
Galaxy     & References             \\
\hline
NGC~224   &  \citet{Bresolin99}, \citet{Blair82}, \citet{Dennefeld81}, \citet{esteban09}, \citet{Esteban2020},\\
          & \citet{Galarza}, \citet{sanders12}, \citet{Zurita12} \\
NGC~300   & \citet{Bresolin09}, \citet{Edmunds84}, \citet{pagel79}, \citet{Toribio16} \\
NGC~598   & \citet{Bresolin10}, \citet{Bresolin99}, \citet{bresolin11_m33}, \citet{Crockett06}, \citet{esteban09}, \\
          & \citet{Kehrig11}, \citet{KwitterAller81}, \citet{lin_m33}, \citet{lopez-hernandez},  \\
	  &  \citet{Magrini10}, \citet{Mccall85}, \citet{Relano10}, \citet{rosolowsky08},\\
	  & \citet{Toribio16}, \citet{vilchez-pagel88} \\
NGC~628   & {\bf \citet{berg13}}, {\bf \citet{Berg15}}, \citet{Bresolin99}, \citet{Castellanos02}, \citet{Ferguson98}, \\
          & \citet{Gusev12}, \citet{Mccall85}, \citet{Rosales-ortega11}, \Citet{vanZee98} \\
NGC~925   & \citet{Castellanos02}, \citet{Martin94}, \citet{Zaritsky94}, \Citet{vanZee98} \\
NGC~1058  & \citet{Ferguson98}, {\bf \citet{Bresolin2019}} \\
NGC~1068  & \citet{Evans87}, \citet{Oey93}, \citet{vanZee98} \\
NGC~1097  & \citet{storchi96} \\
NGC~1232  & \citet{bresolin05}, \citet{Castellanos02}, \citet{vanZee98}\\
NGC~1313  &  \citet{Hadfield07}, \citet{Pagel80}, \citet{Walsh97}\\
NGC~1365  & \citet{Alloin81}, \citet{bresolin05}, \citet{pagel79}, \citet{Roy97} \\
NGC~1512  & {\bf \citet{b12}}, \citet{lopez-sanchez15} \\
NGC~1637  & \citet{Castellanos02}, \citet{vanZee98} \\
NGC~1672  & \citet{storchi96} \\
NGC~2336  & \citet{Gusev12} \\
NGC~2403  & \citet{berg13}, \citet{Bresolin99}, \citet{esteban09}, \citet{Fierro86}, \citet{Garnett97}, \\
          & \citet{Mao18}, \citet{Mccall85}, \citet{smith75}, \citet{vanZee98} \\
NGC~2541  & \citet{Zaritsky94} \\
NGC~2805  & \citet{vanZee98} \\
NGC~2903  & \citet{bresolin05}, \citet{diaz07}, \citet{Mccall85}, \citet{smith75}, \citet{vanZee98}, \\
          & \citet{Zaritsky94}\\
NGC~2997  & \citet{bresolin05}, \citet{Edmunds84}, \citet{Firpo05}, \citet{Mccall85}  \\
NGC~3031  & \citet{arellano-cordova}, \citet{Bresolin99}, \citet{Garnett87}, \citet{Patterson12}, \\
          & \citet{stanghellini10}, \citet{Stauffer84} \\ 
NGC~3184  & \citet{Berg2020}, \citet{Mccall85}, \citet{vanZee98}, \citet{Zaritsky94} \\
NGC~3227  & \citet{Gonzalez-delgado97}, \citet{Lisenfeld08},  \citet{Werk11}\\
NGC~3310  & \citet{Bresolin99}, \citet{Werk11} \\
NGC~3319  & \citet{Zaritsky94} \\
NGC~3344  & \citet{Mccall85}, \citet{Vilchez88a}, \citet{Zaritsky94} \\
NGC~3351  & \citet{fabio02}, \citet{Bresolin99}, \citet{diaz07}, \citet{Mccall85},  \citet{Oey93}\\
NGC~3359  & \citet{Werk11}, {\bf \citet{zahid}} \\
NGC~3521  & \citet{Bresolin99}, \citet{Zaritsky94} \\
NGC~3621  & {\bf \citet{b12}}, \citet{Zaritsky94} \\
NGC~4254  & \citet{Henry94}, \citet{Mccall85}, \citet{Shields91} \\
NGC~4258  & \citet{Bresolin99}, \citet{bresolin11_n4258}, \citet{Diaz00}, \citet{Oey93}, \citet{Zaritsky94} \\
NGC~4303  & \citet{Henry92}, \citet{Shields91} \\
NGC~4321  & \citet{Mccall85}, \citet{Shields91} \\
NGC~4395  & \citet{esteban09}, \citet{Mccall85}, \citet{vanZee98} \\
NGC~4625  & {\bf\citet{Goddard10}} \\
NGC~4651  & \citet{Skillman96} \\
NGC~4654  & \citet{Skillman96} \\
NGC~5194  & \citet{bresolin2004}, \citet{Bresolin99}, \citet{Croxall15}, \citet{diaz91}, \citet{Mccall85}, \citet{smith75}  \\
NGC~5236  & \citet{Bresolin99}, \citet{fabio02}, \citet{bresolin05}, {\bf \citet{bresolin09b}},  \\
          & \citet{Dufour80}, \citet{esteban09}\\
NGC~5248  & \citet{storchi96} \\
NGC~5457  & \citet{b07}, {\bf \citet{croxall16}}, \citet{esteban09}, \citet{Esteban2020}, \citet{Garnett94},  \\
          & \citet{Izotov07}, \citet{Kennicutt96}, \citet{kennicutt03}, \citet{Li13}, \citet{Luridiana02},  \\
	  & \citet{Mccall85}, \citet{Skillman85},  \citet{smith75}, \citet{Torres-peimbert89}, \citet{vanZee98} \\
NGC~6384  & \citet{fabio02}, \citet{Bresolin99}, \citet{Oey93} \\
NGC~6946  & \citet{Ferguson98}, \citet{Mccall85} \\
NGC~7331  & \citet{Bresolin99}, \citet{Gusev12}, \citet{Oey93}, \citet{Zaritsky94} \\
NGC~7518  & \citet{Robertson12} \\
NGC~7529  & \citet{Robertson12} \\
NGC~7591  & \citet{Robertson12} \\
NGC~7678  & \citet{Gusev12} \\
NGC~7793  & \citet{Bibby10}, \citet{Edmunds84}, \citet{Mccall85}, \citet{stanghellini15},\\
          & \citet{Webster83} \\
Milky Way & \citet{Esteban18}, \citet{Esteban17}, \citet{Esteban04},  \citet{Esteban13}, \\
          & \citet{Fernandez-martin17}, \citet{Rojas04}, \citet{Rojas05}, \citet{Rojas06}, \\
	  & \citet{Rojas07}, \citet{Rojas14} \\
\end{tabular}
\end{table*}

\section{References for galaxy sample images}
Table~\ref{referencias_imagenes} shows the photometric bands, sources and references for the images employed in the morphological analysis of the galaxy sample (Sect.~\ref{morfo}), and for the images shown in Appendix~\ref{perfiles} to illustrate the location of the \hii\ regions employed in  the derivation of the radial abundance profiles.
\begin{table*}
\centering
\caption{Photometric bands and references for the images shown in Appendix~\ref{perfiles} and for those used in 
the morphological analysis (Sect.~\ref{morfo}). \label{referencias_imagenes}}
\begin{tabular}{l|ll|ll}
\hline
galaxy & Photometric band            & Source$^{a}$/Reference & Photometric band  &  Source$^{a}$/Reference \\
       &  (Appendix ~\ref{perfiles}) &                  &  (Morphology)     &                  \\
\hline
NGC~224        &      NUV      &      NED/\citet{GildePaz07}  &     -      &      -   \\
NGC~300        &    \ha\       &      NED/\citet{Larsen99}    &     R      &     NED/\citet{Larsen99} \\
NGC~598        &      NUV      &      NED/\citet{GildePaz07}  &     R      &     NED/\citet{Massey06} \\
NGC~628        &     \ha\      &      4.2m WHT/Private image &     r      &     SDSS\\
NGC~925        &     \ha\      &    NED/2007SINGS.5......0:      &     R      &     NED/2007SINGS.5......0: \\
NGC~1058       &     \ha\      &    NED/2011PrivC.U..G....D$^{b}$ &     R      &     NED/\citet{James04} \\
NGC~1068       &     \ha\      &      NED/\citet{Knapen04}       &     r      &     NED/\citet{Brown14} \\
NGC~1097       &     \ha\      &    NED/2007SINGS.5......0:      &     R      &     NED/2007SINGS.5......0: \\
NGC~1232       &   IIIaJ-4680A  &   DSS                          &     R      &     ESO archive/FORS \\
NGC~1313       &     \ha\       &   NED/\citet{Larsen99}         &     R      &     NED/\citet{Larsen99} \\
NGC~1365       &      NUV       &   NED\citet{GildePaz07}        &     I$^{c}$ &     NED/\citet{Kuchinski00}\\ 
NGC~1512       &     \ha\       &   NED/2007SINGS.5......0:      &     R      &     NED/2007SINGS.5......0:\\
NGC~1637       &     \ha\      &    NED/2002SINGG.U.......M      &     R      &     NED/2002SINGG.U.......M\\
NGC~1672       &      NUV      &    NED/\citet{GildePaz07}       &     V      &     NED/\citet{Kuchinski00} \\
NGC~2336       &      r        &    CAHA 2.2m (this work)        &     r      &     CAHA 2.2m (this work)\\
NGC~2403       &     NUV       &    NED/\citet{GildePaz07}       &     r      &     NED/\citet{Brown14} \\
NGC~2541       &     \ha\      &    NED/2011PrivC.U..G....D      &     r      &     NED/SDSS\\
NGC~2805       &     \ha\      &    NED/\citet{Knapen04}         &     r$^{c}$ &     NED/SDSS\\
NGC~2903       &     NUV       &    NED/\citet{GildePaz07}       &     r      &     NED/SDSS\\
NGC~2997       &     \ha\      &    NED/\citet{Larsen99}         &     R      &     NED/\citet{Larsen99} \\
NGC~3031       &     \ha\      &    NED/\citet{Hoopes01}         &     R      &     NED/\citet{Hoopes01} \\ 
NGC~3184       &     \ha\      &    NED/2007SINGS.5......0:      &     R      &     NED/\citet{Knapen04} \\
NGC~3227       &     \ha\      &    NED/\citet{Knapen04}         &     r      &     SDSS \\
NGC~3310       &      r        &          SDSS                   &     r      &     SDSS \\
NGC~3319       &      Hac      &    NED/2011PrivC.U..G....D      &     r      &     SDSS  \\
NGC~3344       &     \ha\      &    NED/\citet{Knapen04}         &     r      &     SDSS \\
NGC~3351       &       NUV     &    NED/\citet{GildePaz07}       &     r      &     SDSS \\
NGC~3359       &     \ha\      &      \citet{n3359_hii}          &     r      &     SDSS \\
NGC~3521       &     \ha\      &    NED/2007SINGS.5......0:      &     r      &     SDSS \\
NGC~3621       &      \ha\     &    NED/2007SINGS.5......0:      &     R      &     NED/2007SINGS.5......0:\\
NGC~4254       &       NUV     &    NED/\citet{GildePaz07}       &     r      &     SDSS \\
NGC~4258       &      NUV      &    NED/\citet{Dale09}           &     r      &     NED/\citet{Frei96} \\
NGC~4303       &       NUV     &    NED/\citet{GildePaz07}       &     r      &     SDSS \\
NGC~4321       &     \ha\      &    NED/\citet{Cheng97}          &     r      &     SDSS \\
NGC~4395       &      NUV      &    NED/\citet{GildePaz07}       &     R      &     NED/\citet{Cook14} \\
NGC~4625       &      NUV      &    NED/\citet{GildePaz07}       &     r      &     SDSS \\
NGC~4651       &      r        &      SDSS                       &     r      &     SDSS\\
NGC~4654       &      r        &      SDSS                       &     r      &     NED/\citet{Baillard11} \\
NGC~5194       &     \ha\      &     NED/\citet{Hoopes01}        &     r      &     SDSS \\
NGC~5236       &     \ha\      &     NED/\citet{Meurer06}        &     R      &     NED/\citet{Dale09} \\   
NGC~5248       &     \ha\      &     NED/\citet{Knapen04}        &     r      &     SDSS \\
NGC~5457       &      \ha\     &     NED/\citet{Hoopes01}        &     r      &     NED/\citet{Brown14} \\
NGC~6384       &      \ha\     &     NED/\citet{Knapen04}        &     r      &     CAHA 2.2m (this work)\\
NGC~6946       &     \ha\      &     4.2m WHT/Private image      &     R \& \ha\ off-band      &     NED/2007SINGS.5......0: \&Private image\\
NGC~7331       &       NUV     &     NED/\citet{GildePaz07}      &     r      &     SDSS \\
NGC~7518       &       r       &      SDSS                       &     r      &     SDSS \\
NGC~7529       &       r       &      SDSS                       &     r      &     SDSS \\
NGC~7591       &      r        &      SDSS                       &     r      &     SDSS \\
NGC~7678       &      r        &      SDSS                       &     r      &     SDSS \\
NGC~7793       &      \ha\     &    NED/2007SINGS.5......0:      &     R      &     NED/\citet{Cook14} \\
\end{tabular}
{\footnotesize
\begin{center}
  \begin{tabular}{l}    
    $^{a}$ NED: The NASA/IPAC Extragalactic Database (\url{https://ned.ipac.caltech.edu/}); CAHA: Calar Alto \\
    observatory (\url{https://www.caha.es}); SDSS: The Sloan Digital Sky Survey (\url{https://www.sdss.org/}); DSS: ESO Online Digitized \\
    Sky Survey (\url{https://archive.eso.org/dss/dss}) \\
    $^{b}$ \url{http://ha-atlas.obs.carnegiescience.edu/FITS/}\\
    $^{c}$ Images only partially useful for our purposes. See Sect.~\ref{morfo} and Table~\ref{tab:sample} for further details.\\ 
  \end{tabular}
\end{center}
}\end{table*}

\section{Comparison of electron temperatures for different ions}
\begin{figure*}
\hspace{-1.5cm}\includegraphics[width=1.1\textwidth]{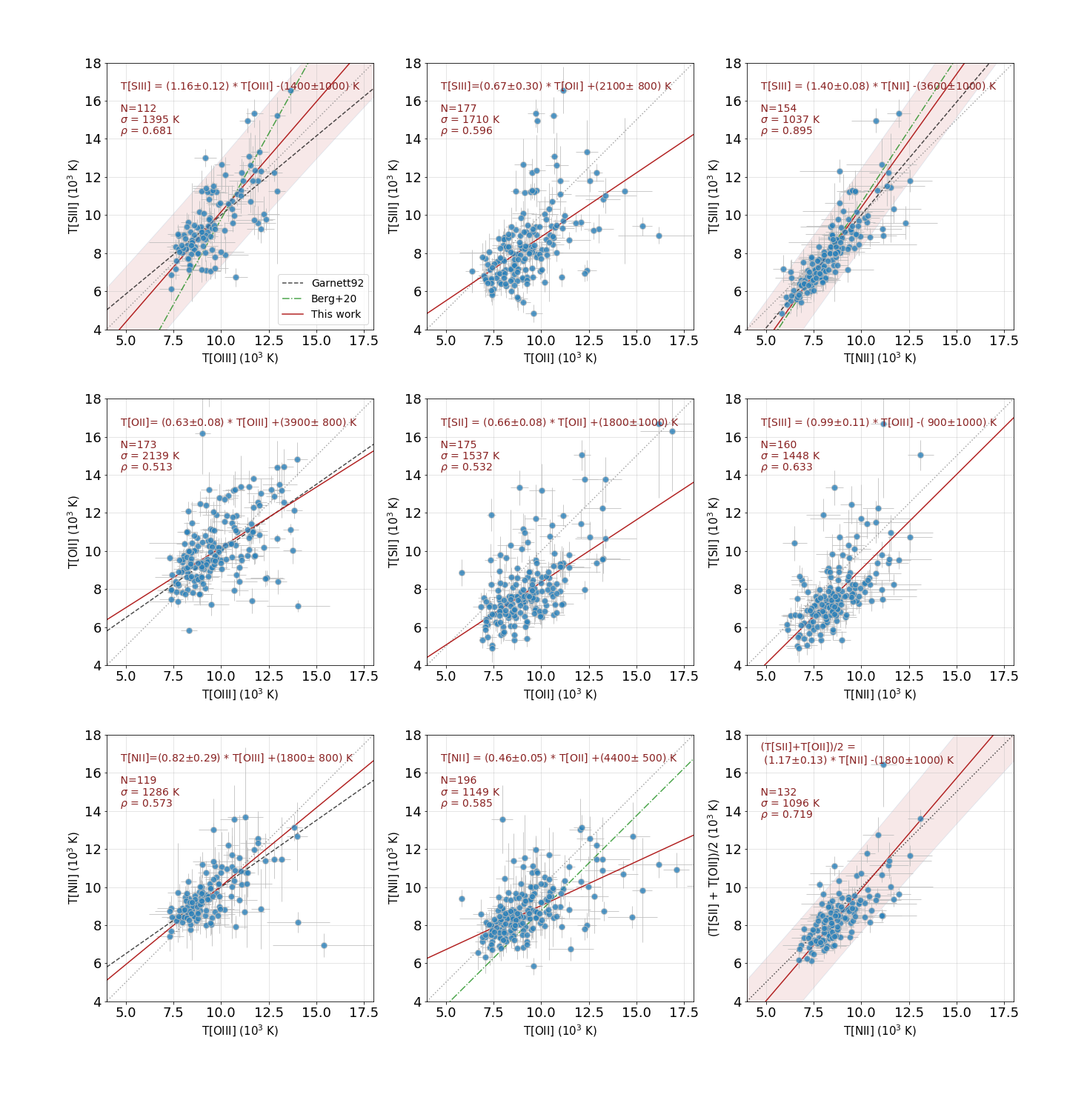}
\vspace{-1.5cm}\caption{Comparison between the electron temperature obtained from different ions. The grey dotted straight lines show the 1:1 temperature relation in each panel. The red straight lines show our best-fitting to the data in each panel, and the shaded area in the panels of the $T_e$[\siii]-$T_e$[\oiii],  $T_e$[\siii]-$T_e$[\nii] and  ($T_e$[\sii]-$T_e$[\oii])/2-$T_e$[\nii]  relations show the area that has a 95\% chance of containing the true regression line. The best-fitting equation, the number of data points (N), the scatter (rms) around the best-fitting ($\sigma$) and the Spearman's Rank correlation coefficient ($\rho$) are shown in the upper-left corner in each panel.  The \citet{garnett92} and   \citet{Berg2020}  relations are marked with grey dashed and green dotted-dashed straight lines, respectively.
\label{temp_rel}}
\end{figure*}

Fig.~\ref{temp_rel} shows the relations  obtained between  the electron temperatures derived for the different ionic species: $T_e$[\siii],  $T_e$[\oiii], $T_e$[\nii],  $T_e$[\sii], and  $T_e$[\oii]. See Sect.~\ref{Te} for details on their derivation.  We obtained values of $T_e$ for 639 \hii\ regions, and for 377 of them we could derive at least two measurements of $T_e$. For a given pair of $T_e$ temperatures we have been able to use a number between 112 and 196 \hii\ regions, to compare the two determinations of $T_e$. 
The plots show with dashed grey straight lines the scaling relations  between the  temperature of different species predicted by the photoionization models of \citet{garnett92}. The relationships derived  by \citet{Berg2020} from observational data are also overplotted with a green dotted-dashed line, and the dotted grey straight lines mark the 1:1 temperature relation in each panel. All sets of data show moderate to strong correlation, with Spearman's Rank correlation coefficients, $\rho$, larger than 0.5, but the dispersion is high in all cases, with values in the range $\sim$1000-2200K. The correlation is remarkably  good for $T_e$[\siii]-$T_e$[\nii], with $\rho$=0.90 (top-right). $T_e$[\siii] and $T_e$[\oiii] ($\rho$=0.68, top-left) are also more strongly correlated than other sets of $T_e$. These relations have been confirmed previously by a number of authors \citep[e.g.][]{Perez-Montero2017,Croxall15, Bresolin09, Berg15}  but to our knowledge, except in \citet{Berg2020}, always with a much smaller data sample. \citet{Berg15} observe  an increasing dispersion in the T[\oiii]-T[\siii] relationship with decreasing temperature for regions in NGC~628 and M101, that is not observed in our derived relations.

We have performed a linear regression to the data\footnote{\hii\ regions with $T_e$ values above 18000~K have not been taken into account for the fits. There are only 3 regions with high  $T_e$[\nii] and $T_e$[\oii], and one with  $T_e$[\oiii] above that value. In any case, due to the methodology employed the fitting results are only marginally affected by these data points if included.} in each panel with the {\tt linmix} {\sc python} code\footnote{\url{https://linmix.readthedocs.io/en/latest/}} that is a hierarchical Bayesian model  \citep{Kelly2007} for fitting a straight line to data with errors in both the x and y directions. The best-fitting is plotted in red for each set of data, and the corresponding fitting parameters are shown in the upper-left corner of each panel, together with the number of data points and the standard deviation from the best-fitting.
A first look at the best-fitting can give the impression of disagreement between our best-fitting and the \citet{garnett92} and/or \citet{Berg2020} relations. However, for the $T_e$ - $T_e$ relations with strongest correlation we show the prediction band of the regression model at the
95\% confidence level or the area  that has a 95\% chance of containing the true regression line. It can be seen that the bands include
\citet{garnett92} and  \citet{Berg2020} relations, and are also compatible with equality. These bands are larger for the rest of relations. Therefore, given the scatter in the data, our modelled temperature relations are in good agreement with \citet{garnett92} and/or \citet{Berg2020} in the range of temperatures covered by the regions of our sample.
We note the reader the good correlation between the average of $T_e$[\oii] and $T_e$[\sii] with  $T_e$[\nii], in the lower-right panel, that is consistent with equality.  In spite that the scatter is high and the correlation not very strong for the $T_e$[\oii]-$T_e$[\nii] and the $T_e$[\sii]-$T_e$[\nii] relations, the average of  $T_e$[\oii] and $T_e$[\sii] resembles rather well the $T_e$[\nii] scale.

\section{Notes on individual galaxies}
\label{notes_galaxies}
\begin{itemize}
\item {\bf Milky Way:} The adopted morphological parameters for the Milky Way disc and bar have been chosen from  published data, and were derived from a different methodology to the one used for the galaxy sample (surface brightness  ellipse fitting, Sect.~\ref{morfo}).  Measurements of the Milky Way disc scale length range from $\sim$2.2~kpc to $\sim$3.6~kpc, depending on  the photometric band and methodology \citep[e.g.][]{Bovy_Rix13,Licquia2016,Sofue2018}. We have adopted the result  by \citet{Licquia2016}, $r_d=(2.71\pm0.2)$~kpc, that considers visible light\footnote{The result come from the analysis of star counts along various lines-of-sight through the disc. It requires assuming a model of the stellar density profile that  matches the observations, and assumptions on the stellar mass and extinction.}  only (being closer than other works to the $r$-band used for our sample). This value is in good agreement   with the statistical analysis by \citet{Bland-Hawthorn2016}  of the main papers ($\sim15$) on the Milky Way radial scale length, that leads to $r_d=(2.6\pm0.5)$~kpc. The adopted value  for $r_d$ implies a disc effective radius of $(4.6\pm0.4)$~kpc. \citet{Licquia2016} also estimate an absolute magnitude $M_B=(-20.07\pm0.44)$.
  
The Milky Way bar semi-major axis, $r_{bar}=(5.0\pm0.2)$~kpc,  was estimated from red clump giant star counts from the combined 2MASS, UKIDSS, VVV, and GLIMPSE surveys by \cite{Wegg2015}. The bar ellipticity was  assumed to be 0.46. This value does not come from observations, but from the prediction of simulations of the Milky Way bar that reproduce well observed star count distributions \citep{Inma2011}.
 
\item {\bf NGC~224 (M31):} The emission-line fluxes from \citet{sanders12} have been extinction-corrected using their published values for A$_V$ and the \citet{ccm} extinction curve. Only \hii\ regions meeting the following criteria have been kept in our sample: the  atmospheric dispersion corrector was functioning during observations (as indicated by the authors in the on-line data), A$_V>0$ and surface brightness class 'normal' or 'bright'. The final sample comprises 114 \hii\ regions (out of 255).

\item {\bf NGC~598 (M33):} 
\begin{itemize}
\item {\bf \citet{rosolowsky08}:} We consider only \hii\ regions for which the relative error in the reported [\oiii]$\,\lambda$4363 flux is lower than 15\%, or the signal-to-noise ratio (S/N) is greater than $\sim$6.5. This cut-off is similar to the one applied by \citet{bresolin11_m33}.  The number of \hii\ regions included in our sample from these authors is therefore 42 (out of 63).
\item {\bf \citet{KwitterAller81}:} We assume a 20\% error in auroral line emission line fluxes, which is the average error quoted  by the authors for the faintest detected lines.
\end{itemize}

\item {\bf NGC~1512:} \hii\ regions of the southern inner arm of NGC~1512 have not been taken into account in the fits (lower abundance possibly due to the interaction with NGC~1510; see  \citet{lopez-sanchez15}). 

\item {\bf NGC~3184:} We have estimated relative errors in the emission-line fluxes from the quoted flux uncertainties in \citet{Berg2020}, and have only used flux measurements with relative errors smaller than 33.3\% (or  signal-to-noise ratio  greater than 3). This criterion has been applied on individual flux measurements of the two lines of doublets, even if the total flux is used (e.g. for [\oii]$\,\lambda\lambda$7320, 7330, or [\sii]$\,\lambda\lambda$4068, 4076). 
 
\item {\bf NGC~3351:} \hii\ region N3351(+014+067) from \citet{Mccall85} has been excluded from the fits,  due to emission-line ratios that yield unrealistic abundances with some of the calibrations. These erroneously affected the fitted radial abundance profile.

\item {\bf NGC~4395:} This is the lowest luminosity strongly barred galaxy in our sample and its derived radial 12+$\log$(O/H) and $\log$(N/O) gradients are very relevant for \citetalias{paperII}. It is not possible to derive reliably the gradients from the  $T_e$-based method. However, the analysis of the slope values derived from strong-line methods indicates a shallow gradient for both the 12+log(O/H) and the log (N/O) profiles in this galaxy: (a) The median of the seven values of $\alpha_{O/H}$ derived from strong-line methods is $-0.005$~dex kpc\me\ with a  standard deviation of 0.007~dex kpc\me. (b) Regarding the slope of the $\log$(N/O) radial profile, the HCM and $R$ methods yield $-0.01\pm0.01$~dex~kpc\me\ and $-0.03\pm0.01$~dex~kpc\me,  respectively. Given the good agreement of the slopes derived from these two methods with the $T_e$-based ones, we expect the $T_e$-based slope to be in the range $-0.04$-$0.00$~dex~kpc\me, that is considerably shallower than the slope of unbarred galaxies of similar luminosity, as shown in \citetalias{paperII}.

\item {\bf NGC~5457 (M101):} For two \hii\ regions the  [\nii]$\,\lambda$5755 fluxes have no error in \citet{kennicutt03}. We assume a $\sim25$\% error, which is a representative value for rest of regions with [\nii]$\,\lambda$5755  detections by these authors. Same for regions number \#143 and \#394 in \citet{lin_m33},  for which an error of 0.1 have been assumed  for the [\nii]$\,\lambda$5755/\hb\ line ratio.
  
\item {\bf NGC~7793 and NGC~4945:} \hii\ regions with $c(H\beta)$=0 or low S/N \citep[or possible PNe according to authors][]{stanghellini15} have not been included in our sample \citet{stanghellini15}.
\end{itemize}

\section{Galaxy images and radial abundance profiles}
\label{perfiles}
\begin{figure*}
\begin{minipage}{1.05\textwidth}
\hspace{-1.2cm}\includegraphics[width=1.12\textwidth]{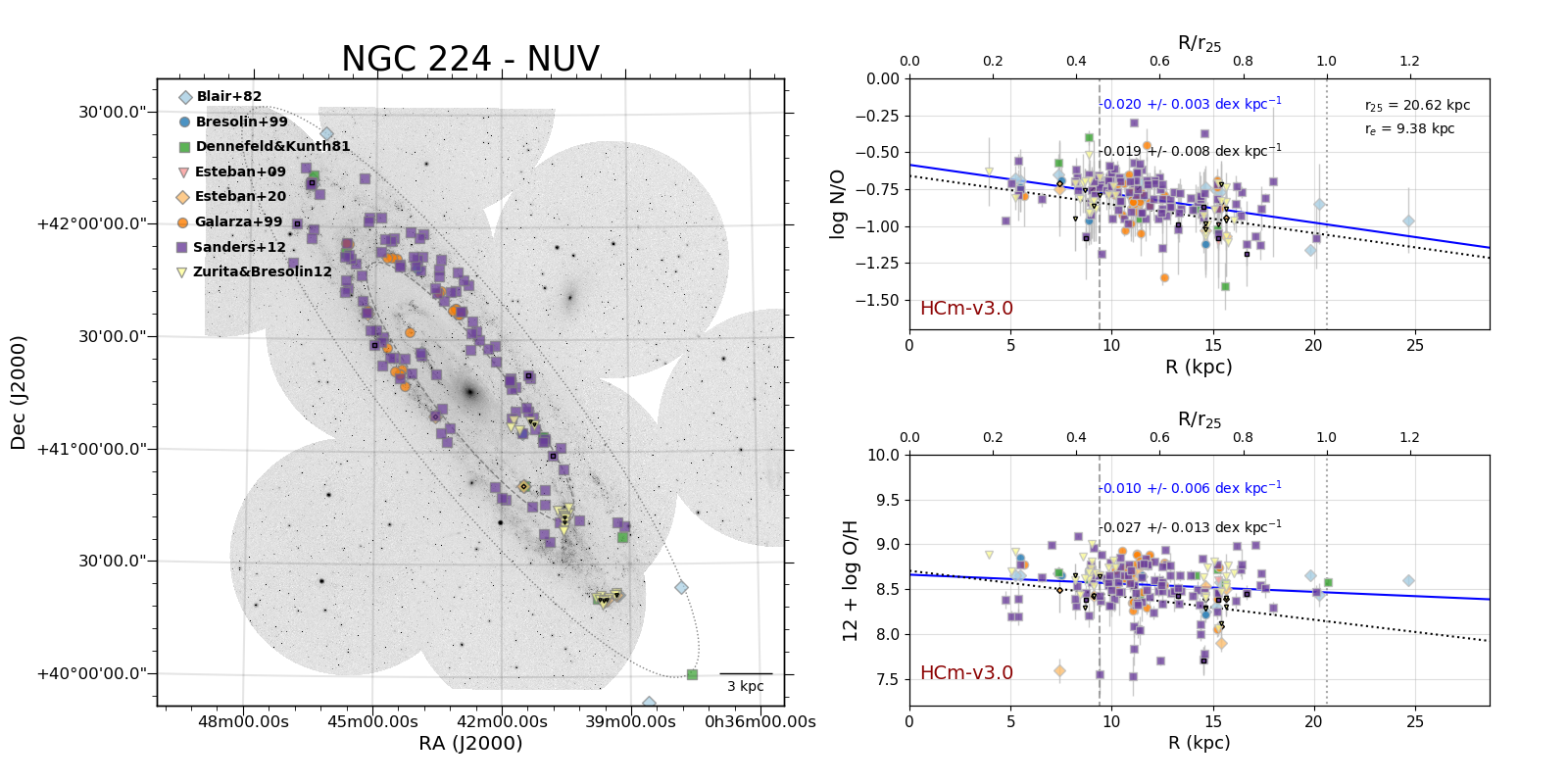}
\end{minipage}
\begin{minipage}{1.05\textwidth}
\hspace{-1.2cm}\includegraphics[width=1.12\textwidth]{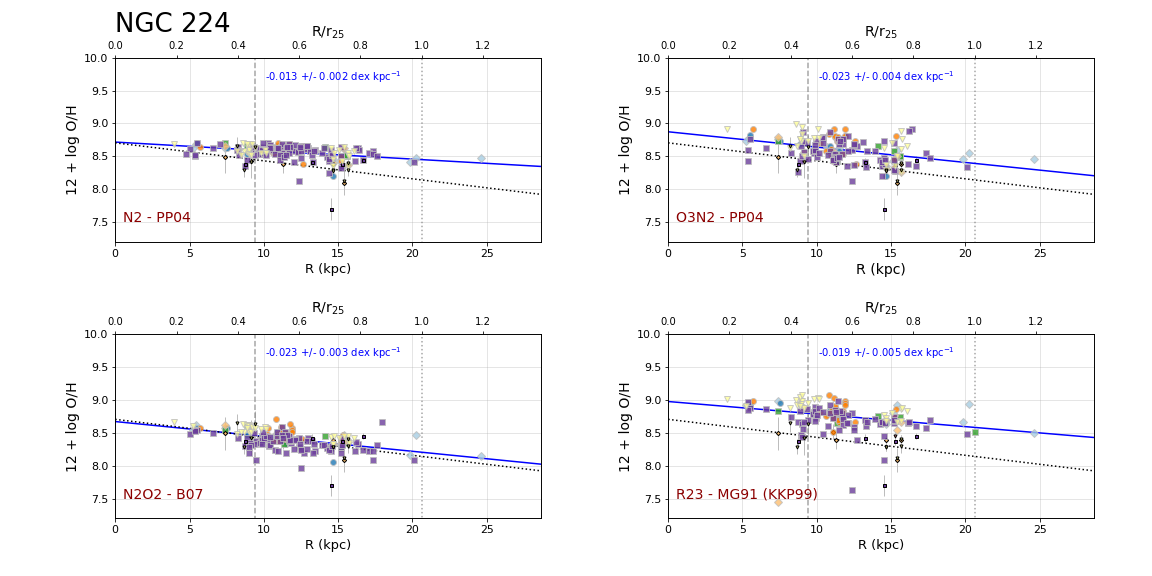}
\end{minipage}
\caption{(Top-left) Image of NGC~224 (M31) showing the location of the \hii\ regions compiled  for this work with different colours and symbols for the different authors. (Top-right) Radial $\log$(N/O) and $12+\log$(O/H) abundance profile for abundances obtained from the HII-CHI-mistry code \citep{epm14}. (Bottom)  Oxygen abundance profile as determined from different {\em strong-line} methods: N2 and O3N2  as calibrated by \citet{pp04}, N2O2 as calibrated by \citet{b07} and R$_{23}$ as calibrated from theoretical  models by \citet{m91} through the \citet{kuzio} parametrization. See the text for further information on the plots.\label{Afig1}}
\end{figure*}

\begin{figure*}
\begin{minipage}{1.05\textwidth}
\hspace{-1.2cm}\includegraphics[width=1.12\textwidth]{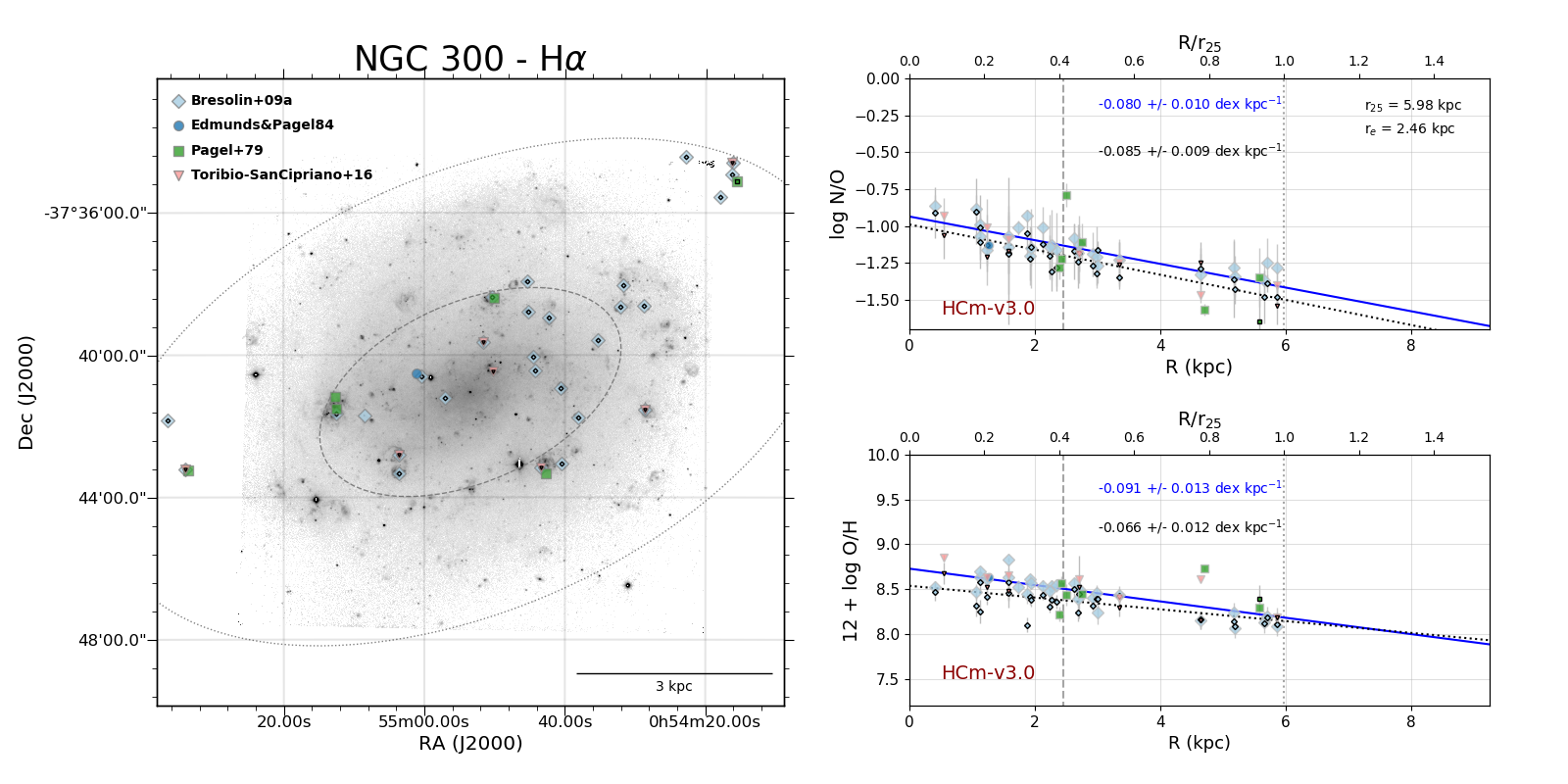}
\end{minipage}
\begin{minipage}{1.05\textwidth}
\hspace{-1.2cm}\includegraphics[width=1.12\textwidth]{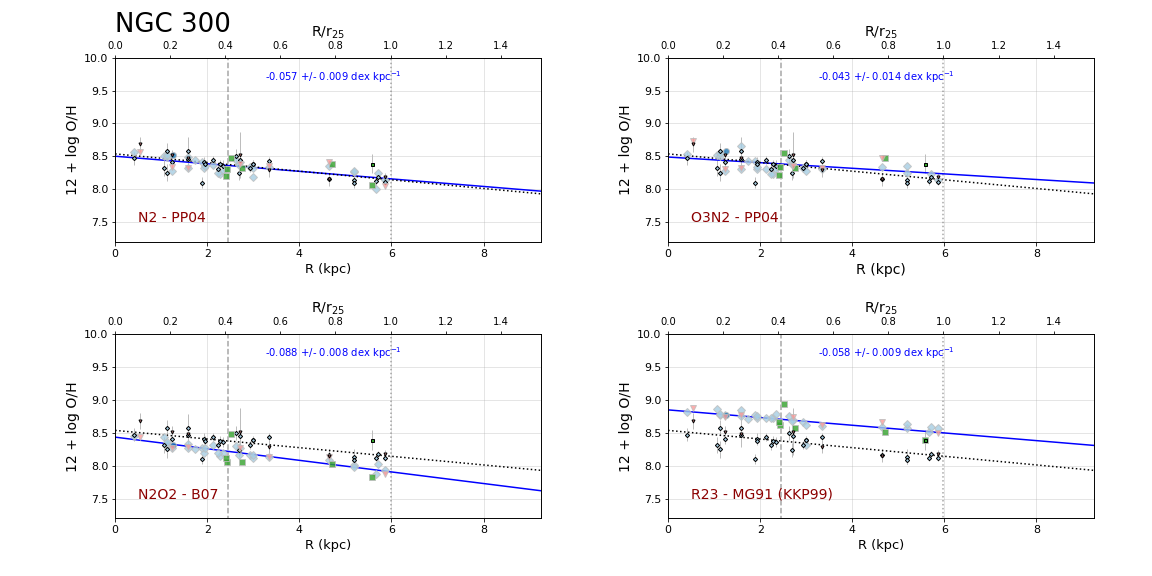}
\caption{\label{Afig2} Same as Fig.~\ref{Afig1} for NGC~300.}
\end{minipage}
\end{figure*}

\begin{figure*}
\begin{minipage}{1.05\textwidth}
\hspace{-1.2cm}\includegraphics[width=1.12\textwidth]{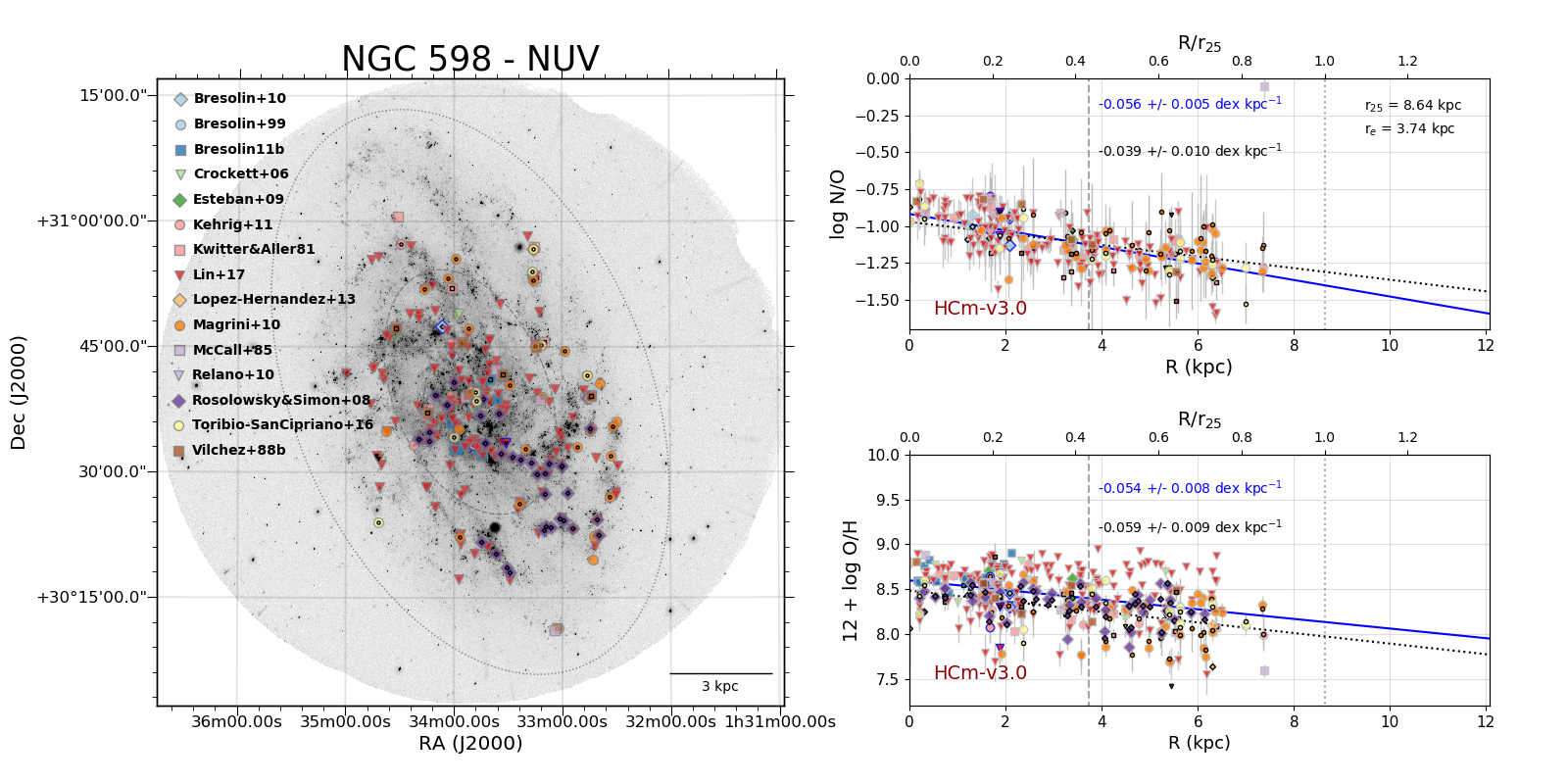}\\
\end{minipage}
\begin{minipage}{1.05\textwidth}
\hspace{-1.2cm}\includegraphics[width=1.122\textwidth]{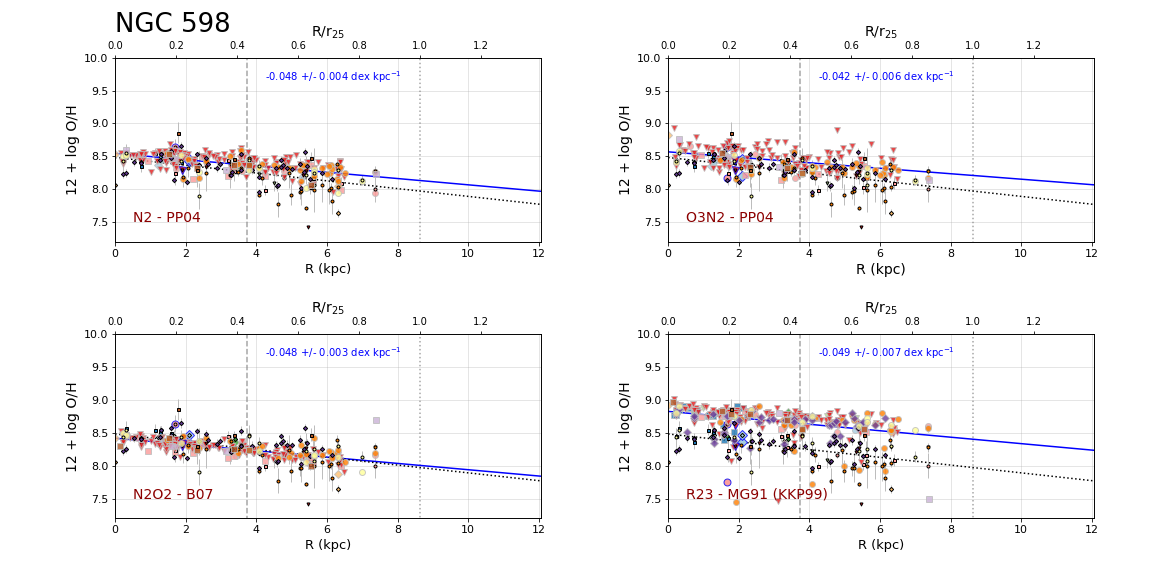}
\caption{\label{Afig3} Same as Fig.~\ref{Afig1} for NGC~598 (M33).}
\end{minipage}
\end{figure*}

\begin{figure*}
\begin{minipage}{1.05\textwidth}
\hspace{-1.2cm}\includegraphics[width=1.12\textwidth]{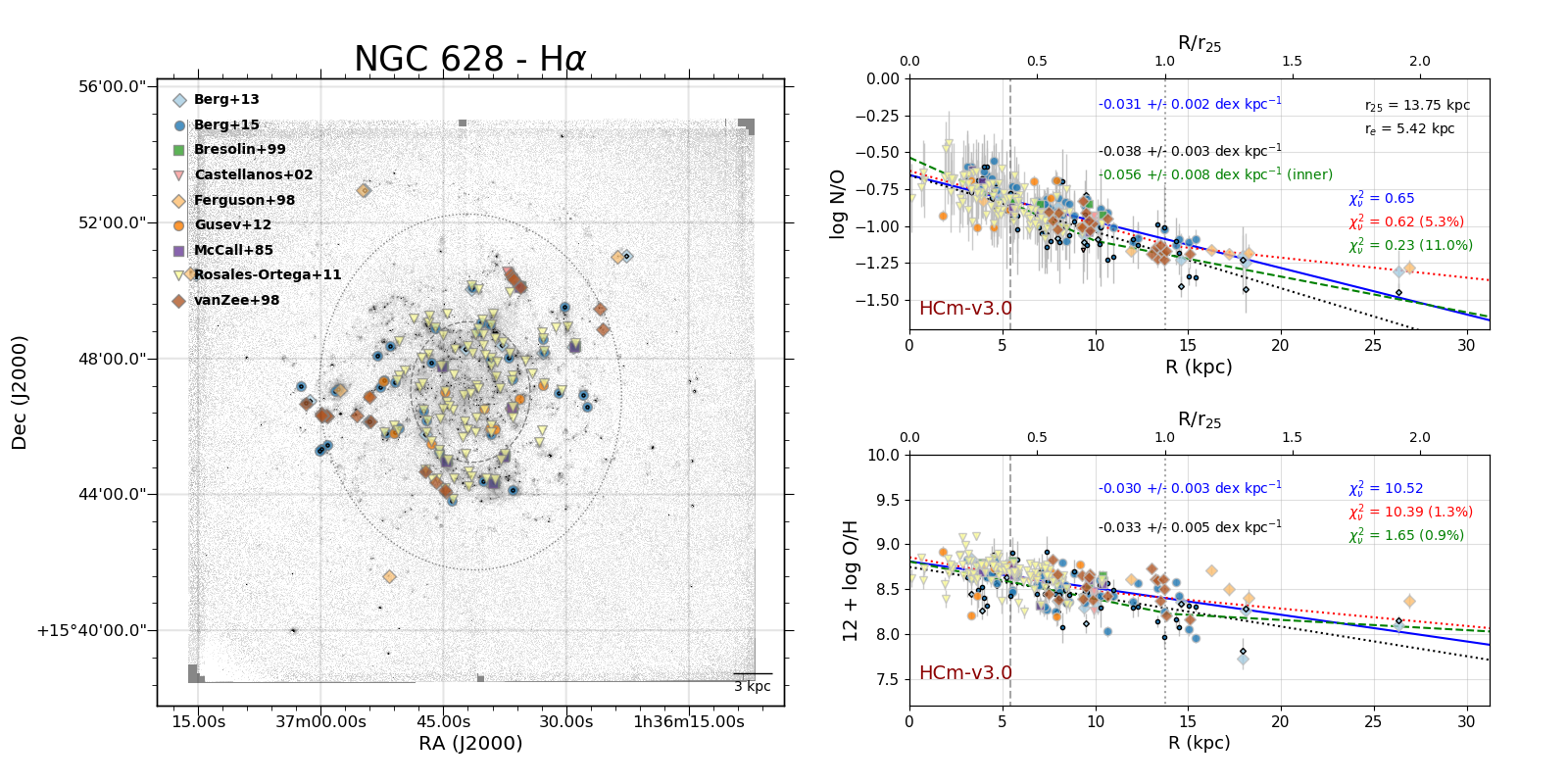}
\end{minipage}
\begin{minipage}{1.05\textwidth}
\hspace{-1.2cm}\includegraphics[width=1.12\textwidth]{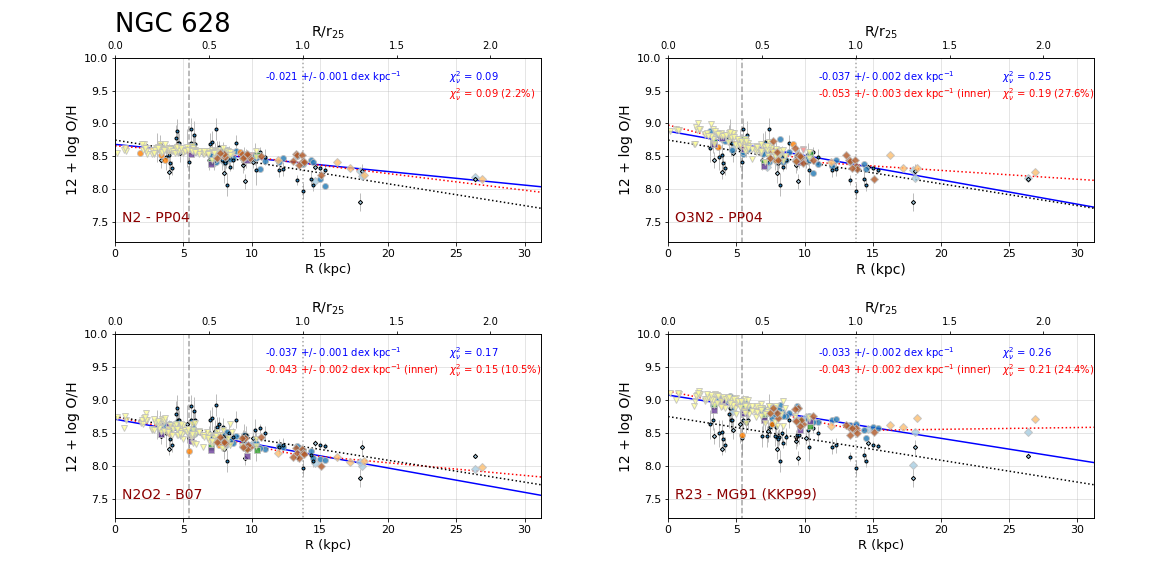}
 \end{minipage}
\caption{\label{Afig4} Same as Fig.~\ref{Afig1} for NGC~628 (M74).}
\end{figure*}

\begin{figure*}
  \begin{minipage}{1.05\textwidth}
  \hspace{-1.2cm}\includegraphics[width=1.12\textwidth]{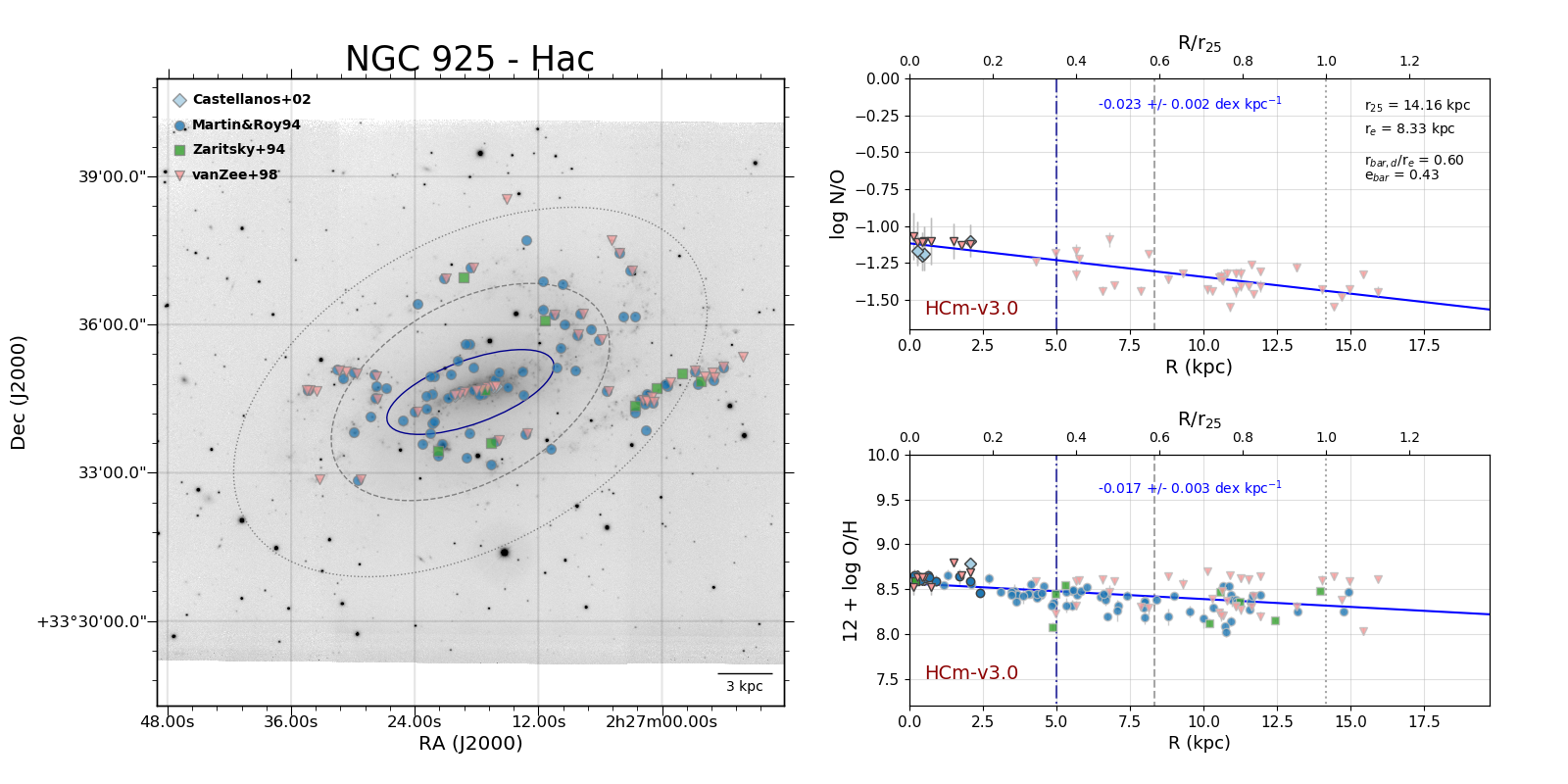}
  \end{minipage}
  \begin{minipage}{1.05\textwidth}
  \hspace{-1.2cm}\includegraphics[width=1.12\textwidth]{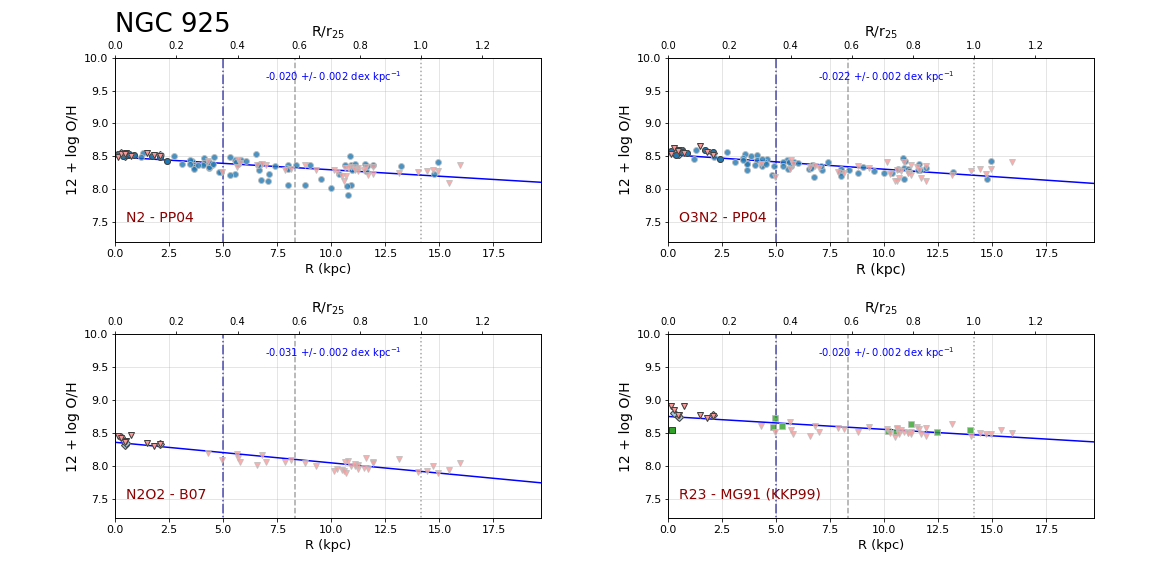}
  \end{minipage}
\caption{\label{Afig6}Same as Fig.~\ref{Afig1} for NGC~925.}
\end{figure*}

\begin{figure*}
  \includegraphics[width=1.12\textwidth]{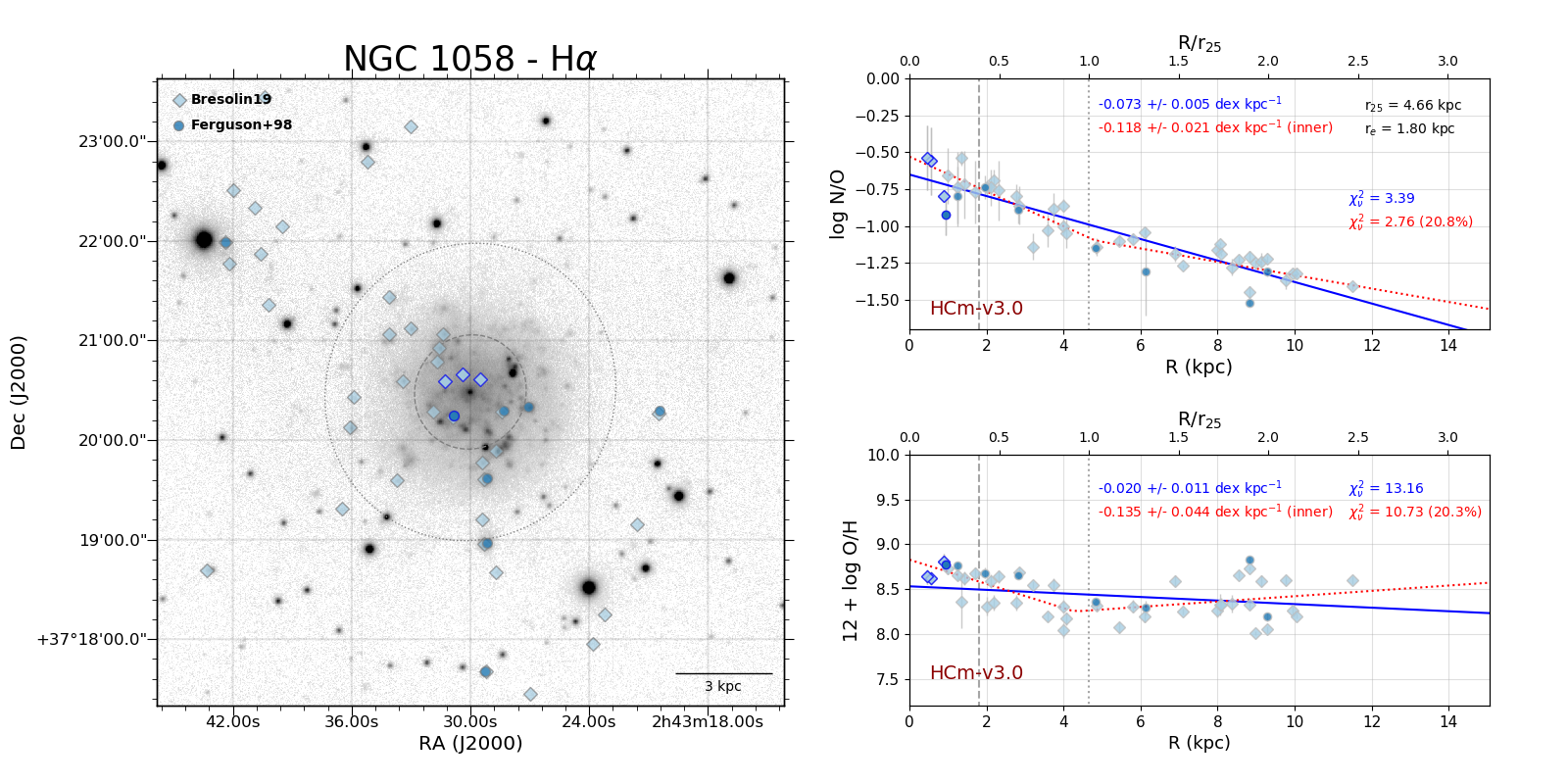}
  \includegraphics[width=1.12\textwidth]{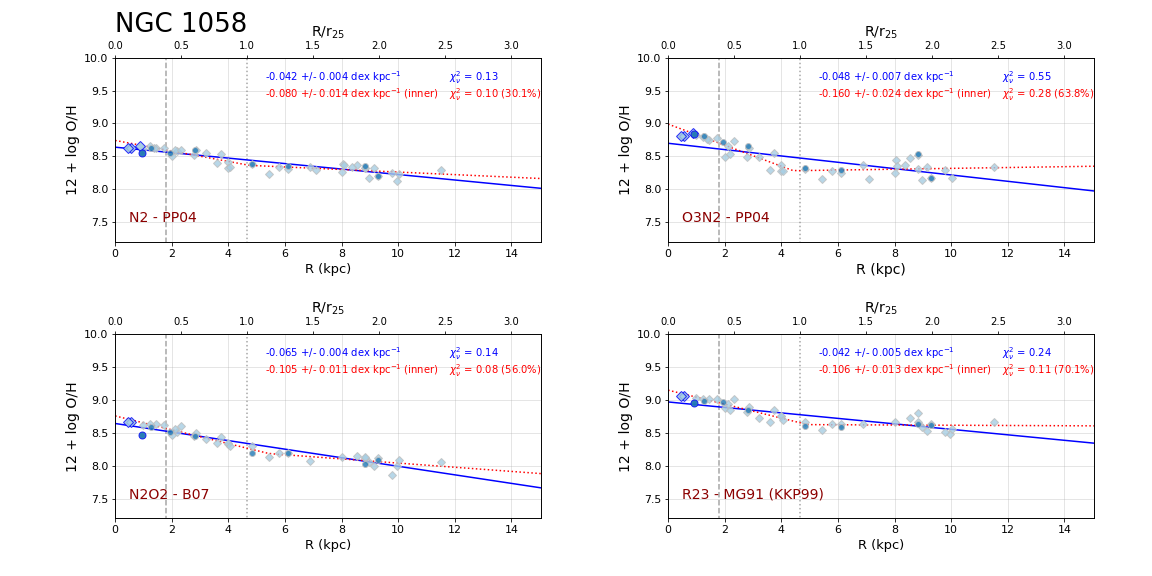}
\caption{}\label{Afig7}
\end{figure*}

\begin{figure*}
\begin{minipage}{1.05\textwidth}
\hspace{-1.2cm}\includegraphics[width=1.12\textwidth]{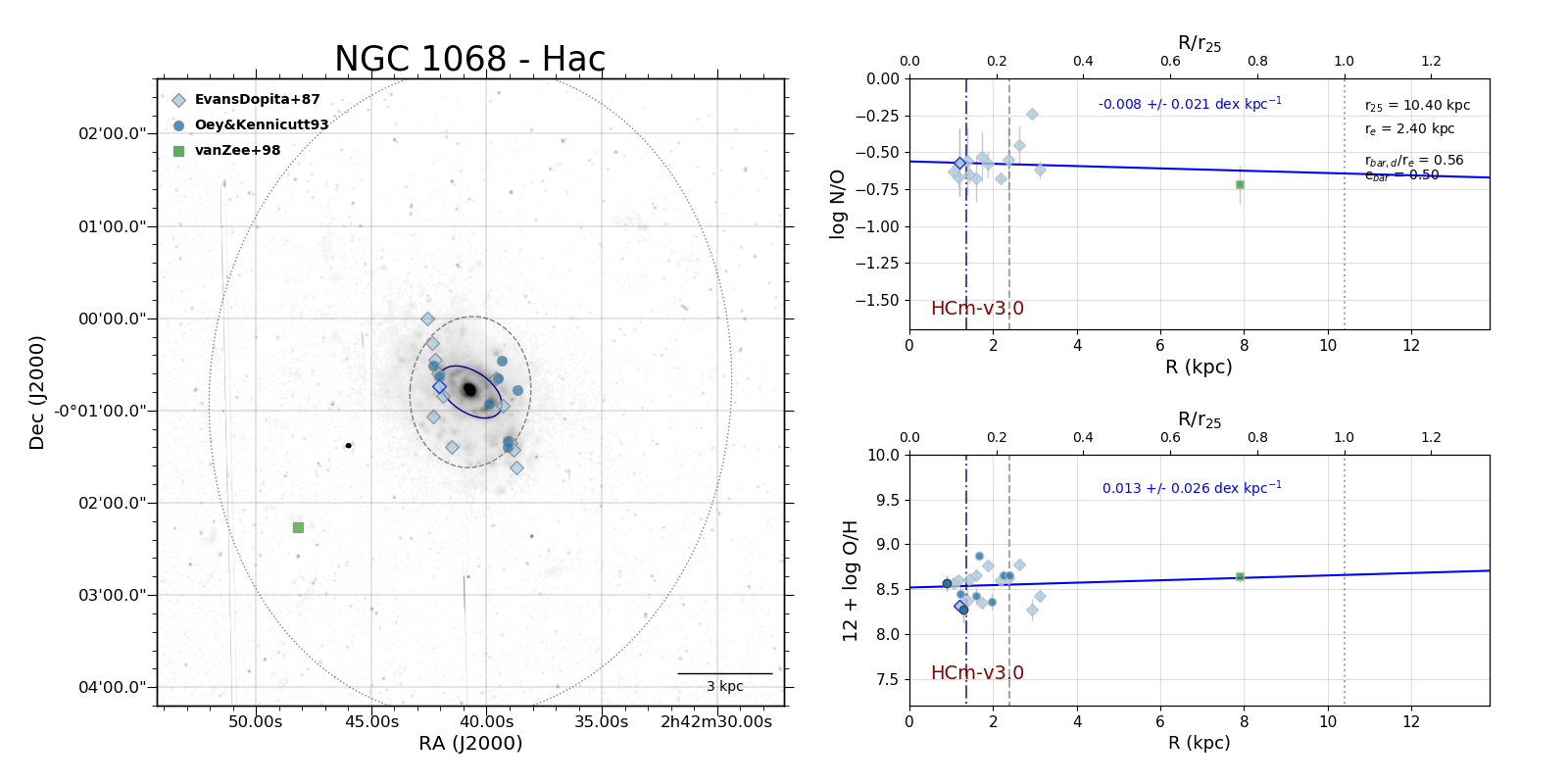}
\end{minipage}
\begin{minipage}{1.05\textwidth}
\hspace{-1.2cm}\includegraphics[width=1.12\textwidth]{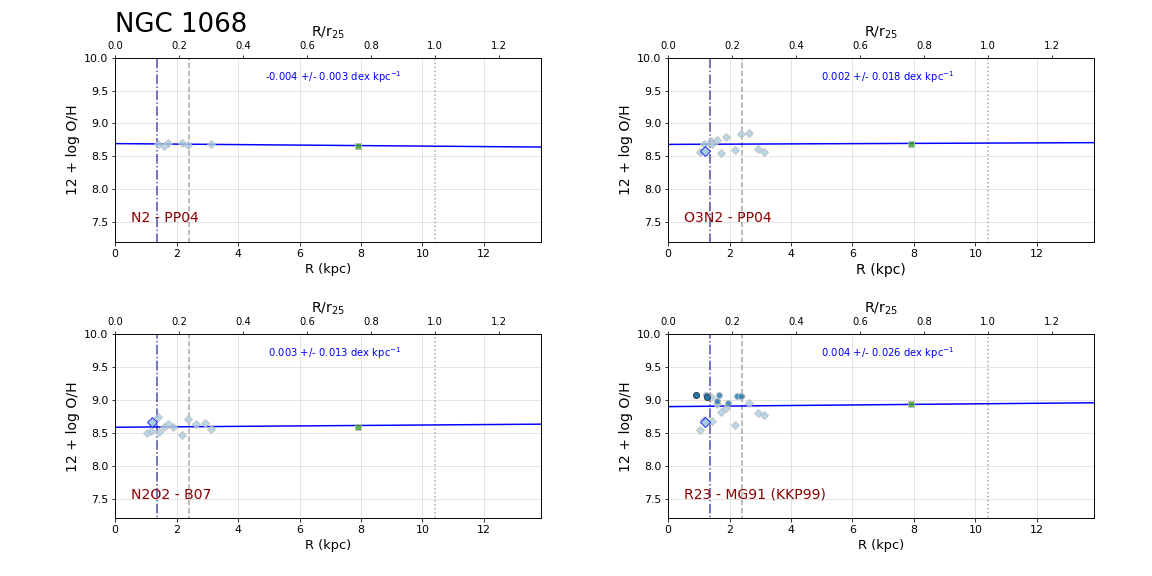}
\end{minipage}
\caption{Same as Fig.~\ref{Afig1} for NGC~1068.}\label{Afig8}
\end{figure*}

\begin{figure*}
\begin{minipage}{1.05\textwidth}
\hspace{-1.2cm}\includegraphics[width=1.12\textwidth]{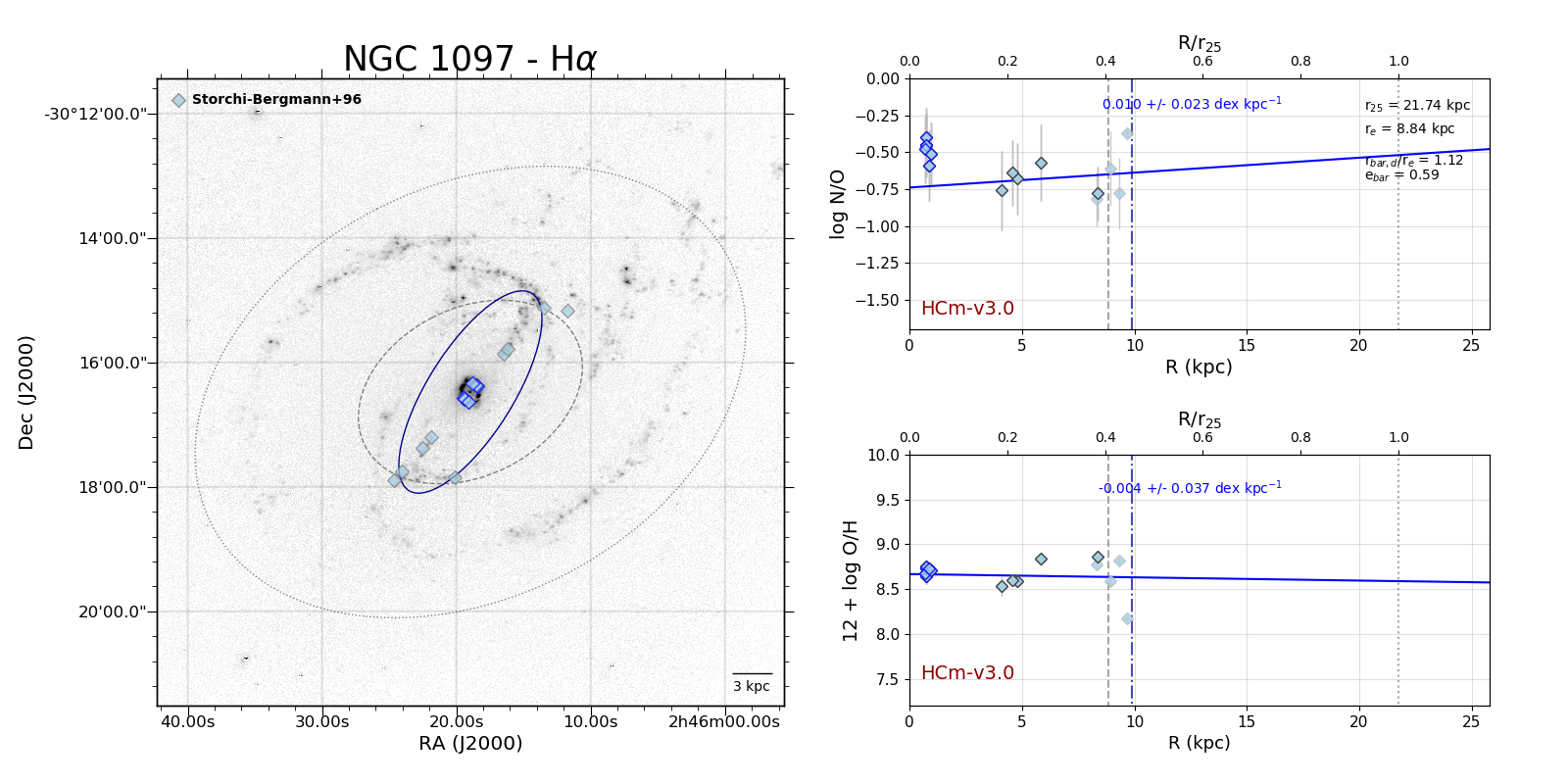}
\end{minipage}
\begin{minipage}{1.05\textwidth}
\hspace{-1.2cm}\includegraphics[width=1.12\textwidth]{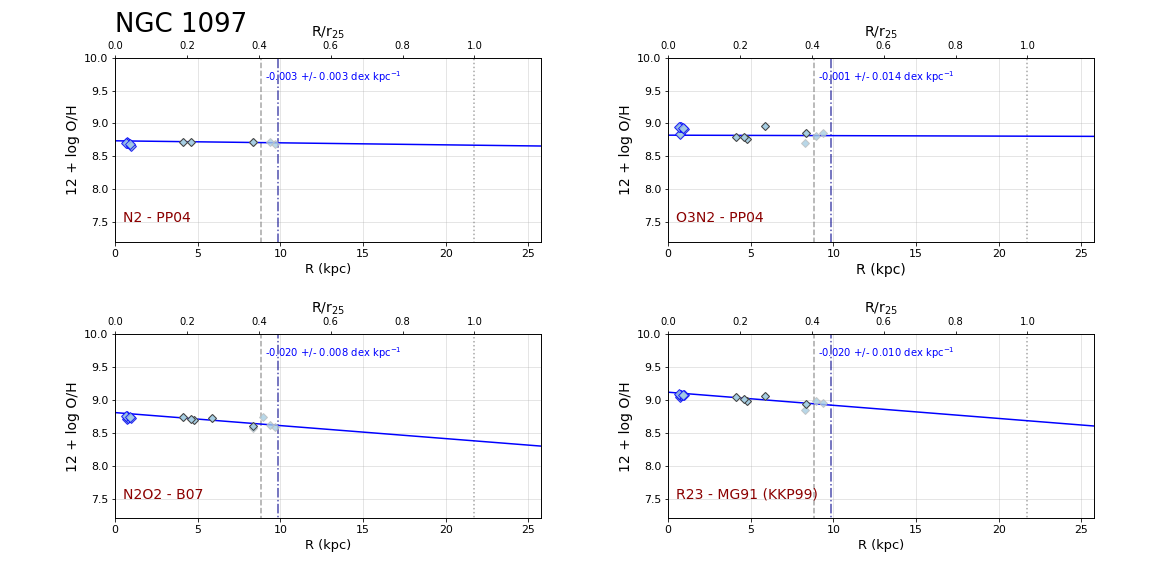}
\end{minipage}
\caption{Same as Fig.~\ref{Afig1} for NGC~1097.}\label{Afig9}
\end{figure*}

\begin{figure*}
\begin{minipage}{1.05\textwidth}
\hspace{-1.2cm}\includegraphics[width=1.12\textwidth]{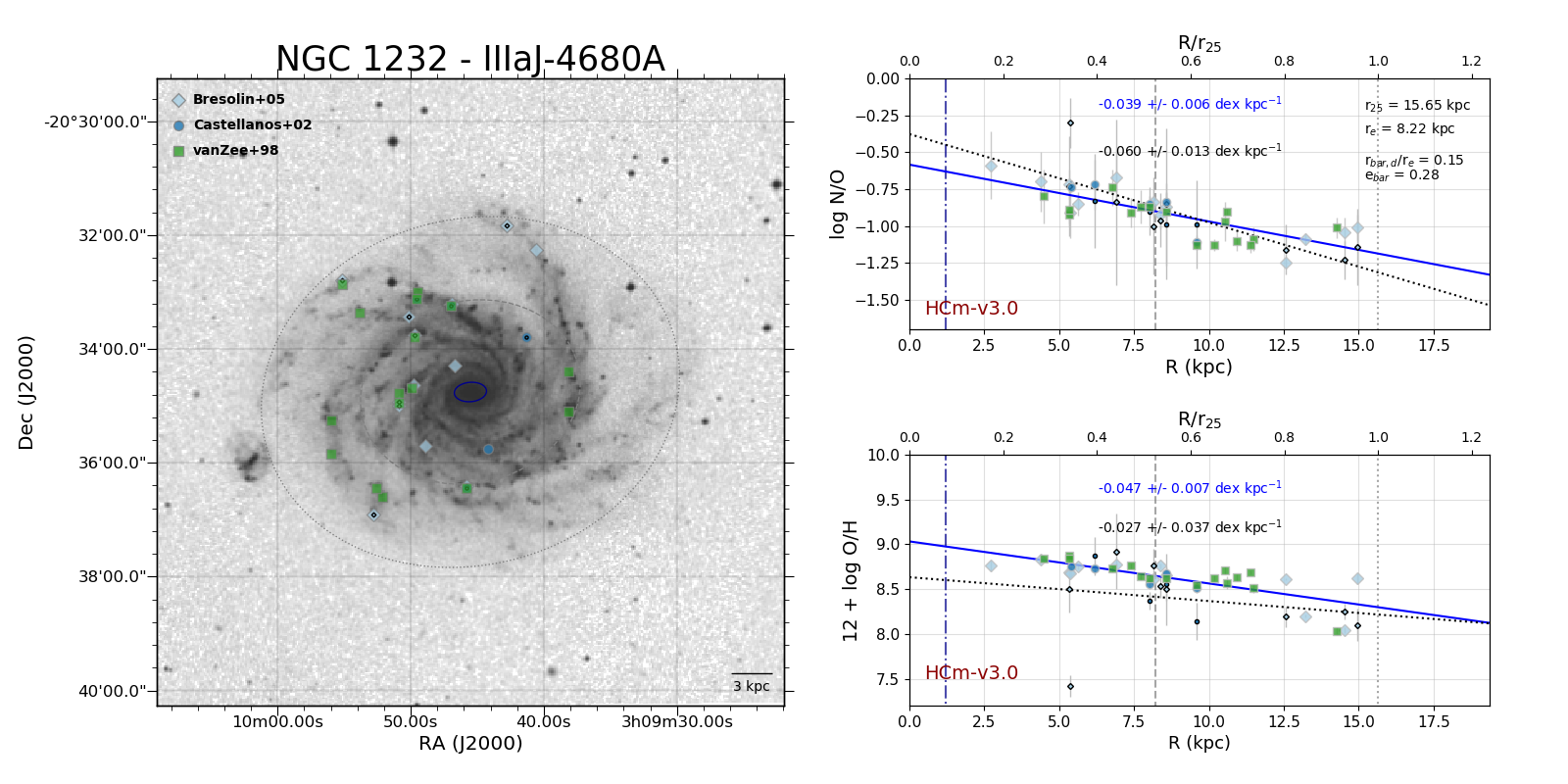}
\end{minipage}
\begin{minipage}{1.05\textwidth}
\hspace{-1.2cm}\includegraphics[width=1.12\textwidth]{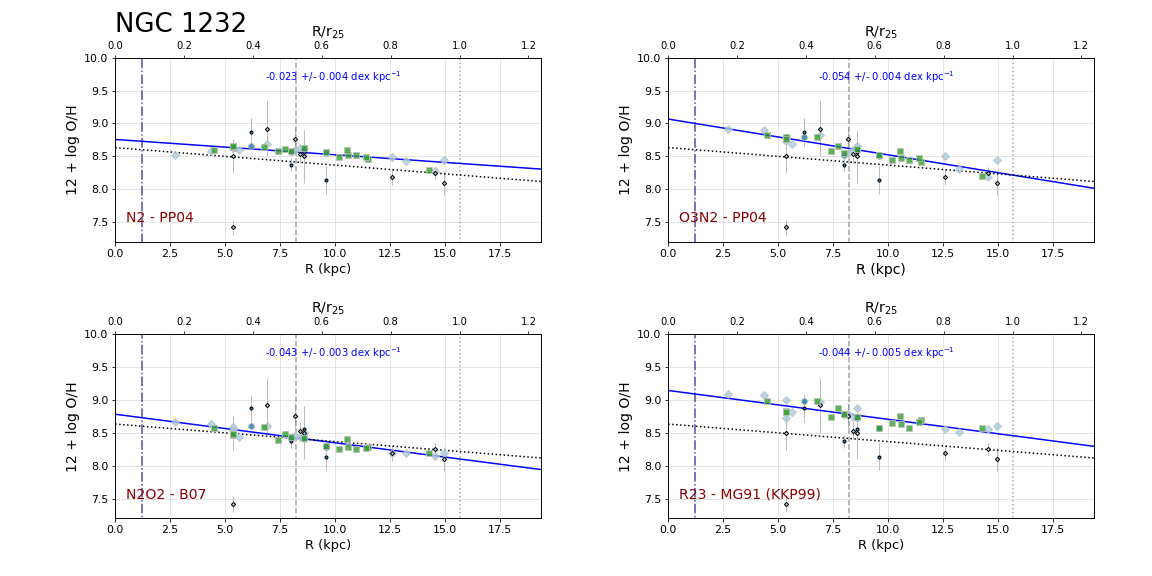}
\end{minipage}
\caption{Same as Fig.~\ref{Afig1} for NGC~1232.}\label{Afig10}
\end{figure*}

\begin{figure*}
\begin{minipage}{1.05\textwidth}
\hspace{-1.2cm}\includegraphics[width=1.12\textwidth]{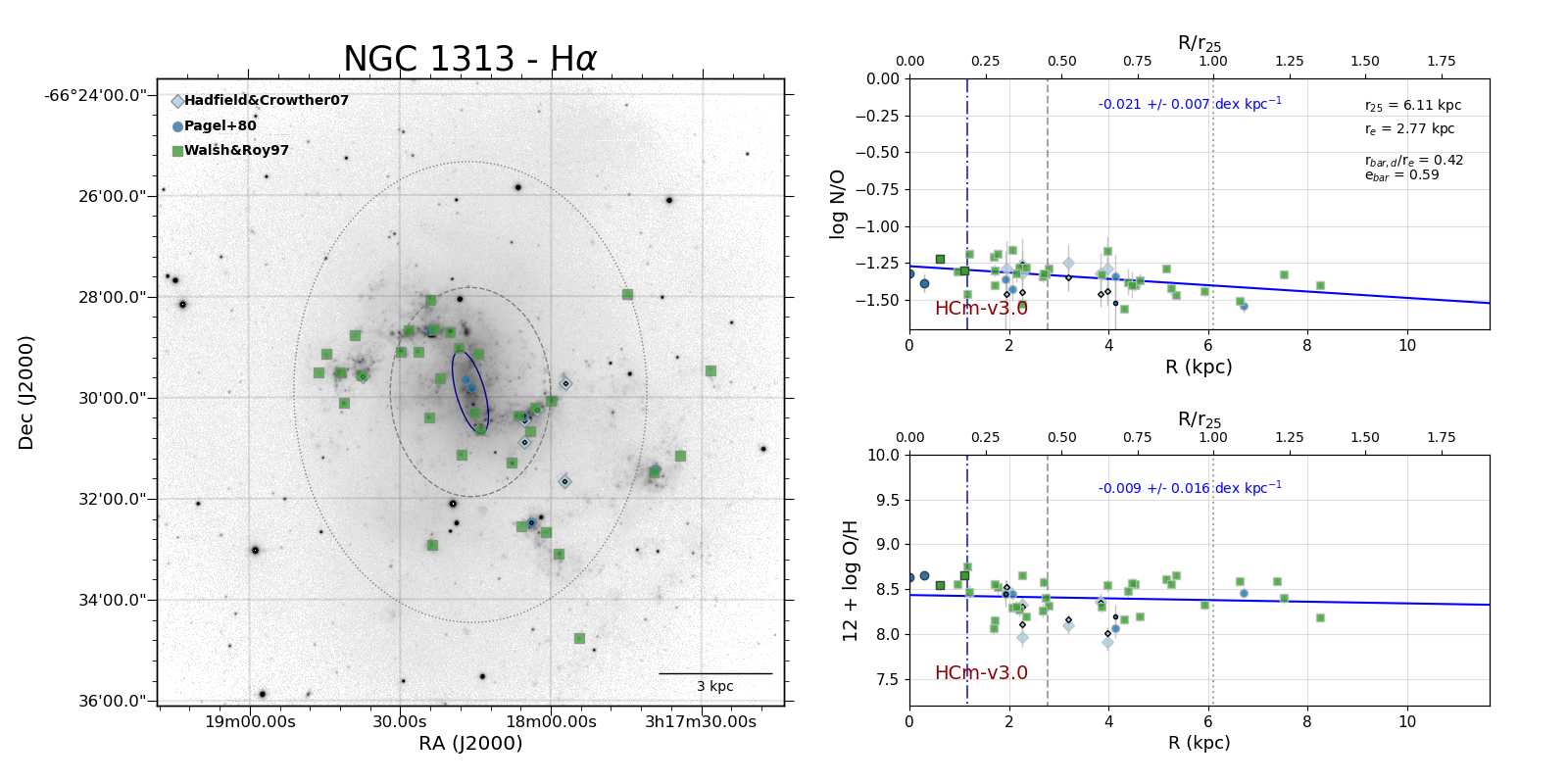}
\end{minipage}
\begin{minipage}{1.05\textwidth}
\hspace{-1.2cm}\includegraphics[width=1.12\textwidth]{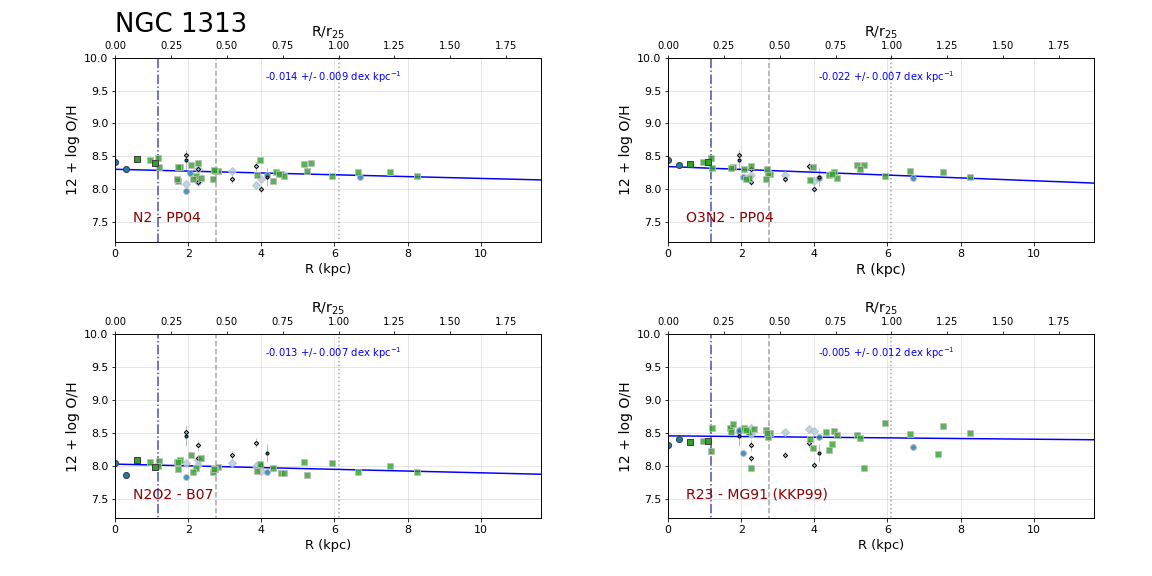}
\end{minipage}
\caption{Same as Fig.~\ref{Afig1} for NGC~1313.}\label{Afig11}
\end{figure*}

\begin{figure*}
\begin{minipage}{1.05\textwidth}
\hspace{-1.2cm}\includegraphics[width=1.12\textwidth]{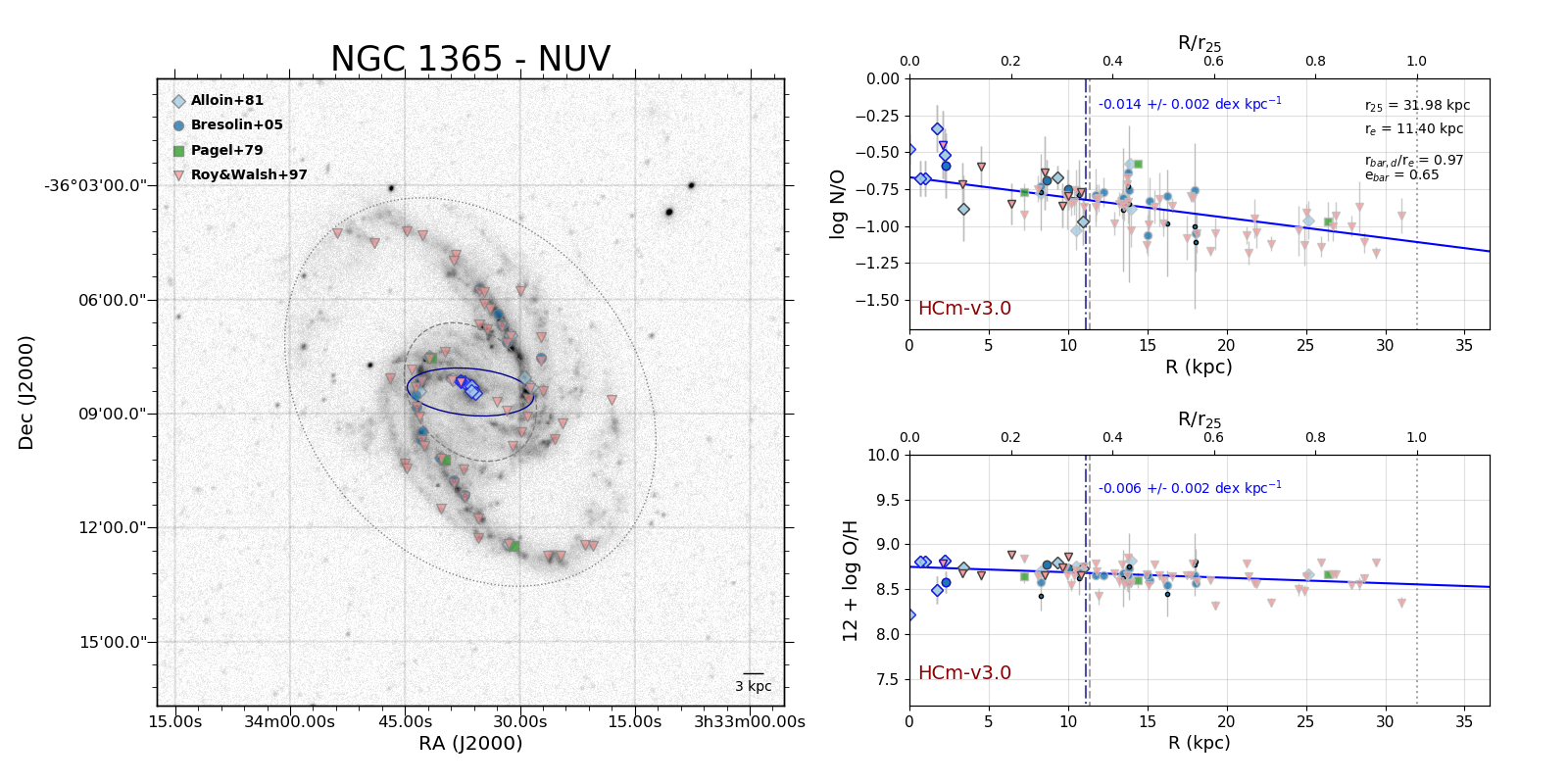}
\end{minipage}
\begin{minipage}{1.05\textwidth}
\hspace{-1.2cm}\includegraphics[width=1.12\textwidth]{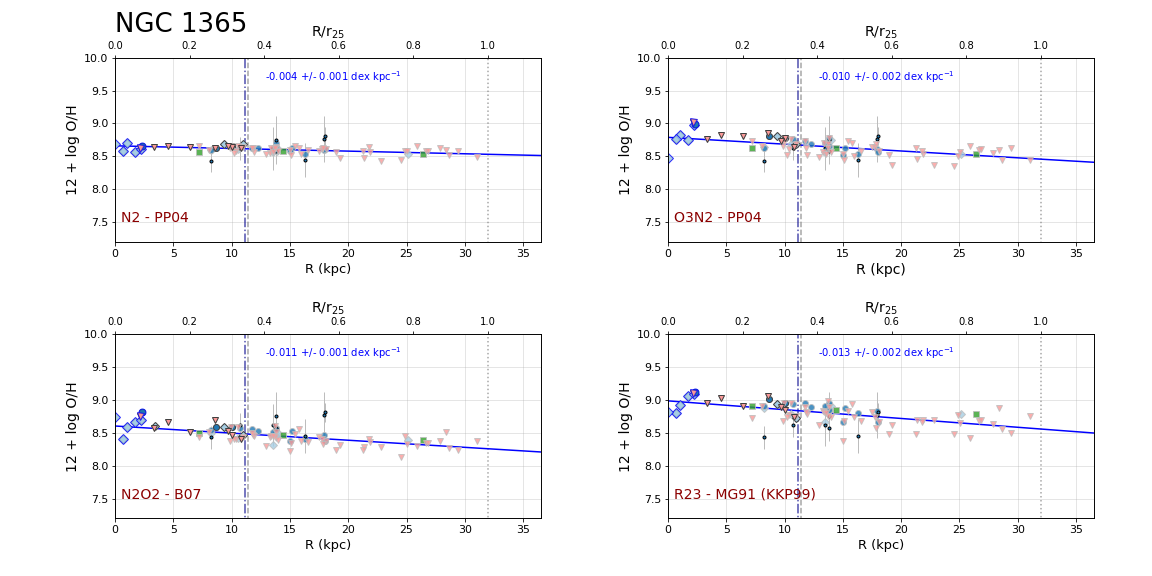}
\end{minipage}
\caption{Same as Fig.~\ref{Afig1} for NGC~1365.}\label{Afig12}
\end{figure*}

\begin{figure*}
\begin{minipage}{1.05\textwidth}
\hspace{-1.2cm}\includegraphics[width=1.12\textwidth]{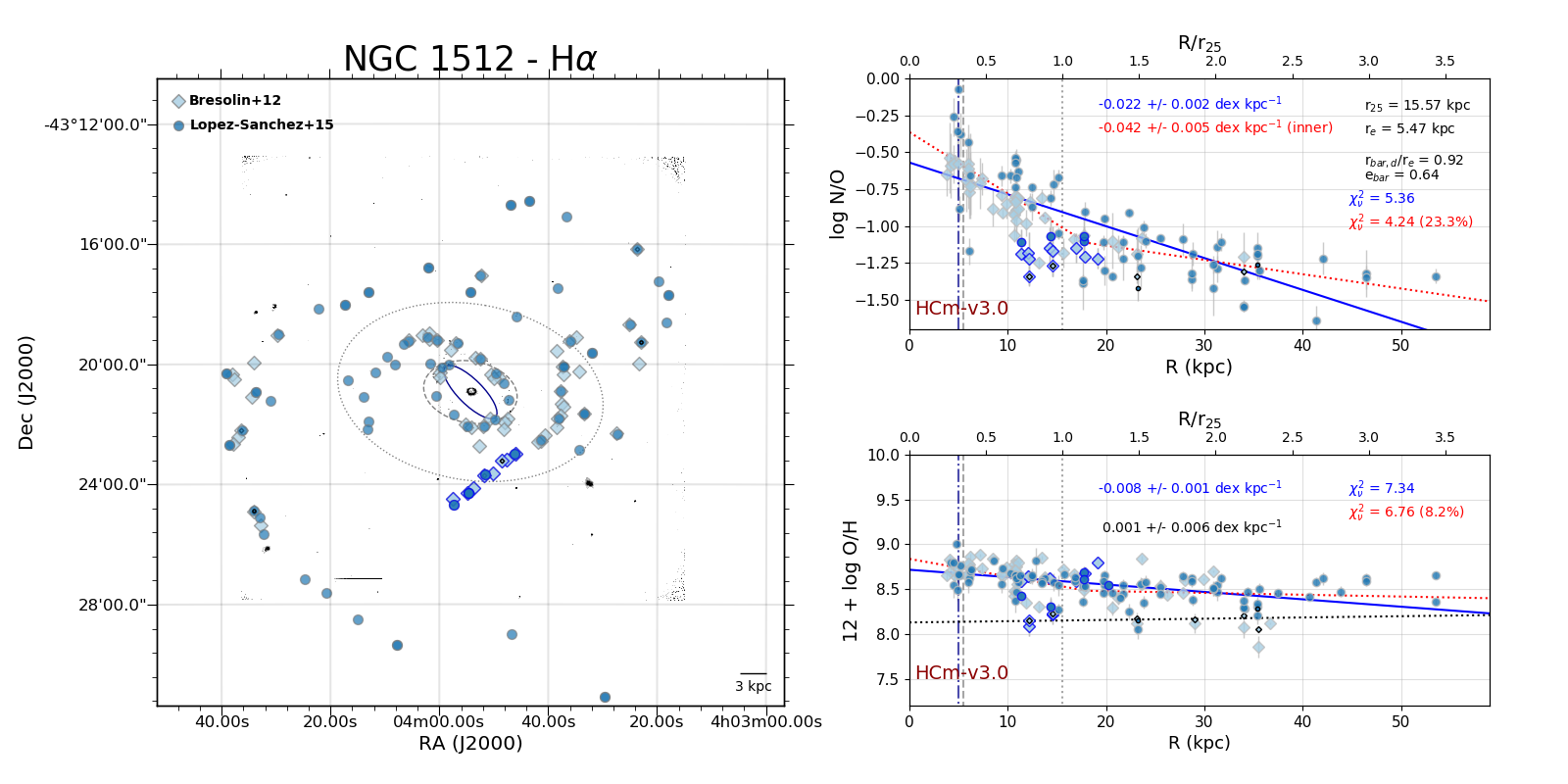}
\end{minipage}
\begin{minipage}{1.05\textwidth}
\hspace{-1.2cm}\includegraphics[width=1.12\textwidth]{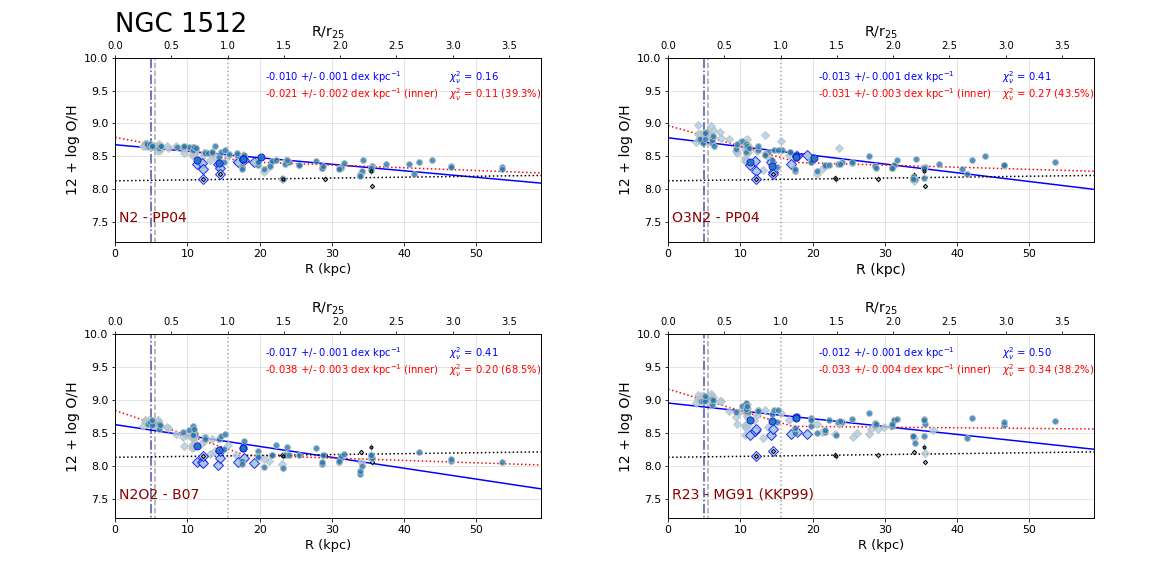}
\end{minipage}
\caption{Same as Fig.~\ref{Afig1} for NGC~1512.}\label{Afig13}
\end{figure*}

\begin{figure*}
\begin{minipage}{1.05\textwidth}
\hspace{-1.2cm}\includegraphics[width=1.12\textwidth]{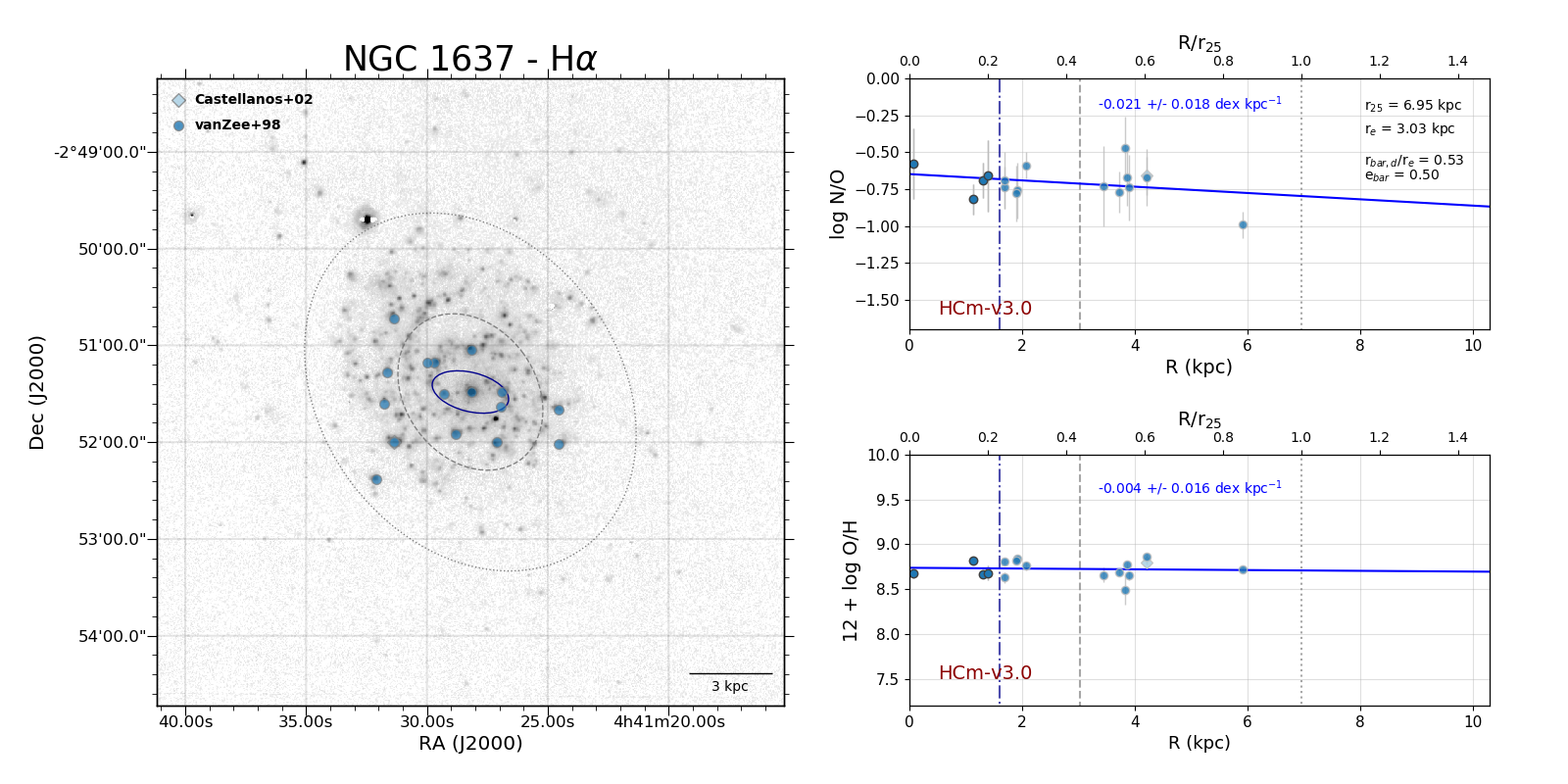}
\end{minipage}
\begin{minipage}{1.05\textwidth}
\hspace{-1.2cm}\includegraphics[width=1.12\textwidth]{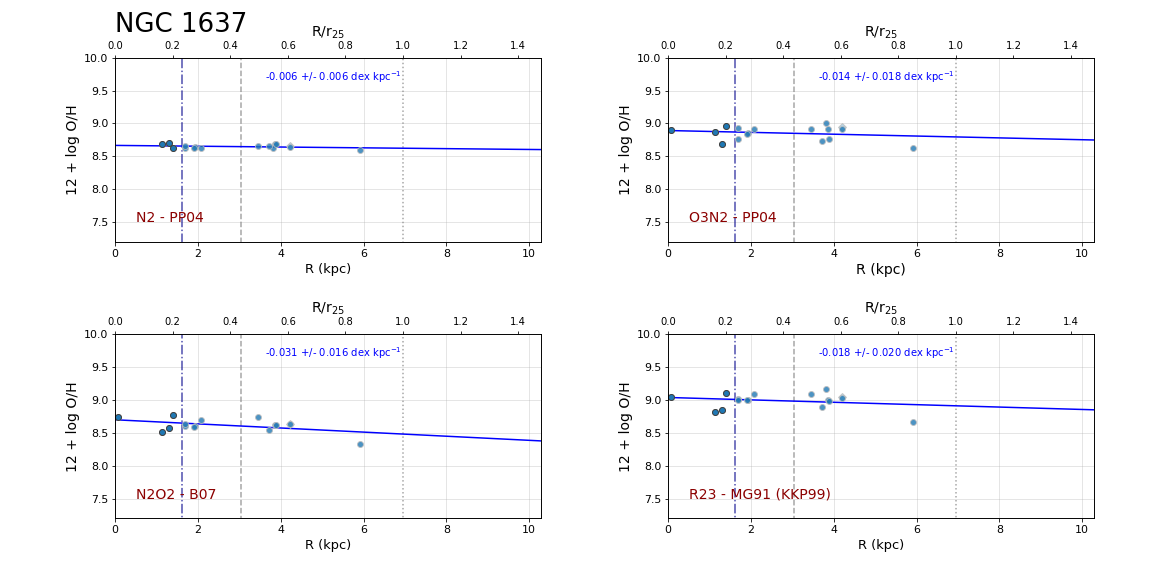}
\end{minipage}
\caption{Same as Fig.~\ref{Afig1} for NGC~1637.}\label{Afig14}
\end{figure*}

\begin{figure*}
\begin{minipage}{1.05\textwidth}
\hspace{-1.2cm}\includegraphics[width=1.12\textwidth]{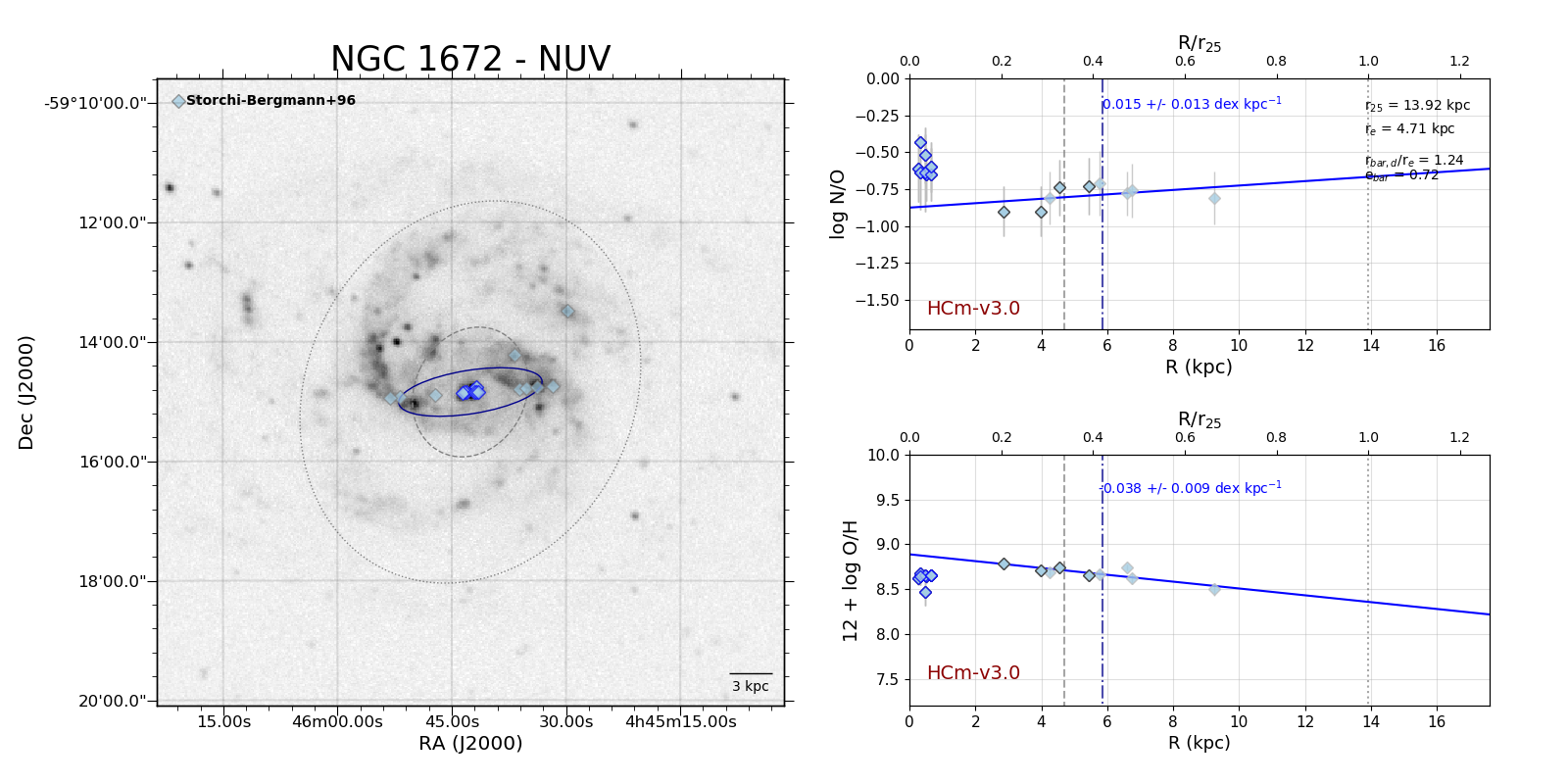}
\end{minipage}
\begin{minipage}{1.05\textwidth}
\hspace{-1.2cm}\includegraphics[width=1.12\textwidth]{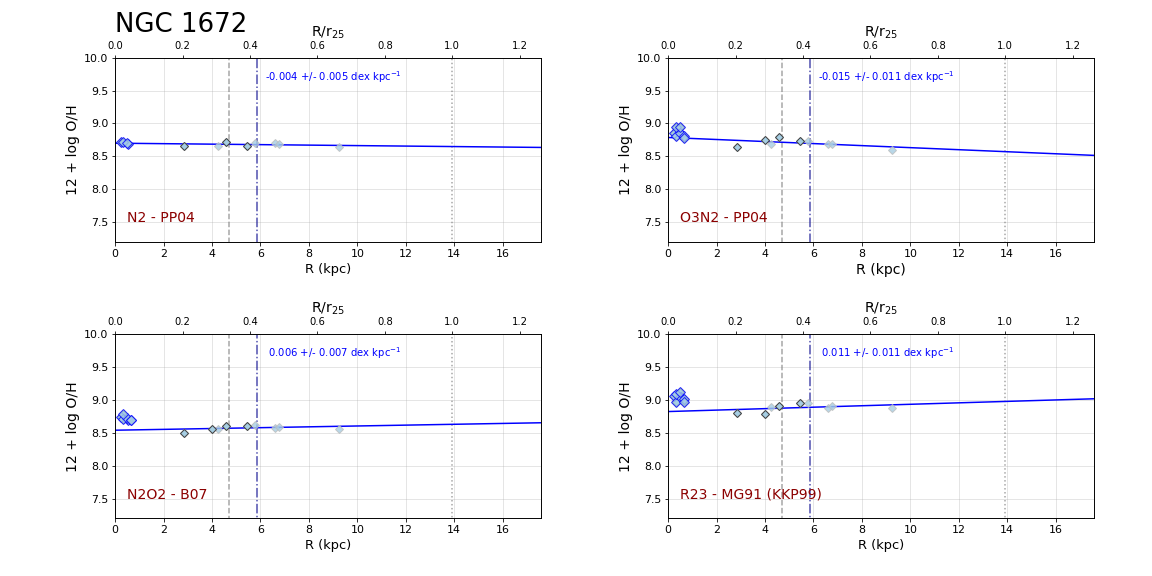}
\end{minipage}
\caption{Same as Fig.~\ref{Afig1} for NGC~1672.}\label{Afig15}
\end{figure*}

\begin{figure*}
\begin{minipage}{1.05\textwidth}
\hspace{-1.2cm}\includegraphics[width=1.12\textwidth]{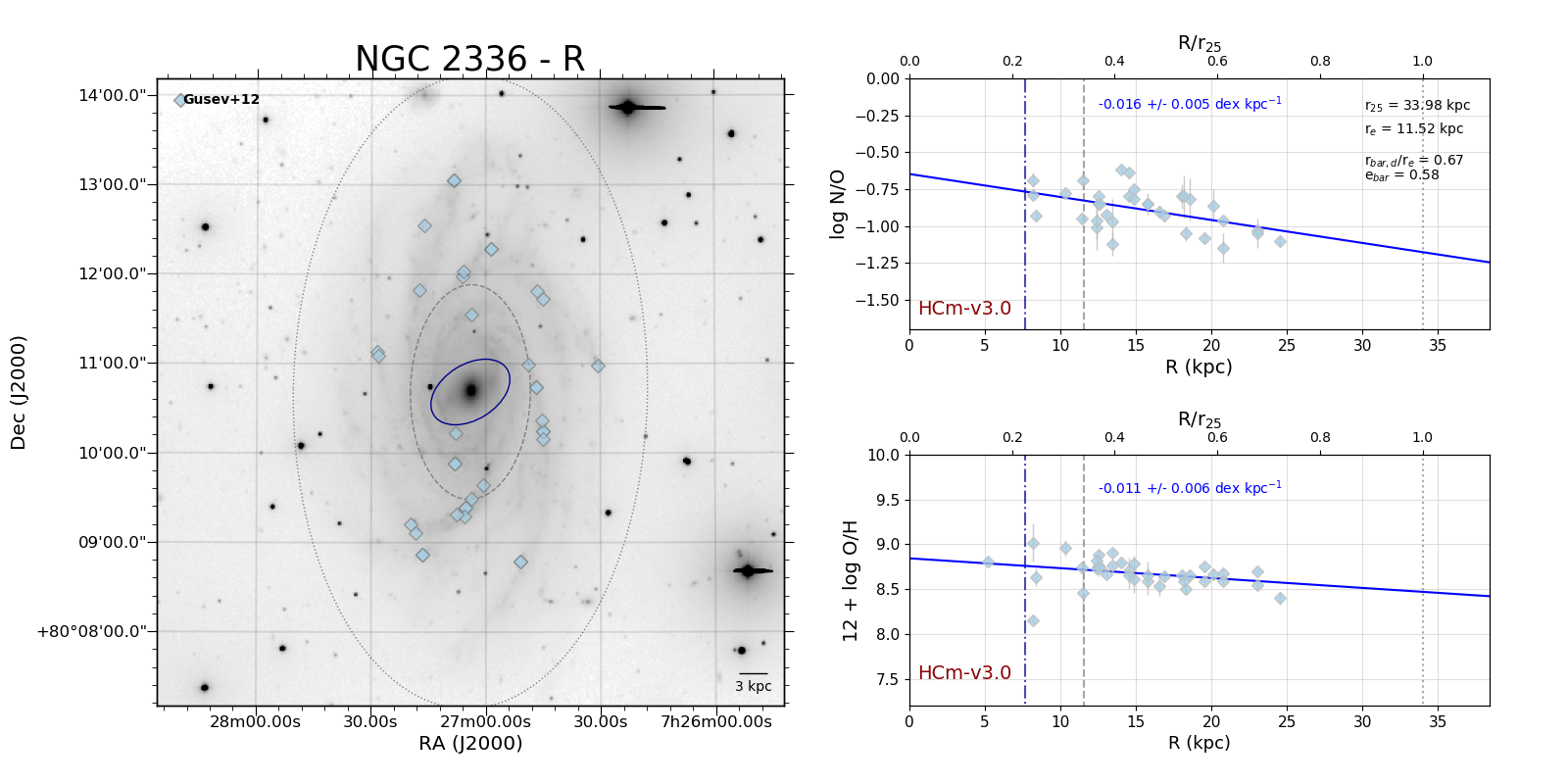}
\end{minipage}
\begin{minipage}{1.05\textwidth}
\hspace{-1.2cm}\includegraphics[width=1.12\textwidth]{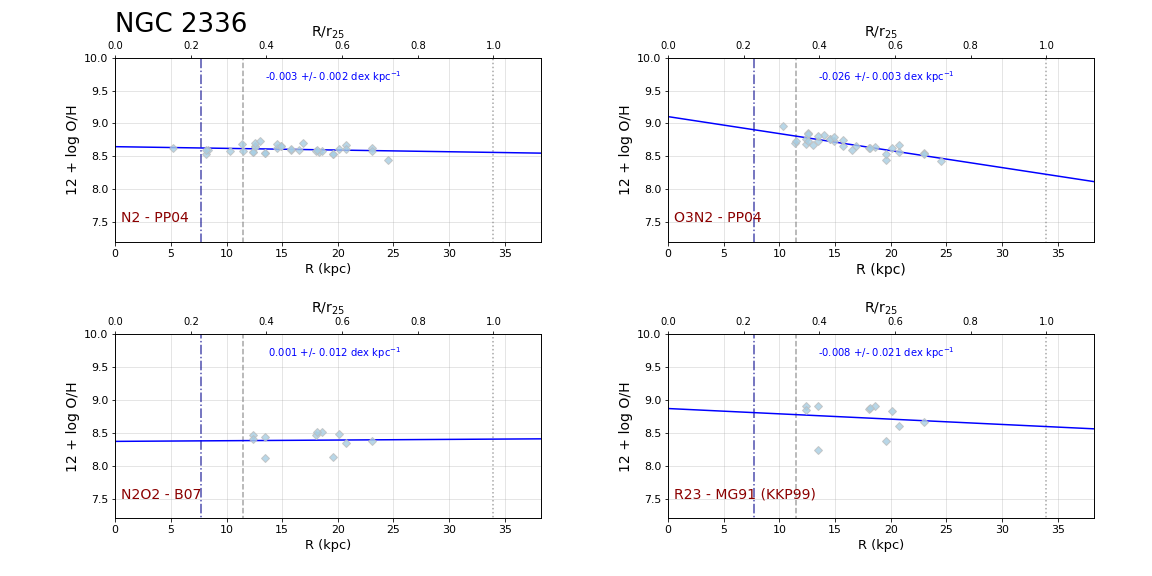}
\end{minipage}
\caption{Same as Fig.~\ref{Afig1} for NGC~2336.}\label{Afig16}
\end{figure*}

\begin{figure*}
\begin{minipage}{1.05\textwidth}
\hspace{-1.2cm}\includegraphics[width=1.12\textwidth]{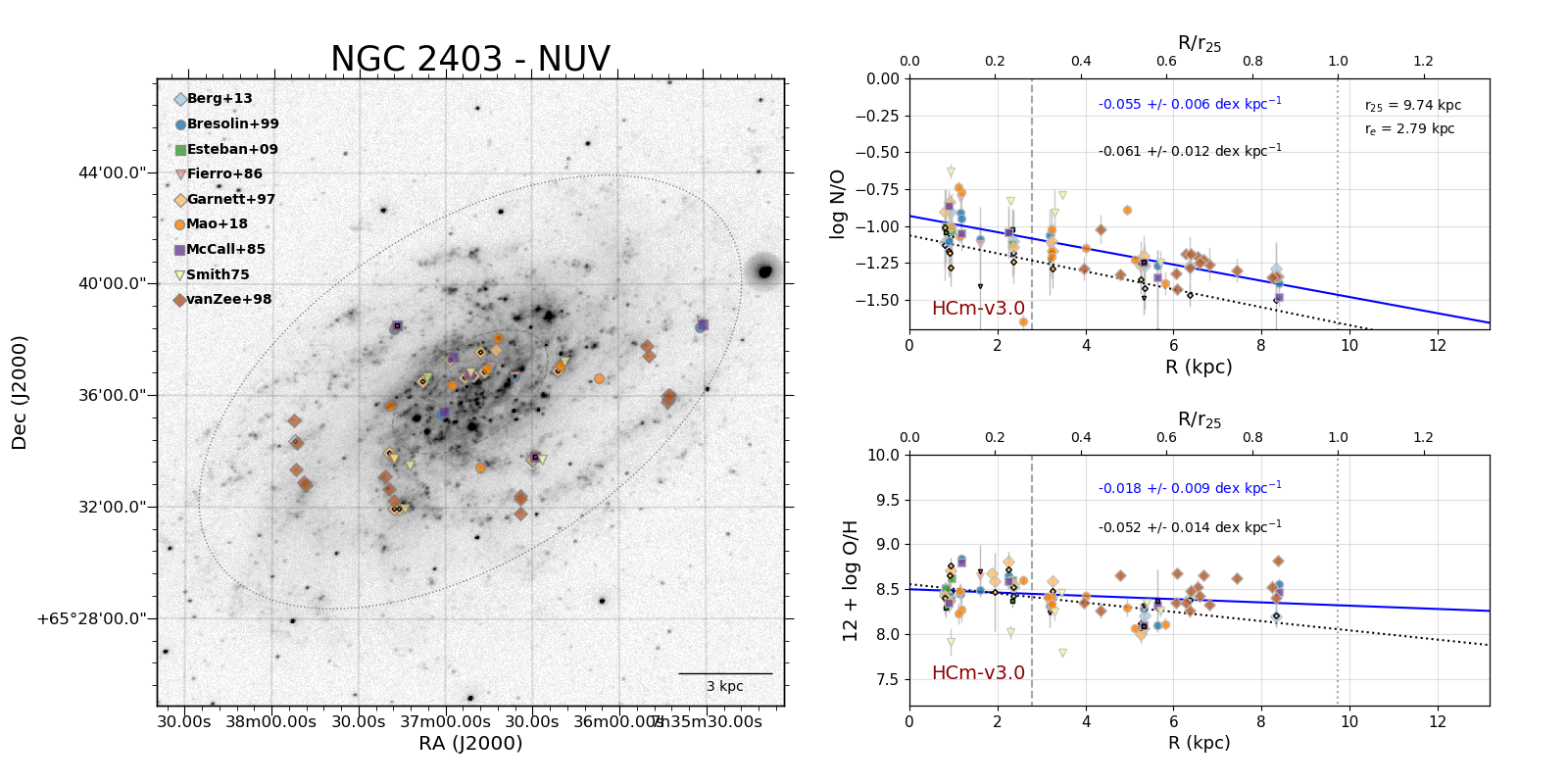}
\end{minipage}
\begin{minipage}{1.05\textwidth}
\hspace{-1.2cm}\includegraphics[width=1.12\textwidth]{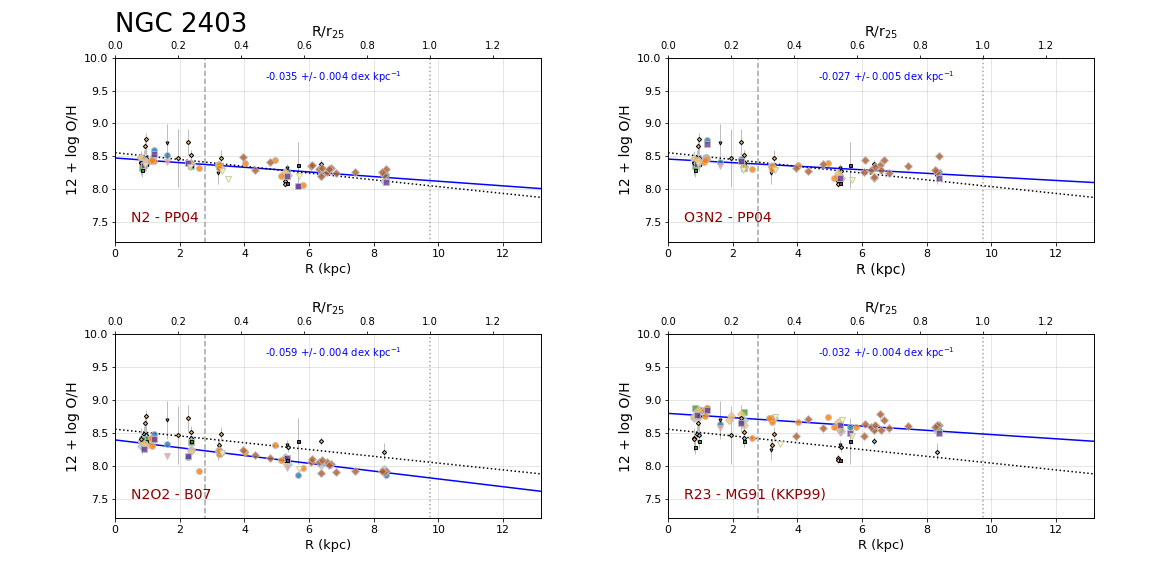}
\end{minipage}
\caption{Same as Fig.~\ref{Afig1} for NGC~2403.}\label{Afig17}
\end{figure*}

\begin{figure*}
\begin{minipage}{1.05\textwidth}
\hspace{-1.2cm}\includegraphics[width=1.12\textwidth]{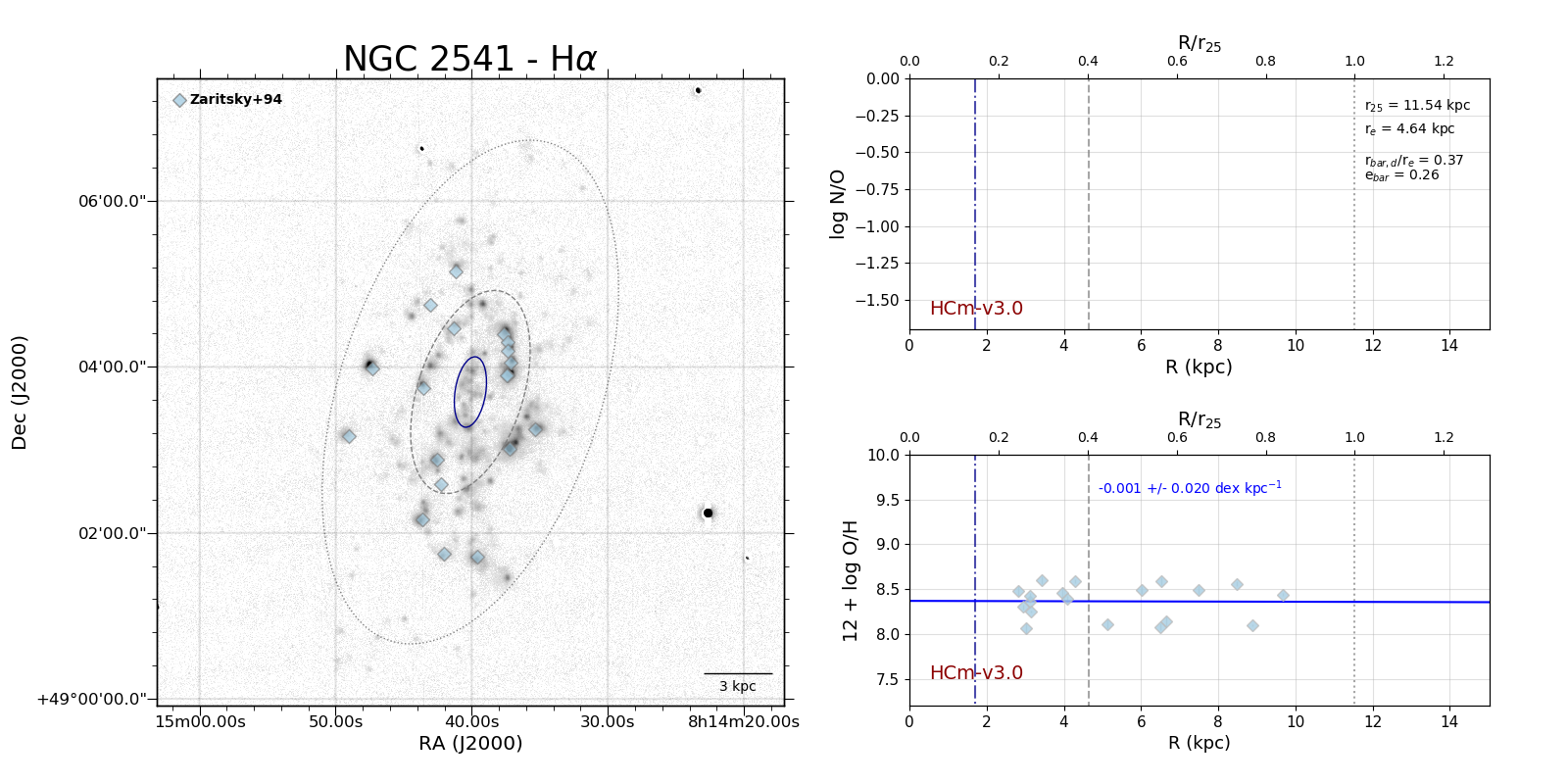}
\end{minipage}
\begin{minipage}{1.05\textwidth}
\hspace{-1.2cm}\includegraphics[width=1.12\textwidth]{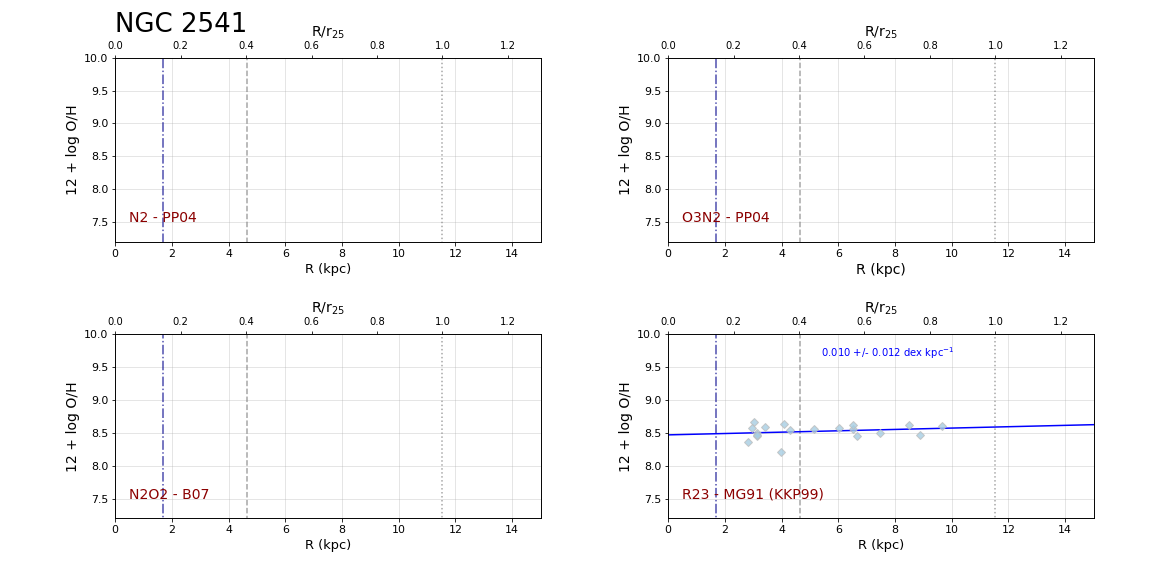}
\end{minipage}
\caption{Same as Fig.~\ref{Afig1} for NGC~2541.}\label{Afig18}
\end{figure*}

\begin{figure*}
\begin{minipage}{1.05\textwidth}
\hspace{-1.2cm}\includegraphics[width=1.12\textwidth]{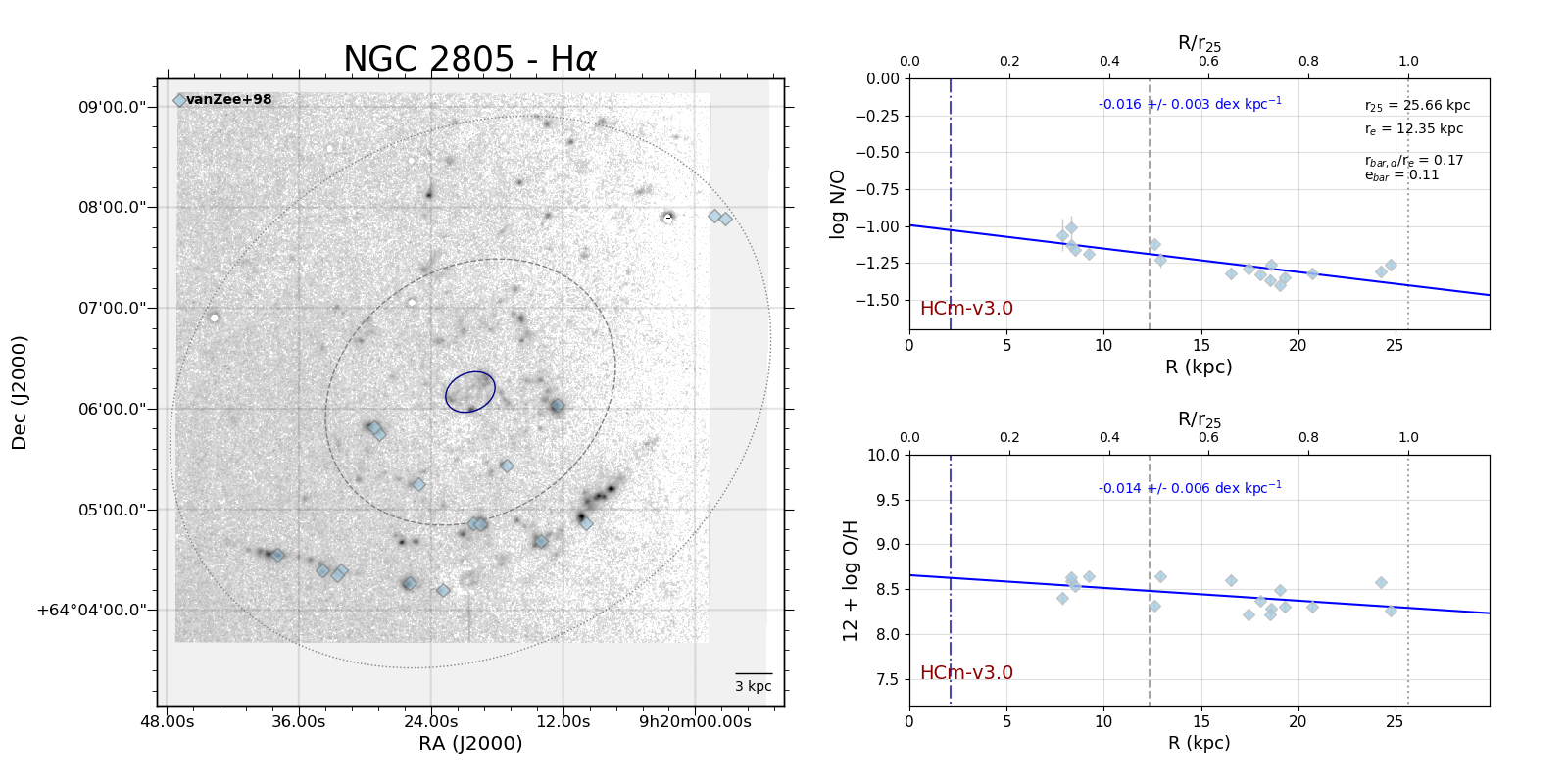}
\end{minipage}
\begin{minipage}{1.05\textwidth}
\hspace{-1.2cm}\includegraphics[width=1.12\textwidth]{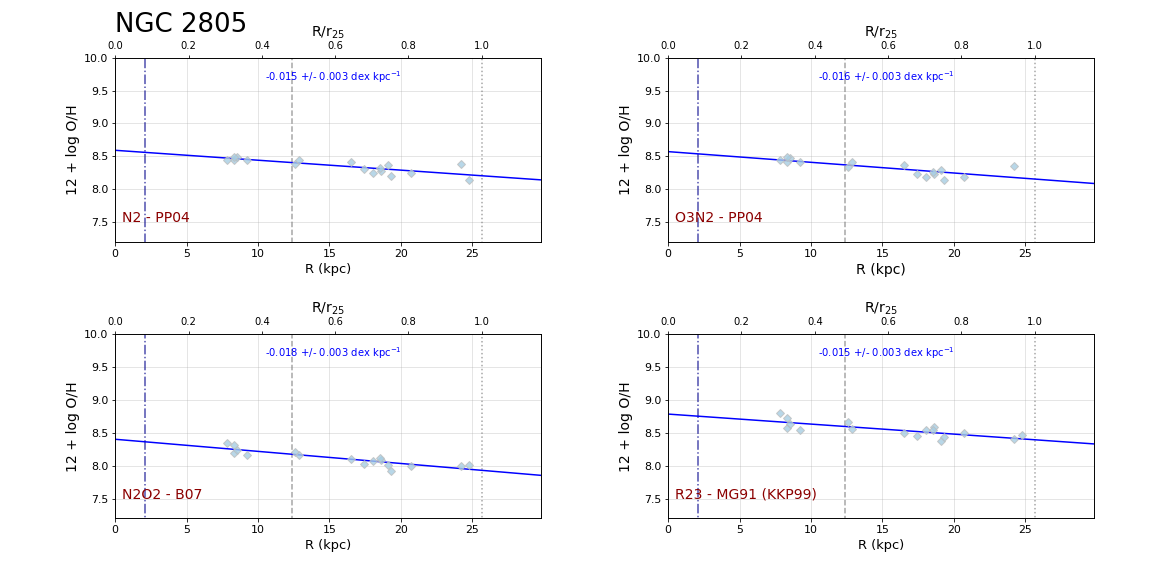}
\end{minipage}
\caption{Same as Fig.~\ref{Afig1} for NGC~2805.}\label{Afig19}
\end{figure*}

\begin{figure*}
\begin{minipage}{1.05\textwidth}
\hspace{-1.2cm}\includegraphics[width=1.12\textwidth]{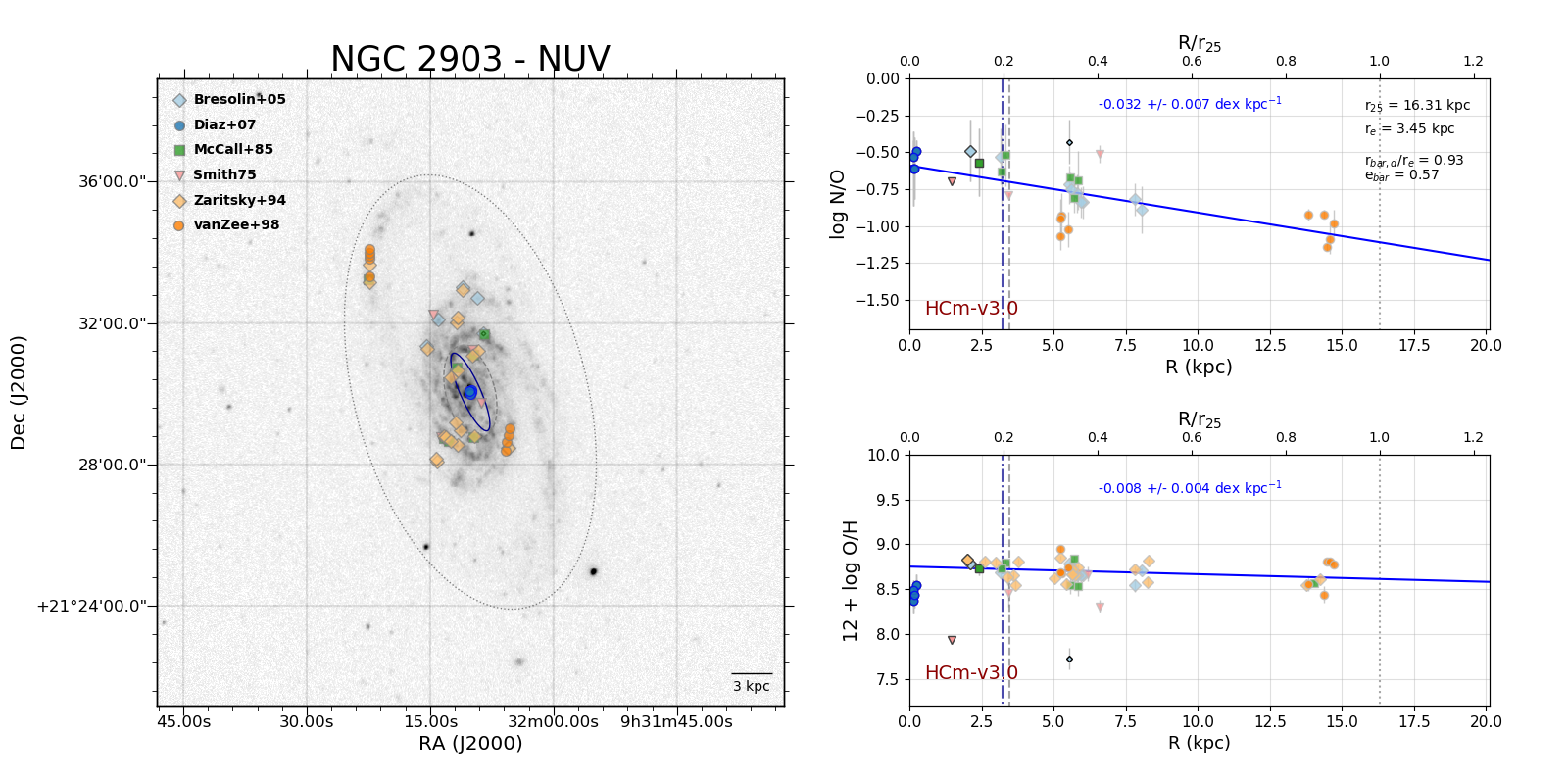}
\end{minipage}
\begin{minipage}{1.05\textwidth}
\hspace{-1.2cm}\includegraphics[width=1.12\textwidth]{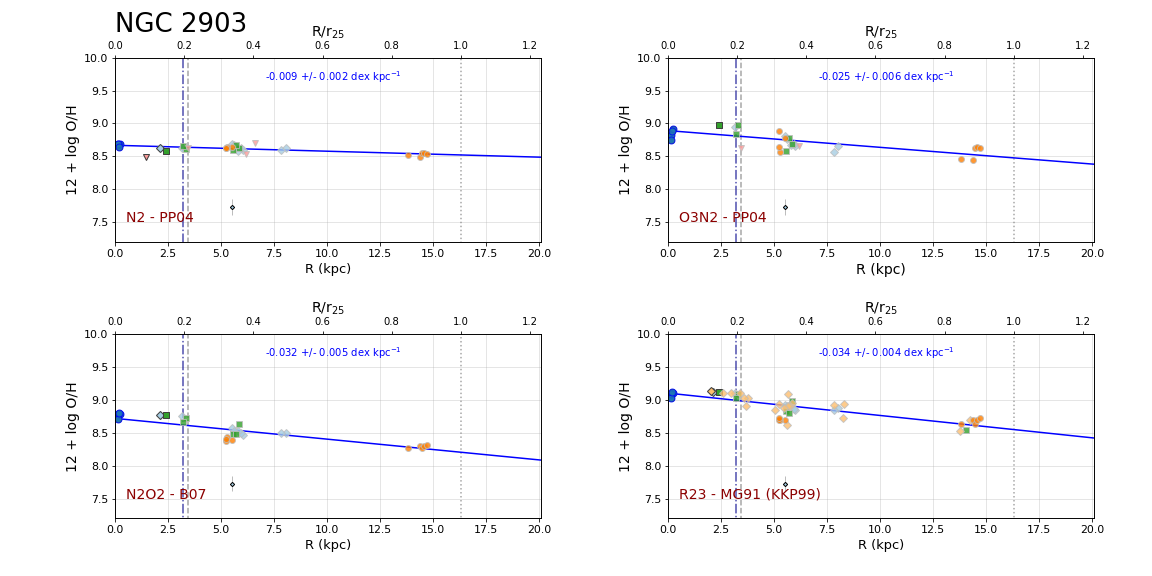}
\end{minipage}
\caption{Same as Fig.~\ref{Afig1} for NGC~2903.}\label{Afig20}
\end{figure*}

\begin{figure*}
\begin{minipage}{1.05\textwidth}
\hspace{-1.2cm}\includegraphics[width=1.12\textwidth]{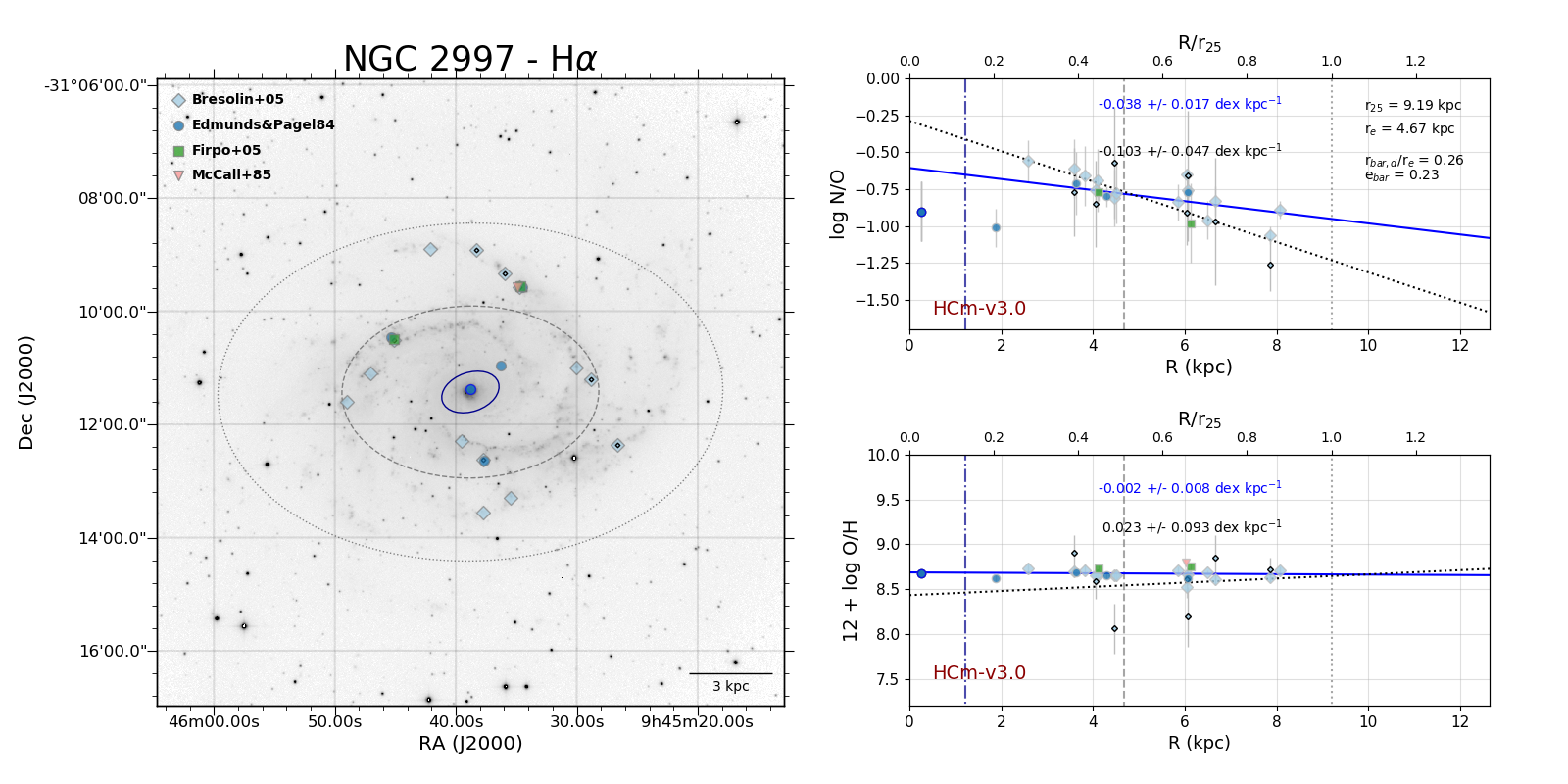}
 \end{minipage}
\begin{minipage}{1.05\textwidth}
\hspace{-1.2cm}\includegraphics[width=1.12\textwidth]{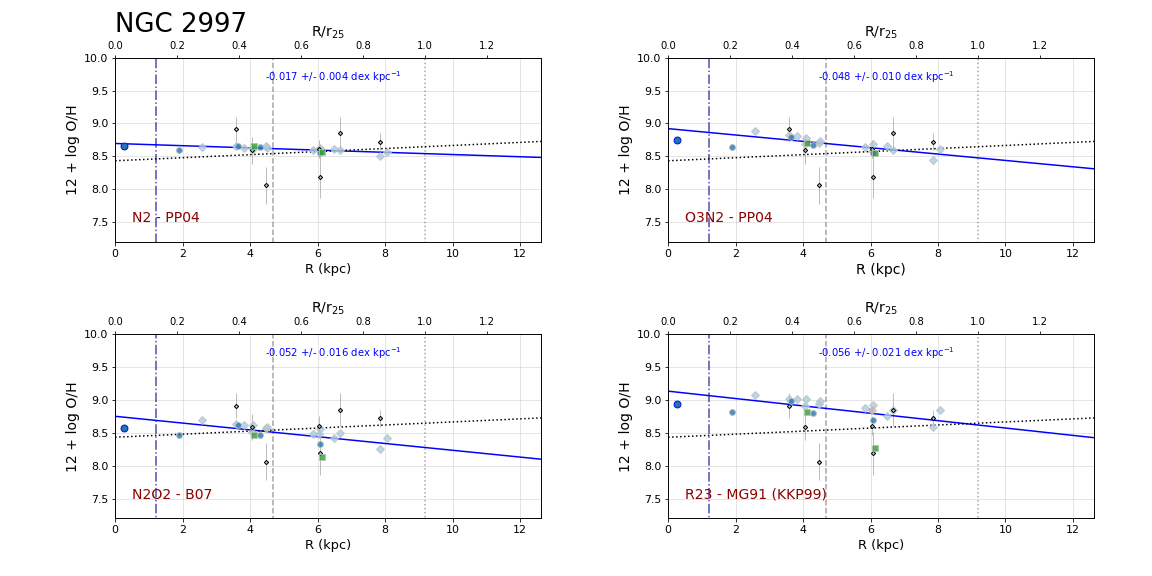}
\end{minipage}
\caption{Same as Fig.~\ref{Afig1} for NGC~2997.}\label{Afig21}
\end{figure*}

\begin{figure*}
\begin{minipage}{1.05\textwidth}
\hspace{-1.2cm}\includegraphics[width=1.12\textwidth]{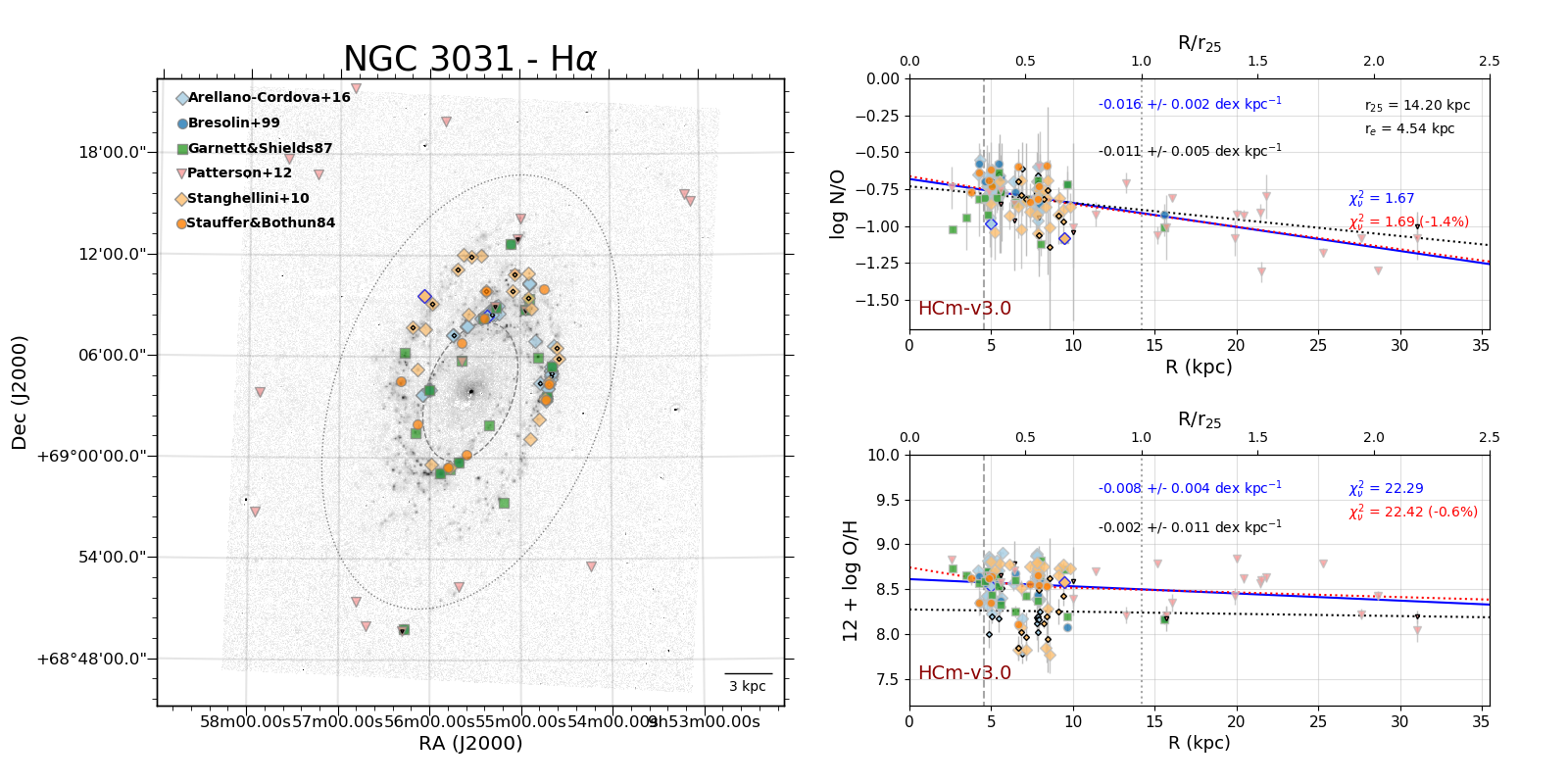}
\end{minipage}
\begin{minipage}{1.05\textwidth}
\hspace{-1.2cm}\includegraphics[width=1.12\textwidth]{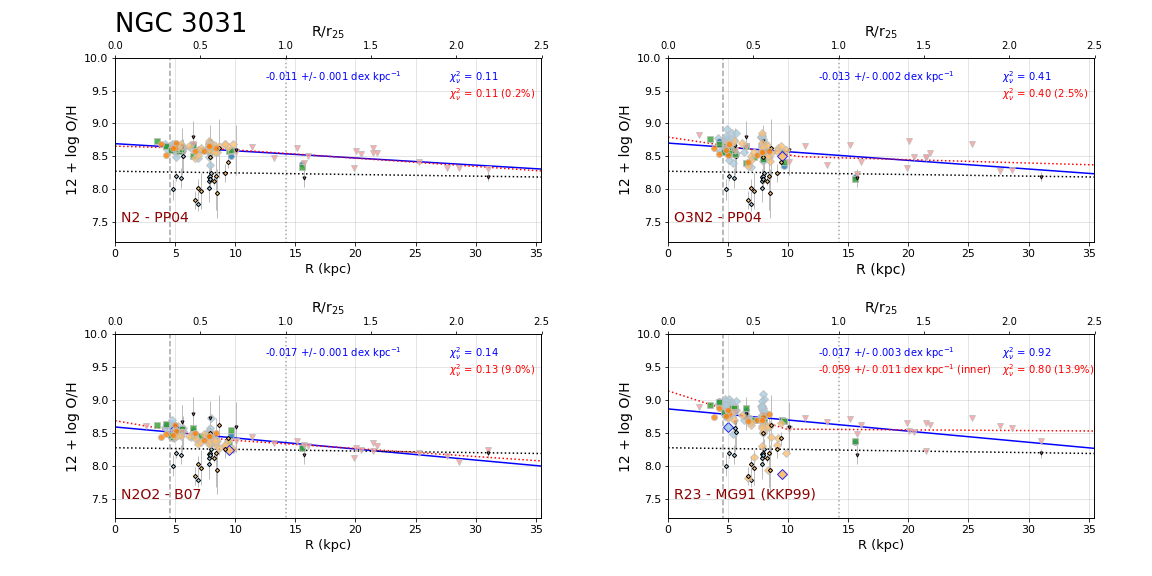}
\end{minipage}
\caption{Same as Fig.~\ref{Afig1} for NGC~3031.}\label{Afig22}
\end{figure*}

\begin{figure*}
\begin{minipage}{1.05\textwidth}
\hspace{-1.2cm}\includegraphics[width=1.12\textwidth]{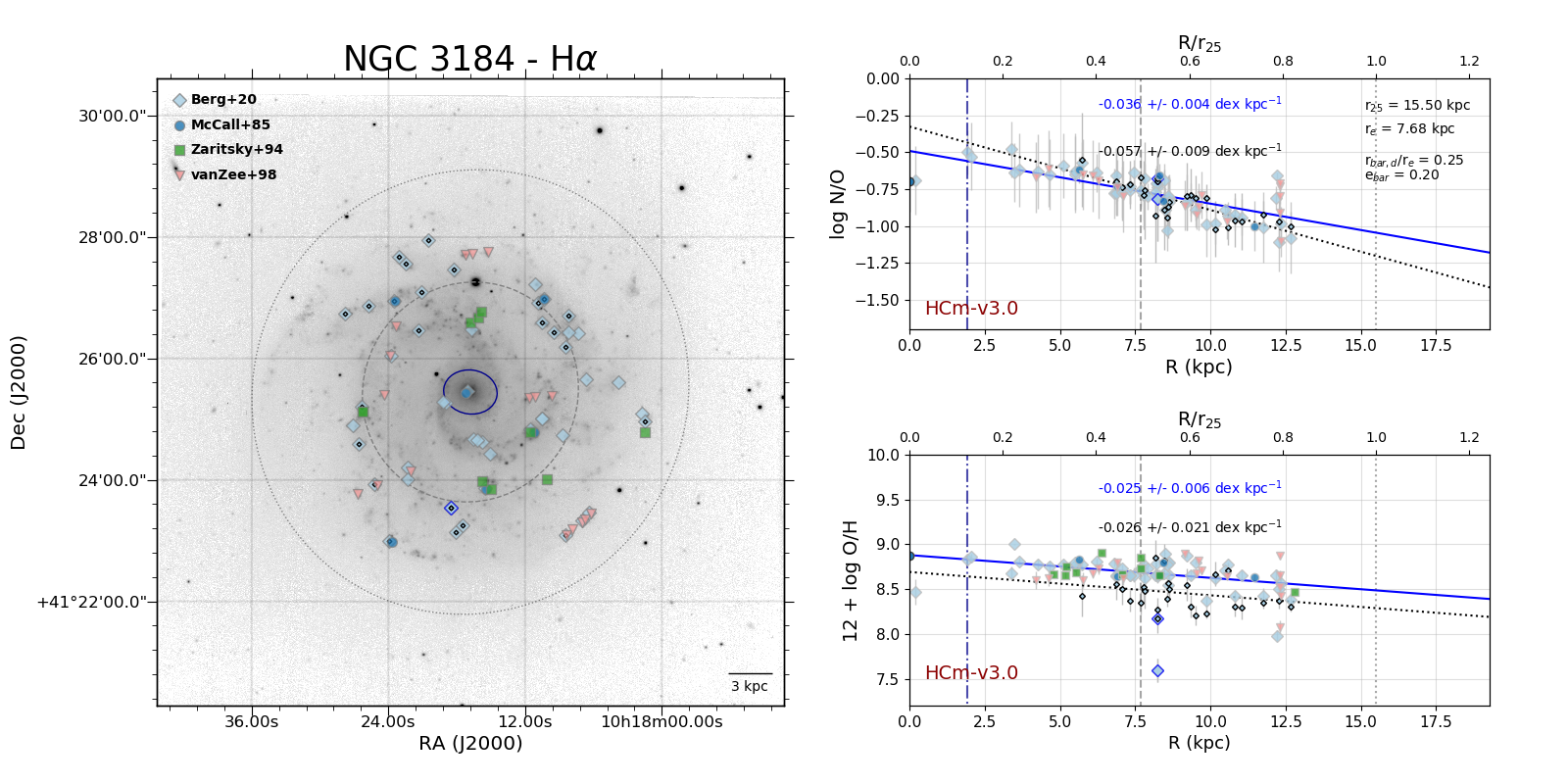}
\end{minipage}
\begin{minipage}{1.05\textwidth}
\hspace{-1.2cm}\includegraphics[width=1.12\textwidth]{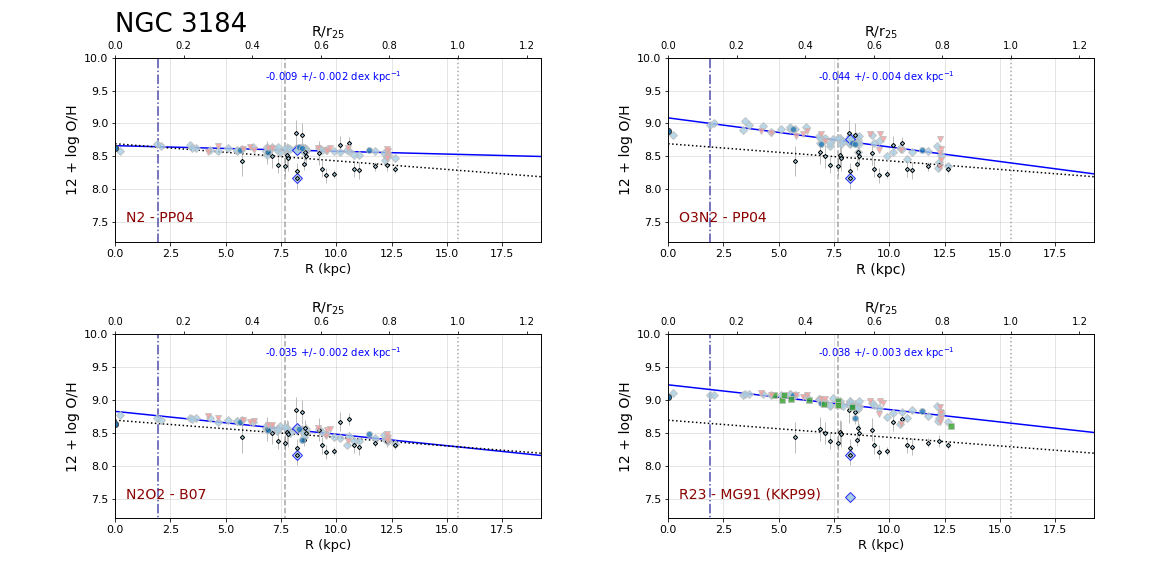}
\end{minipage}
\caption{Same as Fig.~\ref{Afig1} for NGC~3184}\label{Afig23}
\end{figure*}

\begin{figure*}
\begin{minipage}{1.05\textwidth}
\hspace{-1.2cm}\includegraphics[width=1.12\textwidth]{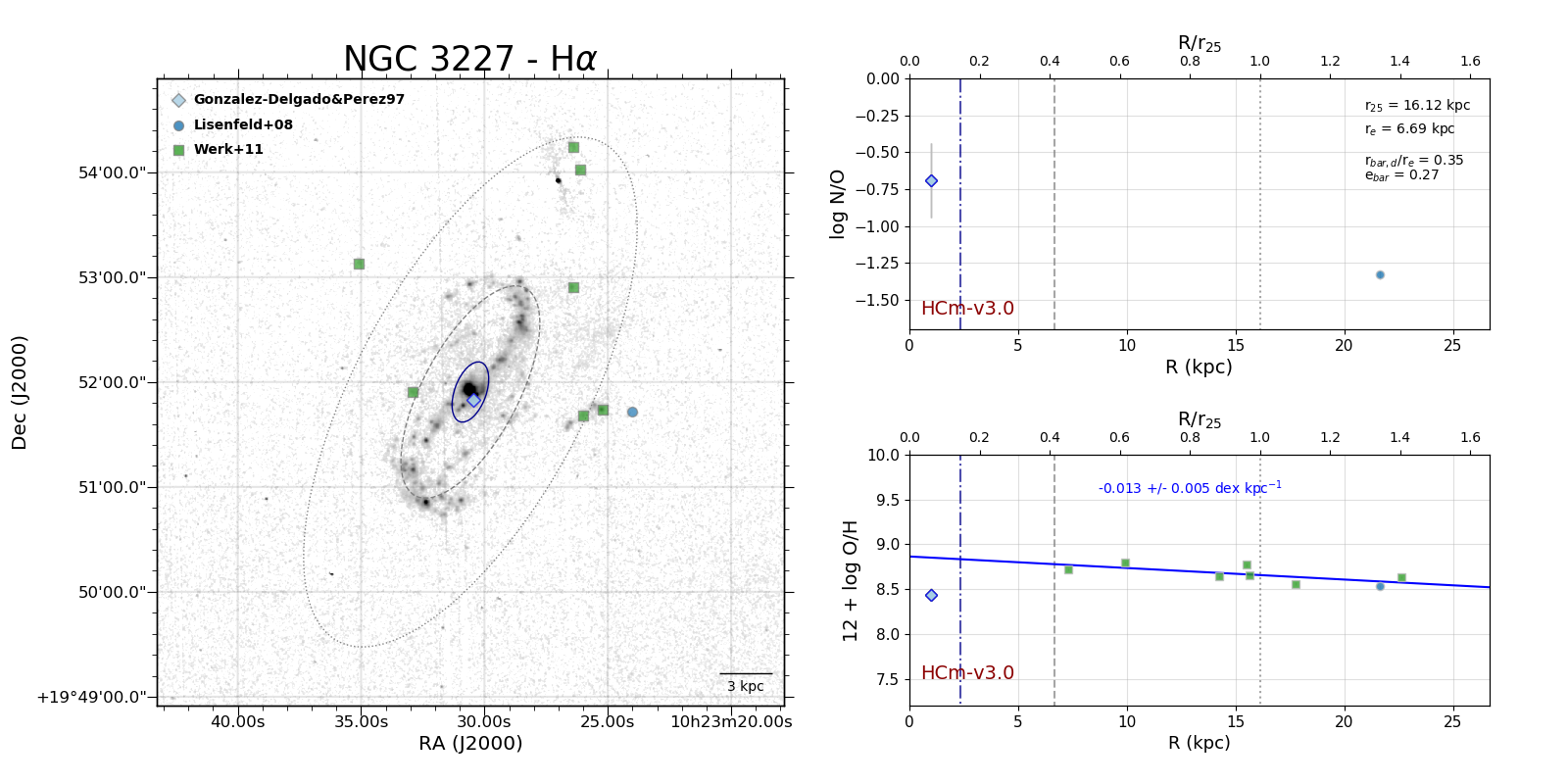}
\end{minipage}
\begin{minipage}{1.05\textwidth}
\hspace{-1.2cm}\includegraphics[width=1.12\textwidth]{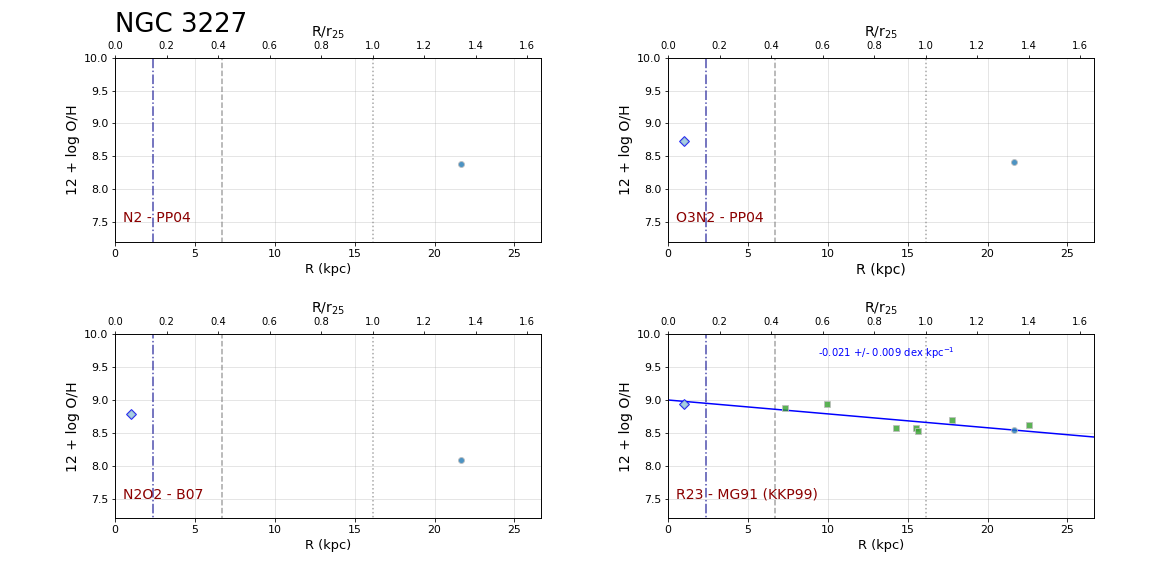}
\end{minipage}
\caption{Same as Fig.~\ref{Afig1} for NGC~3227.}\label{Afig24}
\end{figure*}

\begin{figure*}
\begin{minipage}{1.05\textwidth}
\hspace{-1.2cm}\includegraphics[width=1.12\textwidth]{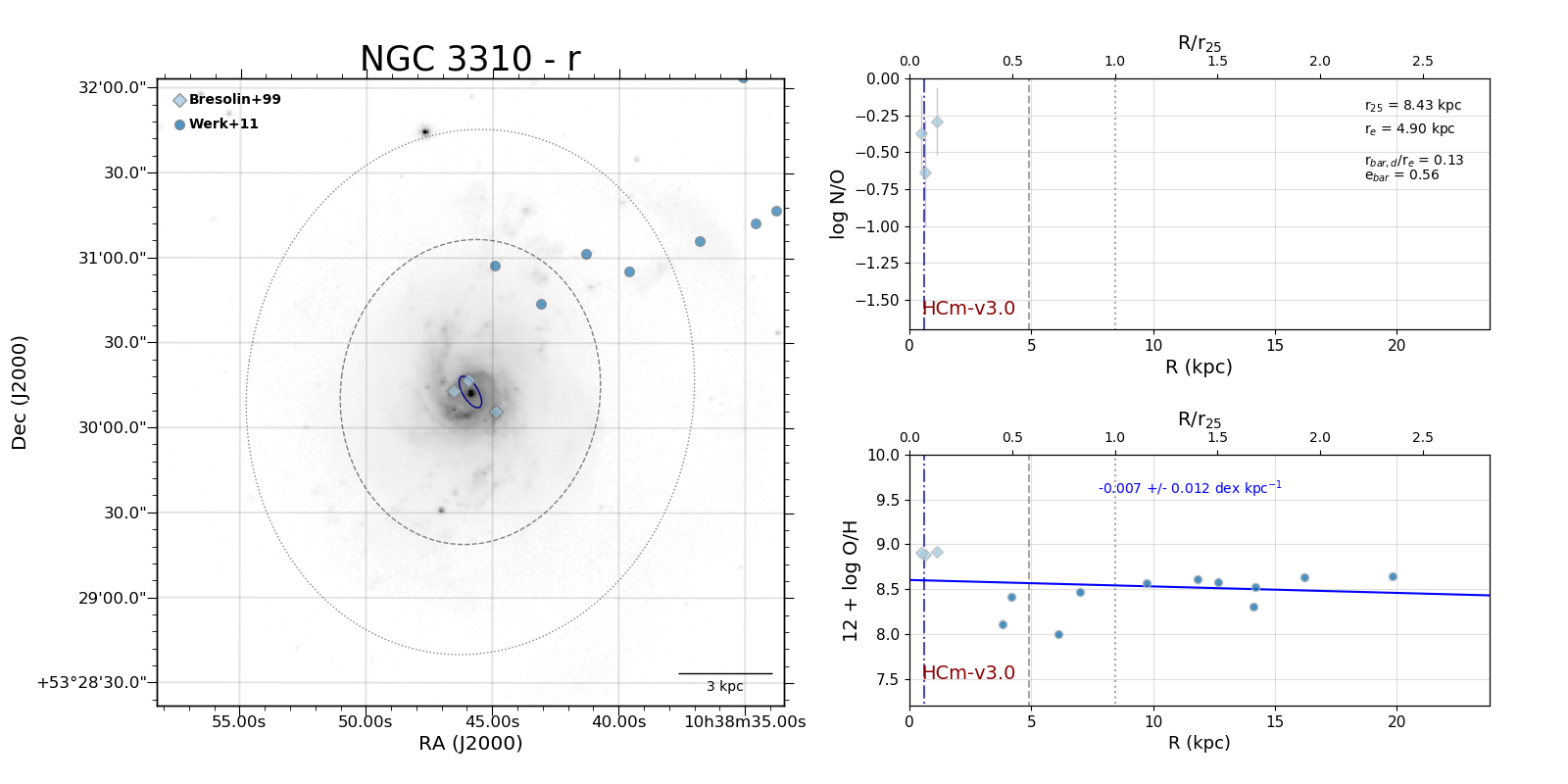}
\end{minipage}
\begin{minipage}{1.05\textwidth}
\hspace{-1.2cm}\includegraphics[width=1.12\textwidth]{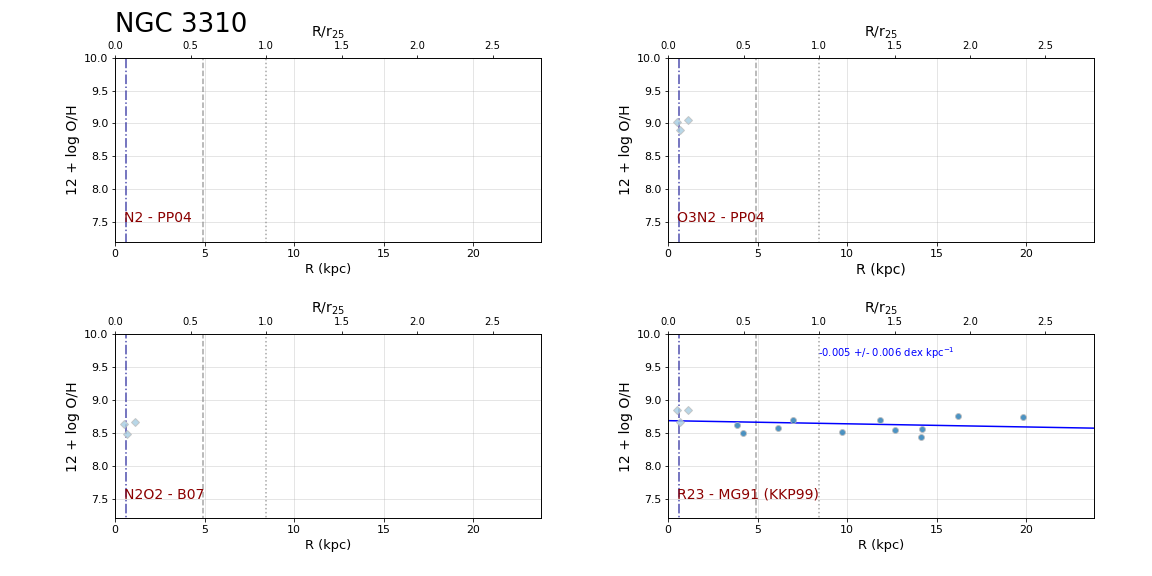}
\end{minipage}
\caption{Same as Fig.~\ref{Afig1} for NGC~3310.}\label{Afig25}
\end{figure*}

\begin{figure*}
\begin{minipage}{1.05\textwidth}
\hspace{-1.2cm}\includegraphics[width=1.12\textwidth]{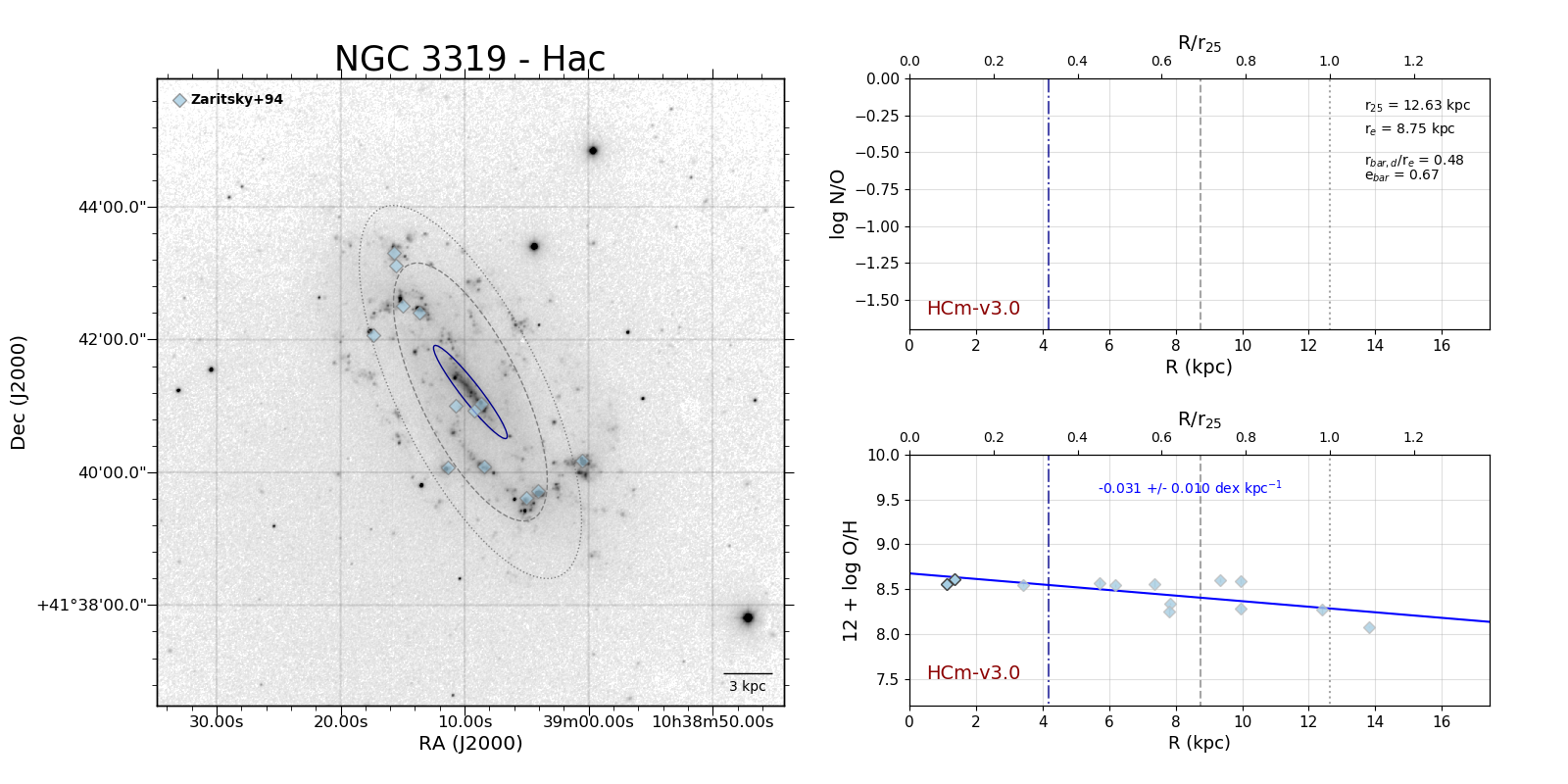}
\end{minipage}
\begin{minipage}{1.05\textwidth}
\hspace{-1.2cm}\includegraphics[width=1.12\textwidth]{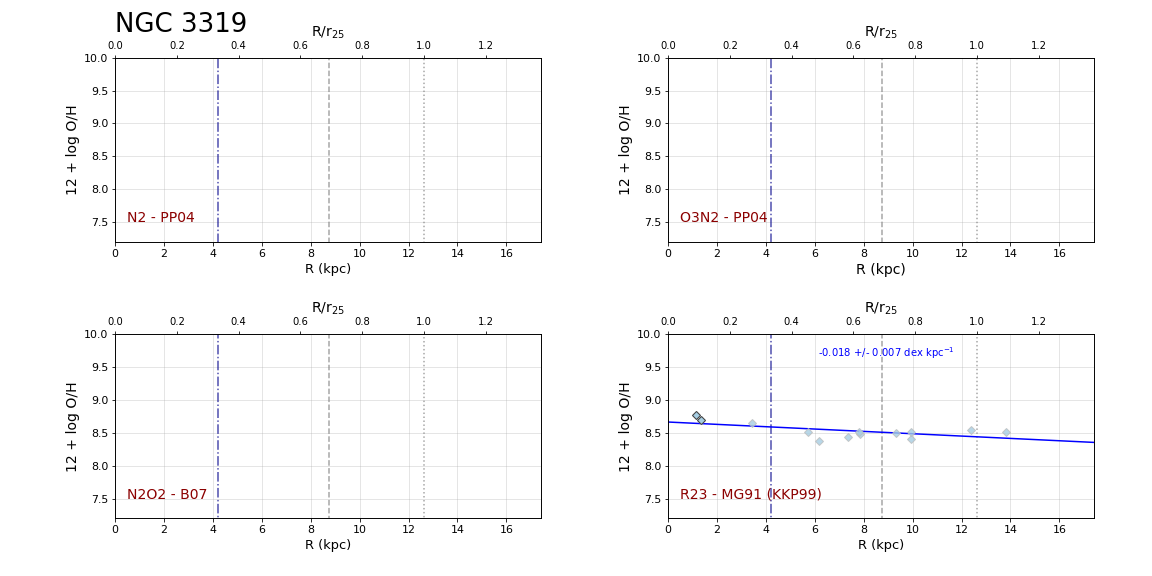}
\end{minipage}
\caption{Same as Fig.~\ref{Afig1} for NGC~3319}\label{Afig26}
\end{figure*}

\begin{figure*}
\begin{minipage}{1.05\textwidth}
\hspace{-1.2cm}\includegraphics[width=1.12\textwidth]{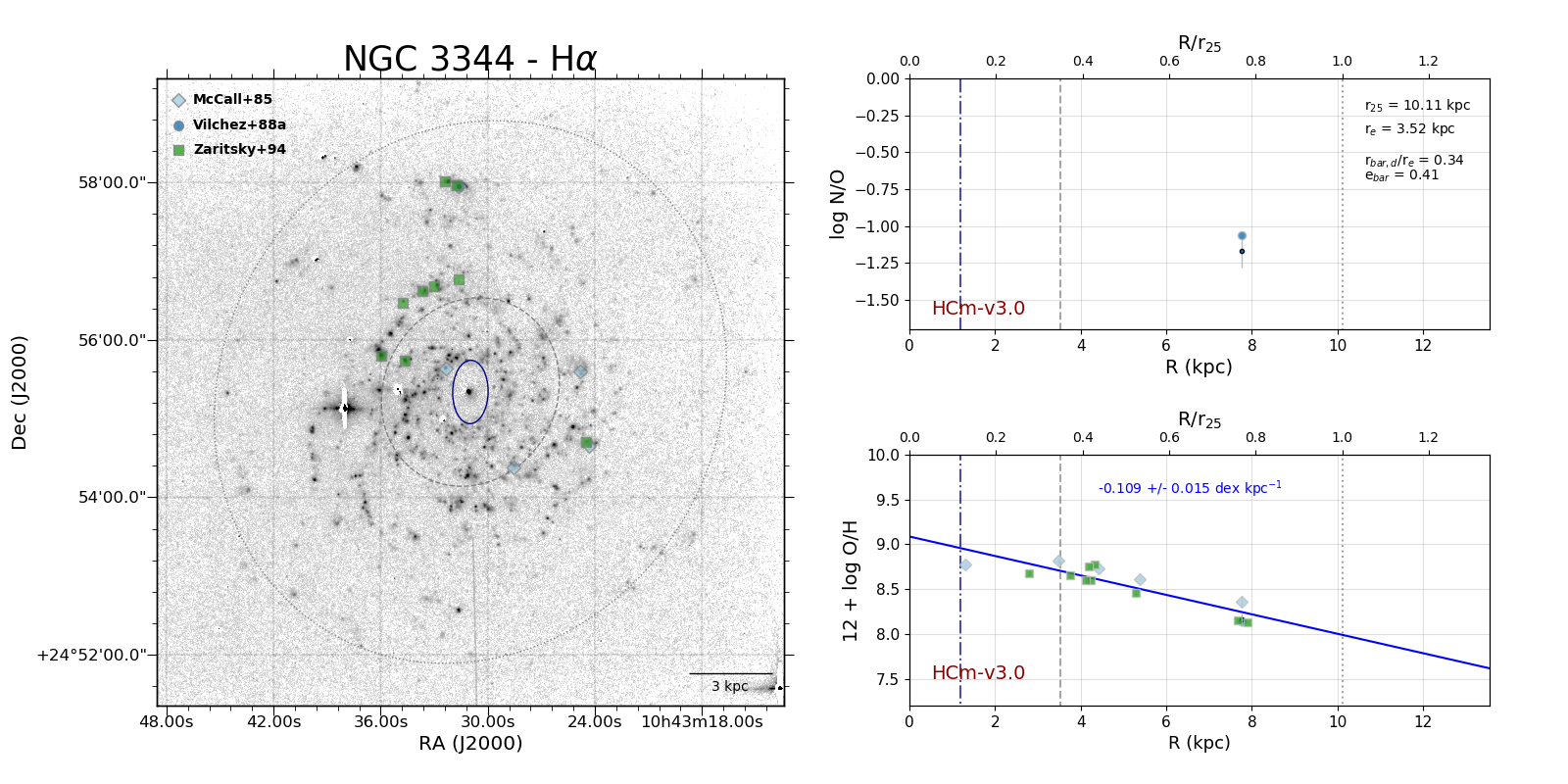}
\end{minipage}
\begin{minipage}{1.05\textwidth}
\hspace{-1.2cm} \includegraphics[width=1.12\textwidth]{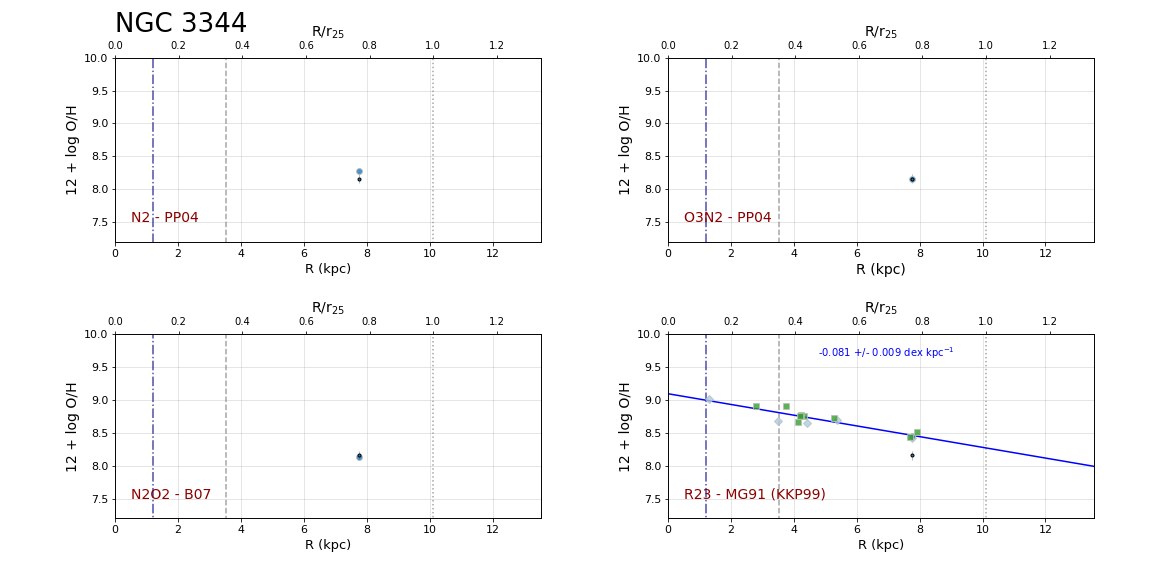}
\end{minipage}
\caption{Same as Fig.~\ref{Afig1} for NGC~3344.}\label{Afig27}
\end{figure*}

\begin{figure*}
\begin{minipage}{1.05\textwidth}
\hspace{-1.2cm}\includegraphics[width=1.12\textwidth]{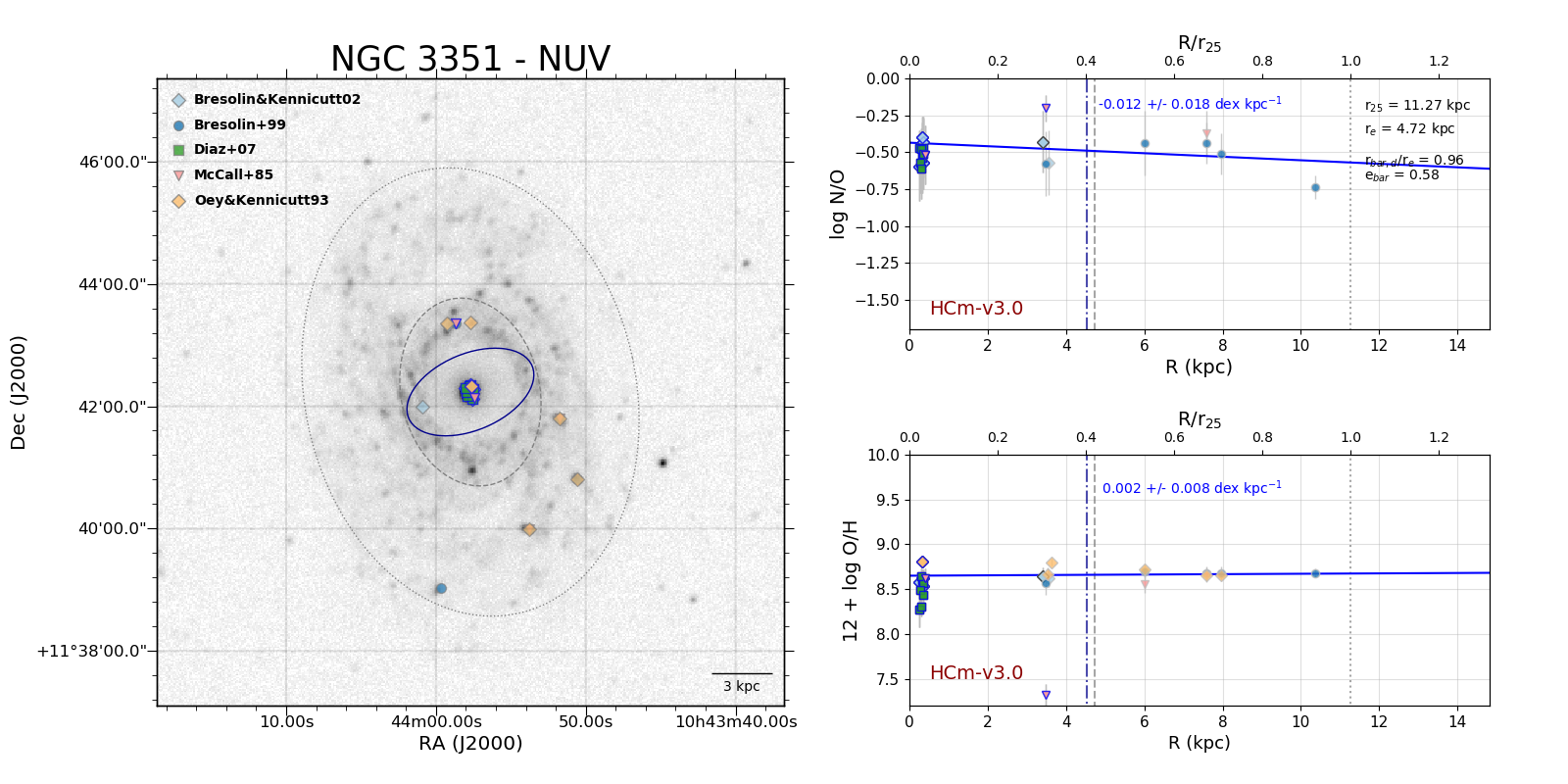}
\end{minipage}
\begin{minipage}{1.05\textwidth}
\hspace{-1.2cm} \includegraphics[width=1.12\textwidth]{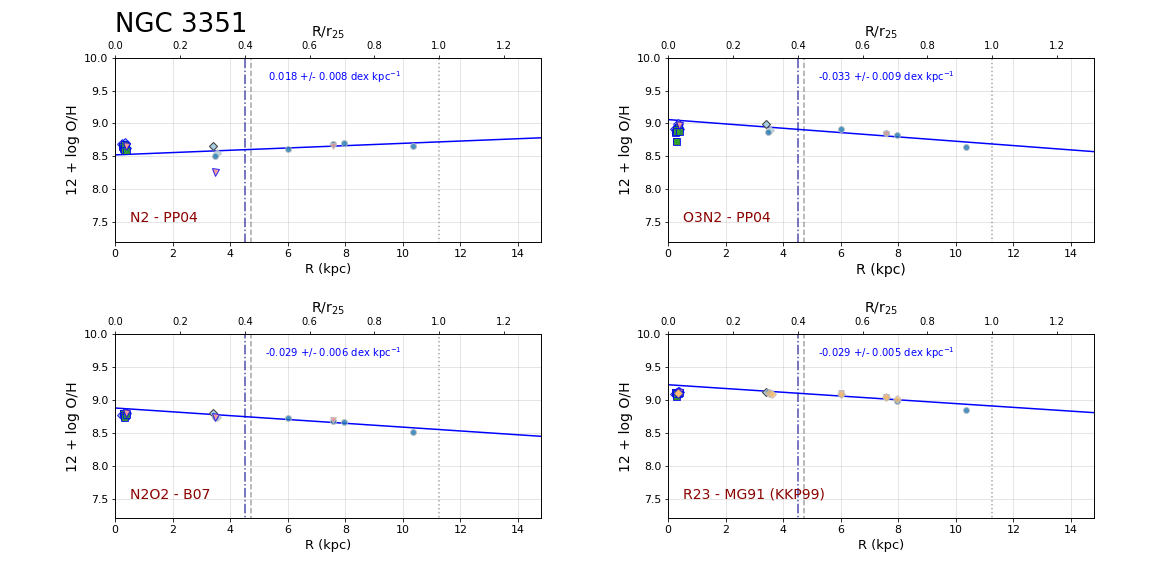}
\end{minipage}
\caption{Same as Fig.~\ref{Afig1} for NGC~3351.}\label{Afig28}
\end{figure*}

\begin{figure*}
\begin{minipage}{1.05\textwidth}
\hspace{-1.2cm}\includegraphics[width=1.12\textwidth]{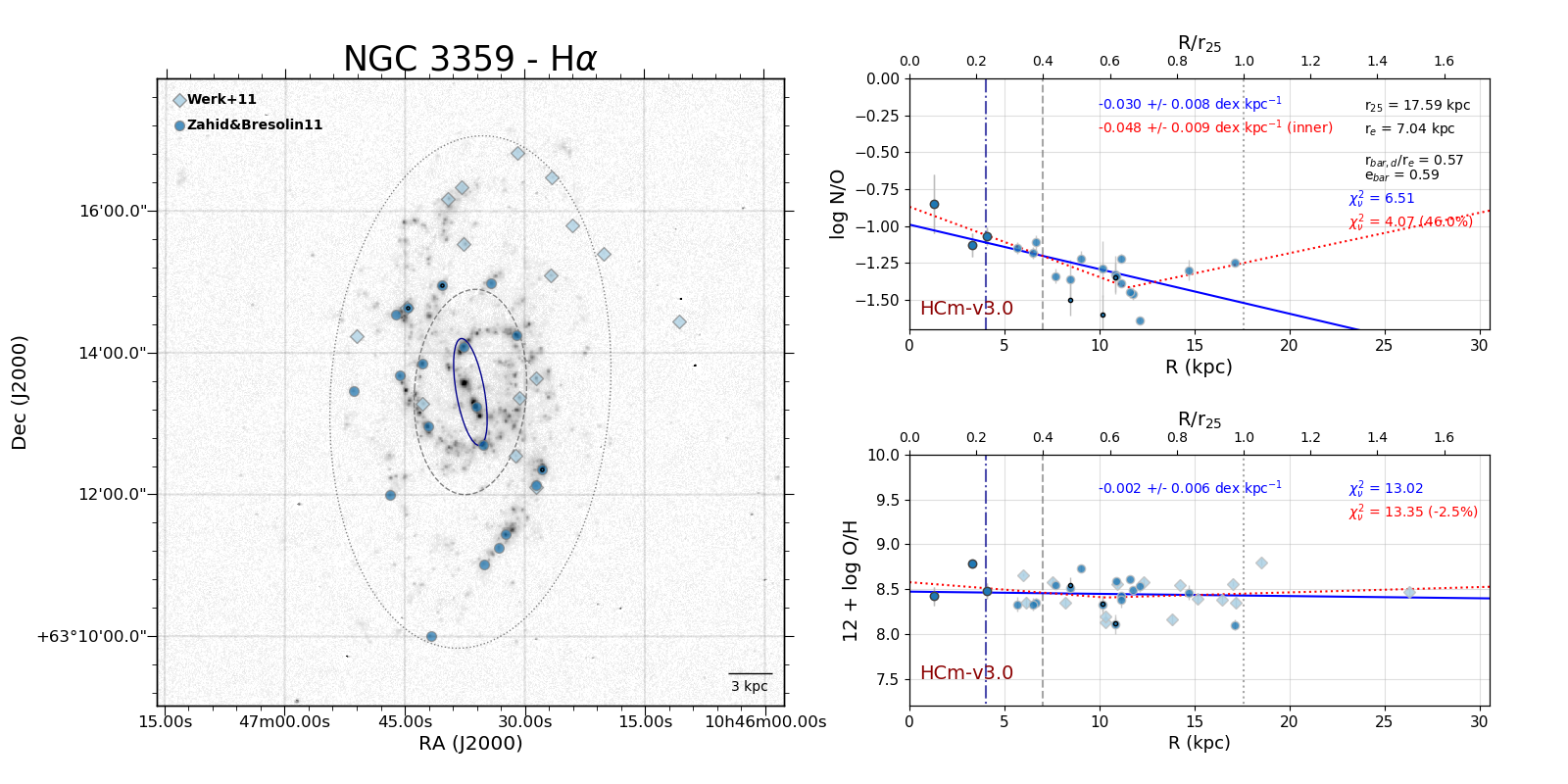}
\end{minipage}
\begin{minipage}{1.05\textwidth}
\hspace{-1.2cm}\includegraphics[width=1.12\textwidth]{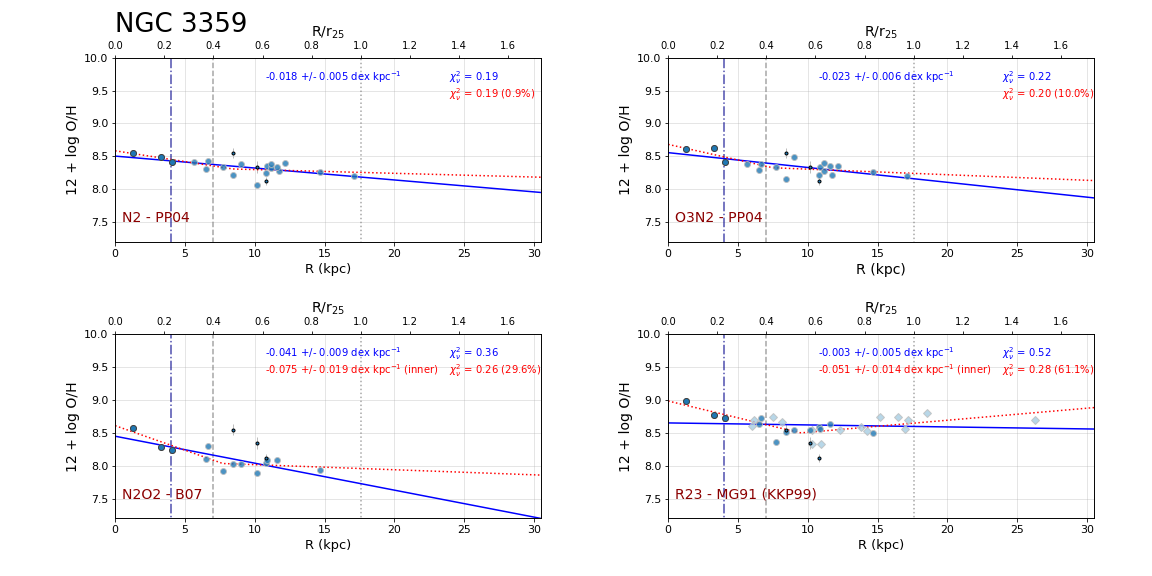}
\end{minipage}
\caption{Same as Fig.~\ref{Afig1} for NGC~3359.}\label{Afig29}
\end{figure*}

Figs.~\ref{Afig1} to \ref{Afig54} contain a mosaic of  all the galaxies of the sample, with the following information:
\begin{itemize}
\item (Top-left) Image of the galaxy (photometric band indicated in the plot title) showing the location of the \hii\ regions compiled for this work with different colours and symbols for the different authors. The dashed and dotted grey ellipses mark the extent of the disc effective radius (r$_e$) and r$_{25}$, respectively. The bar (if present) is marked with a dark blue ellipse.

\item (Top-right) Radial $\log$(N/O) and  $12+\log$(O/H) abundance profile for abundances obtained from the HII-CHI-mistry code \citep{epm14}. 

\item (Bottom) Oxygen abundance profile as determined from different {\em strong-line} methods: N2 and O3N2  as calibrated by \citet{pp04}, N2O2 as calibrated by \citet{b07} and R$_{23}$ as calibrated from theoretical  models by \citet{m91} through the \citet{kkp99} parametrization.
\end{itemize}

Black small open symbols mark regions with T$_e$-based or {\em direct} abundance estimates. Symbols for \hii\ regions located within the bar have a dark grey edge. Regions with a blue edge have not been taken into account for the fitting (as explained in Sec.~\ref{fits}). The vertical lines in the radial abundance profiles mark the location of  r$_e$ (grey dashed line), r$_{25}$ (grey dotted line) and the deprojected (dark blue dashed-dotted) bar radius. Dotted black straight lines mark the best-fitting to T$_e$-based abundances. Solid blue straight lines are linear fits to abundances derived from strong-line methods. Double linear fits  are plotted with red dotted straight lines and green dashed lines for strong-line and T$_e$-based abundances respectively.

In the upper region of the profiles the radial abundance gradient in dex~kpc\me\ is given with the corresponding uncertainty, in the same colour as the corresponding fit (i.e. black for the direct abundances, blue for the strong-line ones). When double linear fittings are perform and these improve the $\chi_\nu^2$ value with respect to the single linear fitting, the inner slope value is also given (in red for strong-line abundances and in green for T$_e$-based ones).
When double linear fits are performed, the $\chi_\nu^2$ value for the single and double linear fits are also shown, with the relative improvement in $\chi_\nu^2$ with respect to the single linear fit in parenthesis.

\clearpage
\clearpage
\clearpage
\clearpage
\clearpage
\clearpage
\clearpage
\clearpage
\clearpage

\clearpage
\clearpage
\clearpage
\clearpage
\clearpage
\clearpage
\clearpage
\clearpage
\clearpage
\clearpage
\clearpage
\clearpage
\clearpage
\clearpage
\clearpage
\clearpage
\clearpage
\clearpage
\clearpage
\begin{figure*}
\begin{minipage}{1.05\textwidth}
\hspace{-1.2cm}\includegraphics[width=1.12\textwidth]{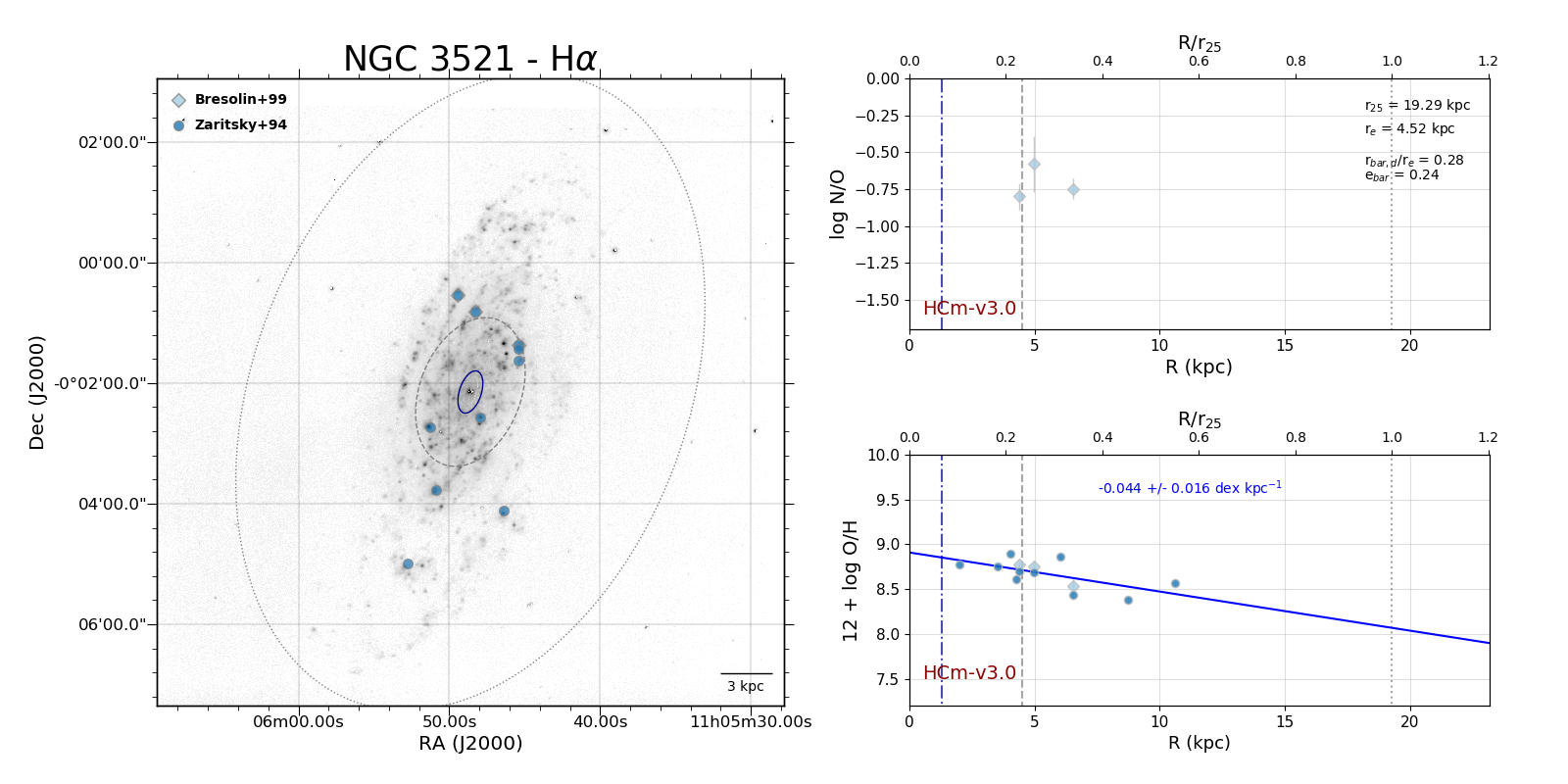}
\end{minipage}
\begin{minipage}{1.05\textwidth}
\hspace{-1.2cm}\includegraphics[width=1.12\textwidth]{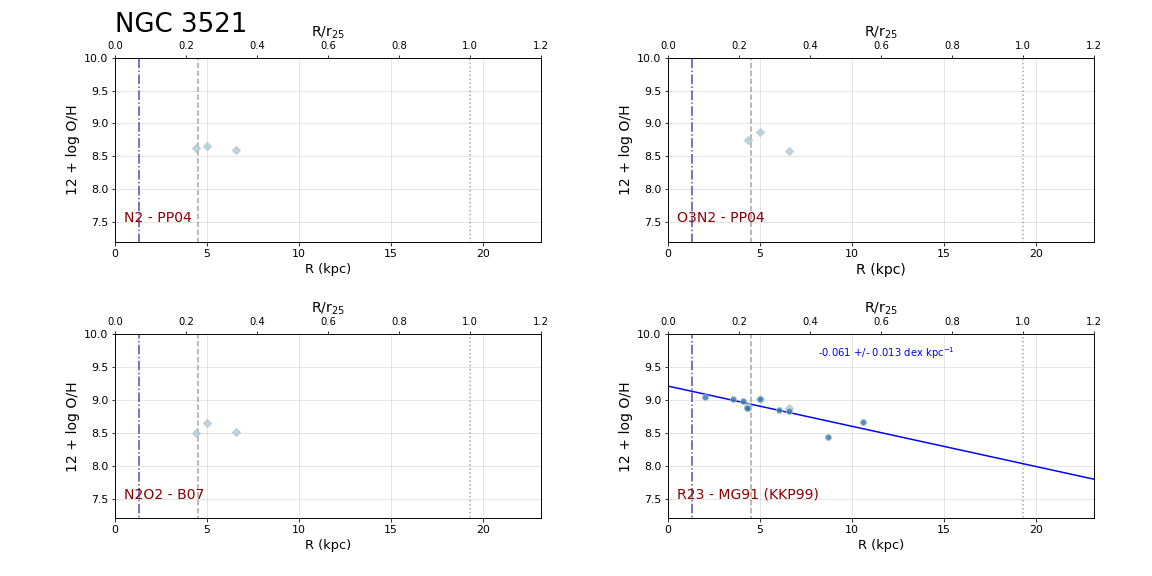}
\end{minipage}
\caption{Same as Fig.~\ref{Afig1} for NGC~3521.}\label{Afig31}
\end{figure*}
\clearpage
\begin{figure*}
\begin{minipage}{1.05\textwidth}
\hspace{-1.2cm}\includegraphics[width=1.12\textwidth]{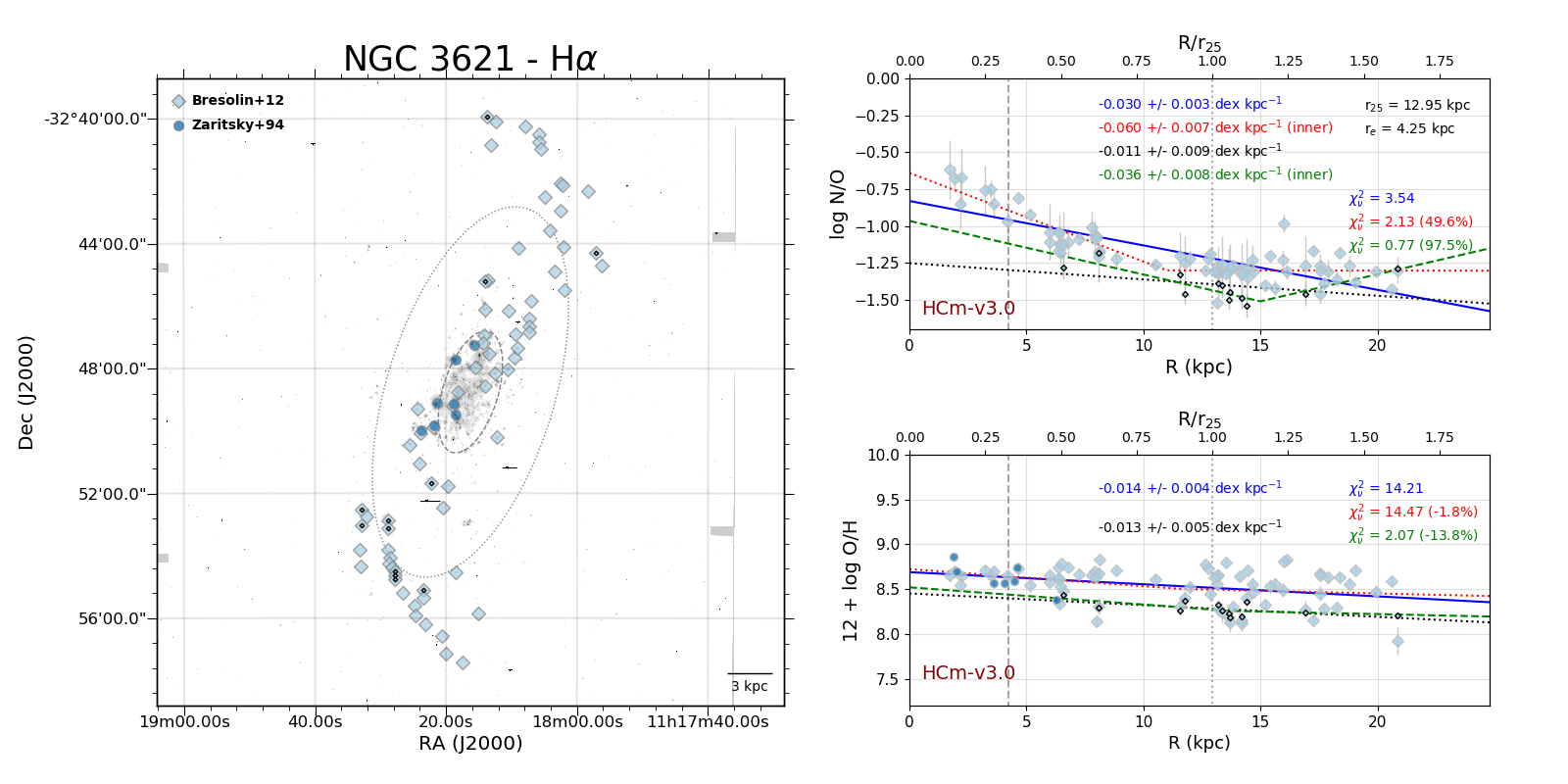}
\end{minipage}
\begin{minipage}{1.05\textwidth}
\hspace{-1.2cm}\includegraphics[width=1.12\textwidth]{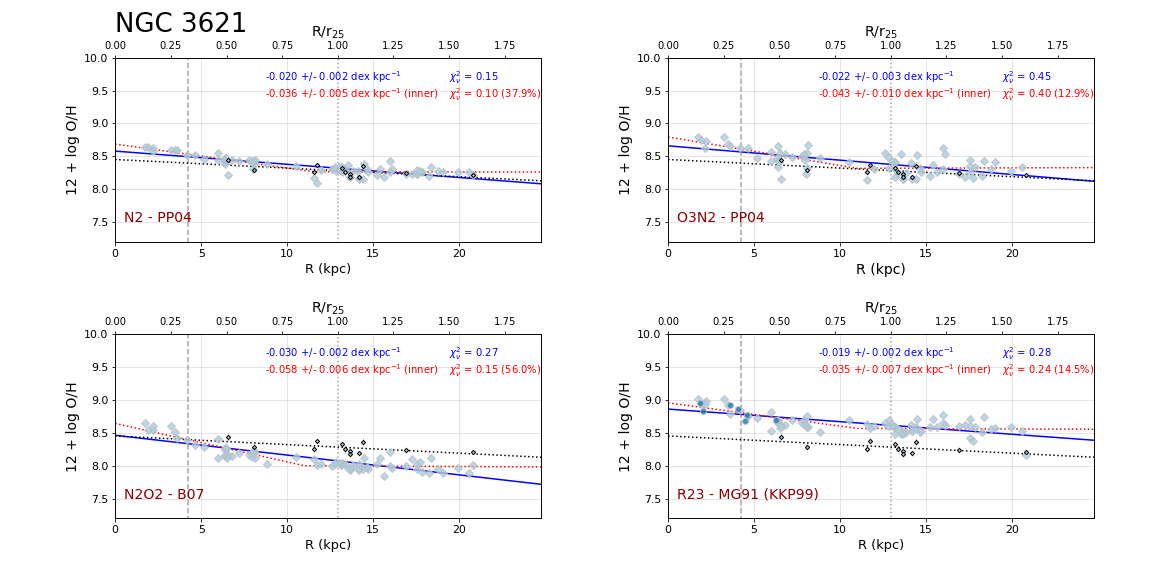}
\end{minipage}
\caption{Same as Fig.~\ref{Afig1} for NGC~3621.}\label{Afig32}
\end{figure*}
\clearpage
\begin{figure*}
\begin{minipage}{1.05\textwidth}
\hspace{-1.2cm}\includegraphics[width=1.12\textwidth]{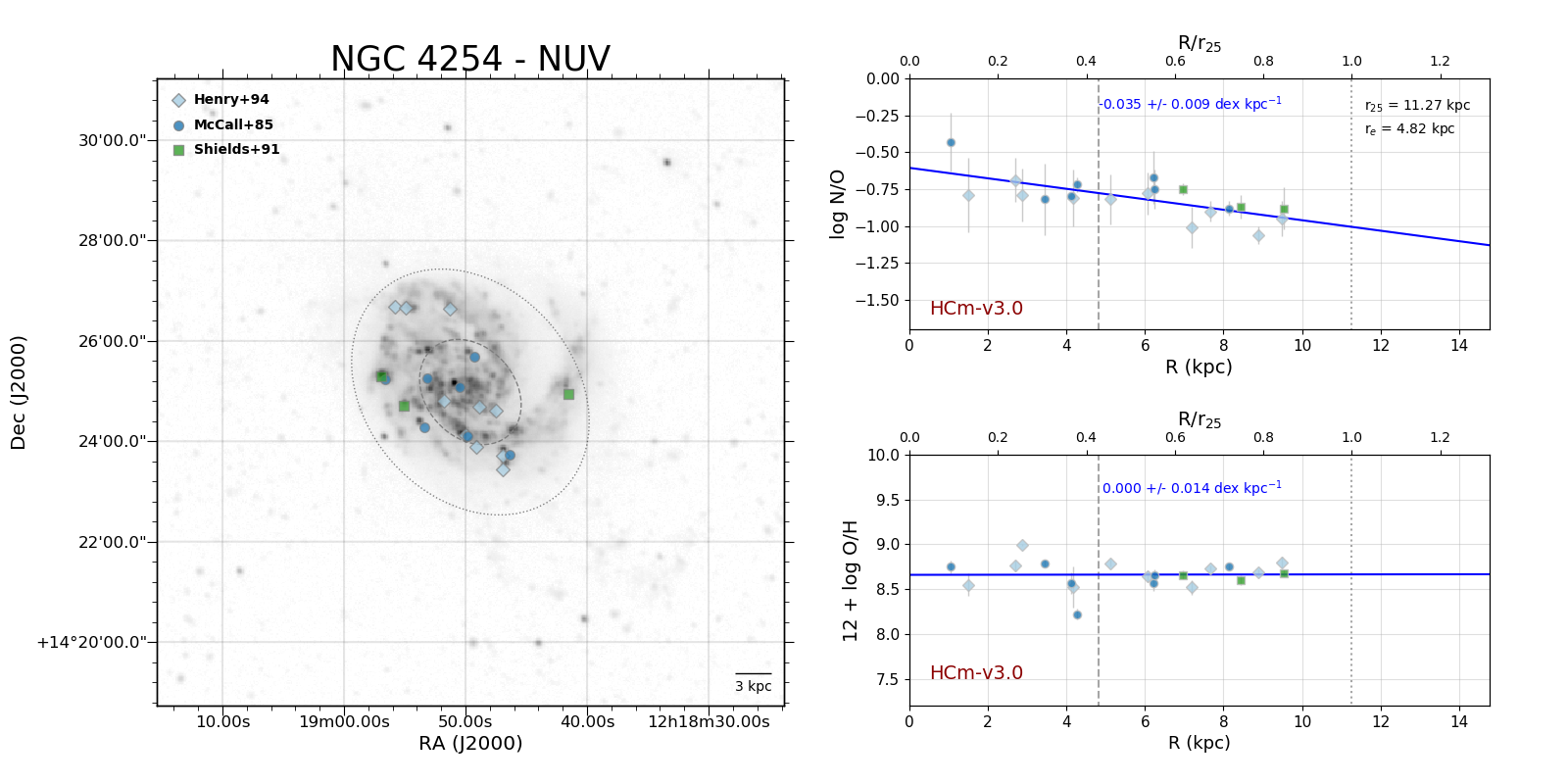}
\end{minipage}
\begin{minipage}{1.05\textwidth}
\hspace{-1.2cm}\includegraphics[width=1.12\textwidth]{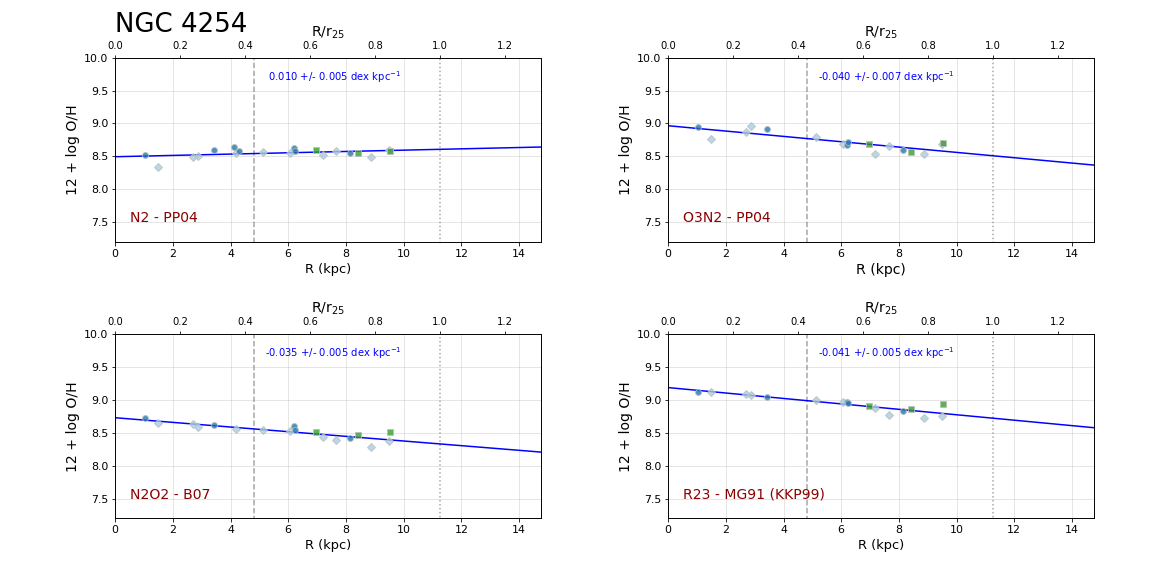}
\end{minipage}
\caption{Same as Fig.~\ref{Afig1} for NGC~4254.}\label{Afig33}
\end{figure*}
\clearpage
\begin{figure*}
\begin{minipage}{1.05\textwidth}
\hspace{-1.2cm}\includegraphics[width=1.12\textwidth]{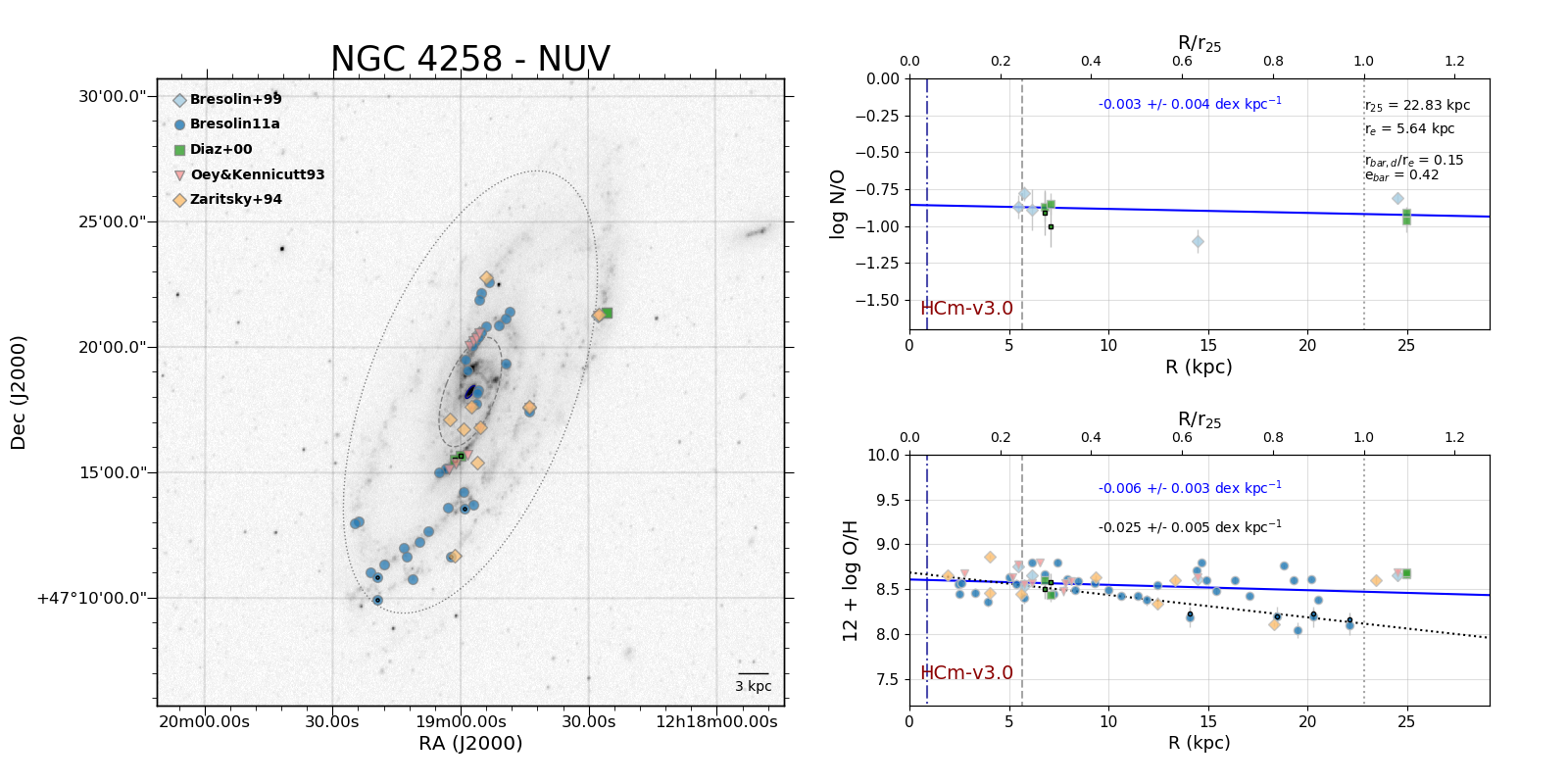}
\end{minipage}
\begin{minipage}{1.05\textwidth}
\hspace{-1.2cm}\includegraphics[width=1.12\textwidth]{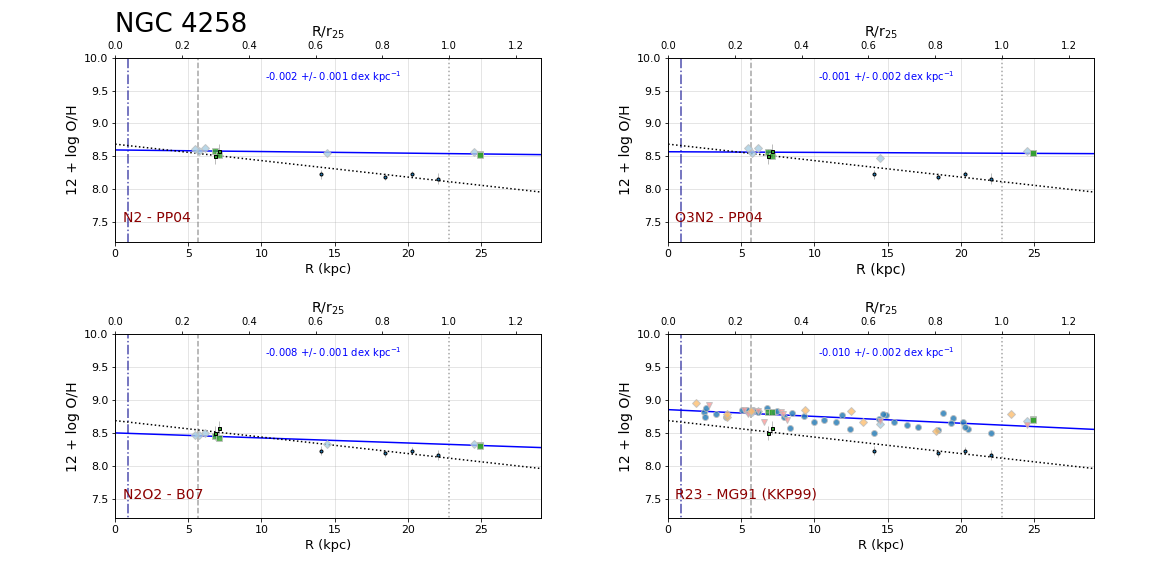}
\end{minipage}
\caption{Same as Fig.~\ref{Afig1} for NGC~4258.}\label{Afig34}
\end{figure*}
\clearpage
\begin{figure*}
\begin{minipage}{1.05\textwidth}
\hspace{-1.2cm}\includegraphics[width=1.12\textwidth]{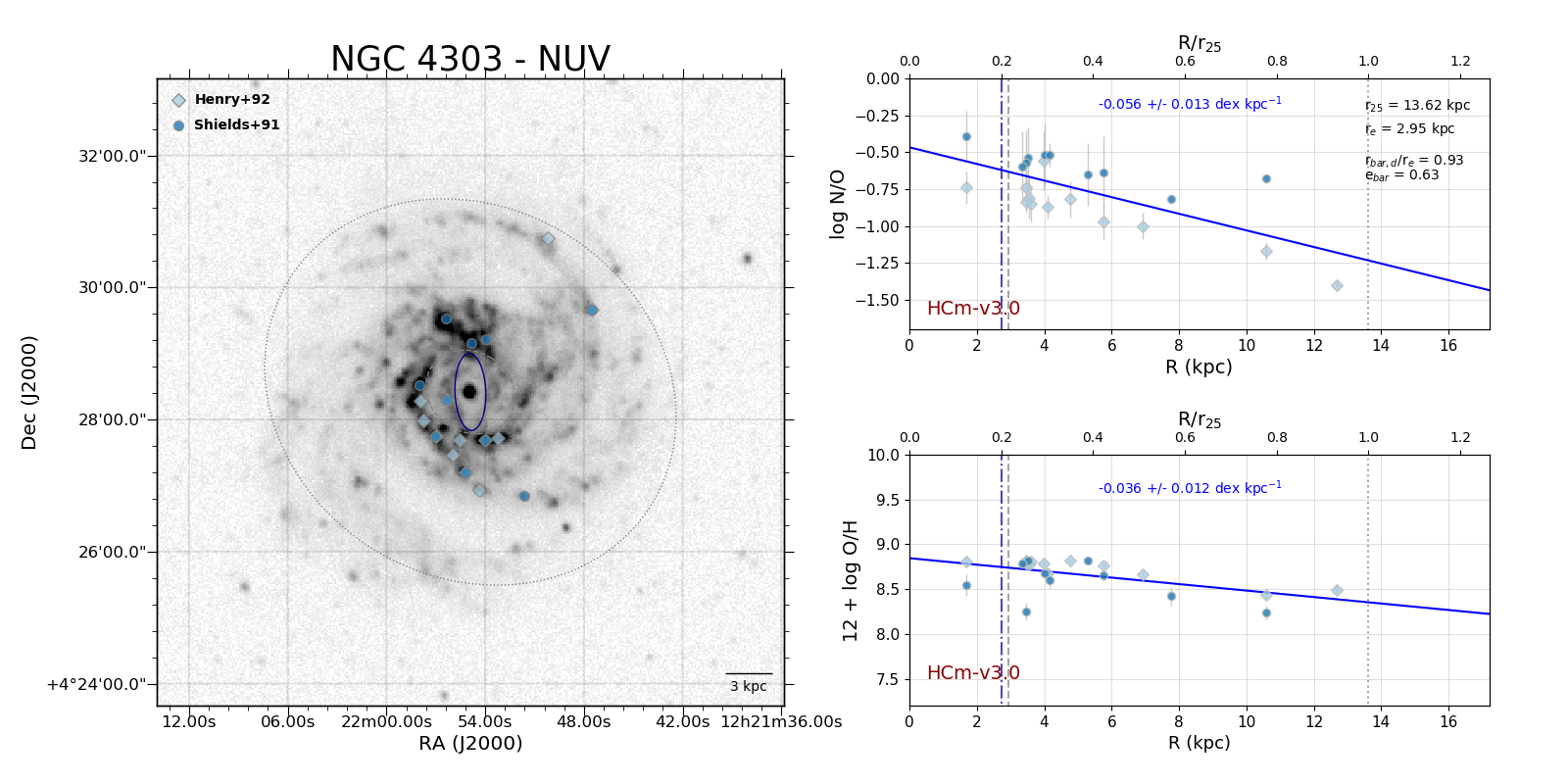}
\end{minipage}
\begin{minipage}{1.05\textwidth}
\hspace{-1.2cm}\includegraphics[width=1.12\textwidth]{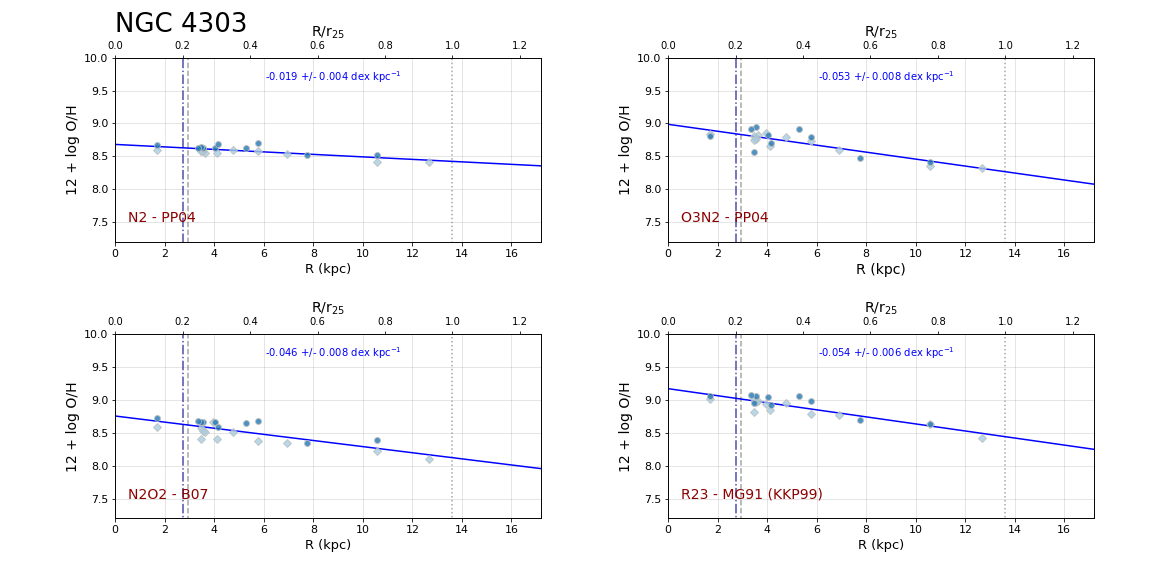}
\end{minipage}
\caption{Same as Fig.~\ref{Afig1} for NGC~4303.}\label{Afig35}
\end{figure*}
\clearpage
\begin{figure*}
\begin{minipage}{1.05\textwidth}
\hspace{-1.2cm}  \includegraphics[width=1.12\textwidth]{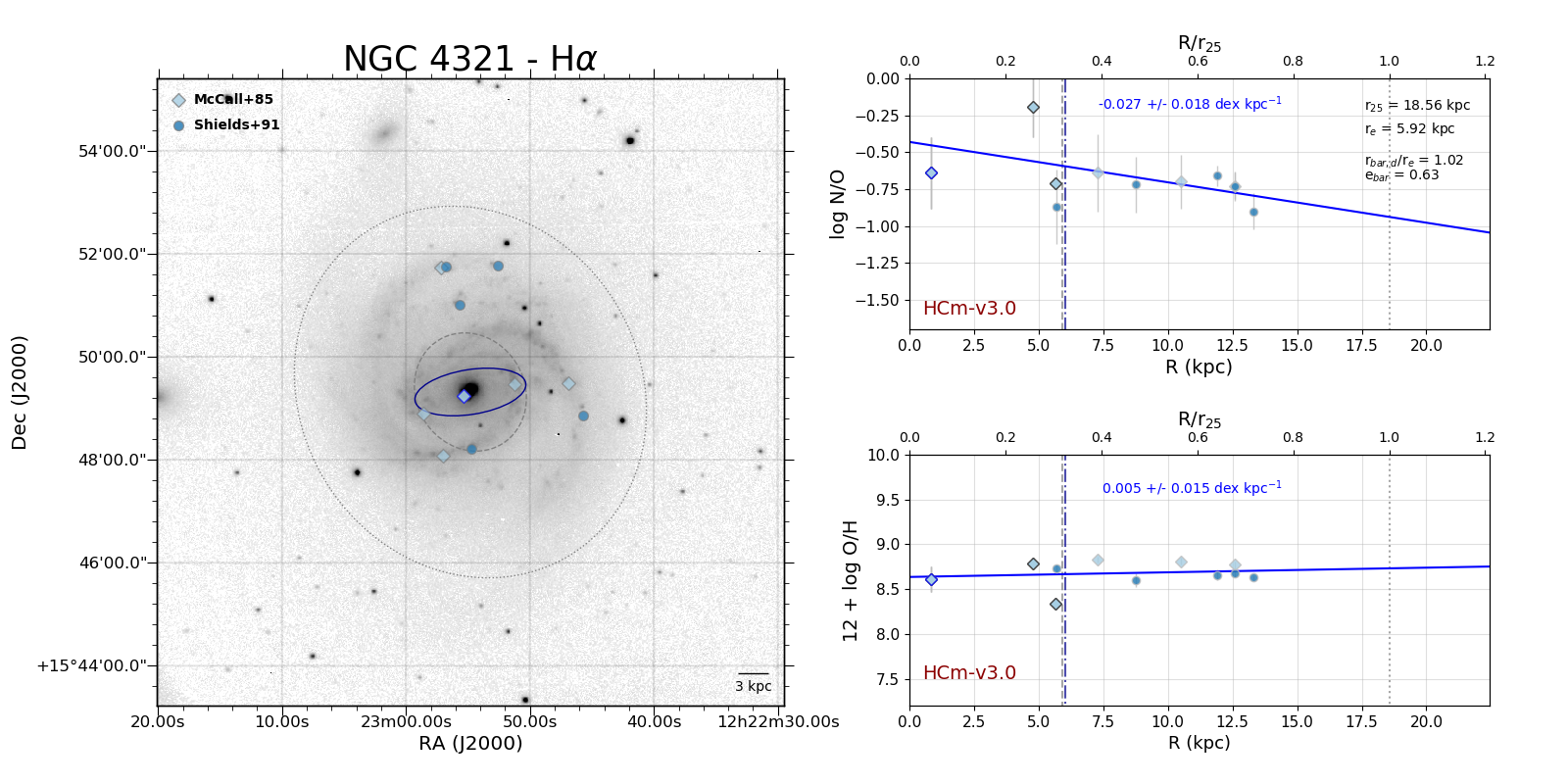}
\end{minipage}
\begin{minipage}{1.05\textwidth}
\hspace{-1.2cm}  \includegraphics[width=1.12\textwidth]{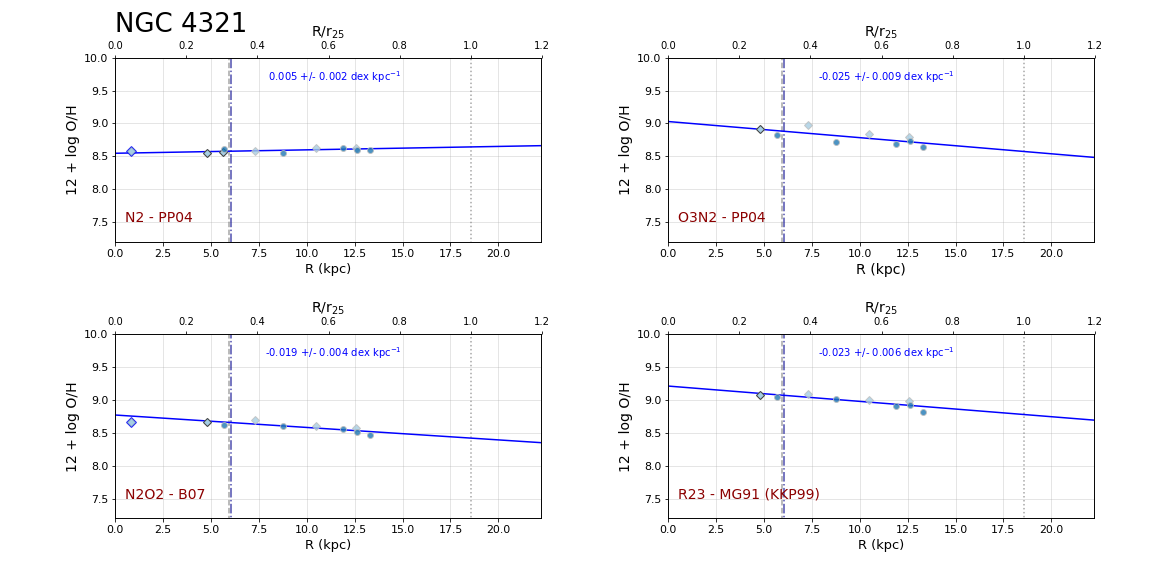}
\end{minipage}
\caption{Same as Fig.~\ref{Afig1} for NGC~4321.}\label{Afig36}
\end{figure*}
\clearpage
\begin{figure*}
\begin{minipage}{1.05\textwidth}
\hspace{-1.2cm}\includegraphics[width=1.12\textwidth]{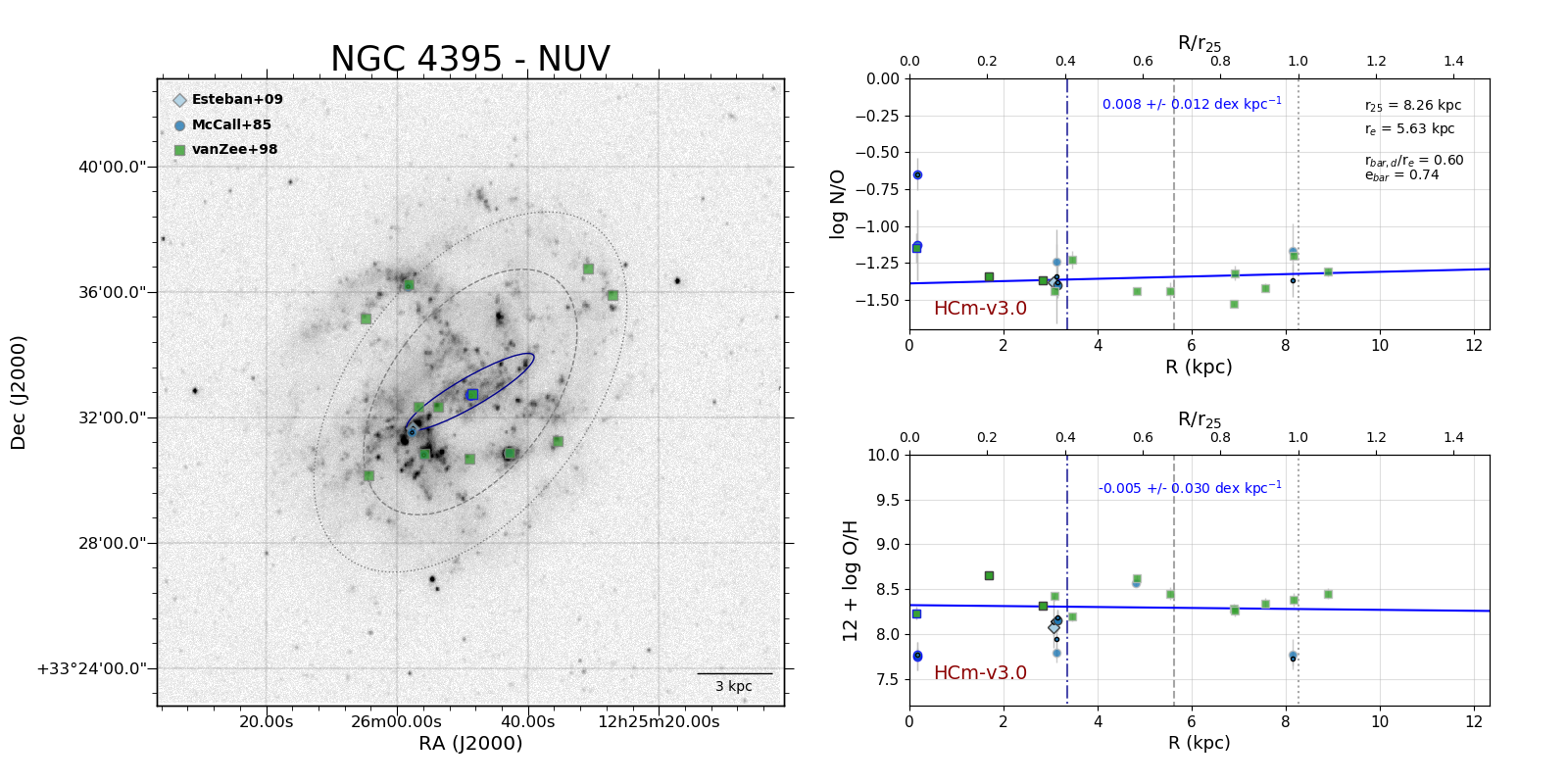}
\end{minipage}
\begin{minipage}{1.05\textwidth}
\hspace{-1.2cm}\includegraphics[width=1.12\textwidth]{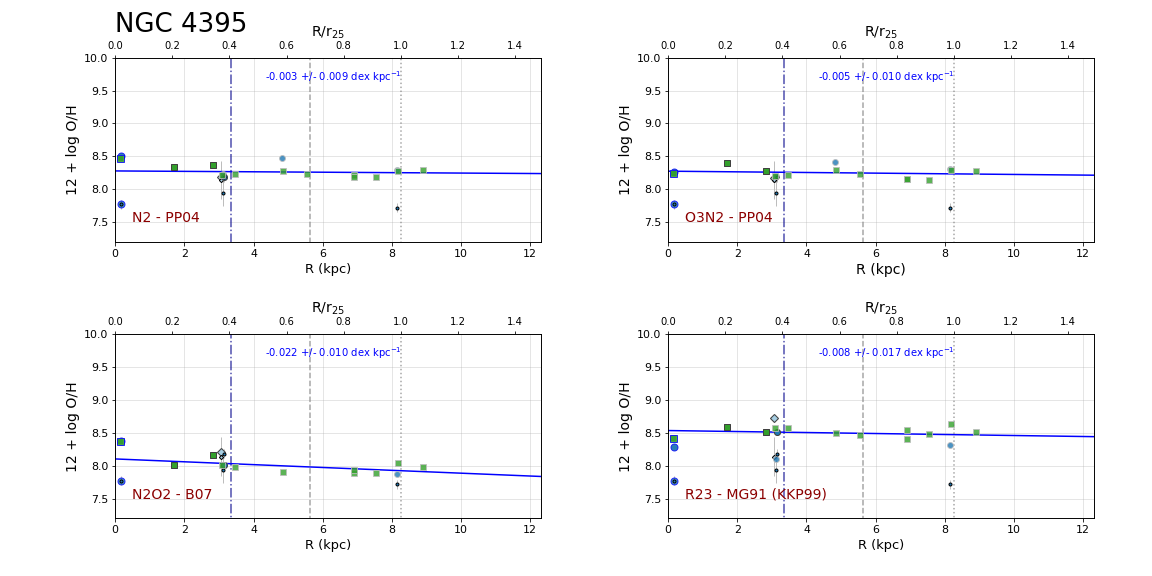}
\end{minipage}
\caption{Same as Fig.~\ref{Afig1} for NGC~4395.}\label{Afig31}
\end{figure*}
\clearpage
\begin{figure*}
\begin{minipage}{1.05\textwidth}
\hspace{-1.2cm}  \includegraphics[width=1.12\textwidth]{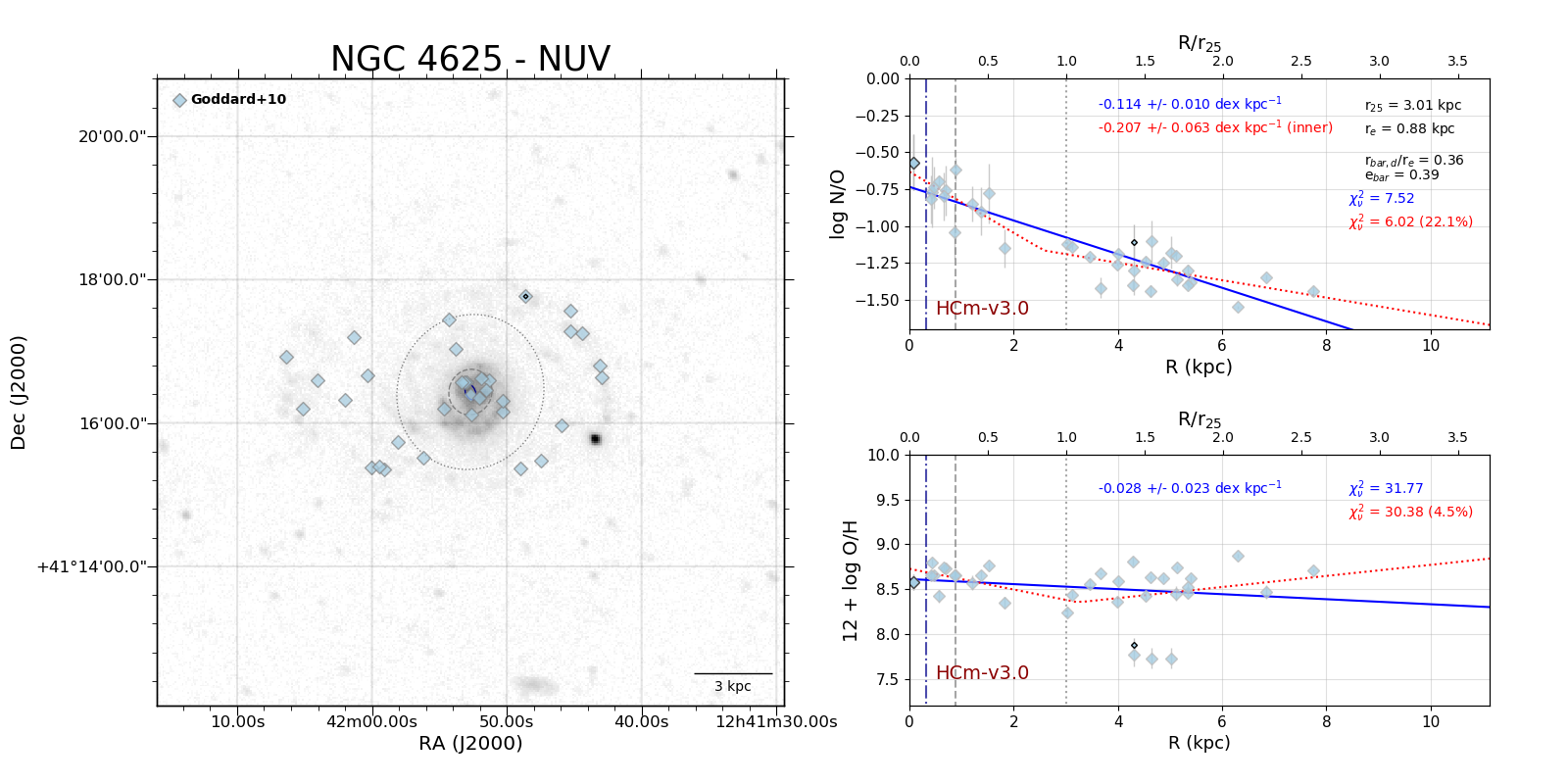}
 \end{minipage}
\begin{minipage}{1.05\textwidth}
\hspace{-1.2cm} \includegraphics[width=1.12\textwidth]{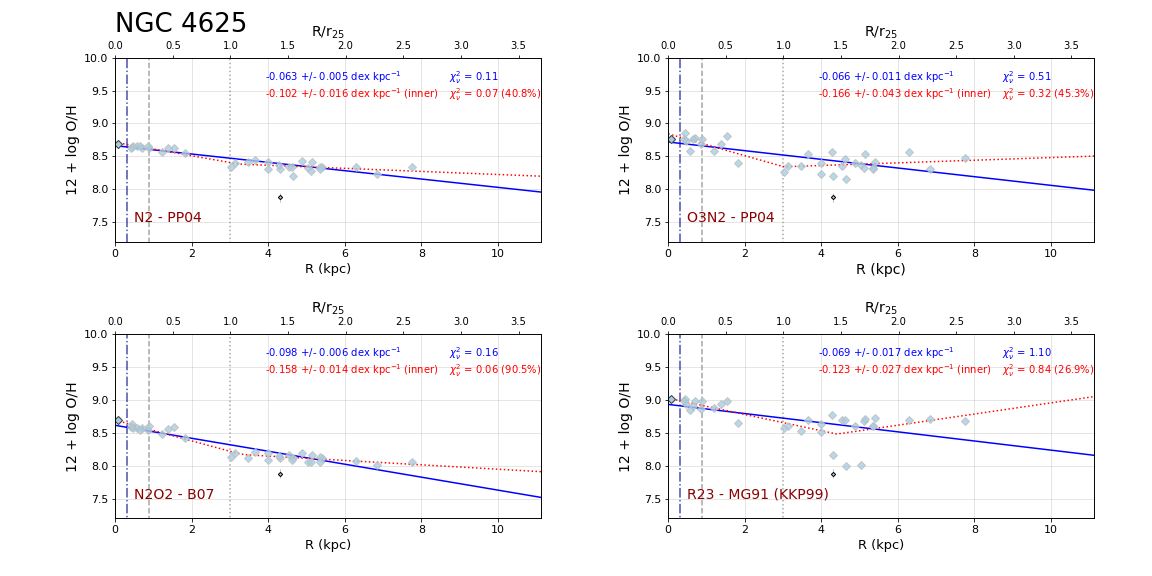}
\end{minipage}
\caption{Same as Fig.~\ref{Afig1} for NGC~4625.}\label{Afig37}
\end{figure*}
\clearpage
\begin{figure*}
\begin{minipage}{1.05\textwidth}
\hspace{-1.2cm}   \includegraphics[width=1.12\textwidth]{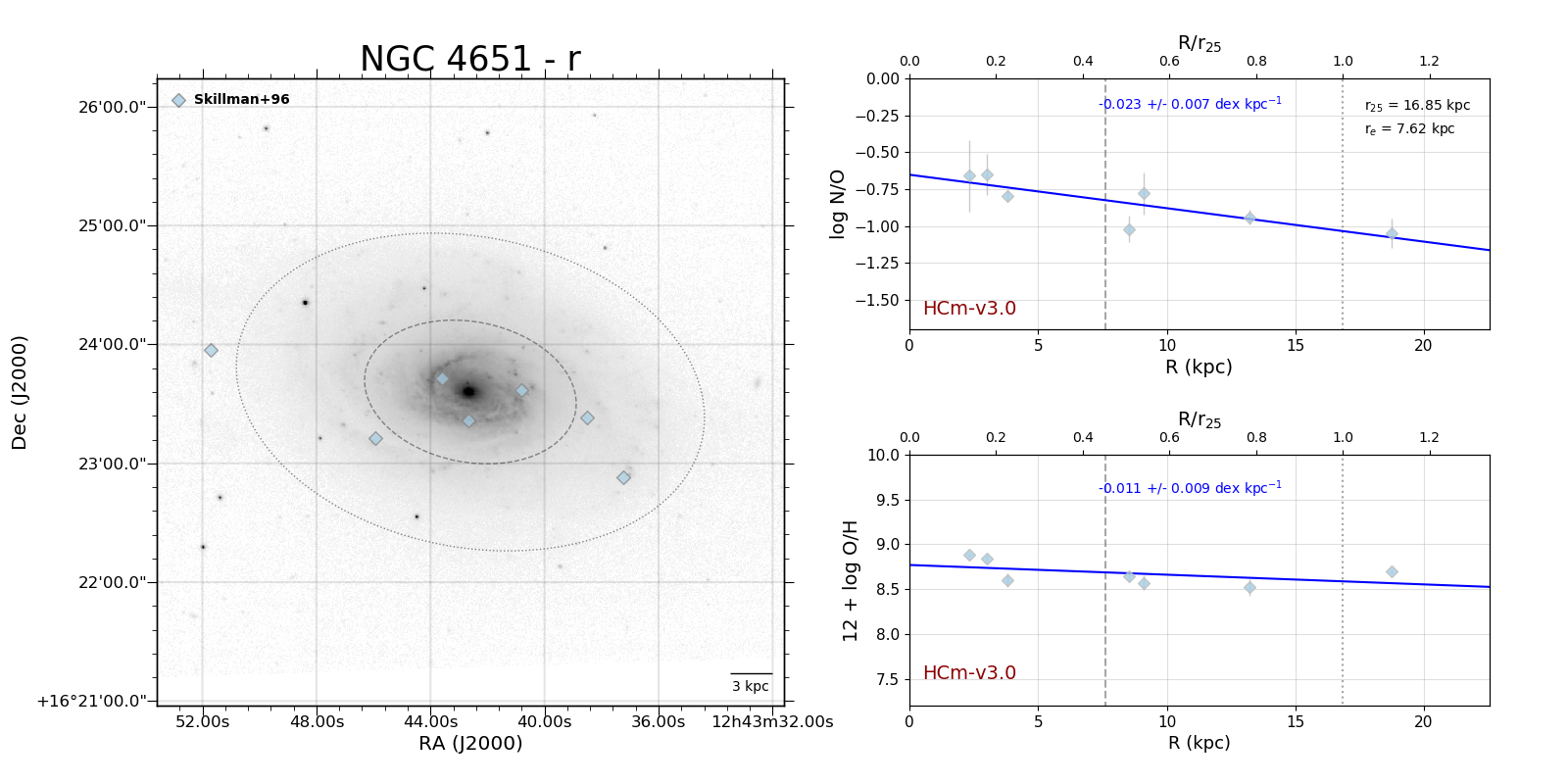}
\end{minipage}
\begin{minipage}{1.05\textwidth}
\hspace{-1.2cm}   \includegraphics[width=1.12\textwidth]{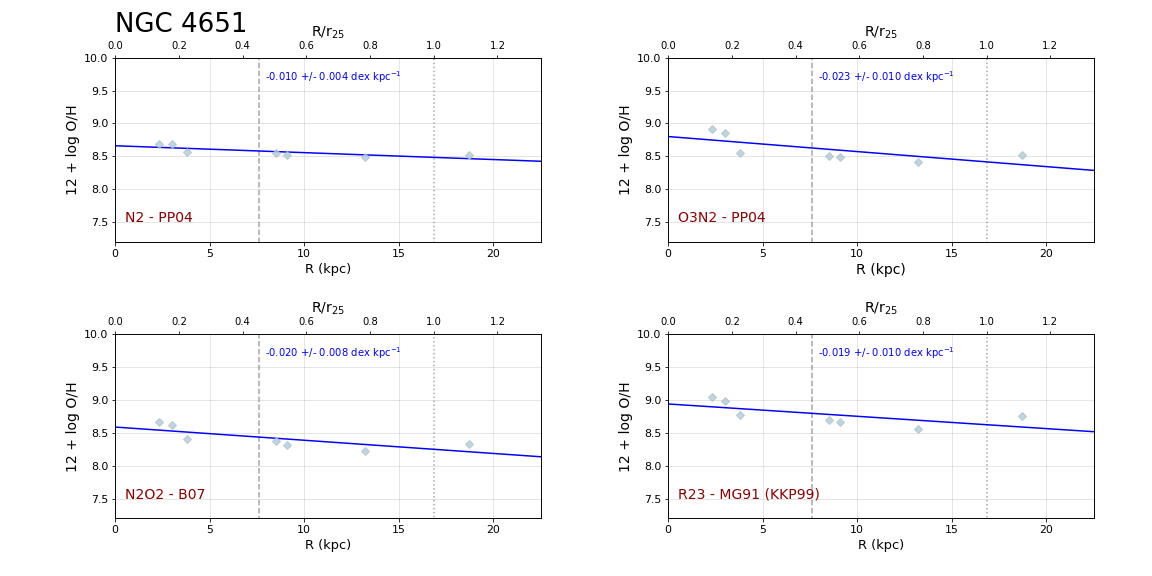}
\end{minipage}
\caption{Same as Fig.~\ref{Afig1} for NGC~4651.}\label{Afig38}
\end{figure*}
\clearpage
\begin{figure*}
\begin{minipage}{1.05\textwidth}
\hspace{-1.2cm}   \includegraphics[width=1.12\textwidth]{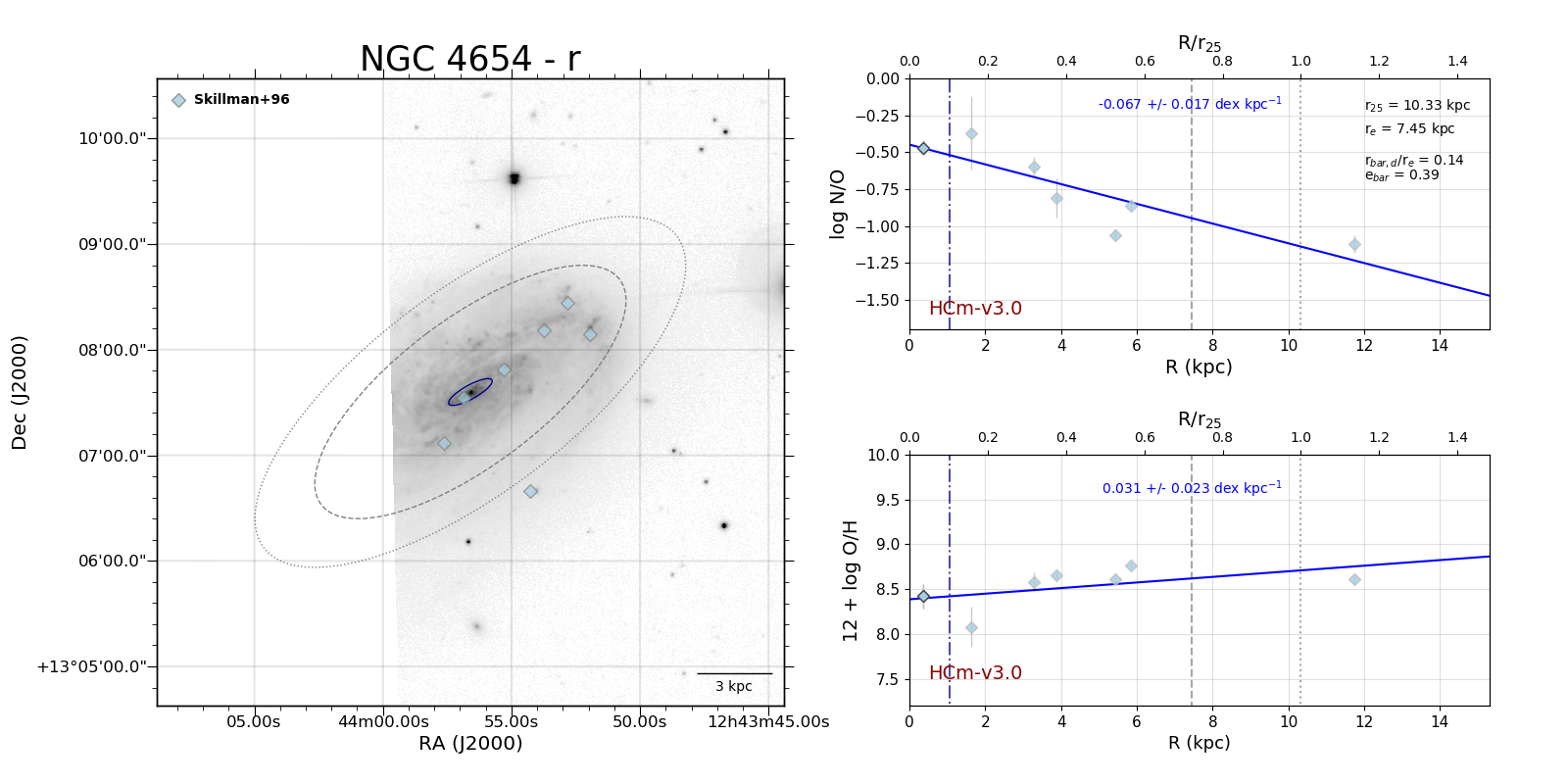}
 \end{minipage}
\begin{minipage}{1.05\textwidth}
\hspace{-1.2cm}   \includegraphics[width=1.12\textwidth]{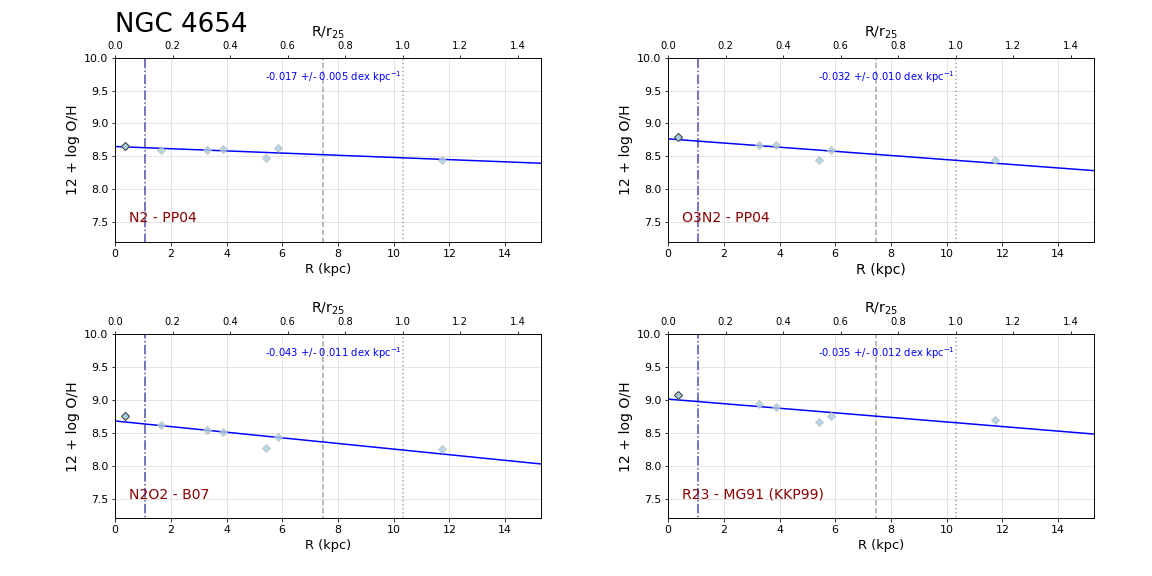}
\end{minipage}
\caption{Same as Fig.~\ref{Afig1} for NGC~4654.}\label{Afig39}
\end{figure*}
\clearpage
\begin{figure*}
\begin{minipage}{1.05\textwidth}
\hspace{-1.2cm}  \includegraphics[width=1.12\textwidth]{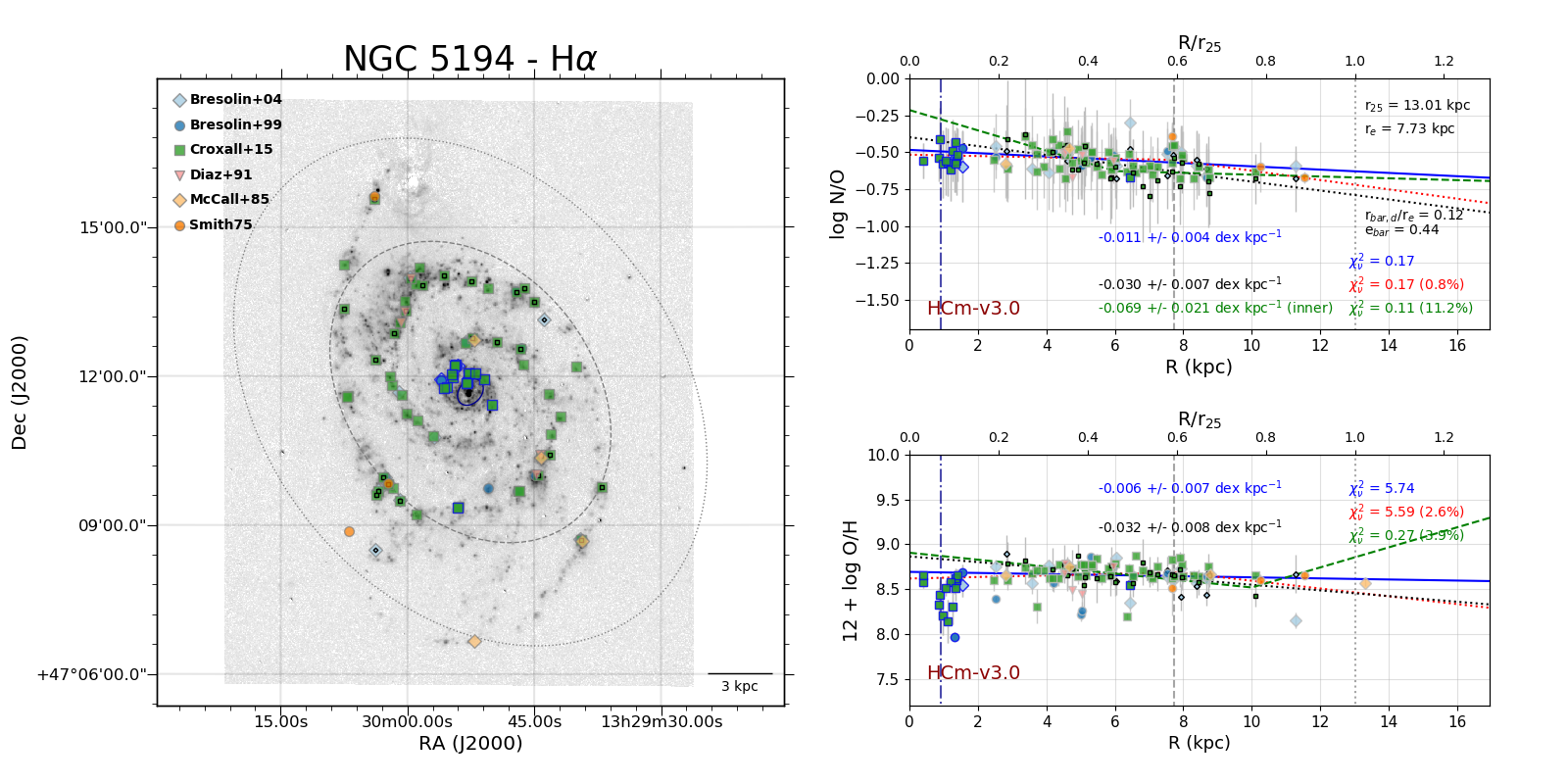}
\end{minipage}
\begin{minipage}{1.05\textwidth}
\hspace{-1.2cm}   \includegraphics[width=1.12\textwidth]{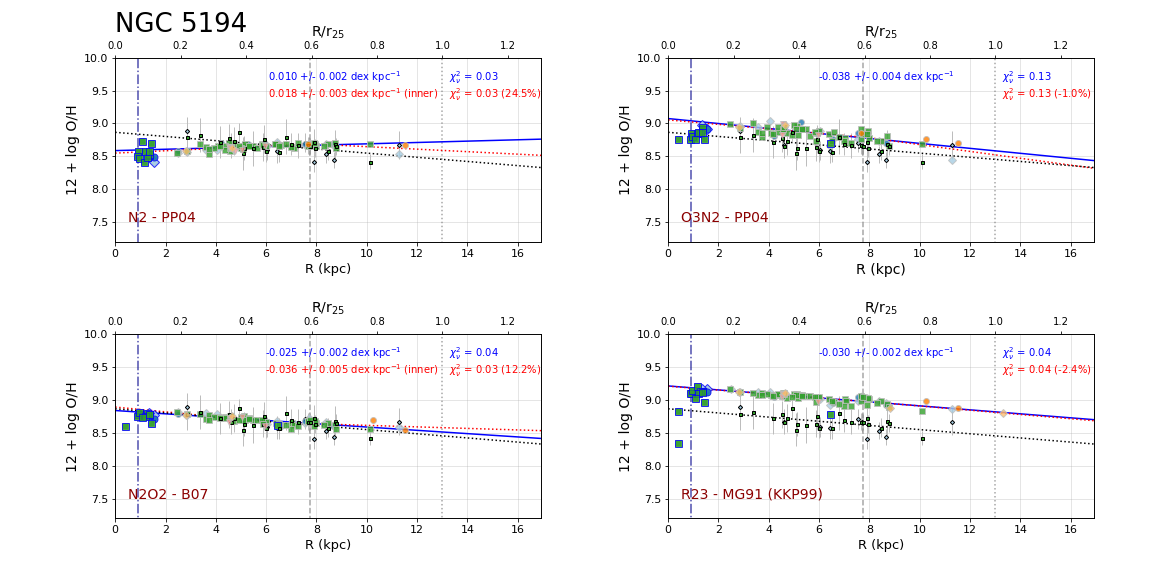}
\end{minipage}
\caption{Same as Fig.~\ref{Afig1} for NGC~5194.}\label{n5194}
\end{figure*}
\clearpage
\begin{figure*}
\begin{minipage}{1.05\textwidth}
\hspace{-1.2cm}\includegraphics[width=1.12\textwidth]{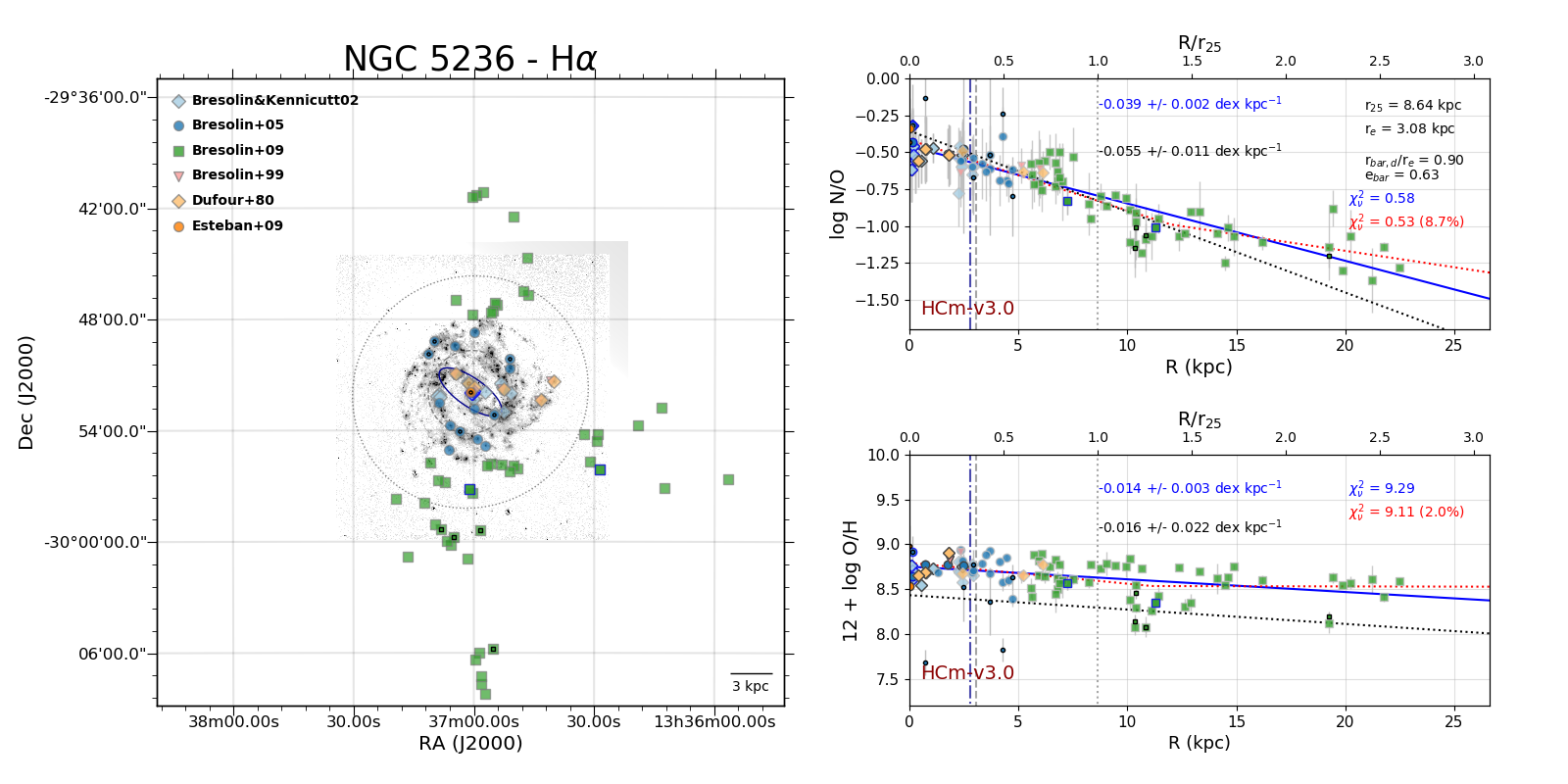}
\end{minipage}
\begin{minipage}{1.05\textwidth}
\hspace{-1.2cm}\includegraphics[width=1.12\textwidth]{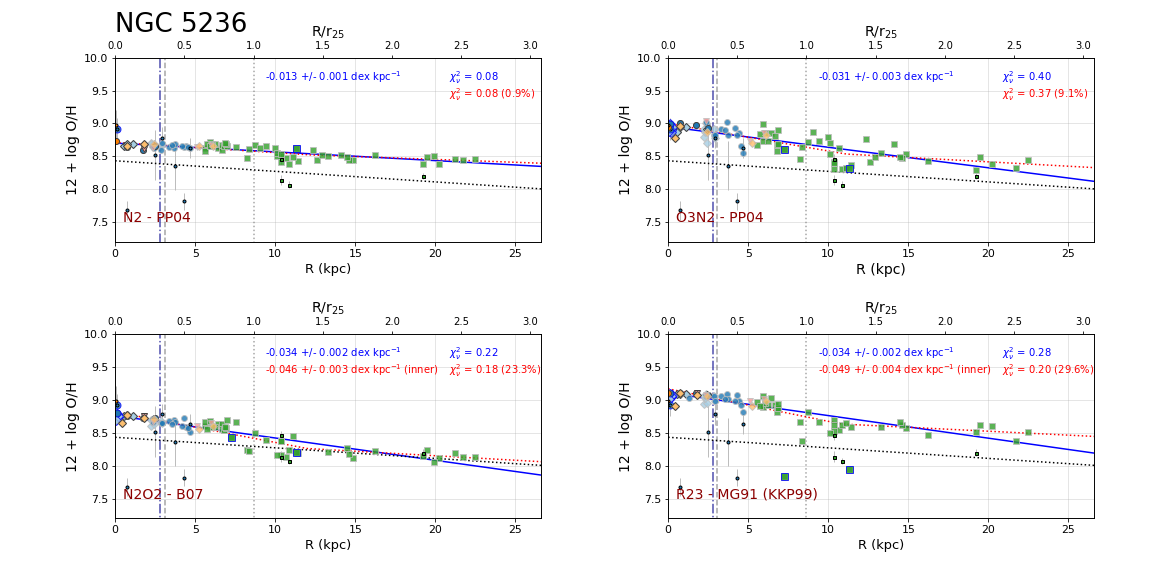}
\end{minipage}
\caption{Same as Fig.~\ref{Afig1} for NGC~5236.}\label{Afig43}
\end{figure*}
\clearpage
\begin{figure*}
\begin{minipage}{1.05\textwidth}
\hspace{-1.2cm}\includegraphics[width=1.12\textwidth]{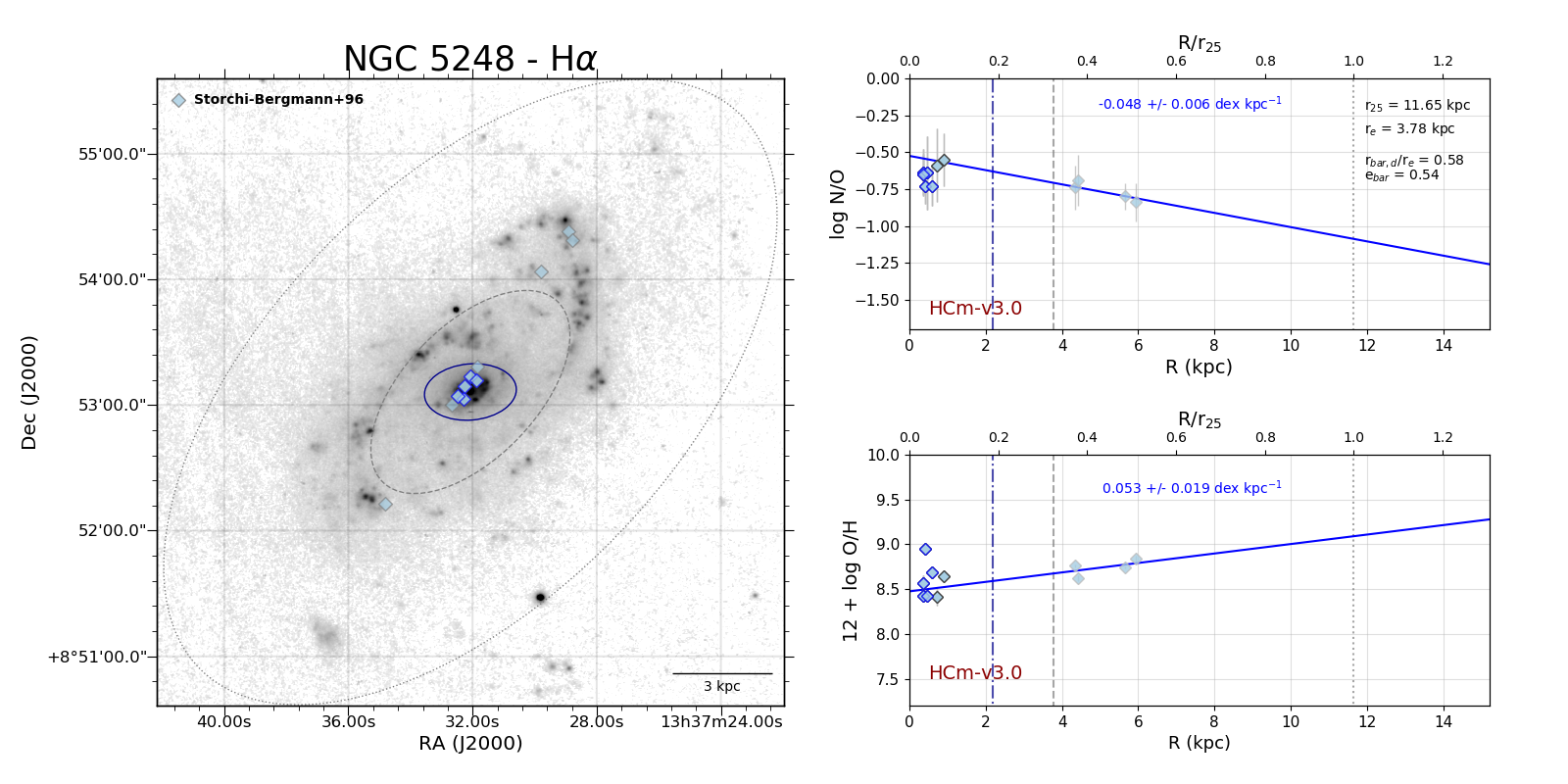}
\end{minipage}
\begin{minipage}{1.05\textwidth}
\hspace{-1.2cm}\includegraphics[width=1.12\textwidth]{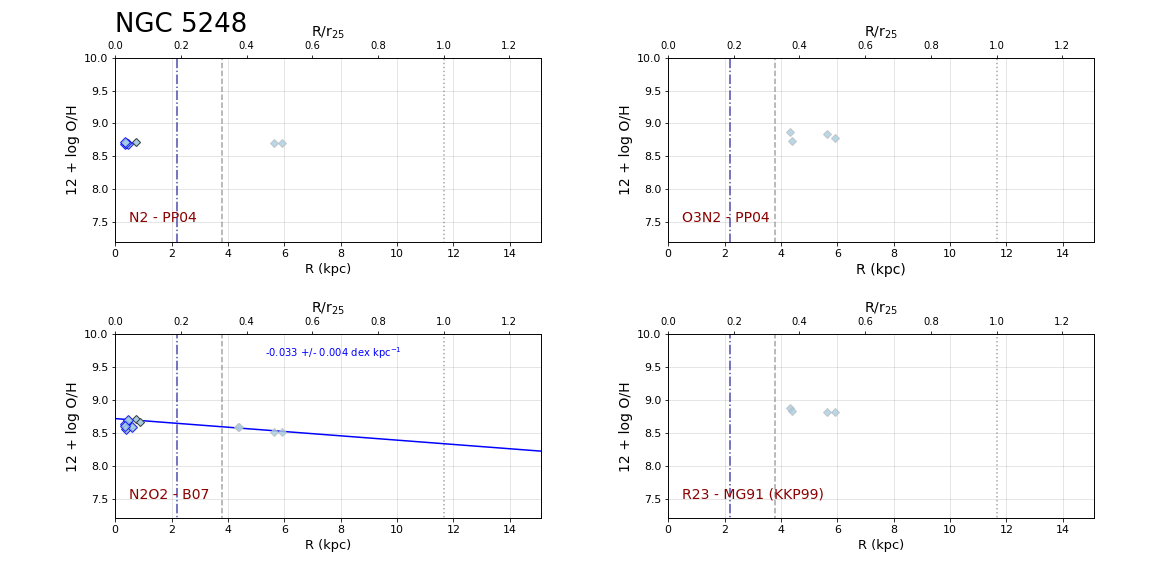}
\end{minipage}
\caption{Same as Fig.~\ref{Afig1} for NGC~5248.}\label{Afig44}
\end{figure*}
\clearpage
\begin{figure*}
\begin{minipage}{1.05\textwidth}
\hspace{-1.2cm}\includegraphics[width=1.12\textwidth]{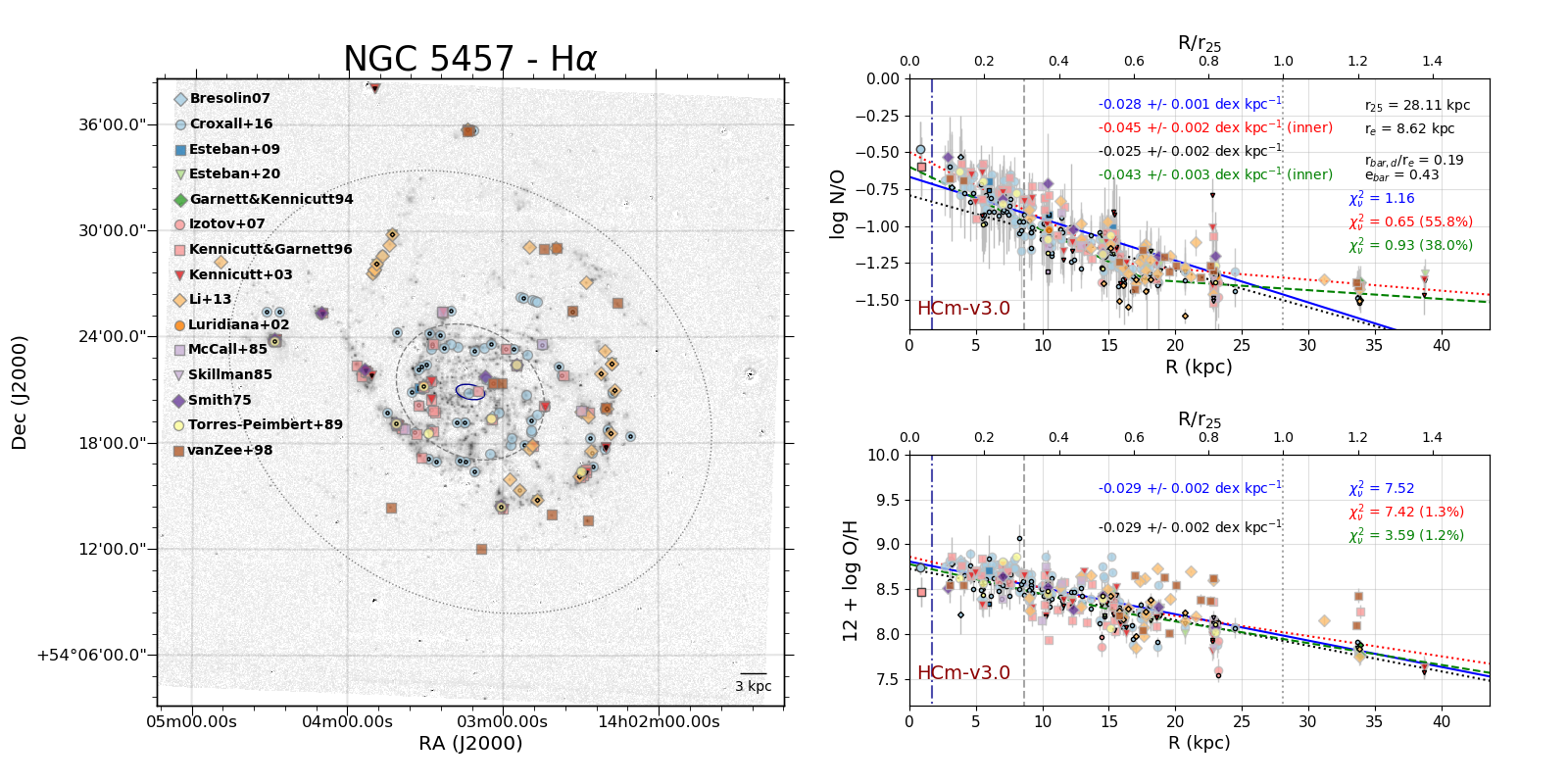}
\end{minipage}
\begin{minipage}{1.05\textwidth}
\hspace{-1.2cm}\includegraphics[width=1.12\textwidth]{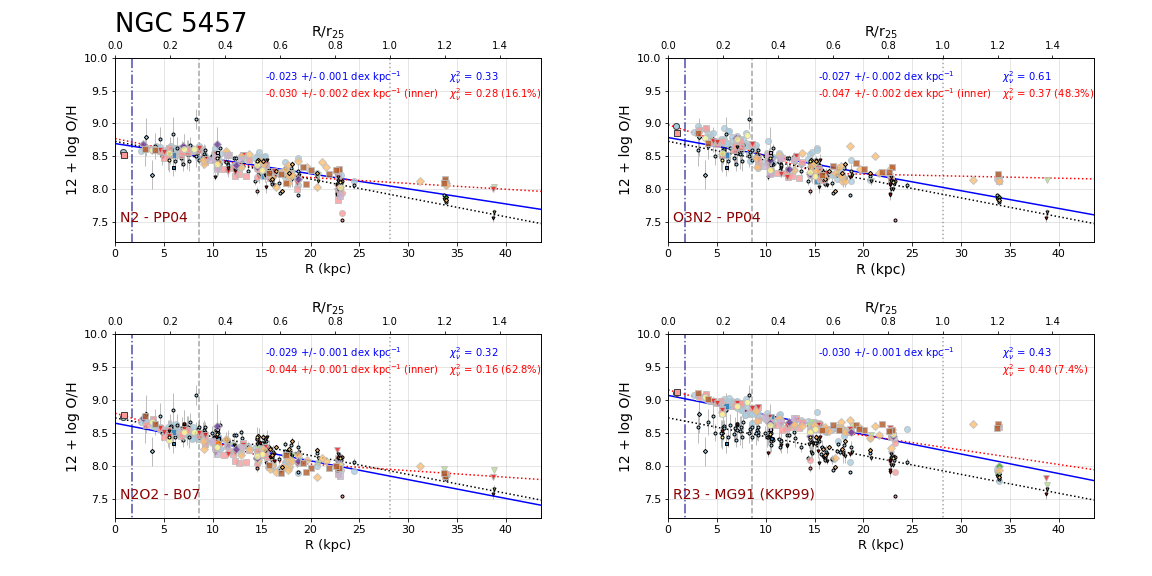}
\end{minipage}
\caption{Same as Fig.~\ref{Afig1} for NGC~5457.}\label{Afig45}
\end{figure*}
\clearpage
\begin{figure*}
\begin{minipage}{1.05\textwidth}
\hspace{-1.2cm}\includegraphics[width=1.12\textwidth]{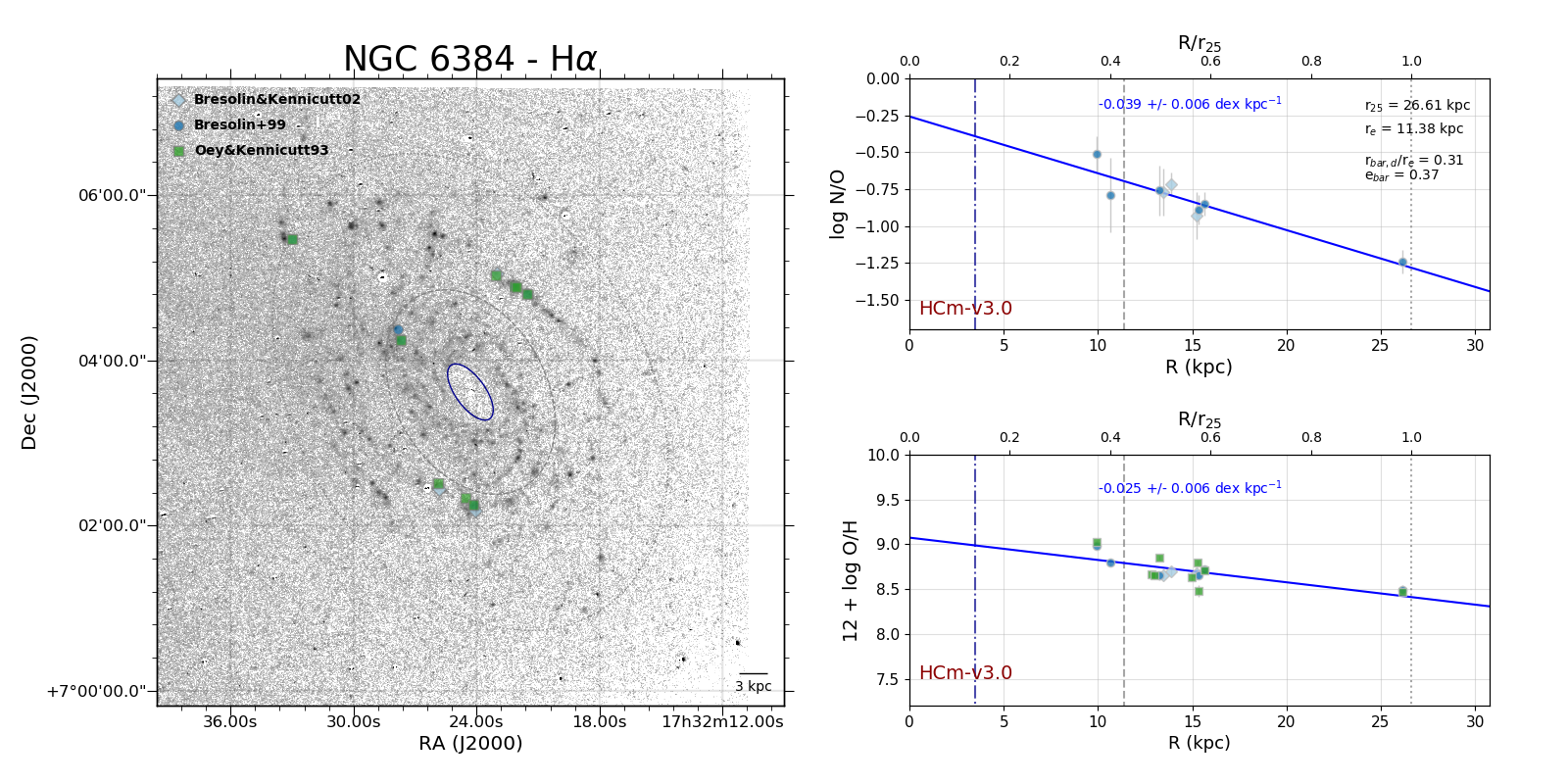}
\end{minipage}
\begin{minipage}{1.05\textwidth}
\hspace{-1.2cm}\includegraphics[width=1.12\textwidth]{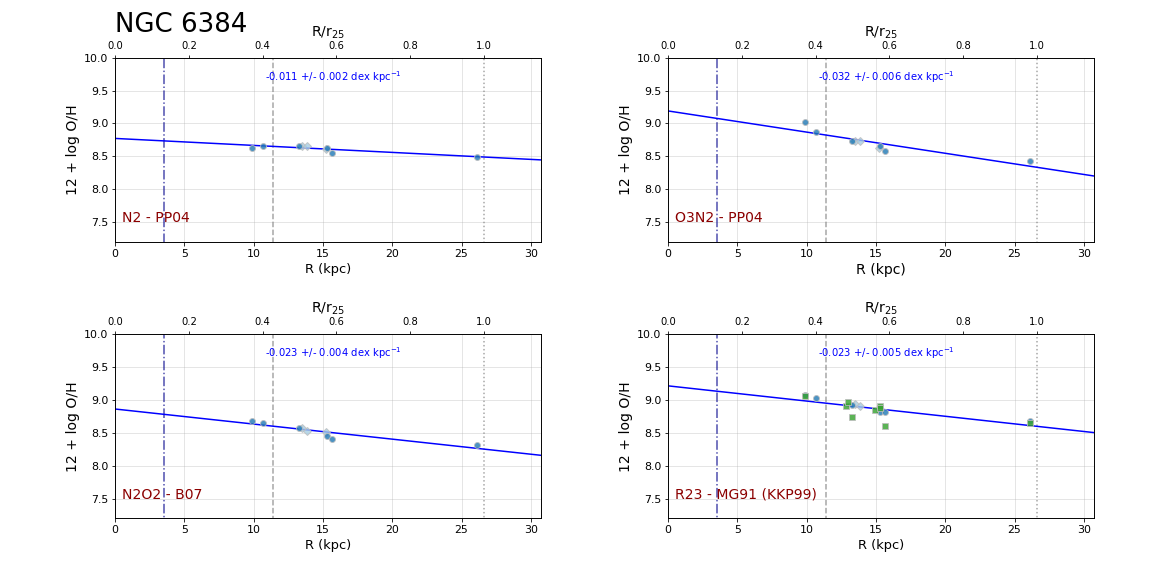}
\end{minipage}
\caption{Same as Fig.~\ref{Afig1} for NGC~6384.}\label{Afig46}
\end{figure*}
\clearpage
\begin{figure*}
\begin{minipage}{1.05\textwidth}
\hspace{-1.2cm}\includegraphics[width=1.12\textwidth]{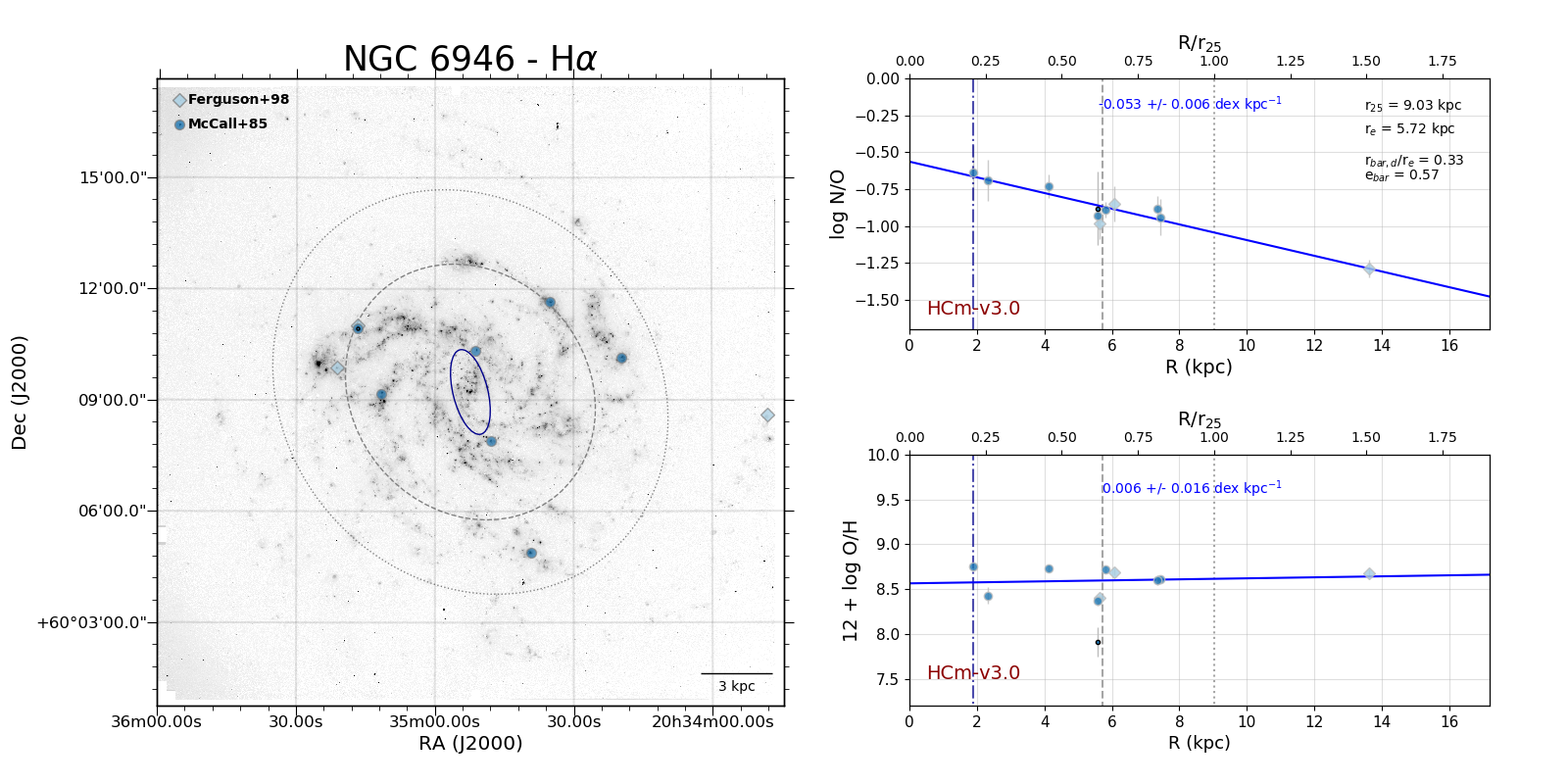}
\end{minipage}
\begin{minipage}{1.05\textwidth}
\hspace{-1.2cm}\includegraphics[width=1.12\textwidth]{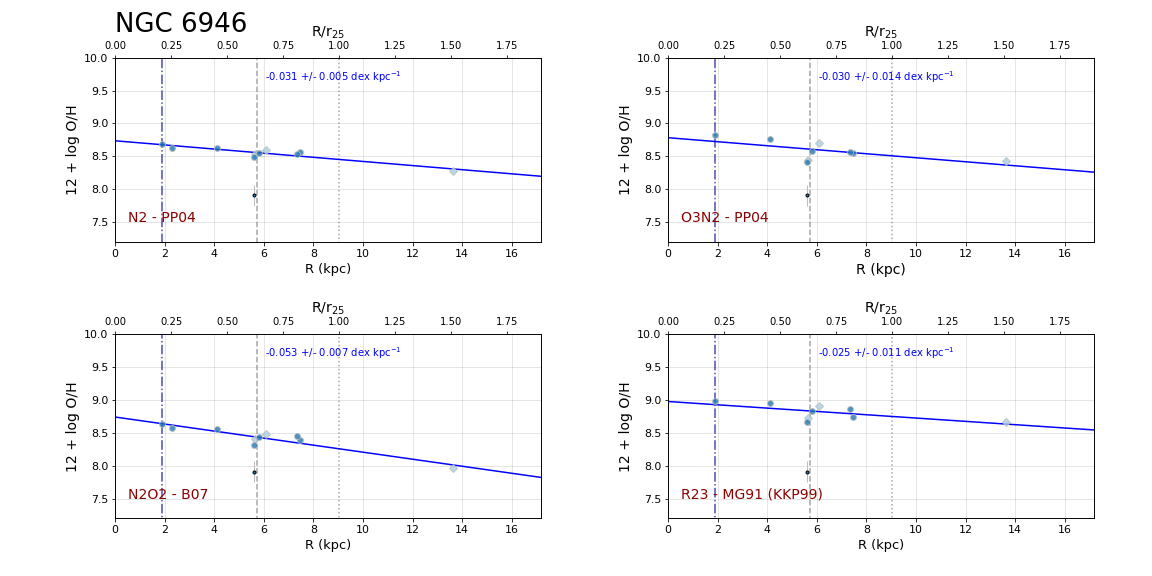}
\end{minipage}
\caption{Same as Fig.~\ref{Afig1} for NGC~6946.}\label{Afig47}
\end{figure*}
\clearpage
\begin{figure*}
\begin{minipage}{1.05\textwidth}
\hspace{-1.2cm}\includegraphics[width=1.12\textwidth]{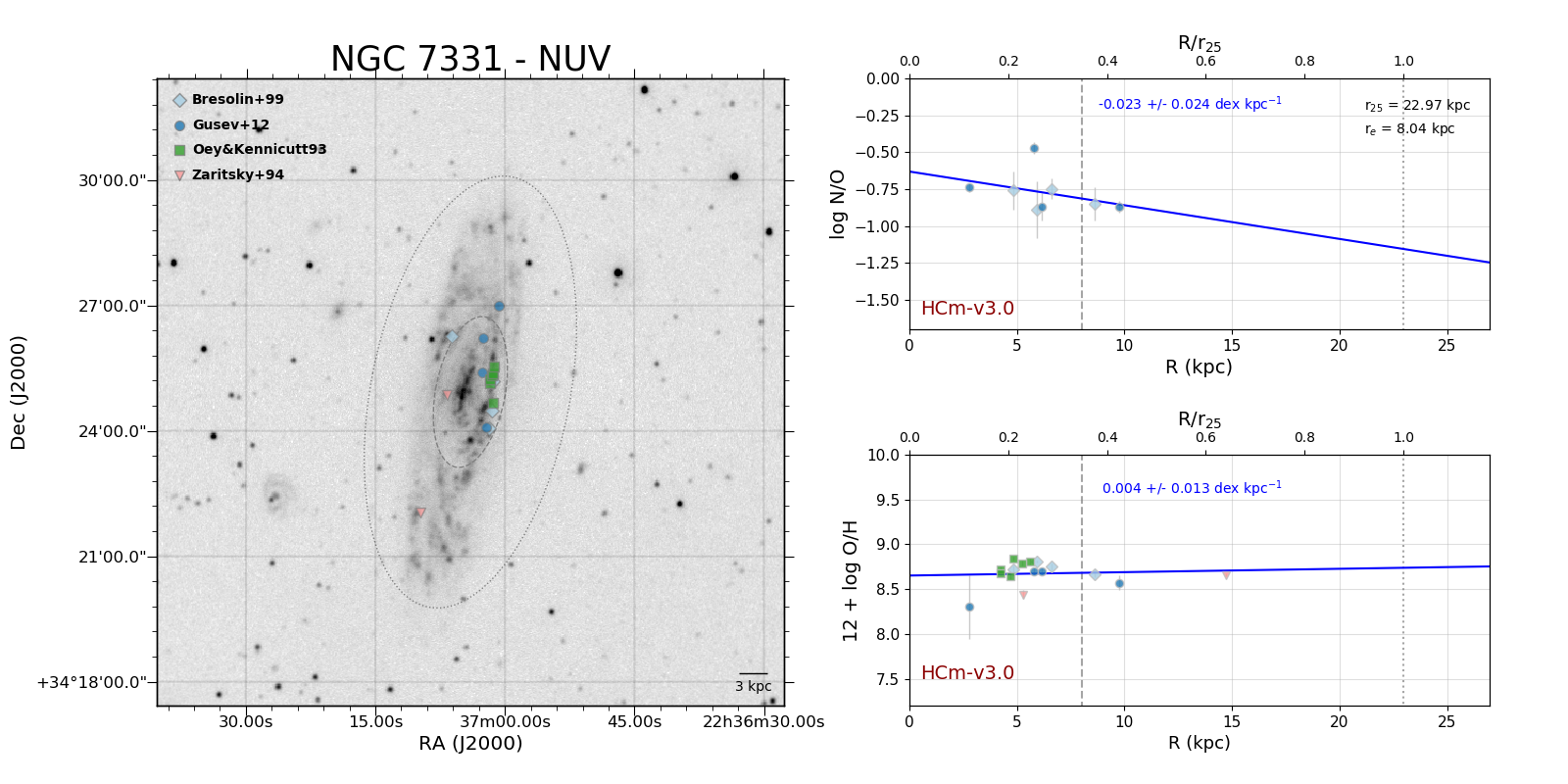}
\end{minipage}
\begin{minipage}{1.05\textwidth}
\hspace{-1.2cm}\includegraphics[width=1.12\textwidth]{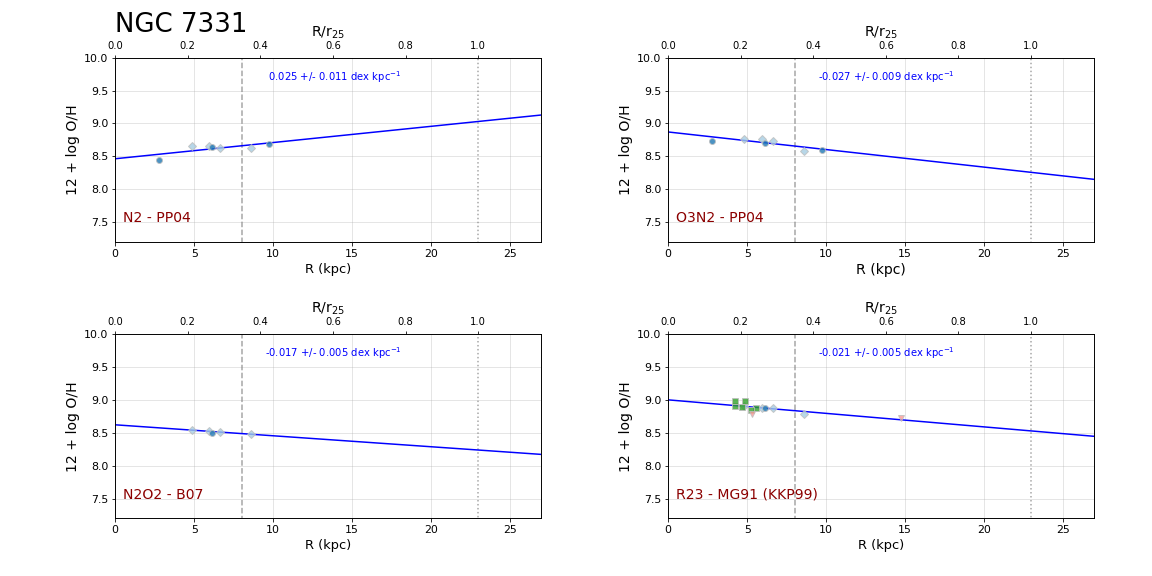}
\end{minipage}
\caption{Same as Fig.~\ref{Afig1} for NGC~7331.}\label{Afig48}
\end{figure*}
\clearpage
\begin{figure*}
\begin{minipage}{1.05\textwidth}
\hspace{-1.2cm}\includegraphics[width=1.12\textwidth]{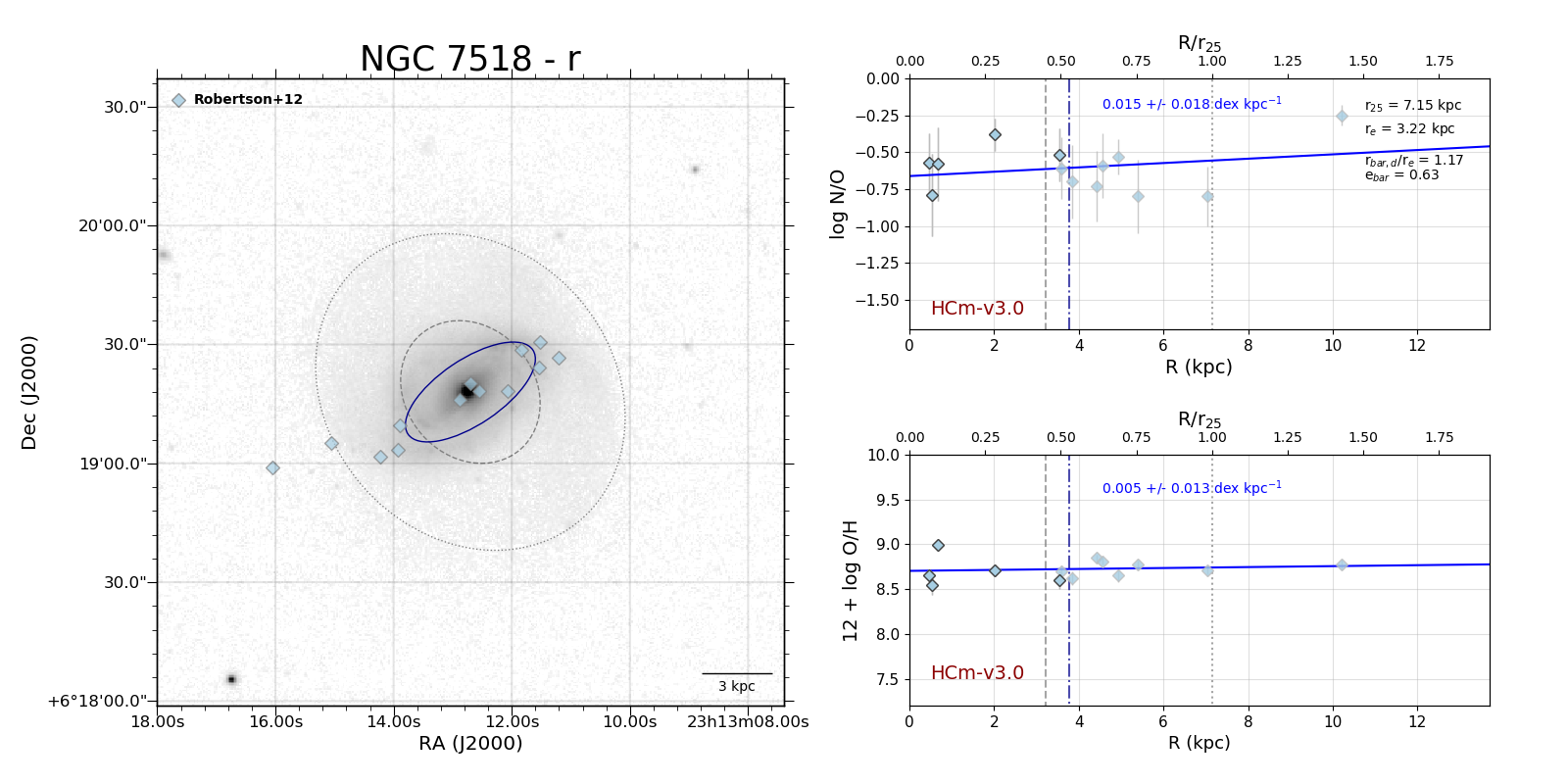}
\end{minipage}
\begin{minipage}{1.05\textwidth}
\hspace{-1.2cm}\includegraphics[width=1.12\textwidth]{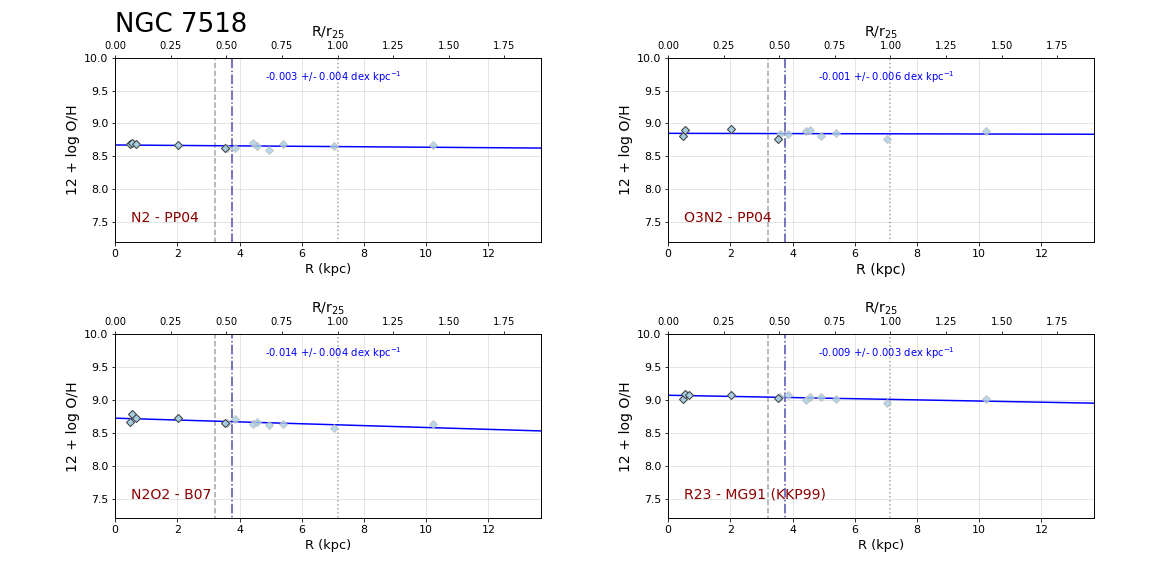}
\end{minipage}
\caption{Same as Fig.~\ref{Afig1} for NGC~7518.}\label{Afig49}
\end{figure*}
\clearpage
\begin{figure*}
\begin{minipage}{1.05\textwidth}
\hspace{-1.2cm}\includegraphics[width=1.12\textwidth]{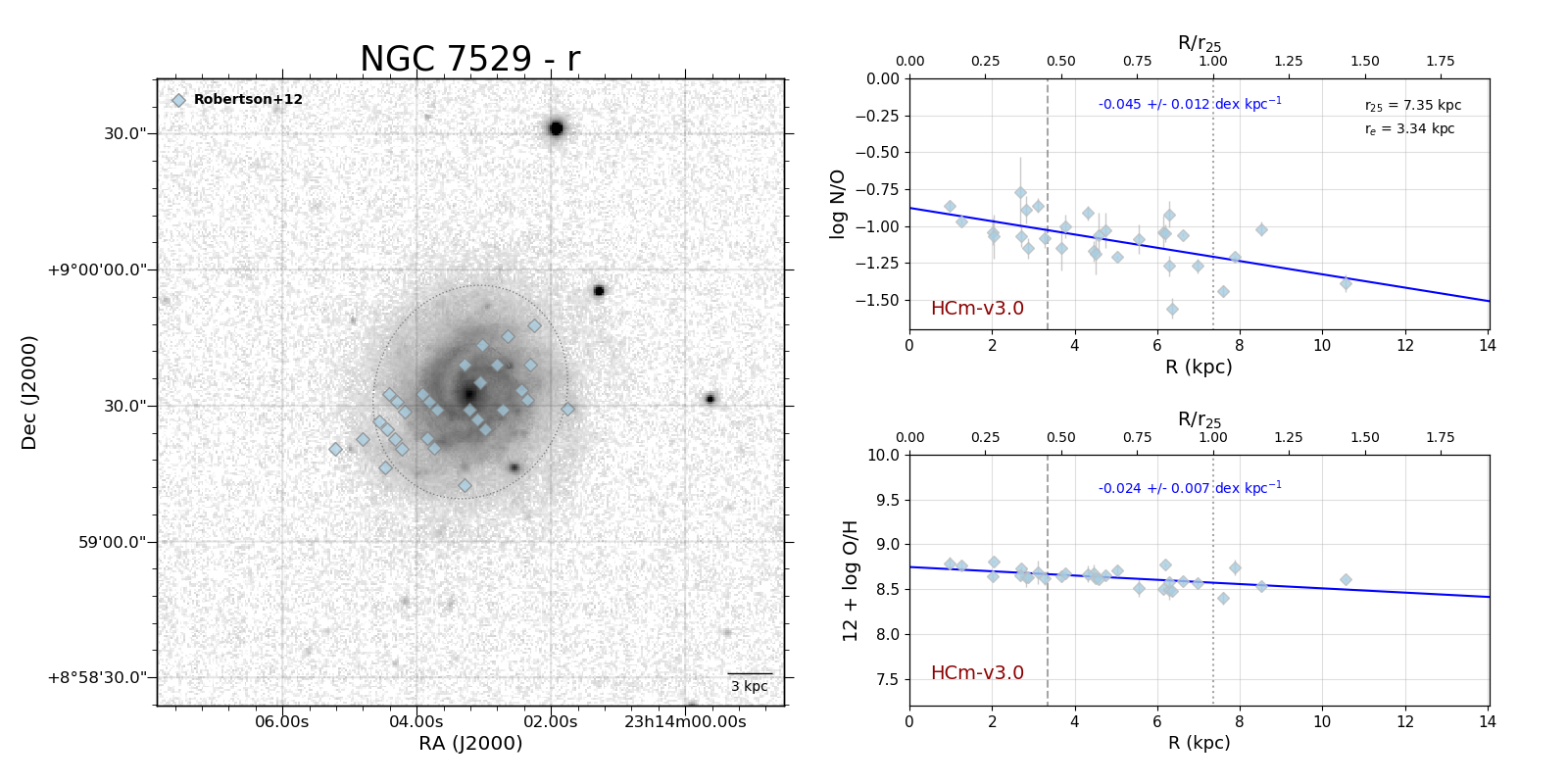}
\end{minipage}
\begin{minipage}{1.05\textwidth}
\hspace{-1.2cm}\includegraphics[width=1.12\textwidth]{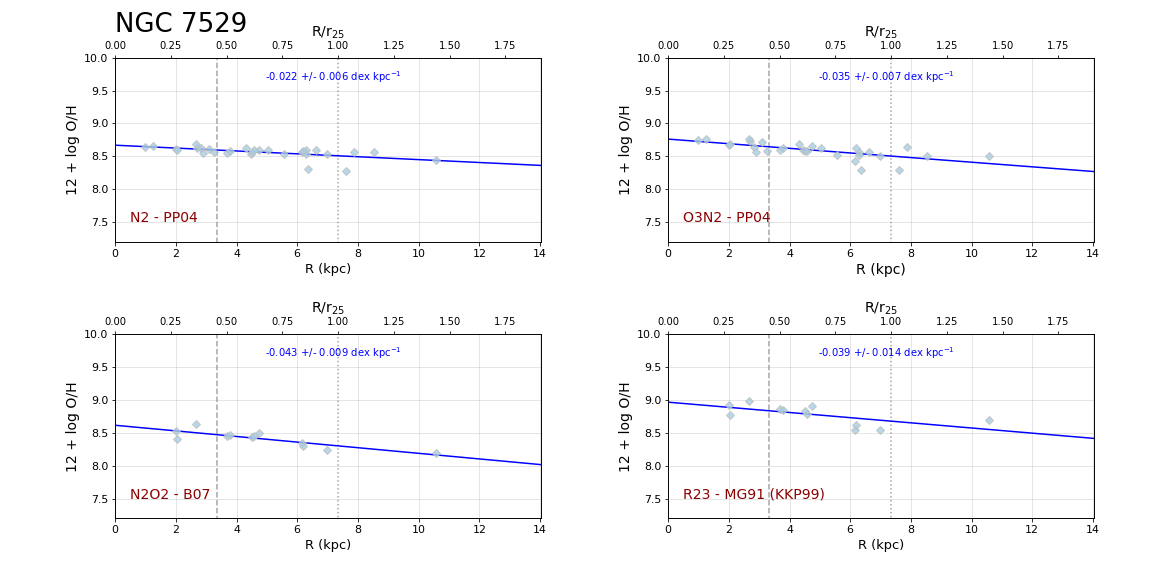}
\end{minipage}
\caption{Same as Fig.~\ref{Afig1} for NGC~7529.}\label{Afig50}
\end{figure*}
\clearpage
\begin{figure*}
\begin{minipage}{1.05\textwidth}
\hspace{-1.2cm}\includegraphics[width=1.12\textwidth]{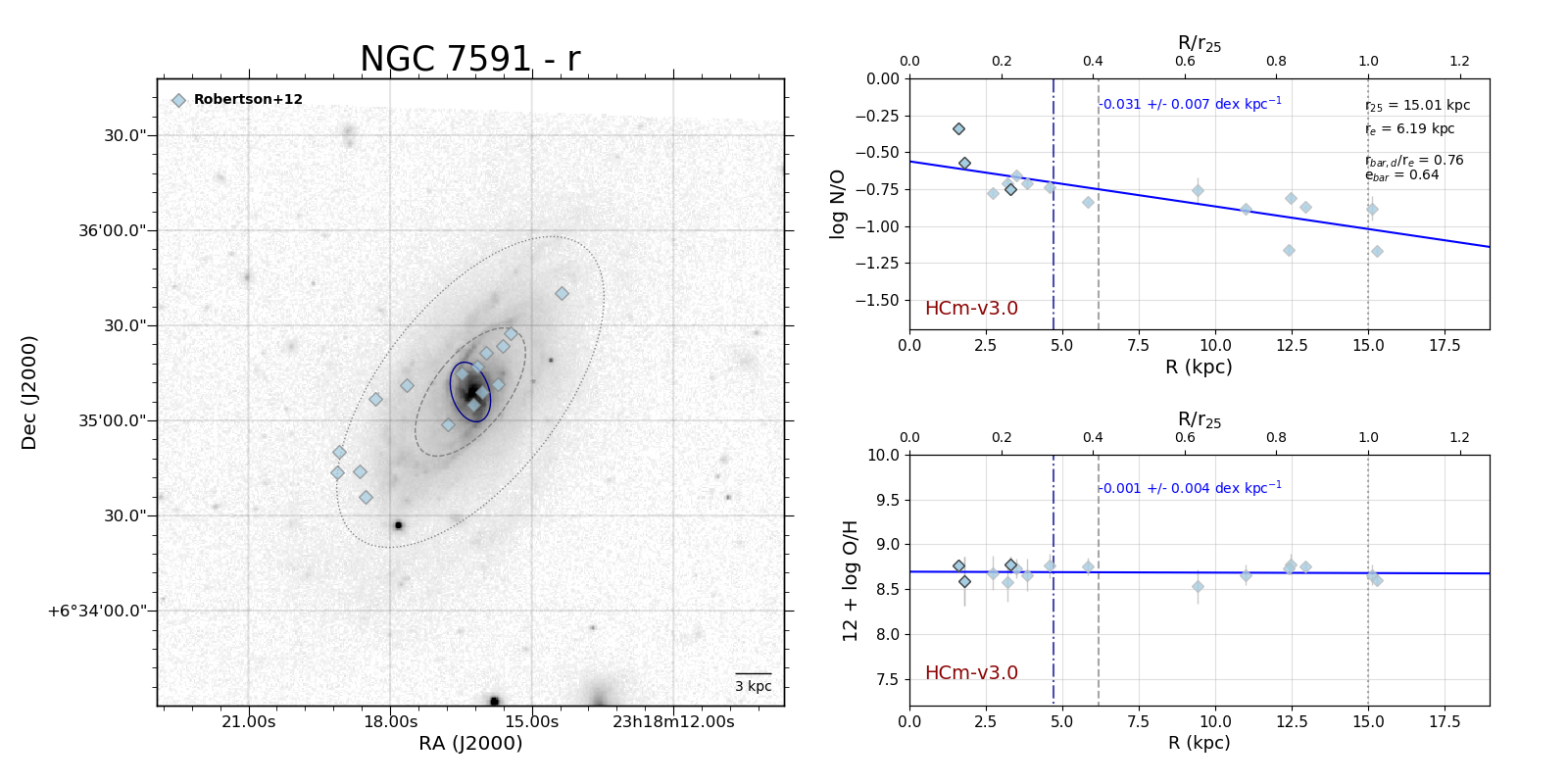}
\end{minipage}
\begin{minipage}{1.05\textwidth}
\hspace{-1.2cm}\includegraphics[width=1.12\textwidth]{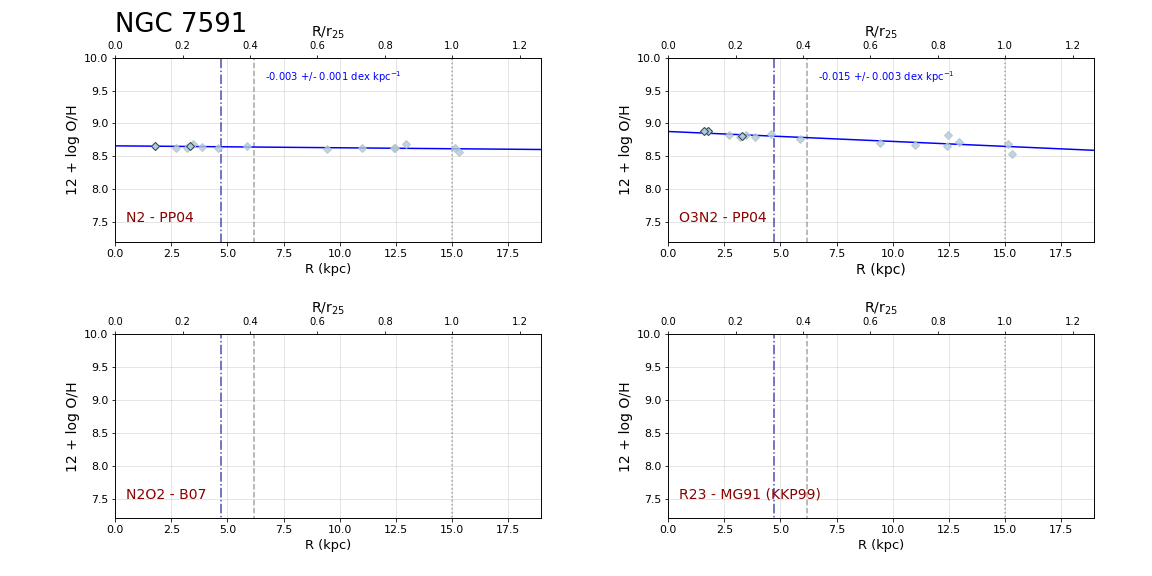}
\end{minipage}
\caption{Same as Fig.~\ref{Afig1} for NGC~7591.}\label{Afig51}
\end{figure*}
\clearpage
\begin{figure*}
\begin{minipage}{1.05\textwidth}
\hspace{-1.2cm}\includegraphics[width=1.12\textwidth]{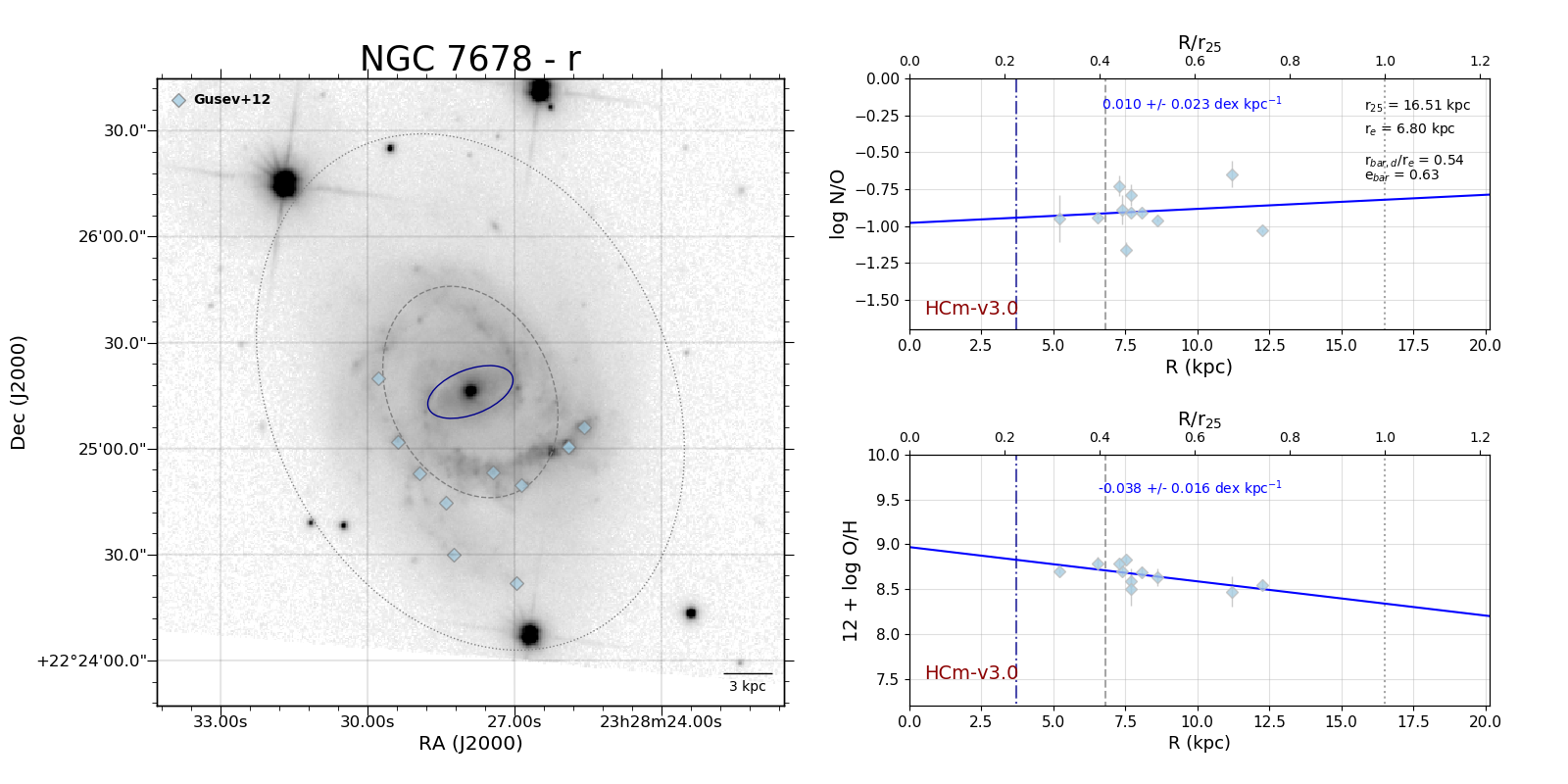}
\end{minipage}
\begin{minipage}{1.05\textwidth}
\hspace{-1.2cm}\includegraphics[width=1.12\textwidth]{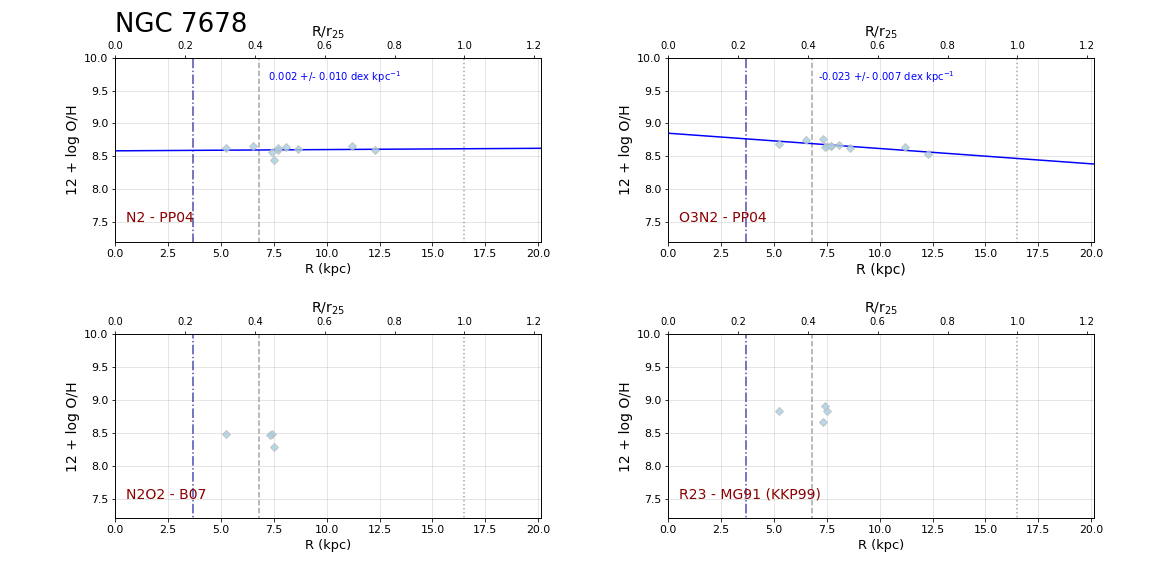}
\end{minipage}
\caption{Same as Fig.~\ref{Afig1} for NGC~7678.}\label{Afig52}
\end{figure*}
\clearpage
\begin{figure*}
\begin{minipage}{1.05\textwidth}
\hspace{-1.2cm}\includegraphics[width=1.12\textwidth]{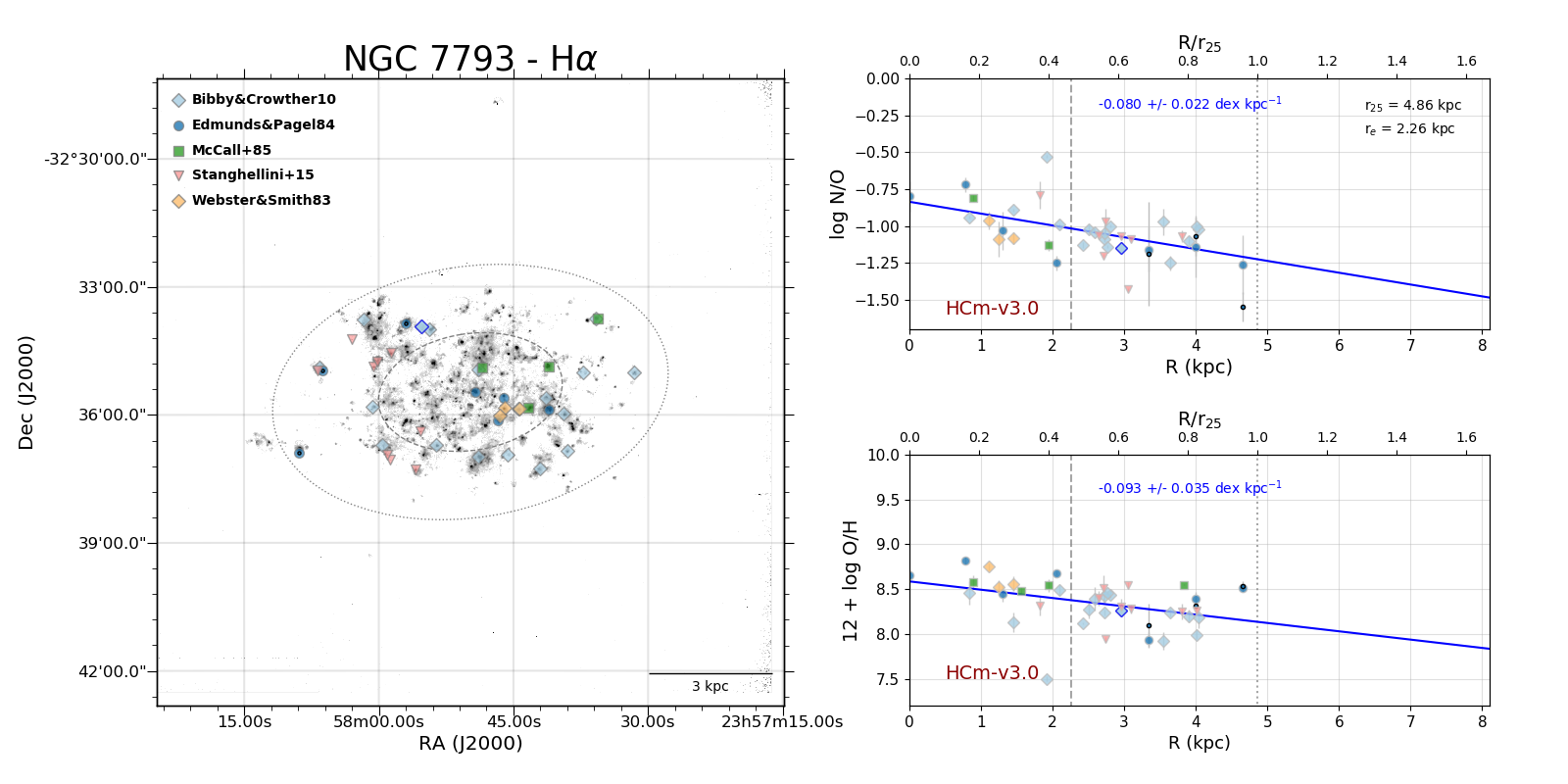}
\end{minipage}
\begin{minipage}{1.05\textwidth}
\hspace{-1.2cm}\includegraphics[width=1.12\textwidth]{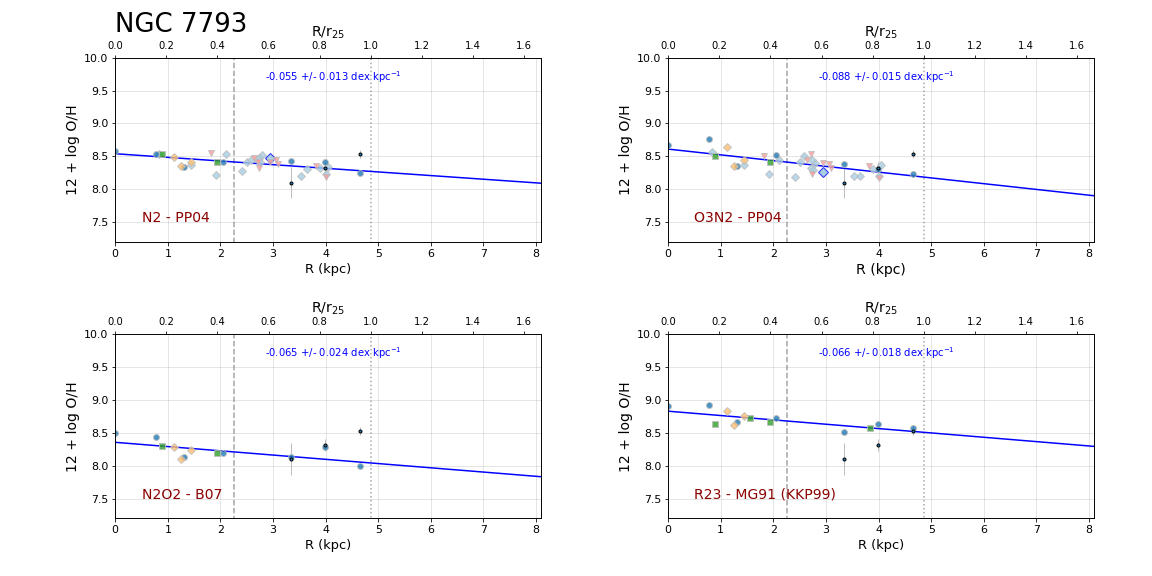}
\end{minipage}
\caption{Same as Fig.~\ref{Afig1} for NGC~7793.}\label{Afig53}
\end{figure*}
\clearpage
\begin{figure*}
\begin{minipage}{1.05\textwidth}
\hspace{-1.2cm}\includegraphics[width=1.12\textwidth]{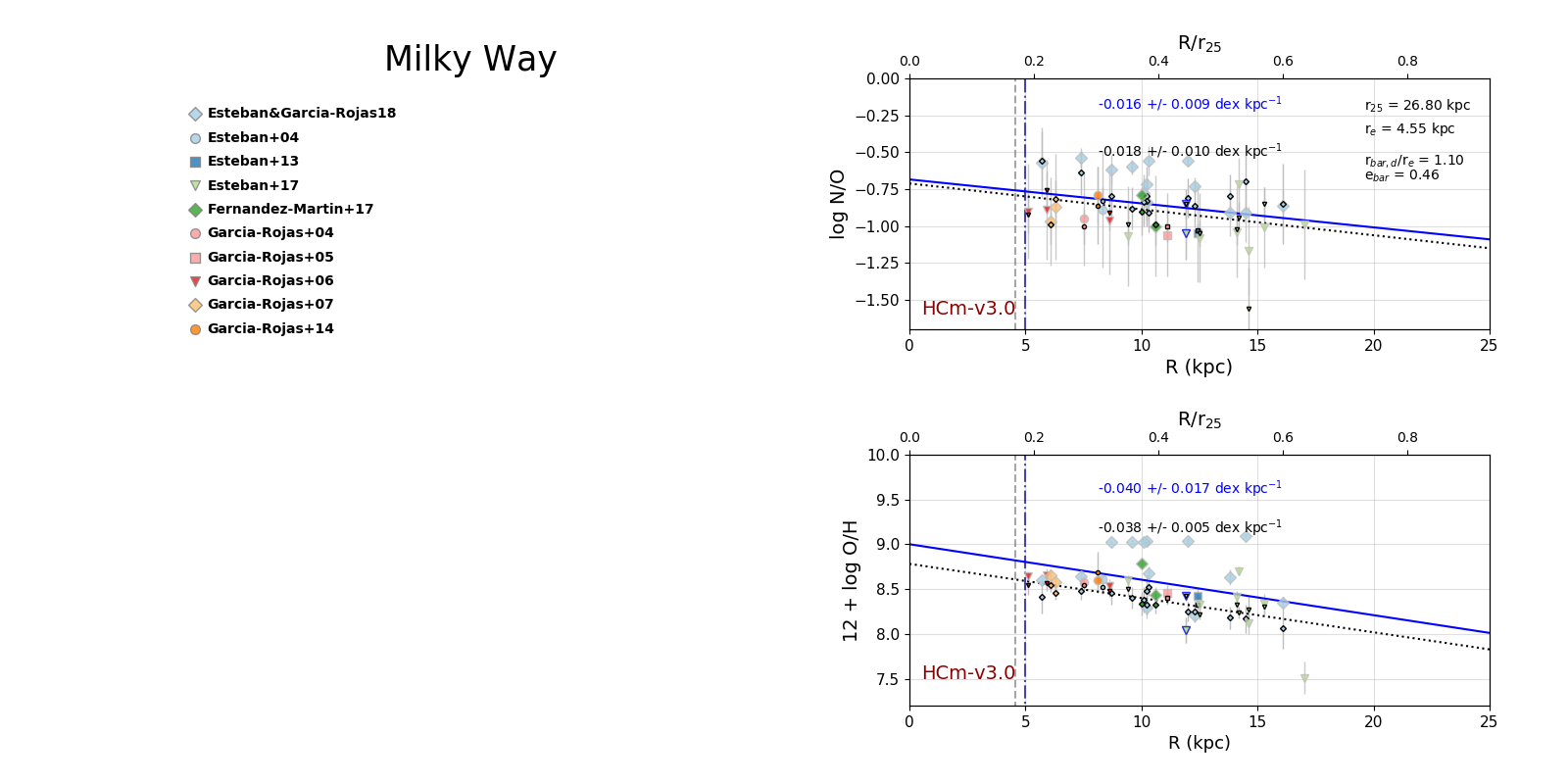}
\end{minipage}
\begin{minipage}{1.05\textwidth}
\hspace{-1.2cm}\includegraphics[width=1.12\textwidth]{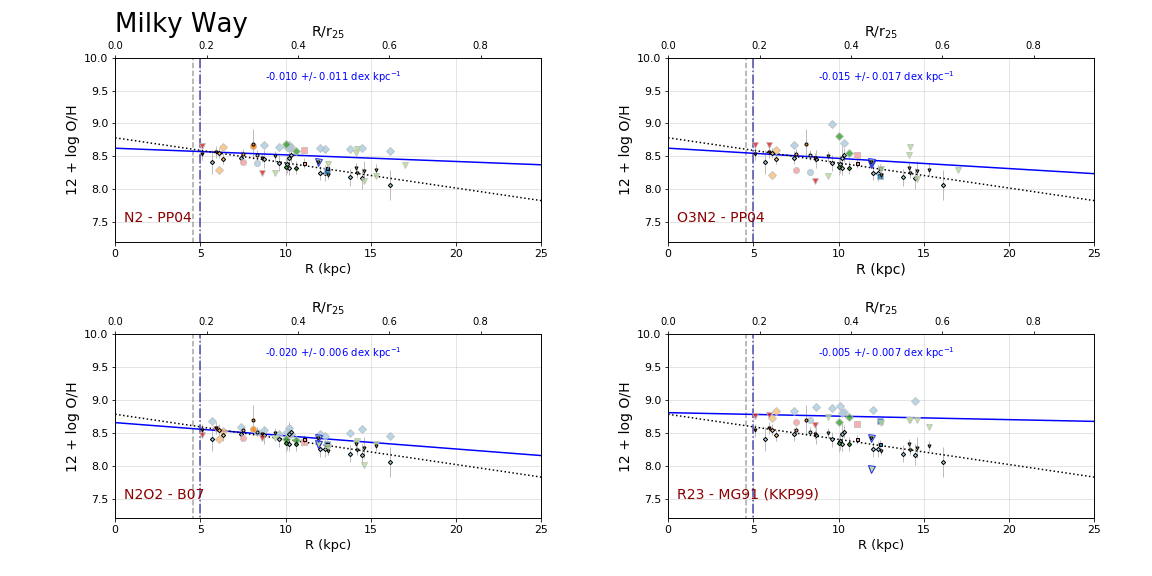}
\end{minipage}
\caption{Same as Fig.~\ref{Afig1} for the Milky Way.}\label{Afig54}
\end{figure*}
\clearpage

\bsp	
\label{lastpage}
\end{document}